\documentclass[hidelinks,pdftex,phd]{pittetd} 
\usepackage{etoolbox}
\makeatletter
\patchcmd{\@caption}{\csname the#1\endcsname}{\csname fnum@#1\endcsname:}{}{}
\renewcommand*\l@figure{\@dottedtocline{1}{1.5em}{4.5em}} % default for 3rd arg: 2.3em
\let\l@table\l@figure % as in article.cls
\makeatother

\usepackage{graphicx}
\usepackage{indentfirst}

\usepackage{threeparttable}

\usepackage{amsmath,amsthm}
\usepackage{pdflscape}
\hypersetup{hidelinks}
\InputIfFileExists{amsmath.pit}{%
            \ClassInfo{pittetd}{Patch for amsmath loaded}}{%
            \ClassInfo{pittetd}{No patch found for amsmath}}
\InputIfFileExists{amsthm.pit}{%
            \ClassInfo{pittetd}{Patch for amsthm loaded}}{%
            \ClassInfo{pittetd}{No patch found for amsthm}}

\usepackage{caption}
\usepackage{bbm}
\usepackage{amsmath,amsfonts,bm}
\usepackage[table]{xcolor}
\def\rvx{{\mathbf{x}}}
\def\rvz{{\mathbf{z}}}
\def\rvy{{\mathbf{y}}}
\def\rvc{{\mathbf{c}}}
\def\rvv{{\mathbf{v}}}
\def\rvw{{\mathbf{w}}}
\DeclareMathOperator{\vect}{vec}
\usepackage{stackengine}

\newcommand\etal {{\it et al.}}
\newcommand{\KL}{D_{\mathrm{KL}}}
\newcommand{\E}{\mathbb{E}}

\def\gL{{\mathcal{L}}}
\makeatletter
\renewcommand{\@biblabel}[1]{[#1]\hfill}
\newcommand\ie {{\it i.e.}, }
\newcommand\eg {{\it e.g.}, }
\newcommand\etc{{\it etc.}}

\newcommand{\nood}[1]{\textcolor[RGB]{196, 78, 82}{#1}}

\newcommand{\aid}[1]{\textcolor[RGB]{34,135, 60}{#1}}

\newcommand{\uid}[1]{\textcolor[RGB]{76,114, 176}{#1}}

\newcommand{\food}[1]{\textcolor[RGB]{221, 132, 82}{#1}}
\newcommand\Tstrut{\rule{0pt}{2.6ex}}         % = `top' strut
\newcommand\Bstrut{\rule[-1.0ex]{0pt}{0pt}}   % = `bottom' strut

\makeatletter
\renewcommand{\@biblabel}[1]{[#1]\hfill}
\makeatother
\makeatletter
\def\@hangfrom#1{\setbox\@tempboxa\hbox{{#1}}%
      \hangindent 0pt%\wd\@tempboxa
      \noindent\box\@tempboxa}
\makeatother

\title[Title displayed in Acrobat Reader's Document Info Dialog Box]{Deep Learning for Medical Imaging \\ From Diagnosis Prediction to its Explanation}
\author{Sumedha Singla}
\degree{Masters in Computing, University of Utah, 2015}

\school{Department of Computer Science}
\date{July 12th 2022}

\keywords{deep learning, explanation, counterfactual, COPD, concepts}
\subject{Dissertation}
\begin{document}
\maketitle
% If moved around the document, it might generate errors and warnings while compiling.
%==========================================================================================%
%==========================================================================================%

%==========================================================================================%
%CREATING THE COMMITTEE PAGE
%==========================================================================================%
% For the committee membership page, you have to provide the names and affiliations of the members. The first one will 

% THESIS ADVISOR (First member of the committee)

% THESIS CO-ADVISOR

\committeemember{Dr. Adriana Kovashka, Assistant Professor, Department of Computer Science, School of
Computing and Information}
\committeemember{Dr. Xiaowei Jia, Assistant Professor, Department of Computer Science, School of
Computing and Information}
\committeemember{Dr. Sofia Triantafillou, Assistant Professor, Department of Mathematics and Applied Mathematics, University of Crete}
\committeemember{Dr. Milos Hauskrecht, Professor, Department of Computer Science, School of Computing and
Information (Dissertation Co-Chair)}
\committeemember{Dr. Kayhan Batmanghelich, Assistant Professor, Department of Biomedical Informatics,
University of Pittsburgh School of Medicine (Dissertation Chair)}

% THESIS CO-ADVISOR
%\coadvisor{Second advisor, Co-advisor Departmental Affiliation}
%Uncomment to add a 'Second Advisor' into the document
% COMMITTEE MEMBERS

% To add more committee members
%\committeemember{Fourth member's name, Departmental Affiliation}
%\committeemember{Fifth member's name, Departmental Affiliation}
%\committeemember{Sixth member's name, Departmental Affiliation}
% To use uncommon the different committee members or add '\committeemember' commands as needed

%Special Option
% For master's theses, the committee may be omitted, naming only the advisor.

% DEPARTMENT INFORMATION 
\school{School of Computing and Information}
\makecommittee
%==========================================================================================%
%==========================================================================================%

%==========================================================================================%
%CREATING THE COPYRIGHT PAGE
%==========================================================================================%

% Create a copyright page with the author and year specified in the 'Title Page'
\copyrightpage                     
%Uncomment to get a copyright page.
%==========================================================================================%
%==========================================================================================%

%==========================================================================================%
%CREATING THE ABSTRACT
%==========================================================================================%
\begin{abstract}
Deep neural networks (DNN) have achieved unprecedented performance in computer-vision tasks almost ubiquitously in business, technology, and science. While substantial efforts are made to engineer highly accurate architectures and provide usable model explanations, most state-of-the-art approaches are first designed for natural vision and then translated to the medical domain. This dissertation seeks to address this gap by proposing novel architectures that integrate the domain-specific constraints of medical imaging into the DNN model and explanation design. 

Prior work on DNN design commonly performs lossy data manipulation to make volumetric data compatible with 2D or low-resolution 3D architectures. To this end, we proposed a novel DNN architecture that transforms volumetric medical imaging data of any resolution into a robust representation that is highly predictive of disease. For DNN model explanation, current explanation methods primarily focus on highlighting the essential regions (where) for the classification decisions. The location information alone is insufficient for applications in medical imaging. We designed counterfactual explanations to visually demonstrate how adding or removing image-features changes the DNN decision to be positive or negative for a diagnosis.

Further, we reinforced the explanations by quantifying the causal relationship between neurons in DNN and relevant clinical concepts. These clinical concepts are derived from radiology reports and are corroborated by the clinicians to be useful in identifying the underlying diagnosis. In the medical domain, multiple conditions may have a similar visual appearance, and it's common to have images with conditions that are novel for the pre-trained DNN. DNN should refrain from making over-confident predictions on such data and mark them for a second reading. Our final work proposed a novel strategy to make any off-the-shelf DNN classifier adhere to this  clinical requirement.

\end{abstract}

% SPECIAL OPTIONS
% Include Keywords
% To include keywords as part of the abstract include the option '[keywords]' (e.g., \begin{abstract}[Keywords:]
% The list comes from the '\keywords' specified in the 'Title Page'

% Include the word 'ABSTRACT'
% Use '\begin{abstract*}' and '\end{abstract*} instead of '\begin{abstract}' and '\end{abstract}
% The word `ABSTRACT' appears on the top of the page
%==========================================================================================%
%==========================================================================================%

%==========================================================================================%
% TABLE OF CONTENTS, FIGURES, AND TABLES
%==========================================================================================%

% Table of contents
\tableofcontents
% Comment (Use the '%' character) to omit

% List of Tables
\listoftables                      
% Comment (Use the '%' character) to omit

% List of Figures

\listoffigures
             
% Comment (Use the '%' character) to omit

% **If no figures and/or tables are included in the document, 'PITETD' will still create an empty page for the figures and/or tables **

% ** LaTex automatically includes all the figures and tables from the figures/tables included in the document **%
%==========================================================================================%
%==========================================================================================%

%==========================================================================================%
% INCLUDING A PREFACE
%==========================================================================================%
\phantomsection
\preface

% To include a preface for the document uncomment the '\preface' command
% Include the text of the preface after the command

First and foremost, I would like to thank my advisor
  Dr. Kayhan Batmanghelich. My research has been guided by his leadership, persistence, and encouragement and I thank him for all the time and energy that he has invested in me. Thank you for your trust in my capacities and your guidance. No matter how busy his schedule was, he was the most available person to me, just a message away. Dr. Kayhan has always given me the best of his mentorship, and  I will be forever grateful to have benefited from his abundance of experience and knowledge. I would also like to thank his postdocs and my close collaborators, Brian Pollock, Junxiang Chen and Mingming Gong. I had a great experience in learning and working with them. 
 
 The final part of this dissertation is the outcome of collaboration with Dr. Sofia Triantafillou. Sofia has been an awesome person to work with. I really enjoyed working and collaborating with her during the last few years, especially discussing research problems through our long meetings. Thank you so much for all your time. I am also grateful to Forough Arabshahi who has been a great industrial collaborator.
 
 I could not thank enough Dr. Stephen Wallace for his time and patience.  As my clinical collaborator he has given me his precious time,  have patiently answers all my questions, and have helped me develop research from a clinical prospective. A special thanks to Dr. Frank Sciurba for his interesting discussions and his passion for small things. I will always remember our meetings and your joyful nature.
 
 I am grateful to Dr. Motahhare Eslami who have rescued me at my time of need by providing her expertise in human computer interaction. Her smiling face and words of encouragement have made our interactions so much  more than  academic collaboration. Thank you so much for making my research human-relevant.

 I am also thankful to former and current members of Batman Lab who have been a great source of support during my PhD years. I am grateful to all of them, particularly Yanwu Xu, Rohit Jena, Ke Yu, Li Sun, Matthew Ragoza, Maxwell Reynolds, Nihal Murali, Sead Nikšić, Shantanu Ghosh and Payman Yadollahpour.

I am also thankful to the staff members in DBMI and Computer Science department  who have been always extremely helpful during the last 6 years. Special thanks goes to Keena Walker and Genine Bartolotta for all their help.

Finally and most importantly, I would like to thank my family, Vasu Gangrade and my beloved Vanya. You both have been the pillars of my strength.  Your support has given me courage and determination over this long journey. With all my heart and soul, thanks to my mother, father and sister Suruchi. You three have been there for me all this time, despite the distance. Thank you mom for sharing your experiences and making sure that I keep making progress. I undoubtedly could not have achieved this without them.

% This is the text of the preface, with acknowledgments, dedication, etc. 
%==========================================================================================%
%==========================================================================================%

%==========================================================================================%
% STARTING THE DOCUMENT
%==========================================================================================%

%The document begins by specifying the first 'chapter' with the command '\chapter{}

\chapter{Overview} 
Decision-making pipelines are inclining towards using deep neural networks (DNN) for high-stake applications especially medical diagnostics~\cite{esteva2017dermatologist,Gonzalez2018DiseaseTomography,Eitel2019TestingClassification,graziani2018regression}. DNNs are sophisticated and complex machine learning (ML) models trained using large datasets and computational resources~\cite{Huang2016DenselyNetworks,Karras2019ANetworks}. Many advancements have been proposed to use different sources of medical data such as imaging, electronic health records~\cite{miotto2016deep,rajkomar2018scalable,malakouti2019predicting}, patient discharge summaries~\cite{icu}, diagnosistic tests, and others to build deep learning (DL) systems for diagnostic applications. The primary focus of this thesis is on using medical imaging information for diagnosis prediction. Further, in medical image analysis,  DL models are used for solving different computer vision-related problems such as classification~\cite{yadav2019deep}, detection~\cite{xie2019automated}, segmentation~\cite{gordienko2018deep}, and registration~\cite{eppenhof2018pulmonary}. Among them, we primarily focus on classification methods that classify an input image or a series of images with diagnostic labels of some predefined diseases.

Over the last few years, convolutional neural networks (CNNs) have become the de-facto backbone of numerous models using medical-imaging data for diagnostic classification~\cite{Eitel2019TestingClassification,Rajpurkar2017CheXNet:Learning,Winkler2019AssociationRecognition,Mitani2020DetectionLearning}. CNNs superior predictive performance compared to simpler counterparts makes them a lucrative option for real-world deployment. But this improvement comes at the cost of decreased model interpretability~\cite{MONTAVON20181}.  They are essentially \emph{Black box} predictive models that  often predict the correct answer for the wrong reasons~\cite{Gal2016UncertaintyID,nguyen2015deep} and are over-confident on out-of-domain data~\cite{guo2017temperaturescaling,pmlr-v48-gal16}. These challenges remain a primary reason for lack of trust and a barrier to broader acceptance of such algorithms in practice.

The clinical deployment of DNN models is contingent on satisfying three essential requirements. First, DNN model design should integrate domain-specific constraints of medical imaging into its architectural design while providing sufficiently high predictive performance~\cite{Singla2018Subject2VecVector,jena2021self,sun2021context}. Second, the decision-making process of these models should be explainable to clinicians to obtain their trust in the model ~\cite{glass2008toward,Gastounioti2020IsAI,Jiang2018ToClassifier}. Third, the model should communicate the uncertainty in its prediction and  raise a flag when it doesn't have sufficient information to make a confident prediction~\cite{Gal2016UncertaintyID,hendrycks17baseline}. This dissertation proposes models to account for subtleties of medical imaging and add support for these clinical needs. At the same time, our foundation is a general approach and applies to different model and dataset domains. Next, we will discuss each requirement in more detail.

\section{DNN model design}

Researchers have proposed numerous DNN models that use medical imaging data for autonomous diagnostics~\cite{Lynch2015CTDefinableSociety}. In this thesis, we primarily focus on lung imaging and associated diseases such as Chronic obstructive pulmonary disease (COPD), pleural effusion, and others. However, our proposed models apply to a wide range of heterogeneous disorders. Having an objective way to characterize local patterns of disease is essential in diagnosis, risk prediction, and sub-typing~\cite{Estepar2013ComputedImplications,Hayhurst1984DiagnosisTomography,Muller1988DensityTomography,Shapiro2000EvolvingDisease.}. Previously, researchers have proposed intensity and texture-based feature descriptors to represent the visual appearance of a disease. However, most image features are generic and not necessarily optimized for a given diagnosis~\cite{Cheplygina2017TransferDisease,Sorensen2012Texture-basedApproach,Yang2017UnsupervisedStudy}. Recent advances in deep learning (DL) have enabled researchers to use raw images directly for predicting clinical outcomes without specifying radiological features~\cite{Gonzalez2018DiseaseTomography}. These clinical outcomes may include diagnosis identification, symptom score, and mortality.  The classical DL methods, which operate on entire volumes or slices, are challenging to interpret and require resizing the input images to a fixed dimension~\cite{Gonzalez2018DiseaseTomography}. My first project proposes an attention-based method that aggregates local image features to a subject-level representation for predicting disease severity. Our proposed method operates on a set of image patches; hence it can accommodate variable-length input volumes without image resizing.

\section{DNN model explanation}
Explainability is essential for auditing DNNs~\cite{Winkler2019AssociationRecognition}; identifying various failure modes~\cite{Eaton-Rosen2018TowardsPredictions,Oakden-Rayner2020HiddenImaging} or hidden biases in the data~\cite{cramer2018assessing} or the model~\cite{Larrazabal2020GenderDiagnosis}; evaluating the model's fairness~\cite{doshi2017towards}; and obtaining new insights from large-scale studies~\cite{Rubin2018LargeNetworks}. There are two general approaches to model explanation: (1) developing \emph{interpretable} models and (2)  \emph{posthoc explaining} a pre-trained model. 
Interpretable predictive models are constrained to make their reasoning processes more understandable to humans, making them much easier to troubleshoot and use in practice~\cite{Rudin2022InterpretableChallenges}. However, such models may impose simplifications to ensure interpretability and thus achieve a lower predictive accuracy~\cite{Azar2013DecisionDiagnosis,Tsumoto2004MiningModel}. Other times, they have a complicated design, making them difficult to train~\cite{Chen2019ThisRecognition}. Another popular line of work builds attention-based interpretability into the DNN. These methods use attention weights to highlight parts of an input that the network focuses on while making its decision~\cite{twoLevelAtten,8099959}. 

 \emph{Post-hoc explanation} aims to improve human understanding of a pre-trained model~\cite{doshi2017towards,guidotti2019survey,kim2015interactive,molnar2018interpretable,Zeiler2013VisualizingNetworks}. Hence, the performance of the model is not compromised. Post-hoc explanation comprises several broad approaches, such as example-based explanations~\cite{KimM18,examplesBeenKim}; approximating DNN models with simpler models~\cite{ribeiro2016should,ribeiro2018anchors}; understanding feature attribution~\cite{Sundararajan2017AxiomaticNetworks,Lundberg2017APredictions} or importance~\cite{Selvaraju2017Grad-cam:Localization,Kindermans2017TheMethods,Bach2015OnPropagation,Fong2017InterpretablePerturbation}. These methods provide a local (image-level) or a global (target label-level) explanation.  They explain by highlighting the critical regions (\emph{where}) for the classification decisions. However, the location information alone is insufficient for applications in medical imaging. My second project focuses on providing posthoc model explanations in the form of counterfactual images. Counterfactual images show \emph{what} image features are important for the classification decision and \emph{how} to modify the critical features to change the classifier's decision.

Recently, researchers have focused on providing explanations that resemble a domain expert's decision-making process. Very often, such explanations are expressed using human-understandable concepts or terminology~\cite{yeche2019ubs,graziani2020concept,gloabl_local}. Existing approaches for concept-based explanation depend on explicit concept annotations. The concept annotations are provided either as a representative set of images~\cite{examplesBeenKim} or as semantic segmentation~\cite{bau2017network}. Such annotations are expensive to acquire, especially in the medical domain. 
 Furthermore, these methods measure correlations between concept perturbations and classification predictions to quantify the concept's relevance~\cite{examplesBeenKim,bau2017network,Zhou2018InterpretableExplanation}.  However, NN may not use the information from learned concepts to arrive at its decision. My third project focuses on using counterfactual images to quantify the causal effect of a concept on the classification decision.

\section{DNN model uncertainty quantification }
Most DNN models are deterministic functions that provide point estimates of parameters and predictions. In the absence of probabilistic distribution, it is essential to convey the uncertainty in the DNN model's decision to the end-user~\cite{nguyen2015deep,guo2017temperaturescaling}. For example, consider a DNN model trained on adult face images, predicting whether the person is young or old. Such a model should refrain from making an overconfident decision on ambiguous images from middle-aged people or out-of-distribution (OOD) images of children and animals~\cite{mukhoti2021deterministic,van2020uncertainty}. Uncertainty in prediction may arise from noisy data or data in high class-overlap regions, leading to \emph{aleatoric uncertainty}~\cite{ABDAR2021243}. The other kind of uncertainty is related to the model, \emph{epistemic uncertainty}, that arises from the model's limited information on unseen data or due to a mismatch between training and testing data distributions~\cite{aleatoricEpistemic}. Communicating predictive uncertainty can help the end-user understand the predictions better, anticipate when uncertainty is irreducible, prioritize gathering more data to reduce model uncertainty, and decide when to discard the model prediction and rely on expert knowledge~\cite{9022324}. 
%A highly accurate, well explained but overconfident model is ill-suited for high-stake applications such as healthcare~\cite{Gal2016UncertaintyID,christian2017}. 

Much of the prior work focused on deriving uncertainty measurements from a pre-trained DNN output~\cite{hendrycks17baseline,guo2017temperaturescaling,liang2018enhancing,liu2020energy}, feature representations~\cite{lin2021mood,Lee2018ASU}, or gradients~\cite{huang2021on}. Such methods use a threshold-based scoring function to identify OOD samples. The scoring function is derived from softmax confidence scores~\cite{hendrycks17baseline}, scaled logit~\cite{guo2017temperaturescaling,lin2021mood}, energy-based scores~\cite{liu2020energy,wang2021can}, or gradient-based scores~\cite{huang2021on}. These methods help in identifying OOD samples but did not address the over-confidence problem of DNN, which made identifying OOD non-trivial in the first place~\cite{Hein2019WhyRN,nguyen2015deep}. My final project focuses on mitigating the over-confidence issue in a pre-trained classifier by efficiently capturing both epistemic or aleatoric uncertainty.  

\section{Explanation Framework}
This dissertation propose an \textbf{explanation framework} to explain the DNN classification model's decision. The explanation framework has two primary models. The first is an \emph{interpretable model} that uses a carefully designed attention mechanism to provide interpretability while achieving high predictive performance. The second is a \emph{progressive counterfactual explainer} (PCE) that provides a posthoc explanation for a pre-trained classifier. A counterfactual image is a perturbation of the input image with an opposite classification decision compared to an input image. It shows what imaging features are present in salient locations and how changing such features modify the classification decision.
The generative explainer is constrained to create natural-looking images as explanations that resemble medical-imaging data, thus ensuring the clinical usability of our explanations. My work presents a thorough human-grounded experiment with diagnostic radiology residents to compare different styles of explanations (no explanation, saliency map, cycleGAN explanation, and our counterfactual explanation) by evaluating different aspects of explanations. The results show that the counterfactual explanations from my proposed method,  were the only explanations that significantly improved  the users' understanding of the classifier's decision compared to the no-explanation baseline.

Further, in my next project, I extended the explanation framework to support two applications. The first application focuses on enriching the explanation with conceptual information. Specifically, I integrated counterfactual explanations with tools from Causal Inference literature~\cite{imai2011commentary} to quantify the causal relationship between the building units of a DNN, neurons, and clinically relevant concepts~\cite{NEURIPS2020_92650b2e}. The weak annotations from radiology reports were used to derive concept annotations. The second application focuses on fixing an over-confident pre-trained classifier. The counterfactual images derived from PCE were used to fine-tune the classifier. The empirical results show that fine-tuning helps in smoothing the decision boundary and helps in preventing the classifier from being over-confident on samples near the decision boundary. Further, the discriminator of the GAN-generator was used to provide a density score to identify OOD samples.

\section{Dissertation structure}
Chapter~\ref{ch0} provides a detail literature review on different deep learning models in medical imaging. It also provides a thorough background on different  paradigms of deep learning model explanation.

Chapter~\ref{ch1} proposes an attention-based DNN model aggregating local image features from volumetric medical images into a compact latent representation. This representation is then used to predict multiple patient-relevant outcomes such as symptom scores and disease severity. The model provides interpretability by learning an attention weight for each anatomical feature that reflects its contribution to the final prediction decision. We evaluated our proposed model in a large clinical study of over 10K participants with chronic obstructive pulmonary disease (COPD). Our results show that our model independently predicted spirometric obstruction, emphysema severity, exacerbation risk, and mortality from CT imaging alone.

Chapter~\ref{ch2} proposes a Progressive Counterfactual Explainer (PCE) to explain the decision of a pre-trained image classifier. The explainer generates a progressive set of perturbations to a query image, such that the classification decision changes from its original class to its negation. We used counterfactual explanations derived from our framework to audit a classifier. We conducted experiments on a natural image dataset of face images and a medical chest x-ray (CXR) dataset. To quantitatively evaluate our explanations, we proposed new metrics that consider the clinical definition of a target disease while comparing counterfactual changes between normal and abnormal populations, as identified by the classifier. 

We conducted a human-grounded experiment with diagnostic radiology residents to compare different styles of explanations (no explanation, saliency map, cycleGAN explanation, and our counterfactual explanation) by evaluating different aspects of explanations: (1) understandability,  (2) classifier's decision justification, (3) visual quality, (d) identity preservation, and (5) overall helpfulness of an explanation to the users. Our results show that our counterfactual explanation  was the only explanation method that significantly improved  the users' understanding of the classifier's decision compared to the no-explanation baseline. Our metrics established a benchmark for evaluating model explanation methods in medical images. Our explanations revealed that the classifier relied on clinically relevant radiographic features for its diagnostic decisions, thus making its decision-making process transparent to the end-user.

Chapter~\ref{ch3} shows an application of PCE to provide concept-based explanations. In this project, we aim to quantify causal associations between the hidden units of the DNN and human-understandable concepts~\cite{examplesBeenKim,NEURIPS2020_92650b2e}. We take advantage of radiology reports accompanying the chest X-ray images to define concepts. First, we solve sparse linear logistic regression to identify hidden units that are positively correlated with the presence of a concept. Next, we viewed these concept units as a mediator in the treatment-mediator-outcome framework~\cite{imai2011commentary} from mediation analysis. Using PCE to define counterfactual interventions, we measure the in-direct causal effect of a concept on the network's prediction. Finally, we present our findings as a low-depth decision tree over causally relevant concepts, providing the global explanation for the model in the form of clinically relevant decision rules.

Chapter~\ref{ch4} demonstrates an application of PCE in improving the uncertainty quantification of an existing pre-trained DNN. Ideally, the DNN model's output should reflect its confidence in its decision. This project proposes fine-tuning an existing pre-trained classifier on counterfactually augmented data (CAD) generated using PCE to improve its uncertainty estimates. Further, the GAN-PCE discriminator helps identify and reject far-OOD samples. In our experiments, we out-performed state-of-the-art methods for uncertainty quantification on multiple datasets with varying difficulty levels. Chapter~\ref{ch5} summarizes this thesis and suggests future extensions.

All chapters in this dissertation address unique DNN challenges motivated by specific clinical requirements. We investigated and explored efficient DNN architectural, explanation and training paradigms while keeping our end-users ``clinicians" in focus. At the same time, the methods developed in this research have broad applicability and have been used by many researchers in different domains~\cite{9694495}.  We have released open-source implementations of all these methods. 

\section{Contributions}
The most notable contributions of this  dissertation are the development of:

\begin{enumerate}
    \item An interpretable, attention-based DNN architecture that processes an entire 3D volumetric image without any resizing and predicts multiple disease outcomes with high predictive accuracy (summarized in Chapter~\ref{ch1}). 
    \item A new posthoc explainability method that provides visual counterfactual explanations. These explanations not only highlight the important regions but also shows how the image features
should be transformed to flip the classification decision (summarized in Chapter~\ref{ch2}).
\item A concept-based explanation method that explains the classification decision in terms of clinically relevant concepts. This method uses the explanations from the technique described in Chapter 2 to quantify the causal effect of a concept on the network's prediction (summarized in Chapter~\ref{ch3}).

\item A methodology  to fine-tune the existing DNN on counterfactually augmented data to improve its  estimates for aleatoric uncertainty. Further, using the discriminator of the GAN-counterfactual explainer as a selection function to identify and reject samples with high epistemic uncertainty (summarized in Chapter~\ref{ch4}).

\end{enumerate}

\section{List of publications}

Material presented in this dissertation has been published in peer-reviewed conferences and journal papers:

\begin{enumerate}
    \item Singla, S.; Gong, M.; Ravanbakhsh, S.; Sciurba, F.; Poczos, B. and Batmanghelich, K.N.;,``Subject2Vec: Generative-Discriminative Approach from a Set of Image Patches to a Vector," MICCAI, Part I, pp. 502-510, September 2018.~\cite{Singla2018Subject2VecVector}
    \item Singla, S.; Gong, M.; Riley, C.; Sciurba, F. and Batmanghelich, K.N.;, ``Improving clinical disease subtyping and future events prediction through a chest CT‐based deep learning approach," Medical physics, 48, no 3, pp. 1168-1181, March 2021.~\cite{singla2021improving}
    \item Singla, S.; Pollack, B.; Chen, J. and Batmanghelich, K.N.;, ``Explanation by Progressive Exaggeration," International Conference on Learning Representations, September 2019.~\cite{Singla2020ExplanationExaggeration}
    \item Singla, S.; Pollack, B.; Wallace, S. and Batmanghelich, K.N.;, ``Explaining the Black-box Smoothly-A Counterfactual Approach," arXiv e-prints, pp.arXiv-2101, Jan 2021.~\cite{singla2021explaining}
    \item Singla, S.; Wallace, S.; Triantafillou, S. and Batmanghelich, K.N.;, ``Using Causal Analysis for Conceptual Deep Learning Explanation," MICCAI, Vol. 12903, pp. 519-528, January 2021.~\cite{singla2021using}
    \item Singla, S.; Murali, N; Arabshahi, F; S.; Triantafillou, S. and Batmanghelich, K.N.;, ``Augmentation by Counterfactual Explanation - Fixing an Overconfident Classifier," under review WACV, 2022.
\end{enumerate}

\chapter{Literature Review} 
\label{ch0}
\section{Deep learning for medical imaging}
With the expanding development of deep learning (DL) techniques, utilizing advanced deep neural networks (DNNs) for medical image analysis has become an active field of research.
DNNs have shown superior performance over clinicians in many tasks, primarily due to the availability of large training datasets and increased computational power. Applications of DL in medical image analysis involve different computer vision-related problems such as classification~\cite{yadav2019deep}, detection~\cite{xie2019automated}, segmentation~\cite{gordienko2018deep}, and registration~\cite{eppenhof2018pulmonary}. Among them, we primarily focus on classification methods that classify an input image or a series of images with diagnostic labels of some predefined diseases~\cite{Gonzalez2018DiseaseTomography,Pasa2019EfficientVisualization}. Traditional computer algorithms for image classification use feature extractors and statistical models that translate human intuition into handcrafted features~\cite{Diaz2010AirwaySmokers,avni2010x,iijima2010aortic}. These features were then used in a supervised setting to train specialized image classifiers. In contrast, DNN models follow a data-driven approach and learn to optimally represent the data for the given classification task with minimum human intervention. The resulting models are complex functions with millions of parameters but are much more accurate, efficient, generalizable, and easier to scale.

The commonly used image modalities for diagnostic analysis in clinic include projection imaging such as X-ray imaging and computed tomography (CT). As a working example, we focus on DNN models that are developed for chest imaging. Chest CT imaging comprises a continuous sequence of 2D slices that vary in depth and resolution with changes in patient and scanner settings. Many DL architectural designs  have been applied to applications in CT analysis to solve specific clinical tasks such as nodule detection~\cite{7163869}, fibrosis~\cite{fibrosisCNN}, emphysema~\cite{dlEmphysema}, COPD~\cite{TANG2020e259}, and cancer diagnosis~\cite{computerAided}. The most common setting is to sub-sample 2D slices from volumetric images and concatenate, join, or crop them in different ways to create a 2D image~\cite{computerAided,Gonzalez2018DiseaseTomography,dlLungNodule,anthimopoulos2016lung}. The primary motivation behind this re-scaling is to make the input images compatible with the  classical DNN architectures, which were originally designed for natural images~\cite{He2016DeepRecognition,Simonyan2015VeryDC}. Extensive pre-processing pipelines are proposed which include sub-sampling spatially aligned CT volumes into three slices in either axial, sagittal or coronal directions, to accommodate for the RGB input~\cite{TANG2020e259}. Distortion of CT imaging may lead to undesired artefacts and information loss, leading to sub-optimal performance.

Further, researchers have explored variants of recurrent neural networks (RNN)~\cite{45452} to process consecutive slices from sub-sampled 3D volumes~\cite{dlEmphysema,raN_Nodule}. These methods include using long short-term memory (LSTM) networks capable of learning dependencies between a sequence of images~\cite{7298878}. Such algorithms can take multiple slices as input and provide better global features for downstream use-cases, such as classification. More recently, efforts are being made to integrate various designs, such as 3D CNN with RNN~\cite{CNNandRNN}; multiple resolution CNNs with two, two-and-a-half, and three-dimensional architectures~\cite{ensembleDL,multiScaleCNN,ciompi2017towards}; and 3D multi-scale capsule networks~\cite{multiScaleCapsuleNet}. These methods aim to better capture information from 3D imaging at different spatial resolutions with minimum information loss. Their primary motivation is high discriminative performance, while little attention is paid to models' interpretability. %This criticism of deep learning methods hinders their clinical deployment. %We propose an attention-based DNN model that processes an entire 3D CT scan without resizing and predicts multiple COPD outcomes.

Although deep learning models have achieved great success in medical image analysis, minimum interpretability is still the main bottleneck in the clinical deployment of these methods~\cite{salahuddin2022transparency}. The key benefit of DNN is that it identifies essential features without human intervention. However, this makes the model opaque as the end-user has no intuition on how the decision was being made. The legal ramifications of black-box functionality could have severe consequences; hence, healthcare professionals may decline to work with such systems.

\section{Interpretable deep learning}

The overarching goal of any deep learning method for medical imaging is to aid clinicians in their workflow by increasing their efficiency by removing redundancies~\cite{lindsey2018deep}. This requires a partnership between the clinical experts and the AI system, which in turn requires the clinical experts' trust. Interpretability, or the ability of a DNN model to explain its outcomes and assist clinicians in rationalizing the model prediction, is critical to establishing trust~\cite{Tonekaboni2019WhatUse}. Interpretable DL models aim to incorporate interpretability during the design process of the DNN and, thus, alter the network structure to encourage interpretability. They learn to provide both prediction, and explanation are gaining the interest of the medical research community.For example, DNN have been designed to perform case-based reasoning~\cite{Chen2019ThisRecognition}, to incorporate logical structures~\cite{wu2019towards}, to incorporate hard attention to do classification~\cite{NEURIPS2019_8dd48d6a}, and to learn a disentangled latent space~\cite{chen2020concept}. One of the early methods modified the CNN architecture to extract prototypical examples~\cite{Chen2019ThisRecognition}. In another attempt, Song \etal proposed a student-teacher network, where one network is optimized for superior interpretability while the other network is trained to achieve high discriminative performance~\cite{lungNodule}. Some methods provide interpretability by performing multi-modality learning by integrating radiology reports~\cite{mdNet} or electronic health record data~\cite{multiFusion,integratingDL}. This data provides additional information for assisting clinical decision-making.

Furthermore, many variants of the attention-based model have been proposed that learn an attention mechanism to highlight the most relevant part of the input for the prediction decision~\cite{SCHLEMPER2019197}. For instant, Choi \etal proposed a multi-level attention model on time series data for detecting influential past visits, and clinical variables while predicting diagnosis. Another example is interpretable R-CNN~\cite{wu2019towards}, which is an object detection-based DNN that provides a classification score and a bounding box on the region of interest. Recently, a concept whitening approach was proposed that learns a DNN where the latent space of each layer is aligned with a known set of concepts~\cite{chen2020concept}. Creating an interpretable model is much more complicated than a black-box model, as it involves solving a complex optimization problem while satisfying the interpretability constraints. Nevertheless, the benefits of having an explanation built into the model have far better deployment prospects than a highly accurate but opaque model.

\section{Post-hoc deep learning model explanation}
Post-hoc explanation methods provide explanations for the predictions after the DL model has been trained. Such methods can provide local or global explanations. Local explanations provide explanation for individual data point. It identifies attributes or features in a particular image that are important for the DNN model's prediction. On the other hand, global explanations aim at providing an overall summarization of the model behaviour for a particular class. Post-hoc explanation methods can be model-specific \ie they are applicable to only certain types of models and require access to model-specific information. On the other hand, they can be model-agnostic methods, that is they are applicable to any DNN model in general.  

\subsection{Feature attribution-based explanation}
Feature attribution methods provides an explanation as a saliency map that  reflects the importance of each input component (\eg pixel) to the classification decision. Saliency-based methods are the most common form of post hoc explanations for neural networks. Gradient-based methods for obtaining saliency maps is mostly DNN-specific and provides local explanation~\cite{Shrikumar2017LearningDifferences,Sundararajan2017AxiomaticNetworks,Lundberg2017APredictions}. Some earlier work in this direction~\cite{Simonyan2013DeepMaps,Springenberg2015StrivingNet,Bach2015OnPropagation} focuses on computing the gradient of the target class with respect to input image and considers the image regions with large gradients as most informative. Building on this work, the class activation map (CAM)~\cite{Zhou2016LearningLocalization} and its generalized version Grad-CAM~\cite{Selvaraju2017Grad-cam:Localization}  uses the gradients of the target class, flowing into the final convolutional layer to produce a saliency map. The Layer-Wise Relevance Propagation (LRP)~\cite{Bach2015OnPropagation} method back-propagates a class specific error signal through the DNN and considered its product with each convolutional layer's activation to derive the saliency map. DeepLift~\cite{Shrikumar2017LearningDifferences} is a version of LRP  method that back-propagates the contribution back to every feature of the input. The above gradient-based methods are not model-agnostic and require access to intermediate layers. Recently,~\cite{Adebayo2018SanityMaps} have showed that some saliency methods are independent both of the model and of the data generating process.  The saliency maps are also prone to adversarial attacks as shown by~\cite{Ghorbani2019InterpretationFragile} and~\cite{Kindermans2017TheMethods}.  

In another line of work, perturbation-based methods provide interpretation by showing what minimal changes are required in input image to induce a desirable classification output. Some methods employed image manipulation via the removal of image patches~\cite{Zhou2014ObjectCnns,Zeiler2013VisualizingNetworks}, masking with constant values~\cite{Dabkowski2017RealClassifiers,Petsiuk2018RISE:Models} or the occlusion of image regions~\cite{Zhou2014ObjectCnns} to change the classification score. Recently, the use of influence function, as proposed by ~\cite{Koh2017UnderstandingFunctions} are applied as a form of data perturbation to modify a classifier's response. The authors in~\cite{Fong2017InterpretablePerturbation} proposed the use of optimal perturbation, defined as removing the smallest possible image region that results in the maximum drop in classification score. In another approach,~\cite{Chang2019ExplainingGeneration}  proposed a generative process to find and fill the image regions that correspond to the largest change in the decision output of a classifier.  To switch the decision of a classifier, \cite{Goyal2019CounterfactualExplanations} suggested generating counterfactuals by replacing the image regions  with patches from images with a different class label. All of the aforementioned works perform pixel- or patch-level manipulation to input image, which may not result in natural-looking images. Especially for medical images, such perturbations may introduce anatomically implausible features or textures.

 Another interesting approach is to use game theory to compute the Shapley value of each pixel as its marginal contribution to the final prediction decision~\cite{Lundberg2017APredictions,sundararajan20b}. The idea of Shapely values is that all features cooperate to produce model prediction. This is a local interpretation method that can be either model agnostic or model specific depending on the formulation. Classical SHAP method required repeated predictions from the model, as it exhaustively try all possible configuration of the features. This is computationally expensive and hence, multiple approximations are been proposed~\cite{chen2021explaining}.

Saliency map-based methods are frequently applied to the medical imaging studies, \eg chest x-rays~\cite{Rajpurkar2017CheXNet:Learning}, skin imaging~\cite{Young2019DeepDermatologist}, brain MRI~\cite{Eitel2019TestingClassification} and retinopathy~\cite{Sayres2019UsingRetinopathy}. Saliency maps lack a clear interpretation and provide incomplete explanation especially when different diagnoses affect the same regions of the anatomy. Although objects in natural images have a distinct appearance and are easier to identify and isolate by humans, the visual variations in different diagnoses, in medical images, are very subtle and require expert observation. Thus, very similar explanations are given for multiple diagnosis, and often none of them are useful explanations~\cite{rudin2019stop}.

\subsection{Counterfactual explanation}
Recently, researchers have explored generative models that provides explanation by modifying existing examples~\cite{Goyal2019CounterfactualExplanations} or generating new examples~\cite{SamangoueiPouyaandSaeedi2018ExplainGAN:Transformations,Joshi2019TowardsSystems}. A popular direction is to generate counterfactual explanations. Counterfactual explanations are a type of contrastive~\cite{Dhurandhar2018ExplanationsNegatives} explanation that are generated by perturbing the real data such that the classifier's prediction is flipped. Similar to our method, generative models like GANs and variational autoencoders (VAE) are used to compute interventions that generate realistic counterfactual explanations~\cite{ SamangoueiPouyaandSaeedi2018ExplainGAN:Transformations,Joshi2019TowardsSystems,Liu2019GenerativeLearning,Mahajan2019PreservingClassifiers,VanLooveren2019InterpretablePrototypes,ParafitaMartinez2019ExplainingAttribution,Agarwal2019ExplainingModels}. Much of this work is limited to simpler image datasets like MNIST, celebA~\cite{Liu2019GenerativeLearning,Mahajan2019PreservingClassifiers,VanLooveren2019InterpretablePrototypes} or simulated data\cite{ParafitaMartinez2019ExplainingAttribution}. An extension of these methods on large datasets will actually show their scalability and generalizability strengths. This work is yet to be explored by the community in general and provides a great venue for future exploration.

For more complex natural images, previous studies~\cite{Chang2019ExplainingGeneration,Agarwal2019ExplainingModels} focused on finding and in-filling salient regions, in order to generate counterfactual images. In contrast, at inference time, our explanation model doesn't require any re-training for generating explanations for a new image. In another line of work~\cite{Wang2020SCOUT:Explanations,Goyal2019CounterfactualExplanations} provide counterfactual explanations that explains both the predicted and the counter class. Recently~\cite{Narayanaswamy2020ScientificTranslation,DeGrave2020AISignal} used a cycle-GAN~\cite{Zhu2017UnpairedNetworks} to perform image-to-image translation between normal and abnormal images. While images generated by such independently trained GANs may look realistic, these generative models are not explicitly coupled to the classifier that they are aiming to explain. Hence, cycle-GAN may end up learning features that do not reflect the true behaviour of the classifier. In contrast, our model uses the classifier’s predicted probabilities and gradients during the training of the GAN-generator, and hence the generated images are tied to the classifier. 
 
 Since the inception of our work, various extensions to our counterfactual generation process have been proposed. These include adding support for creating diverse and multiple counterfactual explanations~\cite{Rodriguez_2021_ICCV,ghandeharioun2021dissect}, enhancing compatibility with  smaller datasets~\cite{KATZMANN2021141} and inducing a bijective transformation through normalizing flow~\cite{dombrowski2021diffeomorphic}.

\subsection{Concept-based explanation}
Concept learning was used in traditional machine learning to identify and classify samples based on a list of concepts. A concept, in this case is a feature that is discriminative and whose presence is highly associated with the presence of a class label. To summarize, a concept is semantically meaningful attribute that is visually coherent across images and is important for the prediction of a given class \ie its presence is a necessary condition for the classification decision to be true for a given class~\cite{Ghorbani2019TowardsExplanations}. Another benefit of concept-based explanation is, usually concepts are high level attributes that are mentioned in human-friendly manner.
Recent studies have thus focused on bringing such concept-based explainability to DNNs.

Concept-based explanation methods aim to recover concept information from the intermediate DNN activations and then relate them to the classification decision and the data. The essential first step towards deriving concept-based explanations is defining concepts. Some methods used human-labelled supervised data to mark the salient concepts~\cite{Kim2017InterpretabilityTCAV,Zhou2018InterpretableExplanation}, while others used purely unsupervised approaches such as clustering of the DNN activations~\cite{Ghorbani2019TowardsExplanations}. TCAV~\cite{Kim2017InterpretabilityTCAV} learns a concept classifier by training a linear classification model on the activations of an arbitrary intermediate layer, using the ground truth labels for each concept. Gohorbani \etal extended TCAV to used self-supervised labels obtained from automatically super-pixel segmentation followed by k-means clustering. Zhou \etal~\cite{Zhou2018InterpretableExplanation} decomposes
the prediction of one image into multiple human-interpretable conceptual components.
Concept activation vectors (CAV) are used in medical imaging analysis for solving particular tasks such as retina disease diagnosis~\cite{9286420}, skin lesion classification~\cite{lucieri2020interpretability}, breast tumor detection~\cite{Rodriguez-Ruiz2019Stand-AloneRadiologists}, cardiac MRI classification~\cite{gloabl_local}, tumor segmentation in liver CT~\cite{tumor_liver} and radiomics~\cite{yeche2019ubs}.

In another line of work, researchers explore training both a classification model and a concept classifier to obtain an inherently explainable model~\cite{bouchacourt2019educe}.  Similar to this, concept bottleneck method~\cite{cbm} first learns to predict the concepts, then uses only those predicted concepts to make a final prediction~\cite{sabour2017dynamic}. Such approach are popular in medical domain, with applications in lung nodule malignancy classification~\cite{shen2019interpretable}. Goyal \etal measures the
causal effect of concepts by using a conditional VAE model~\cite{Goyal2019Explainingcace}. Measuring the causal effect is essential, as presence of concept information in the latent space of the DNN doesn't necessarily means the network is using that information to make its decision.  To provide causal explanation, Harradon \etal  build a bayesian causal model using these extracted concepts as variables in order to explain image classification~\cite{harradon2018causal}.

\section{DNN model uncertainty quantification}
To facilitate the real-world deployment of a DNN model, it is essentially important to understand what a DNN model does not know. State-of-the-art classification models are mostly DNN such as DenseNet, ResNet and more. These models are deterministic, in the sense they only provide point estimates for the posterior. 
The gold standard for UQ is \textit{Bayesian Neural Network} (BNN)~\cite{neal2012bayesian}. BNN are an alternative to DNN, as they provide a distribution over the model parameters which helps in quantifying model uncertainty. However, computing this information comes at an extra computational cost while also increasing the inference complexity [37, 151]. Moreover, training a BNN is often intractable, and they arguably result in sub-optimal accuracy as compared to deterministic approaches. This is perhaps due to the difficulty in tuning their hyper-parameters~\cite{wilson2020case}. 

Alternatively, approaches such as \textit{Deep Ensembles}~\cite{NIPS2017_9ef2ed4b} and \textit{MC Dropout}~\cite{pmlr-v48-gal16} has been introduced as an approximation of BNN that are compatible with the deterministic DNN architecture with minimal changes at the inference time. Deep ensembles require training multiple copies of the DNN with either  random initialization of the weights or the training data or both. In MC-dropout, weights are randomly dropped at training as well as inference time. Deep ensembles, however, require multiple DNN models to be trained using different initialization seeds, making them computationally expensive to train. MC Dropout is computationally less expensive, but cannot be used for UQ in pre-trained models that are trained without dropout.

The recent interest in single forward pass UQ techniques~\cite{van2020uncertainty,van2021feature} has led to less expensive alternatives for MC Dropout. However, they require a DNN to be trained from scratch using specific constraints or loss functions, and hence cannot be used to fix pre-trained DNNs with poor uncertainty estimates. Further, [118] proposed a novel method that learns the observation noise parameter, which enables it to model both epistemic and aleatoric uncertainty in a single forward pass. Model uncertainty or epistemic uncertainty~\cite{Gal2016UncertaintyID}, measure the uncertainty in estimating the DNN model parameters given the training data. Epistemic uncertainty measures how well the model learns the data. It is reducible as the size of the training data increases. Data uncertainty, or aleatoric uncertainty~\cite{Gal2016UncertaintyID}, is irreducible uncertainty that arises from the natural complexity of the data, such as class overlap or label noise. Data uncertainty is also considered as a `known-unknown' \ie the DNN model understands (knows) the data distribution and can confidently predict whether a given input is difficult to classify \i.e an unknown~\cite{NEURIPS2018_3ea2db50}. However, epistemic uncertainty may also arise when there is a mis-match between the training and testing data distribution. This is  `unknown-unknown' as the model is unfamiliar with the test data and hence, cannot confidently make predictions.

\subsection{Uncertainty quantification in pre-trained DNN models}
Much of the prior work focused on deriving uncertainty measurements from a pre-trained DNN output~\cite{hendrycks17baseline,guo2017temperaturescaling,liang2018enhancing,liu2020energy}, feature representations~\cite{lin2021mood,Lee2018ASU} or gradients~\cite{huang2021on}. Such methods use a threshold-based scoring function to identify OOD samples. A baseline method
for OOD detection was introduced by Hendrycks \etal. They showed that simple statistics derived from softmax distributions provide an effective way to identify out of distribution (OOD) data~\cite{hendrycks17baseline}. Guo\etal extended this work by demonstrating that  a single-parameter variant of Platt scaling, also known as temperature scaling is an effective method to obtain calibrated probabilities, which in turn helps in better  OOD detection~\cite{guo2017temperaturescaling}. Very recently, researchers have proposed energy-based scores for OOD detection~\cite{liu2020energy,wang2021can}. The energy score
helps in mitigating a critical problem with softmax confidence that assigns arbitrarily high values for OOD examples. Further, several works are been proposed that attempts to improve the OOD uncertainty quantification by using ODIN score~\cite{liang2018enhancing} and its variant~\cite{9156473}. Specifically, ODIN proposed adding small perturbations to the input and gradually increasing the softmax score of any input by reinforcing the model’s belief in the predicted label. Further, proposed to use Mahalanobis distance-based confidence score to identify and reject OOD samples~\cite{Lee2018ASU}.

In another attempt, Huang \etal proposed to use  GradNorm, a simple and for detecting OOD inputs by utilizing information extracted from the gradient space. Gradient norm uses the vector norm of gradients, backpropogated from the KL divergence between the softmax output and uniform probability distribution. All these methods help in identifying OOD samples but did not address the over-confidence problem of DNN, that made identifying OOD non-trivial in the first place~\cite{Hein2019WhyRN,nguyen2015deep}. Our work focuses on mitigating the over-confidence issue by fine-tuning a pre-trained classifier on counterfactually augmented data (CAD). Further, we used the discriminator of the GAN-generator to provide a density score to identify OOD samples.

%Our method not only enjoys the computational efficiency of deterministic approaches, but can also be used to fix pre-trained DNNs with minimal fine-tuning steps without incurring loss in predictive accuracy. While these approaches mainly focus on far-OOD data, we also show comparisons on near-OOD and AiD samples. 

\subsection{DNN designs for improved uncertainty estimation}
Designing generalized DNN that provides robust uncertainty estimates has gained significant research attention. 
The Bayesian neural networks are the gold standard for reliable uncertainty quantification~\cite{neal2012bayesian}. Multiple approximate Bayesian approaches have been proposed to achieve tractable inference and to reduce computational complexity~\cite{NIPS2011_7eb3c8be,Weight_Uncertainty,NIPS2015_bc731692,pmlr-v48-gal16}.  Popular non-Bayesian methods include deep ensembles~\cite{NIPS2017_9ef2ed4b} and their variant~\cite{Snapshot,loss_surface}. However, most of these methods are computationally expensive and requires multiple passes during inference. An alternative approach is to modify DNN training~\cite{label_smoothing_szegedy,zhang2017mixup,manifold_mixup}, loss function~\cite{Mukhoti2020CalibratingDN}, architecture~\cite{sun2021react,Liu2020SimpleAP,Geifman2019SelectiveNetAD} or end-layers~\cite{van2020uncertainty,9156473} to support improved uncertainty estimates in a single forward-pass. Further, methods such as DUQ~\cite{van2020uncertainty} and DDU~\cite{mukhoti2021deterministic} proposed modifications to enable the separation between aleatoric and epistemic uncertainty. Unlike these methods, our approach improves the uncertainty estimates of any existing pre-trained classifier, without changing its architecture or training procedure. We used the discriminative head of the fine-tuned classifier to capture aleatoric uncertainty and the density estimation from the GAN-generator to capture epistemic uncertainty.  

\subsection{Uncertainty estimation using GAN}
A popular technique to fix an over-confident classifier is to regularize the model with an auxiliary OOD data which is either realistic~\cite{hendrycks2018deep,Mohseni2020SelfSupervisedLF,PAPADOPOULOS2021138,chen2021atom,liang2018enhancing} or is generated using GAN~\cite{ren2019likelihood,lee2018training,Mandal_2019_CVPR,Xiao2020LikelihoodRA,Serr2020Input}. 
Such regularization helps the classifier to assign lower confidence to anomalous samples, which usually lies in the low-density regions. On of the earlier methods proposed outlier exposure (OE) that leverages diverse, realistic datasets for exposing the model training to OOD distribution~\cite{hendrycks2018deep}. Chen \etal showed that randomly selecting outlier samples for training may yield uninformative samples. They proposed an adversarial training with informative outlier mining (ATOM) technique to selectively collect auxiliary outlier data for estimating  a tight decision boundary between ID and OOD data, which leads to robust OOD detection performance~\cite{chen2021atom}.

Another line of researchers investigate deep generative model based approaches
for OOD detection. Such methods use generative modeling  to detect OOD samples by setting a threshold on the likelihood. An application of generative model such as GAN in OOD detection is the use of entropy loss in the construction of an OD detector for generalized zero-shot action recognition~\cite{Mandal_2019_CVPR}. They learn an OOD detector using real and GAN-generated features from seen and unseen categories, respectively. In another attempt, Ren \etal  propose the use of a likelihood-ratio test by taking the ratio between the likelihood obtained from the model and from a background model which is trained on random perturbations of input data~\cite{ren2019likelihood}.  Further,~\cite{Serr2020Input} proposed to offset the bias of the generative models by a factor that measures the input complexity, such as the length of lossless compression of the image. Further,~\cite{ren2019likelihood,Serr2020Input,salimans2017pixelcnn} obtain high OOD detection performances with Glow, VAE and Pixel-CNN generative models.

Defining the scope of OOD a-priori is generally hard and can potentially cause a selection bias in the learning. Alternative approaches resort to estimating in-distribution density~\cite{Subramanya2017ConfidenceEI}.
 Our work fixed the scope of GAN-generation to CAD~\cite{Singla2020ExplanationExaggeration}. Rather than merging the classifier and the GAN training, we train the GAN in a post-hoc manner to explain the decision of an existing classifier. This strategy defines OOD in the context of pre-trained classifier’s decision boundary. Previously, training with CAD have shown to improved generalization performance on OOD samples~\cite{Kaushik2021ExplainingTE}. However, much of this work is limited to Natural Language Processing, and requires human intervention while curating CAD~\cite{Kaushik2020Learning}. In contrast, we train a GAN-based counterfactual explainer~\cite{singla2021explaining,explaining_in_style} to derive CAD.
 
 \subsection{Data augmentation for improving uncertainty estimation} 
 There is a rich literature on data augmentation (DA) for improving the classification performance of DNNs~\cite{cutout_terrance,autoaugment_da,random_erasing_da,da_survey}. However, most of the classical DA literature is task agnostic and focused on improving accuracy. While GAN-based DA is popular, they are mainly used to generate samples that are consistent with the underlying distribution without taking the DNN into account. In contrast, our GAN-based augmentation network is closely coupled with the pre-trained DNN, and generates samples in ambiguous regions of the distribution to enhance the uncertainty characteristics of the pre-trained model. We take inspiration from recent works~\cite{Singla2020ExplanationExaggeration,singla2021explaining,explaining_in_style} on counterfactual explanations which focus on explaining a DNN. However, they do not explore whether the generated samples can improve a downstream task. Additionally, there is research showing that models trained on counterfactually augmented data have improved generalization performance on out-of-domain samples~\cite{Kaushik2021ExplainingTE}. However, much of this work is limited to Natural Language Processing, and our work differs in terms of both the application and the architecture we use for our proposed method.

\chapter{Improving Clinical Disease Sub-typing and Future Events Prediction through a Chest CT based Deep Learning Approach} 
\label{ch1}
\section{Introduction}
Chronic obstructive pulmonary disease (COPD) is characterized by persistent respiratory symptoms and irreversible airflow obstruction~\cite{Vogelmeier2017GlobalSummary}. The measurement of spirometric obstruction, while traditionally used to define COPD severity, is not sufficient to explain the many critical dimensions required to characterize and manage COPD~\cite{Coxson2014Usingsub1/sub}. Airflow obstruction can result from varying combinations of emphysematous parenchymal destruction~\cite{ODonnell2006PhysiologyCOPD}, chronic airway remodelling~\cite{Grzela2016AirwayMetalloproteinase-9.}, and other poorly characterized imaging patterns, including fibrotic changes common in smokers~\cite{Washko2011LungAbnormalities}. Hence, clinicians must adopt a comprehensive approach while assessing patients with COPD, including identifying risk factors, standardized assessment of symptoms and comorbidities, estimating exacerbation risk~\cite{Soler-Cataluna2005SevereDisease.}, and prognostication of survival. Other established tools for assessing COPD symptoms are the modified Medical Research Council (mMRC) dyspnea scale and prognostication of survival using the body mass index, obstruction, dyspnea and exercise capacity (BODE) index~\cite{Celli2004TheDisease, Martinez2006PredictorsObstruction}. Though radiography has not been historically utilized in routine diagnosis or management of COPD~\cite{ostridge2016present}, the growing use of CT imaging for pulmonary nodule assessment and cancer screening~\cite{tammemagi2017participant,ostrowski2018low}, provides a novel opportunity to leverage imaging data to investigate patients with COPD.

 Despite much interest in using CT imaging in subtyping COPD~\cite{Lynch2015CTDefinableSociety},  stratification of patients as obstructed or non-obstructed is currently based on spirometric pulmonary function testing findings according to the Global Initiative for Chronic Obstructive Lung Disease (GOLD) guidelines~\cite{Vogelmeier2017GlobalSummary}. Much of the clinical workflows rely heavily on qualitative visual assessment for characterizing COPD. Visual assessment includes identifying image features highlighting air trapping in small airways~\cite{matsuoka2008quantitative}, characterizing local patterns for emphysema~\cite{Hayhurst1984DiagnosisTomography,Muller1988DensityTomography}, bronchial wall thickening, or endobronchial mucus~\cite{Kim2021MucusDisease}, and calculating the percentage of low attenuation area (LAA)~\cite{Nishio2017AutomatedRegion}, blood vessel volume\cite{Estepar2013ComputedImplications}, or airway counts\cite{Diaz2010AirwaySmokers}. Also, various intensity and texture-based feature descriptors are proposed to characterize the visual appearance of COPD~\cite{Cheplygina2017TransferDisease,Sorensen2012Texture-basedApproach,Yang2017UnsupervisedStudy}. But most of these image features are generic and are not necessarily optimized for characterizing COPD. Furthermore, some of these methods rely on manual segmentation methods and are thus both labour-intensive and prone to operator error\cite{Lynch2015CTDefinableSociety, Lynch2018CTbasedStudy, MohamedHoesein2012ComputedDecline, Muller1988DensityTomography}.

 While visual CT analysis remains the mainstay of clinical imaging interpretation, there has been growing research interest in quantitative image analysis techniques to quantify abnormalities on CT and characterize disease subtypes~\cite{dlEmphysema}.
Recent advances in deep learning (DL) enable researchers to go directly from raw images to clinical outcomes without specifying radiological features~\cite{Gonzalez2018DiseaseTomography}.
 However, most of the existing work concentrate on some aspect of COPD disease like only spirometry or only emphysema or COPD sub-typing~\cite{TANG2020e259}. There is  room for improvement to bring the prediction of multiple patient-centred outcomes to quantify COPD. Further, much impact can be made by predicting patients' future exacerbation or survival, thus providing helpful input to construct personalized treatment plans.

This paper proposes a novel DL model that takes an entire 3D volumetric image as input and provides a holistic view of a patient's health in terms of multiple COPD outcomes. Our novel DL model followed a data-driven approach and directly analyzed raw HRCT data without manually segmenting or specifying radiological features. Previous, DL approaches~\cite{Gonzalez2018DiseaseTomography} processed slices (three orthogonal slices) of CT images and hence may not be able to characterize the volumetric impact of the disease. In contrast, our proposed method views each subject as a \emph{set} of image patches from the lung region. It can analyze the entire 3D CT scan and requires no image distortion due to resizing or cropping. Previously,~\cite{Cheplygina2017TransferDisease,Schabdach2017AStudies} also viewed CT images as a set and extracted handcrafted image features from each input element. In contrast, the \emph{discriminative} part of our model uses a deep learning approach and directly extracts features from the volumetric patches. Further, we use an attention mechanism~\cite{Xu2015ShowAttention} to adaptively weigh local features and build the subject level representation, which is predictive of the disease severity. Our model is inspired by the Deep Set~\cite{Zaheer2017DeepSets}. We extend it by adapting \emph{generative} regularization, which prevents the redundancy of the hidden features. Furthermore, the \emph{attention mechanism} provides interpretability by quantifying the relevance of a region to the disease.

We predict multiple patient-relevant outcomes such as symptom scores, emphysema severity and pattern, exacerbation risk, and mortality. When compared to other DL model~\cite{Gonzalez2018DiseaseTomography}, our method improved the prediction of important clinical variables, such as COPD disease severity and exacerbation risk. Furthermore, it can distinguish between centrilobular and paraseptal emphysema and quantify the future risk of exacerbation based on the current CT image. Estimating these clinically relevant features using only CT images has a potential application both to clinical care and research.

\section{Method}

\begin{figure}[!ht]
   \includegraphics[width=0.95\linewidth ]{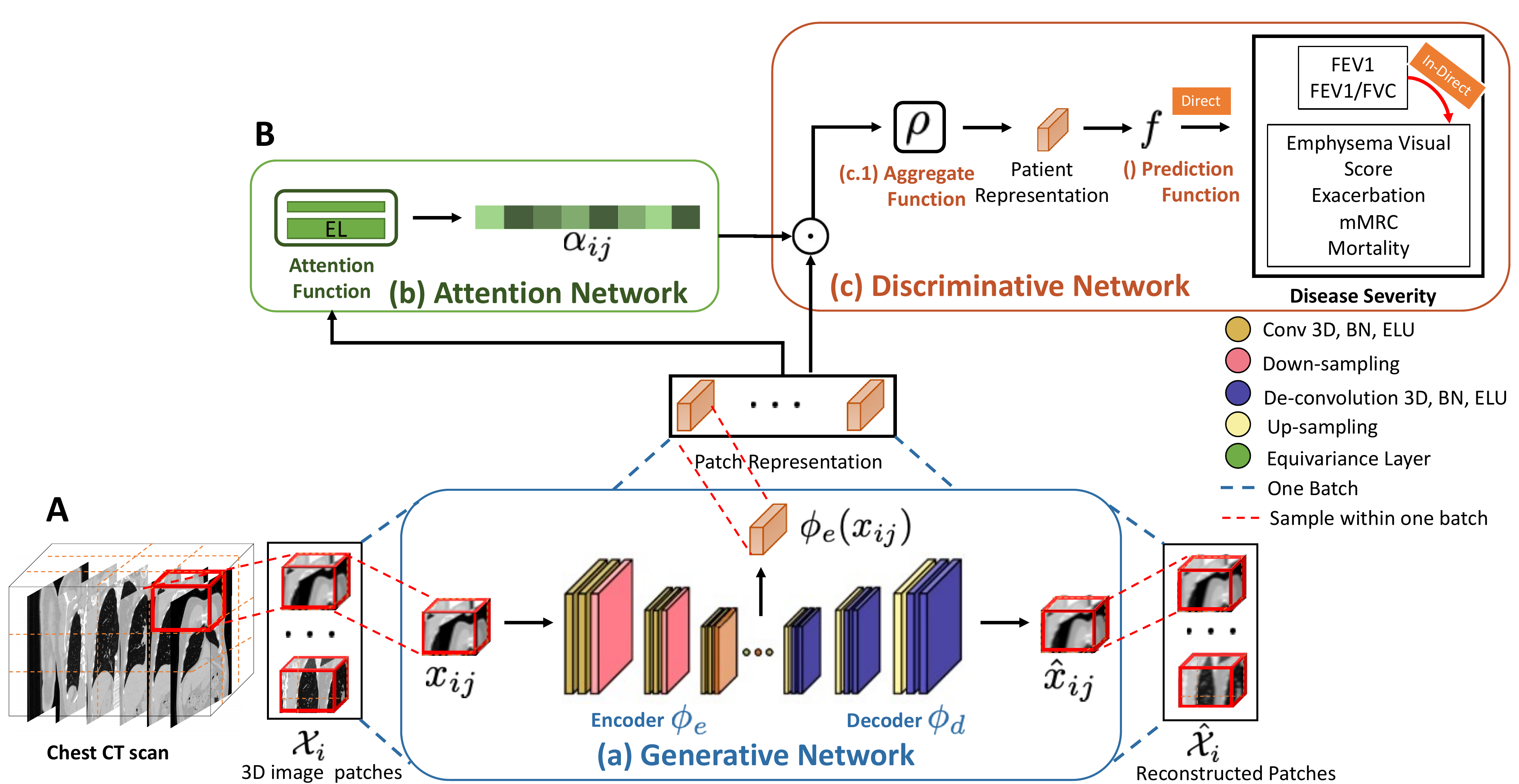}

   \caption[The schematic of our model.]{The schematic of our model. \textbf{A.} The input to our model is a 3D CT scan of the lung. The lung is divided into a set of equally sized, overlapping 3D image patches. (a) The \textbf{Generative Network} is a convolutional auto-encoder (CAE). The encoder function projects the raw image patch to a latent space and the decoder function reconstructs the image patch from the extracted latent features. \textbf{B.} The \textbf{Attention Network} provides interpretability by weighting the patches based on their importance in predicting the disease severity. \textbf{C.} The \textbf{Discriminative Network} (c.1) aggregates the local patch-level information  information, based on their attention weights, to create a patient-level representation, and (c.2) uses it to predict disease severity. }
    \label{fig_1} 
\end{figure}

We represent each subject as a set (bag) of volumetric image patches extracted from the lung region $\mathcal{X}_i = \{x_{ij}\}_{j=1}^{N_i}$, where $N_i$ is the number of patches for subject $i$, which varies with subject.The model learned to extract informative regional features from these patches $x_{ij}$, and then adaptively weight these features to form a fix-length representation for each patient. This patient-representation is then used to predict disease severity ($y_i$). The general idea of our approach is shown in Figure.\ref{fig_1}.

The method consists of three networks that are trained jointly: (1) a \emph{discriminative} network, that aggregates the local information from patches in the set $\mathcal{X}_i$ to predict the disease severity $y_i$, (2) an \emph{attention} mechanism, that helps discriminative network to selectively focus on patch-features by assigning weights to the patches in $\mathcal{X}_i$, and (3) a \emph{generative} network, that regularizes the discriminative network  to avoid redundant representation of patches in the latent space. The model is trained end to end, by minimizing the below objective function:
\newline
\begin{equation}
\label{eq:general}
\min_{\omega, \theta_e , \theta_d, \theta_a} \sum_i \mathcal{L}_d\left(y_i, \hat{y}_i ( \mathcal{X}_i ) ; \theta_e, \omega\right) +
\lambda_1  \mathcal{L}_g\left(  \mathcal{X}_i, \hat{\mathcal{X}}_i ; \theta_e , \theta_d \right) + \lambda_2 \mathcal{R} \left( \mathcal{X}_i ; \theta_e, \theta_a \right),
\end{equation}
\newline
where $\mathcal{L}_d(\cdot, \cdot)$ and  $\mathcal{L}_g(\cdot, \cdot)$ are the discriminative and generative loss functions respectively and $\mathcal{R}(\cdot)$ is a regularization over the attention. The $\theta_e$, $\theta_d$, $\theta_a$ and $\omega$ are the parameters of each term. $\lambda_1, \lambda_2$ controls the balance between the terms. The sum is over number of subjects. Next, we discuss each term in more detail. 
\newline

\vspace{-0.4cm}

\subsection{Generative Network}
The Generative Network is a convolutional auto-encoder (CAE)\cite{Masci2011StackedExtraction}. CAE consists of an encoder $\phi_e(\cdot)$, that extracts local image features from each patch $\left(i.e.,  \phi_e(x_{ij}; \theta_e) \in \mathbb{R}^d \right)$. These features are a summarization of the information in the raw image patch (or region) in a low dimensional “feature space”. To regularize the feature extraction process, CAE have a decoder $\phi_d(\cdot)$. The decoder  recovers the input patch back from the low dimensional feature space as $\hat{x}_{ij}= \phi_d(\phi_e(x_{ij}; \theta_e); \theta_d)$. In the absence of the decoder function, the feature extractor $\phi_e$ will be forced to retain only information that is sufficient for the underlying task of predicting $y$. If $y$ is low dimensional as compared to $d$, $\phi_e$ learns a highly redundant latent space representation for each patch.  To prevent this information loss, we regularize the auto-encoder using a distance loss defined as, $\mathcal{L}_g(  \mathcal{X}_i, \hat{\mathcal{X}}_i ; \theta_e , \theta_d ) = \frac{1}{\left|\mathcal{X}_i \right|}\sum_{x_{ij}\in \mathcal{X}_i} || x_{ij} - \hat{x}_{ij} ||_2$.

\subsection{Attention Network}
\label{att}
The goal of our proposed model is twofold: first to provide a prediction of the disease severity and secondly, to provide a qualitative assessment of our prediction. Here, it is reasonable to assume that different regions in the lung contribute differently to the disease severity. We model this contribution by adaptively weighting the patches. The weight indicates the importance of a patch in predicting the overall disease severity of the lung. This idea is similar to attention mechanism in Computer Vision~\cite{Xu2015ShowAttention} and Natural Language Processing~\cite{Luong2015EffectiveTranslation} communities.

The goal of the attention network is to learn a weight for each of the input image patches, such that the weight indicates the importance of a patch in predicting the overall disease severity of the lung.  We used another neural network to learn these weights for $i^{th}$ subject as   $\left( \bm{\alpha_i} = \{ \alpha_{i1}, \cdots , \alpha_{i,N_i}\} \right)$ 
where
$\bm{\alpha_{i}} = A\left( \phi_e \left( \mathcal{X}_i ; \theta_e\right); \theta_{a} \right)$. We formulated the attention network $A(\cdot)$ as a feed-forward network, consisting of multiple equivariant layers (EL)\cite{Zaheer2017DeepSets}. Assuming $\mathbf{H}_i \in \mathbb{R}^{N_i, d}$ where $k^{th}$ row is $\phi(x_{ik} ; \theta_e) \in \mathbb{R}^d$, an equivariant layer is defined as
\newline
\begin{equation}\label{eq:EL}
\left[ \mathbf{H}_i\right]_k = \mathbf{W}\left(\left[ \mathbf{H}_i\right]_k - \max(\mathbf{H}_i, 1)\right) + \mathbf{b},
\end{equation}
\newline
where $\left[ \mathbf{H}_i \right]_k $ denotes $k^{th}$ row of $\mathbf{H}_i$ and $\max(\mathbf{H}_i,1)$ is the max over rows. $\mathbf{W} \in \mathbb{R}^{L \times d}$, $\bm{b} \in \mathbb{R}^L$ are the parameters of the EL. Such formulation ensures that the weight of any patch depends not only on the corresponding patch feature but also on the features of all the other patches in a patient. Next, we pass the output of the EL layers to a softmax function, to obtain a distribution of weights over the patches. This ensures that the weights $(\bm{\alpha_i})$ are non-negative numbers that sums to one. 

To enable interpretability, the weight vector, $\bm{\alpha_i}$, should follow a sparse distribution. Increased sparsity pushes some weights terms, $\alpha_{ij}$, to zero, and hence, it increases the interpretation by focusing on only the patches relevant for the prediction task. In our formulation, the weights $\alpha_{ij}$, have non-negative values that sum to 1 \ie $(||\bm{\alpha_{i}}|| = \sum_j \alpha_{ij} = 1)$. Hence, its derivative is zero, and using an $\ell_1$ norm over the weight vector will not result in a sparse solution. To ensure high sparsity, we use a log-sum function as a regularizer. Minimizing $\sum_j \log \alpha_{ij}$ is equivalent of maximizing KL-divergence from the uniform distribution. The uniform distribution assigns the same weight to all the patches within one subject, $i.e., \max_{\bm{\alpha_{i}}} \textnormal{KL}([\frac{1}{N_i}, \cdots, \frac{1}{N_i}], \bm{\alpha_i}) = \max_{\bm{\alpha_{i}}} \sum_{j} \frac{1}{N_i} \log \frac{1}{N_i} - \sum_{j} \frac{1}{N_i} \log 
\alpha_{ij} \equiv \min_{\alpha_i} \sum_j \log \alpha_{ij}$.
 We defined the regularization term as, $\mathcal{R} \left( \mathcal{X}_i ; \theta_e, \theta_a \right) = \sum_{j=1}^{N_i} \log (\alpha_{ij} + \epsilon)$ and add it to the loss function in Equation.~\ref{eq:general}.

\subsection{Discriminative Network}
The discriminative network predicts the disease severity  as 
\newline
\begin{equation}\label{eq:4}
\hat{y}_i(\mathcal{X}_i) = f\left( \rho \left(\phi_e \left(\mathcal{X}_i, \theta_e \right)\right), \omega \right).
\end{equation}

\vspace{0.5cm}

The discriminative network takes the patch-level features $\left(i.e.,  \phi_e(x_{ij}; \theta_e)\right)$  extracted by the encoder as input. It transforms the patch-level features using composition of two functions: (1) The aggregate function $\rho(\cdot)$. It is a permutation invariant function that aggregates the patch-level features to form a fixed length patient-representation. (2) A prediction function $f(\cdot;\omega)$, parameterized by $\omega$. It takes the patient representation extracted by $\rho(\cdot)$ as input, and estimates the disease severity. Finally, $\mathcal{L}_d\left(y_i, \hat{y}_i ( \mathcal{X}_i ) ; \theta_e, \omega\right)$ is a regression or classification loss function between predicted and true value.

Conceptually, the aggregate function makes the prediction of disease severity less sensitive to the precise location within an image. It does so by aggregating the information from the local patches. One possible formulation of aggregate function is an average function, defined as $\rho(\cdot) = \frac{1}{N_i} \sum_{j=1}^{N_i} \phi_e(x_{ij})$. It considers all the feature values and hence, spread out the volume of the latent space evenly.  The average function assumes an equal contribution of all the local patches towards final disease severity. However,  COPD disease is often attributed to the diffuse air-sacks obstruction spread unevenly throughout the lung. To incorporate the disease's diffused effect, we adaptively weight the patch-level features to create the patient-representation as, $\rho(\cdot) = \sum_{j=1}^{N_i} \alpha_{ij} \phi_e(x_{ij})$.  An attention network, described in Section~\ref{att}, learns the weights ($\alpha_{ij}$).  
  
\subsection{Architecture Details}
The architecture of the encoder function consists of stacked convolutional layers which down-sampled the patches while doubling the number of channels. The decoder function consists of  transposed convolutional layer (or deconvolutional layer) which up-sample the features while cutting the number of channels to half. Each convolutional layer employs batch-normalization for regularization, followed by an exponential linear unit (ELU)~\cite{Clevert2015FastELUs} for non-linearity. The attention network has 2 equivalence layers with sigmoid activation function, followed by a softmax layer. The model is trained using Adam optimizer~\cite{Kingma2014Adam:Optimization} with hyper-parameters $\beta_1 = 0$ and $\beta_2 = 0.999$ and a fixed learning rate of 0.001. The dimension of the feature vector is 128. The trade-off hyper-parameters are $\lambda_1 = 10$ and $\lambda_2 = 1$. The experiments are performed on two NVIDIA p100 GPUs, each with 16GB GPU memory. The source code is available at
\href{https://github.com/batmanlab/Subject2Vec}{https://github.com/batmanlab/Subject2Vec}.

\section{Experiments and Results}
\subsection{Study cohort}
We evaluated our method on a dataset from the COPDGene study; an NIH funded multi-center clinical trial focused on the genetic epidemiology of COPD~\cite{Regan2010GeneticDesign}. COPDGene includes 10,300 baseline participants, all of which were either current or former smokers. Each participant performed spirometry and had a high resolution inspiratory and expiratory CT scan, using a standardized protocol~\cite{Regan2010GeneticDesign}. The acquired CT scan images were assessed by trained experts to provide a visual quantification of the centrilobular and paraseptal emphysema severity. Survival information was collected using the Social Security Death Index (SSDI) search and the COPDGene longitudinal follow-up (LFU) program. 

\subsection{Experimental setup}

In our analysis, we used full-inspiration CT images, which were re-sampled to isotropic 1 mm$^3$. We worked on the fixed range of intensity values between -1024 HU and 240 HU, as suggested by Bhalla et al.~\cite{Ash2017DensitometricFibrosis}. We represented each subject as a set of equally sized 3-dimensional patches. To extract these patches, we first segmented the chest  using Chest Imaging Platform (CIP)~\cite{ross2015chest}, open-source software for quantitative CT imaging assessment. Next, we extracted 3D overlapping patches from parenchyma region of the chest. The number of patches in a subject ($N_i$) may vary between subjects. A large patch size or a high overlap between the patches increases the $N_i$ for a subject. All the patches of a subject must be processed in the same batch, as they are required to learn the patient-representation, which is then used to predict the disease severity. The available GPU memory restricts the maximum number of patches that can be processed in a single batch.  We experimented with different values and finally used a patch-size of 32$\times$32$\times$32 with a 40\% overlap and an upper limit of 1000 patches per batch in our experiments. The average $Ni$ for this setting is 700 patches per subject. %We consider one subject per batch as shown in Figure~\ref{fig_1}.

We presented an analysis of the performance of our model for predicting patient-centered outcomes related to COPD. We trained two versions; 1) \textbf{Direct}: the model was trained to predict forced expiratory volume in 1 second (FEV1) and the FEV1/forced vital capacity (FVC) ratio, along with a clinical outcome of interest to represent disease severity. We separately trained one such model for each of the target outcomes. 2) \textbf{Indirect}: the model was trained only once, to predict FEV1 and FEV1/FVC as disease severity. The patient-representations from such model were then used in a separate regression analysis to predict other clinical outcomes of interest. The idea is to learn generalized patient-representations by training the model for one clinical variable (spirometry) and testing on another clinical output (emphysema score) which the models haven't seen previously. If two clinical variables are correlated, we should be able to capture much variance. Ofcourse, training directly for the clinical variable, as in direct version, will achieve better results. For all results, we reported average test performance in five-fold cross-validation. 
We compared the performance of our method against 
\begin{enumerate}
    \item Baseline: The low attenuation area (LAA) features. \textbf{LAA-950} is defined as the total percentage of both lungs with attenuation values less than -950 Hounsfield units on inspiratory images. LAA-950 signifies radiographic emphysema~\cite{Nishio2017AutomatedRegion}.
    \item The \textbf{non-parametric} method proposed by Schabdac et al.~\cite{Schabdach2017AStudies}. In this method, hand-crafted image features were extracted for each patient, and non-parametric density estimation was performed to assign a characteristic vector to each patient.
    \item The classical \textbf{k-means} algorithm applied to image features extracted from local lung regions~\cite{Schabdach2017AStudies}. A similar approach was suggested by Ash \etal~\cite{Ash2017DensitometricFibrosis}.
    \item The previous state-of-the-art method based on \textbf{CNN} also, applied to the COPDGene~\cite{Gonzalez2018DiseaseTomography}.
\end{enumerate}

\begin{table}[!ht]
\centering
\caption{ Summarization of the clinical outcomes considered in the experiments and their numerical type and values.}
\label{ch2-table-1}
\begin{tabular}{|c|c|c|l|}
\hline
 \bf Clinical Outcomes &   \bf Type &  \bf Values &  \bf Description\\ 
\hline
\multicolumn{4}{ |c|  }{\bf{Spirometry Measures - Section ~\ref{sm}}} \\
\hline
FEV1 & Continuous &  & Percentage predicted forced expiratory\\
& & &   volume  in  1  sec. \\
FEV1/FVC & Continuous & & FEV1 ratio with forced vital capacity\\
& & &  (FVC) \\
COPD & Binary & 0 or 1 & True if FEV1/FVC $>$ 0.7 \\
GOLD stages & Categorical & 0 - 4 & GOLD stages 0 (non-obstructed) \\
& & & through 4 (severely obstructed).\\
\hline
\multicolumn{4}{ |c|  }{\bf{Visual Emphysema Score - Section~\ref{ves}}} \\
\hline
Centrilobular  & Categorical & 0 - 5 & CLE emphysema severity score: \\
Emphysema (CLE) & & & none (0) to advanced destruction (6).\\
Paraseptal Emphysema  & Categorical & 0 - 2 & Three severity scores: none, mild \\
& & &  and substantial. \\
\hline
\multicolumn{4}{ |c|  }{\bf{Acute Exacerbation - Section~\ref{ae}}} \\
\hline
Historic Exacerbation & Binary & 0 or 1 & True if patient have experienced \\
& & & exacerbation in the last 1 year.\\
Future Exacerbation & Binary & 0 or 1 & True if patient reported an  \\
& & & exacerbation by the 5th year followup.\\
\hline
\multicolumn{4}{ |c|  }{\bf{Others - Section~\ref{mds}}} \\
\hline
mMRC Dyspnea Scale & Categorical & 0 - 4 & Dyspnea with strenuous exertion (0) \\
& & & to dyspnea in daily activities (4) \\
Mortality & Binary & 0 or 1 & Vital status\\
 \hline
\end{tabular}
\end{table}

We perform three experiments: (1) \textit{Predicting COPD outcomes:} we compare the performance of our method against the sate-of-art for different prediction tasks, (2) \textit{Generative regularizer ($\lambda_1$):} we study the effect of the generative regularizer (\ie $\lambda_1$) in terms of prediction accuracy and information preserved in latent space, (3) \textit{Visualization:} we visualize the interpretation of the model on the subject and population level. 

\subsection{Predicting COPD outcomes}

We evaluated our proposed model over multiple COPD outcomes. These outcomes are summarized in Table~\ref{ch2-table-1}. Next, we discuss each COPD outcome in more details and summarize our results.

\subsubsection{Spirometry Measures} 
\label{sm}
As part of the pulmonary function test, following spirometry values were evaluated for all the participants in COPDGene: forced expiratory volume in 1 second (FEV1) and the FEV1/forced vital capacity (FVC) ratio. All spirometric values were expressed as percentage of predicted values. Participants were classified as obstructed or non-obstructed under the 2019 Global Initiative for Chronic Obstructive Lung Disease (GOLD) guidelines using a fixed FEV1/FVC ratio of 0.7~\cite{Vogelmeier2017GlobalSummary}. We defined the disease severity as  the  GOLD stages of 0 (non-obstructed) through 4 (very severely obstructed). Following the GOLD guidelines, in our experiments, we first train the model to predicted FEV1 and FEV1/FVC ratio, and then use these values to diagnose and stage COPD. 

\vspace{-0.1cm}

\begin{table}[h!]
\centering
\caption{ Results for predicting spirometry measurements and using them to diagnose and stage COPD.}
\footnotesize
\label{ch-2-table-2}
\begin{tabular}{|c|c|c|ccc|cc|}
\hline
Method & FEV1 & FEV1/FVC   & \multicolumn{3}{c|}{COPD Diagnosis} &  \multicolumn{2}{c|}{GOLD}\\ 
\cline{1-8}
& R-Square & R-Square & AUC & AUC & Recall & \%   & \%  Accuracy \\
& & & ROC& PR& & Accuracy & \textit{one-off} \\
\hline
Ours (direct) & \textbf{0.67$\pm$0.03} & \textbf{0.74$\pm$0.01} & 0.82 & \textbf{0.72} & \textbf{0.80} & \textbf{65.44} & \textbf{89.14} \\
CNN\cite{Gonzalez2018DiseaseTomography} & 0.53 & - & \textbf{0.86} & - & - & 51.10 &	74.90 \\
Non-Parametric~\cite{Schabdach2017AStudies} &
0.58$\pm$0.03 &	0.70$\pm$0.02 &	0.79&	0.70&	\textbf{0.80} &	58.85	&84.15\\
K-Mean &
0.56$\pm$0.01&	0.68$\pm$0.02&	0.77&	0.68&	\textbf{0.81}&	57.27&	82.28\\
LAA-950&	0.45$\pm$0.02&	0.60$\pm$0.01&	0.75&	0.64&	0.70&	55.75&	75.69\\
 \hline
\end{tabular}
\end{table} 

\vspace{0.5cm}

\begin{figure}[!ht]
   \includegraphics[width=0.9\linewidth]{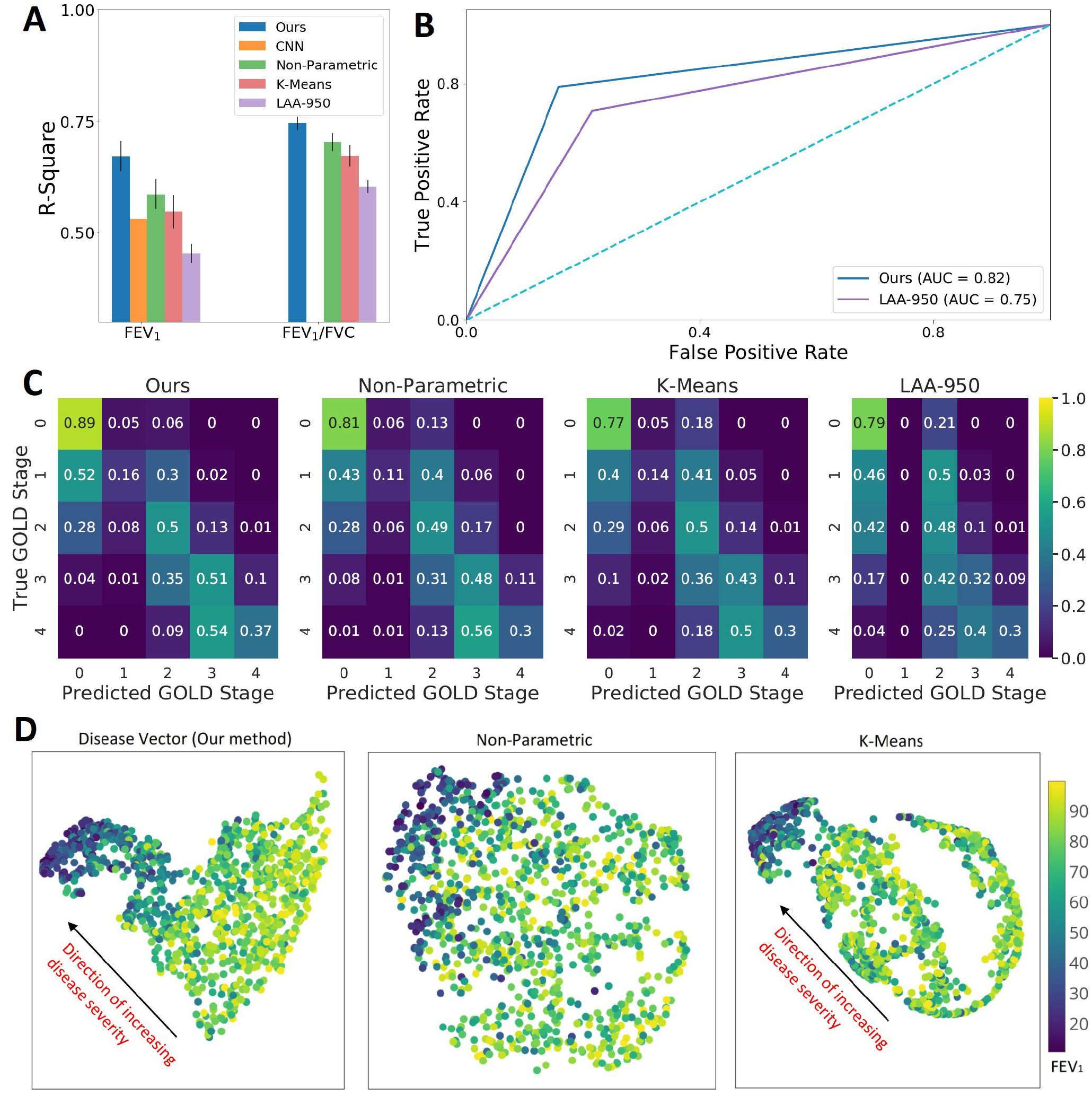}
  
   \caption[Comparing different methods in predicting spirometry measurements, and COPD diagnosis and staging.]{\small
   \textbf{A.} Bar graph comparing the r-square, coefficient of determination, for regression analysis of FEV1 and FEV1/FVC. \textbf{B.} Receiver Operating Characteristic (ROC) curve for prediction of COPD. Higher AUC-ROC suggests better classification. \textbf{C.} Confusion matrix plot for staging subjects using the GOLD stage. Following the GOLD guidelines~\cite{Vogelmeier2017GlobalSummary}, we used the model predicted FEV1 and FEV1/FVC ratio to diagnose and stage COPD. \textbf{D.} Visualizing the population by projecting the patient-level representations to 2D space using a dimensionality reduction method called UMAP~\cite{McInnes2018UMAPReduction}. Each dot represents one subject colored by percentage predicted FEV1. The relative position of a subject can be used to monitor the progression.  We use two dimensions for the sake of visualization; it is straightforward to use a higher dimension and improve patient characterization. Figure is best viewed in color.}
     %note label inside caption
     \label{ch_2_fig_2} 
\end{figure}

\textbf{Results:}  Our model attained an r$^2$ of 0.67 $\pm$ 0.03 for the FEV1 and 0.74 $\pm$ 0.01 for the FEV1/FVC ratio, which is significantly better than previously reported approaches (see Table~\ref{ch-2-table-2}, Figure.~\ref{ch_2_fig_2}). Next, we used the model-predicted FEV1/FVC ratio to diagnose COPD which achieved an AUC-ROC of $0.82$. For the GOLD stage severity classification, our model achieved 65.4\% and 89.1\% exact and one-off accuracy's, respectively. Figure.~\ref{ch_2_fig_2} shows the confusion matrix for the COPD-GOLD stage classification.

\subsubsection{Visual Emphysema Score}
\label{ves}

In the COPDGene cohort, radiographic centrilobular (CLE) and paraseptal emphysema were scored on inspiratory scans by a trained research analysts using the Fleischner Society classification system. Detailed methods for emphysema visual quantification are provided by Lynch \etal~\cite{Lynch2018CTbasedStudy}. They grade the severity of CLE parenchymal emphysema on a scale of zero to five using labels: none, trace, mild, moderate, confluent, and advanced destructive emphysema. While paraseptal emphysema was scored using three labels: none, mild and substantial.

\begin{figure}[!ht]
\centering
   \includegraphics[width=0.9\linewidth]{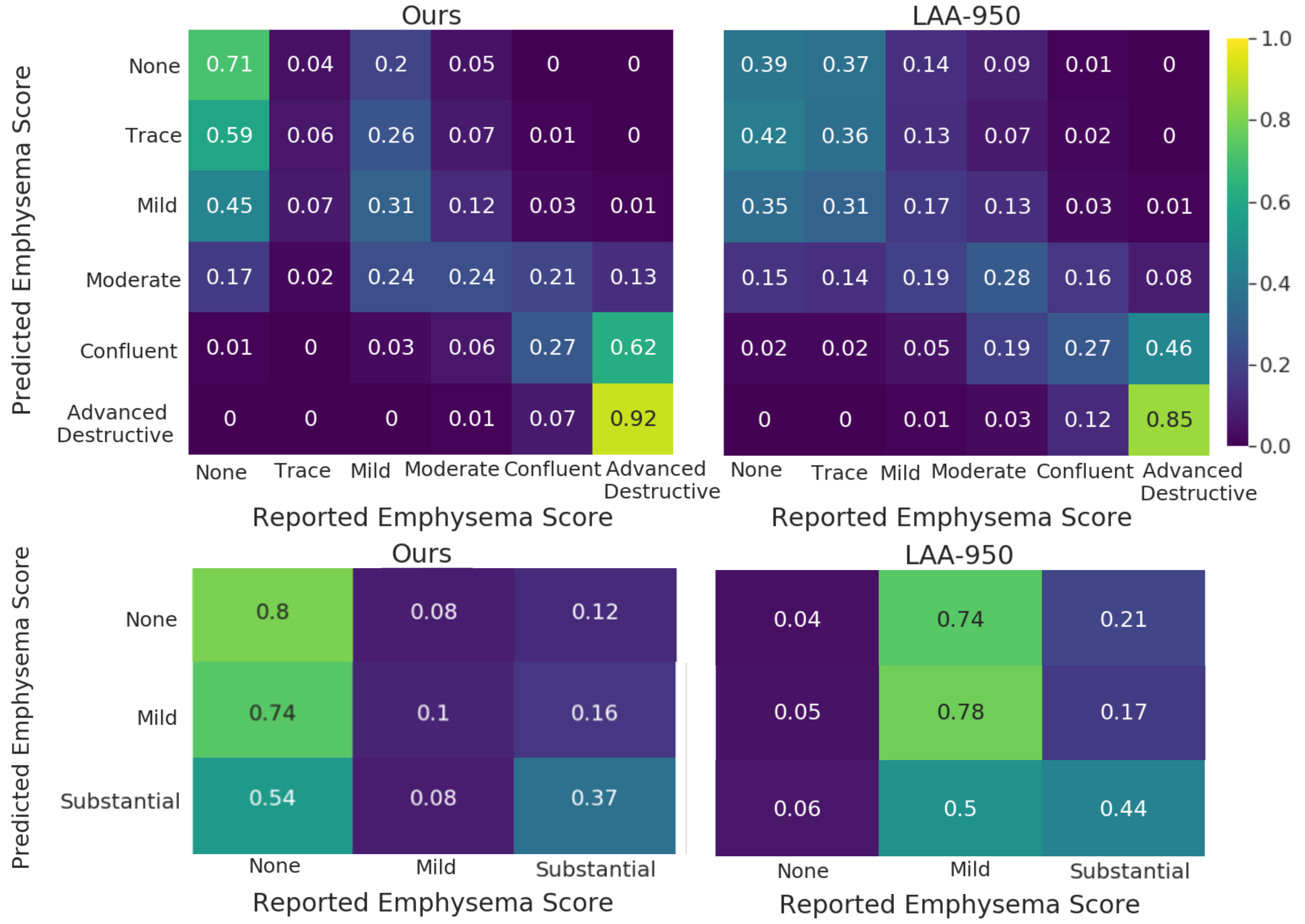}

   \caption[Comparing our method against traditionally used CT quantification measures (LAA-950).]{Comparing our method against traditionally used CT quantification measures (LAA-950).  
   We stratify the population based on centrilobular and paraseptal emphysema severity score. Ours (direct) model is trained to predict spirometry measures and emphysema visual score together in a single loss function. The emphysema visual score is predicted in ordinal multi-class classification analysis. \textbf{A.} Confusion matrix plot for grouping the COPDGene population-based on \textbf{centrilobular emphysema} and \textbf{B.} \textbf{paraseptal emphysema}. Our proposed method performed better than LAA features and created a more significant separation between little and substantial emphysema. Figure is best viewed in color.}
   \label{ch2-fig-3}
\end{figure}

\textbf{Results:} Our model can identify subjects with different degrees of visual emphysema severity. The model correctly identified CLE visual emphysema score in 40.6\% of the subjects in the COPDGene cohort and was within $\pm$ one score 74.8\% of the time.  Figure. \ref{ch2-fig-3} compares the confusion matrices of our method and LAA-950 features. In staging Paraseptal emphysema, the proposed model has an exact and on-off accuracy of 52.8\% and 82.99\% respectively. Results are summarized in Table~\ref{ch-2-table-3}, and the confusion matrix for Paraseptal emphysema prediction is shown in Figure.~\ref{ch2-fig-3}. Application of the Hosmer-Lemeshow~\cite{Lemeshow1982AModels} test did not suggest evidence of poor calibration (p-value 0.079). 

\vspace{-0.2cm}

\begin{table}[!ht]
\centering
\caption{ Results classifying subjects based on their emphysema visual score.}
\label{ch-2-table-3}
\begin{tabular}{|c|c|c|c|c|}
\hline
Method& \multicolumn{2}{c|}{CLE} &  \multicolumn{2}{c|}{Para-septal}\\ 
\cline{2-5}
&  \% Acc.  & \% Acc. \textit{one-off}  & \% Acc.  & \% Acc. \textit{one-off} \\
\hline
Ours (direct) & \textbf{40.61} & \textbf{74.68}  & \textbf{52.82} & 82.99  \\
Ours (in-direct) & 36.30 & 61.33 & 46.87 & 75.97 \\
Spirometry (FEV1)&	33.52&	63.96&	44.64&	72.77 \\
LAA-950&	31.89&	77.74&	33.32&	\bf{87.64}\\
 \hline
\end{tabular}
\end{table}

\subsubsection{Acute Exacerbations} 
\label{ae}

In the COPDGene study, the exacerbations of COPD were self-reported and were quantified by the subject recall on questionnaires. A participant recorded a positive experience of an acute exacerbation if, in the last year, they had experienced at least one episode of increased dyspnea, cough or sputum production, resulting in admission to the hospital or changing of their treatment plan. Approximately 20\% of the subjects reported experiencing at least one exacerbation before enrolling in the study. We used the HRCT acquired at the baseline visit to predict both historical and future exacerbations. The future exacerbation prediction used  exacerbations reported by the longitudinal follow-up participants  at the subsequent 5-year follow-up visit.

\begin{figure}[!ht]
   \includegraphics[width=0.95\linewidth]{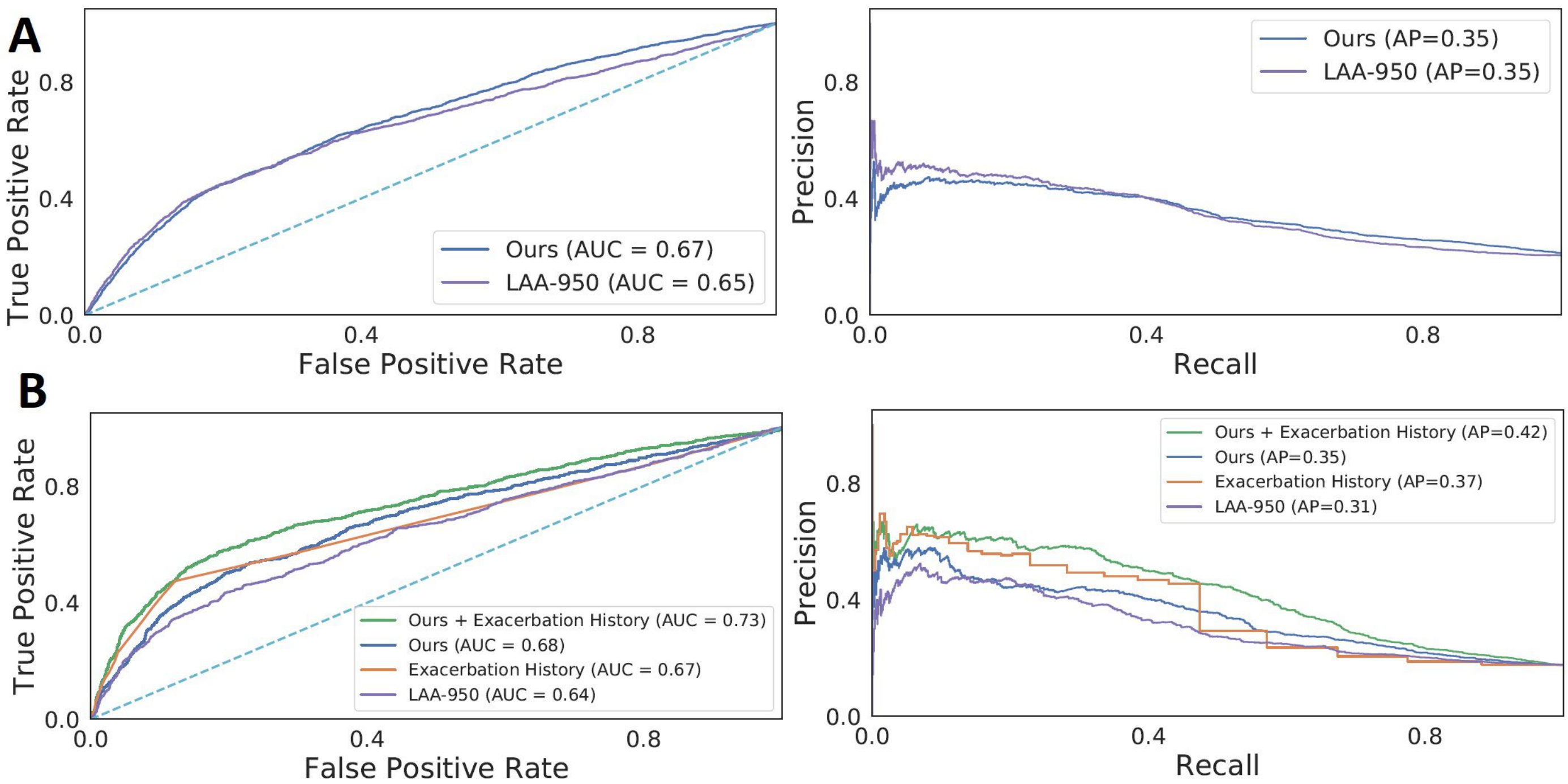}
  
   \caption[Receiver Operating Characteristic (ROC) curve and Precision-Recall (PR) curves.]{Receiver Operating Characteristic (ROC) curve and Precision-Recall (PR) curves. Identifying subjects with \textbf{A. exacerbation history} and \textbf{B. future exacerbation} as given in longitudinal follow up. The ROC curve shows how the true positive vs. false positive relationship changes as we vary the threshold of the positive class. In the top row, the positive class represents those subjects in COPD Cohort who reported experiencing at least one exacerbation before enrolling in the study. In the bottom row, the positive class represents those subjects who reported experiencing at least one exacerbation at the 5-year longitudinal follow up. Higher AUC-ROC number indicates better classification performance. Higher average precision (AP) in the PR curve means the better ability of the model in identifying subjects in a positive class. The plot shows that combining the history of past exacerbation with deep learning features from our model improves the prediction of future exacerbation. Figure is best viewed in color.}
   \label{chp-2-fig-4} 
\end{figure}

\textbf{Results: }Our model achieved an AUC-ROC of 0.70 in identifying the subjects who reported experiencing at least one exacerbation before enrolling in the study. We compared our performance against the intensity-based LAA feature in Figure.~\ref{chp-2-fig-4} (see Table~\ref{chp-2-table-3}). 
\newline

\vspace{-0.5cm}

We also evaluated the performance of our model in identifying the population who reported subsequent exacerbations at the time of the 5-year follow-up. Our model achieved an AUC-ROC of 0.68. Our experiments show that the previous exacerbation history, together with imaging features from our method performs better (AUC-ROC 0.73), in predicting future exacerbation events than using exacerbation history alone (AUC-ROC 0.67). A quantitative comparison between different methods is shown in Table~\ref{chp-2-table-3}. Figure.~\ref{chp-2-fig-4} shows the ROC curve and the PR curve for binary classification. The p-value of the null hypothesis, using the Hosmer-Lemeshow test, is 0.08, suggesting no evidence of poor calibration.

\vspace{-0.2cm}

\begin{table}[!ht]
\begin{center}
\caption{ Results for identifying subjects with exacerbation risk and dyspnea.}
\label{chp-2-table-3}
\begin{tabular}{|c|c|c|c|c|}
\hline
Method & \multicolumn{4}{c|}{\bf Exacerbation History (EH)} \\
\cline{2-5}
& ROC-AUC& PR-AUC& Recall & \% Accuracy\\
\hline
Ours (direct) & 0.68$\pm$0.02 & \textbf{0.38$\pm$0.03} & 0.27$\pm$0.14 & \textbf{76.93} \\
Ours (in-direct) & \textbf{0.73$\pm$0.01} & \textbf{0.43$\pm$0.03} & \textbf{0.59$\pm$0.03} & 74.75 \\
CNN~\cite{Gonzalez2018DiseaseTomography} & 0.643 & - & 0.18 & 	60.40 \\
LAA-950&	0.65$\pm$0.01&	0.35$\pm$0.02&	0.43$\pm$0.02&73.78\\
 \hline
 & \multicolumn{4}{c|}{\bf Future Exacerbation in longitudinal followup} \\
 \cline{2-5}
& ROC-AUC& PR-AUC& Recall & \% Accuracy\\
\hline 
Ours (direct) & 0.65$\pm$0.01 & 0.32$\pm$0.02 & 0.43$\pm$0.01 & 68.30 \\
Ours (in-direct) & 0.70$\pm$0.02 & 0.35$\pm$0.02 & \textbf{0.57$\pm$0.02} & 73.87 \\
LAA-950 &	0.64$\pm$0.01&	0.31$\pm$0.02&	0.43$\pm$0.04&73.80\\
EH&	0.67$\pm$0.02&	0.37$\pm$0.02&	0.47$\pm$0.04&	80.60\\
Ours (in-direct) + EH	&\textbf{0.73$\pm$0.01}&	\textbf{0.42$\pm$0.02}&	0.47$\pm$0.04	&\textbf{80.83}\\
\hline
 &  \multicolumn{4}{c|}{ \bf mMRC Dyspnea  Score}\\ 
\cline{2-5}
&  \multicolumn{2}{c|}{\% Accuracy }  &   \multicolumn{2}{c|}{\% Accuracy \textit{one-off} }  \\
\hline
Ours (direct) & \multicolumn{2}{c|}{\textbf{46.40} }  & \multicolumn{2}{c|}{67.04 }  \\
Ours (in-direct)& \multicolumn{2}{c|}{38.94} & \multicolumn{2}{c|}{59.86 } 	\\
Spirometry (FEV1)& \multicolumn{2}{c|}{42.63} & \multicolumn{2}{c|}{	\textbf{69.07}} 	\\
LAA-950& \multicolumn{2}{c|}{41.52} & \multicolumn{2}{c|}{63.45 } 	\\
\hline
\end{tabular}
\end{center}
\end{table}

\subsubsection{mMRC Dyspnea Scale}
\label{mds}

Subjects completed the modified Medical Research Council (mMRC) dyspnea scale during their baseline visit. The scale ranges from zero (dyspnea only with strenuous exertion) to four (dyspnea with daily activities) 
\newline

\textbf{Results: }Our proposed model was successful in classifying subjects in the COPDGene cohort based on their mMRC dyspnea scale with an accuracy of 43.5\% and was within one score, 64.3\% of the time (Table~\ref{chp-2-table-3}).  Dyspnea scale is used to guide therapeutic strategies in patients with COPD~\cite{Qureshi2014ChronicImplications, Pe15}.

\subsubsection{Mortality}
We used the vital status and censoring time information provided in the mortality dataset to perform survival analysis. In the COPDGene cohort, the mean time between phase 1 data and the censoring time is approximately five years. Nearly 13\% of subjects were reported deceased either in the SSDI search or in the COPDGene LFU. We used Cox proportional hazards (PH) model~\cite{Lin1994CoxApproach} to predict survival utilizing the probability of death predicted by patient-representation against age, gender, smoking status and center of enrollment as fixed covariates. Next, we used Kaplan-Meier plots stratified by quantile of predicted probabilities of death to visualize the results. Kaplan-Meier plot shows the probability of survival plotted against time.

\begin{table}[!ht]
\centering
\begin{threeparttable}
\caption{ Results of Cox Proportional-Hazard (PH) model for survival analysis. The probability of death, learned from binary classification of mortality, is used as covariate in Cox regression. }
\label{ch-2-table-5}
\small
\begin{tabular}{|p{3.5cm}|c|c|c|c|c|}
\hline
Method& Hazard  & Quantile   & Concor- & Global  & PH- \\ 
& Ratio\tnote{b} & p-value\tnote{c} &dance\tnote{d} & statistical  & Assumption\\
& & & & significance\tnote{e} &(Global\\
& & & & Max p-value (LR,& p-value)\tnote{f} \\
& & & &  Wald, log Rank) &  \\
\hline
Ours (direct)&	1.04 &	$<$2e-16	&0.590&	p$=<$2e-16&	0.514\\
&[CI: 0.09, 1.87]&&&&\\
Ours (in-direct)  &	1.54&	$<$2e-16	&0.615&	p$=<$2e-16&	0.598\\
&[CI: 1.09, 2.17]&&&&\\
CNN~\cite{Gonzalez2018DiseaseTomography} &
2.69&	0.017&	0.72&	-&	-\\
&[CI: 1.19, 6.05]&&&&\\
Spirometry (FEV1)&	1.20&	6.91e-07&	0.525&	p$=$4e-06&	-\\
&[CI: 0.94, 1.54]&&&&\\
BODE Index~\cite{Celli2004TheDisease}\tnote{a}& 1.68&	$<$2e-16&	0.568&	p$=<$2e-16&	0.462\\
& [CI: 1.21, 2.31]&&&&\\
LAA-950&	1.13&	6.35e-07&	0.537&	p$=$4e-06&	0.391\\
&  [CI: 0.93, 1.37]&&&&\\
 \hline
\end{tabular}
\begin{tablenotes}
\footnotesize
\item PH: proportional hazards; BODE = Body-mass index, airflow Obstruction, Dyspnea and Exercise index; CI = Confidence interval;
\item All the models have age, gender, smoking pack-years, and center of enrollment as covariates.
\item [a] BODE index is the clinical index used to predict the mortality rate from COPD~\cite{Martinez2006PredictorsObstruction}.
\item [b]The Hazard ratio is the exponential coefficient (exp($\beta$)) of the covariate. A covariate is positively associated with the event probability when the hazard ratio is above one and, thus, is negatively associated with the length of survival. We also report 95\% confidence intervals for the hazard ratio.
\item [c] A significant p-value with $>$ 1 hazard ratio indicates a strong relationship between the covariate and increased risk of death.
\item [d] The concordance shows the fraction of pairs, where the observations with higher survival time have a higher probability of survival predicted by the model. It is analog to the area under the ROC curve in classification analysis. 
\item [e] The Global statistical significance of the model is tested using three alternative tests namely the likelihood-ratio (LR) test, the Wald test, and the score log-rank statistics. p $<$0.001 indicates that the model fits significantly better than the null hypothesis. The null hypothesis states that all the betas ($\beta$) are 0. 
\item [f] We used scaled Schoenfeld residuals to check the proportional hazards assumption. A non-significant p-value shows no evidence of violation of PH assumption by survival model. 
\end{tablenotes}
\end{threeparttable}
\end{table}

We tested the PH assumption by performing a correlation between each of the covariates and their corresponding set of scaled Schoenfeld residuals with time~\cite{Schoenfeld1980ChiSquaredModel}. A non-significant p-value for this test supported the PH assumption. In another test, we checked the global statistical significance of the Cox model. The test validated the null hypothesis that the variables have no association with survival. If the test failed to reject the null hypothesis, this would suggest that removing the variables from the model will not substantially harm the fit of that model. This global test is performed using three alternative tests: the likelihood-ratio test, the Wald test, and the score log-rank statistic. The survival analysis was performed using the lifelines library in Python\cite{DavidsonPilon2019Lifelines} and the survival package in R\cite{Therneau2000ModelingModel}. We also compared the performance of our survival model against the uni-variate Cox regression model using intensity features (LAA-950) and the BODE index. The multidimensional BODE index has been shown to predict survival in cohort studies of COPD~\cite{Martinez2006PredictorsObstruction}. For the Cox PH model, we reported the results in terms of concordance, similar to  AUC-ROC in binary classification.

\textbf{Results: }Our proposed method achieved a concordance of 0.61 in Cox regression\cite{Lin1994CoxApproach} analysis compared to 0.56 for the BODE index and 0.53 for LAA-950 features (Table~\ref{ch-2-table-5}). In testing the proportional hazard (PH) assumption of our model using scaled Schoenfeld residues, we achieved a p-value $>$ 0.3 for all the covariates and a global p-value of 0.59 for the model. A significant p-value for this test provided no evidence for the violation of the PH assumption made by the Cox model. 
\newline

\begin{figure}[!ht]
\begin{center}
   \includegraphics[width=0.95\linewidth ]{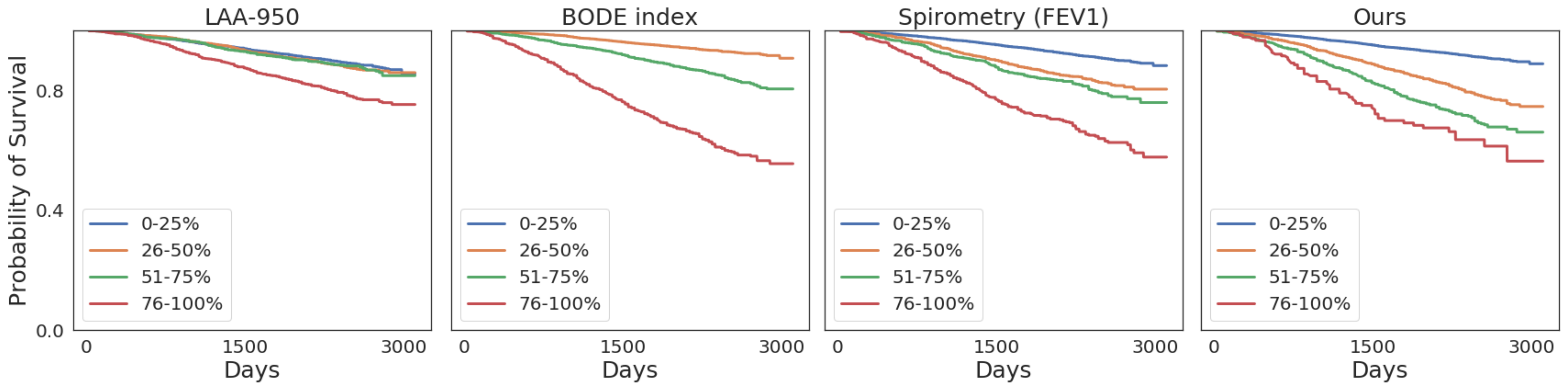}
 
   \caption[Kaplan Meier plot for visualizing the results of survival analysis.]{
   Kaplan Meier plot for visualizing the results of survival analysis.
   The plot is obtained by performing Cox regression analysis stratified on the quantile of predicted probability of mortality in binary classification. A good Kaplan Meier plot has large separations between the groups. BODE index is the Body-mass index, airflow Obstruction, Dyspnea and Exercise index which is highly correlated with mortality~\cite{Martinez2006PredictorsObstruction}. Our model performed better than the conventional emphysema quantification, the BODE index, and spirometry measures for mortality assessment. }
      %note label inside caption
    \label{fig_KM} 
    \end{center}
\end{figure}

\vspace{-0.4cm}

Next, we tested the global statistical significance of the Cox model using three alternative tests: the likelihood-ratio test, the Wald test, and the score log-rank statistic. We achieved a p-value of $<$ 0.001 in all three tests. Hence, we can reject the null hypothesis that all the coefficients are 0, with high confidence. Figure.~\ref{fig_KM} shows the Kaplan Meier (KM) plots to visualize the subjects grouped by quantile of predicted probability of 5-year survival. The KM plot for our method has a large separation between different quantile groups. Thus, our model can divide the population into distinct groups based on their survival risk.

\subsection{Generative regularizer} 
\begin{figure}[!ht]
   \includegraphics[width=1.0\linewidth ]{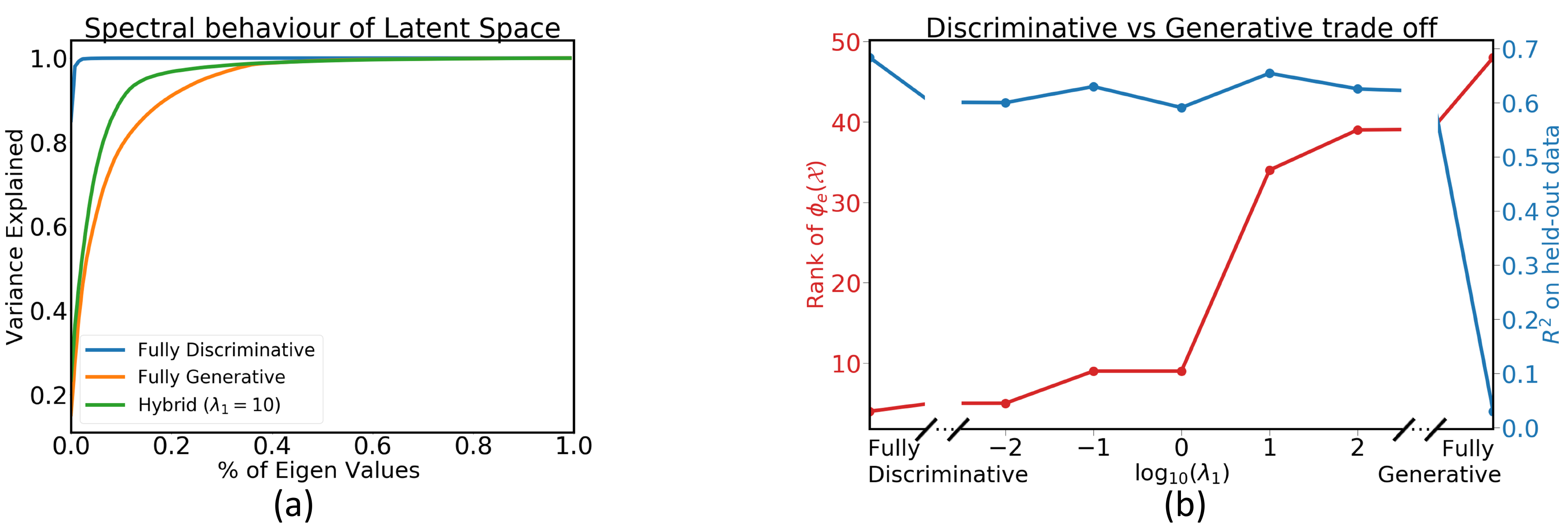}

   \caption[Evaluating generative regularizer.]{Evaluating generative regularizer
    (a) Spectral properties of patch-level features for different values of $\lambda_1$. (b) The trade-off between rank of latent space (red, $y$-axis on left) and predictive power (blue, $y$-axis on right) for different values of $\lambda_1$. Left represents fully discriminative and right represents fully generative models.}
     %note label inside caption
    \label{fig_GR} 
\end{figure}

Hyper-parameter $\lambda_1$ in our overall loss function in Eq.~\ref{eq:general} balances between the discriminative and the generative setting. $\lambda_1 = 0$ represents a fully discriminative setting, in which  the decoder of the auto-encoder is not trainer. Hence, there is no reconstruction of the input patch back from the low dimensional embedding learned by the encoder. In the absence of the decoder function, the feature extractor (encoder) is forced to retain only information that is sufficient for the underlying discriminative task. The Figure.~\ref{fig_GR}(a) reports the spectral behaviors of the latent features (\ie $\phi_e(\mathcal{X}_i)$) for varying $\lambda_1$. For fully discriminative setting with $\lambda_1 = 0$, we observe a highly redundant latent space, with  almost similar patch-level features $\phi_e(x_{ij})$, with attention  weights $\alpha_{ij}$ converging to $\frac{1}{|\mathcal{X}_i|}$.
\newline

\vspace{-0.4cm}

As $\lambda_1 \rightarrow \infty$, the network mostly focuses on the generative task of reconstructing the patch back from patch-level features. In such fully generative setting, the encoder features are not optimized for the downstream prediction task. 
Hence, though the patch-level features are much diverse, as seen in Figure.~\ref{fig_GR}(a), there is a significant drop in $R^2$ for predicting FEV1, as shown in the Figure.~\ref{fig_GR}(b).

We demonstrate the effect of regularization through Figure.~\ref{fig_GR}(b). It shows the trade-off between effective rank  of the latent feature and $R^2$ for predicting FEV1. Although, the $R^2$ drops a little, the rank, which represents the diversity of the latent features, improves drastically. The gap between accuracy's of $\lambda_1=0$ and $\lambda_1>0$ is the price we pay for the interpretability.

\subsection{Visualization} 
\begin{figure}[!ht]
\centering
   \includegraphics[width=0.7\linewidth ]{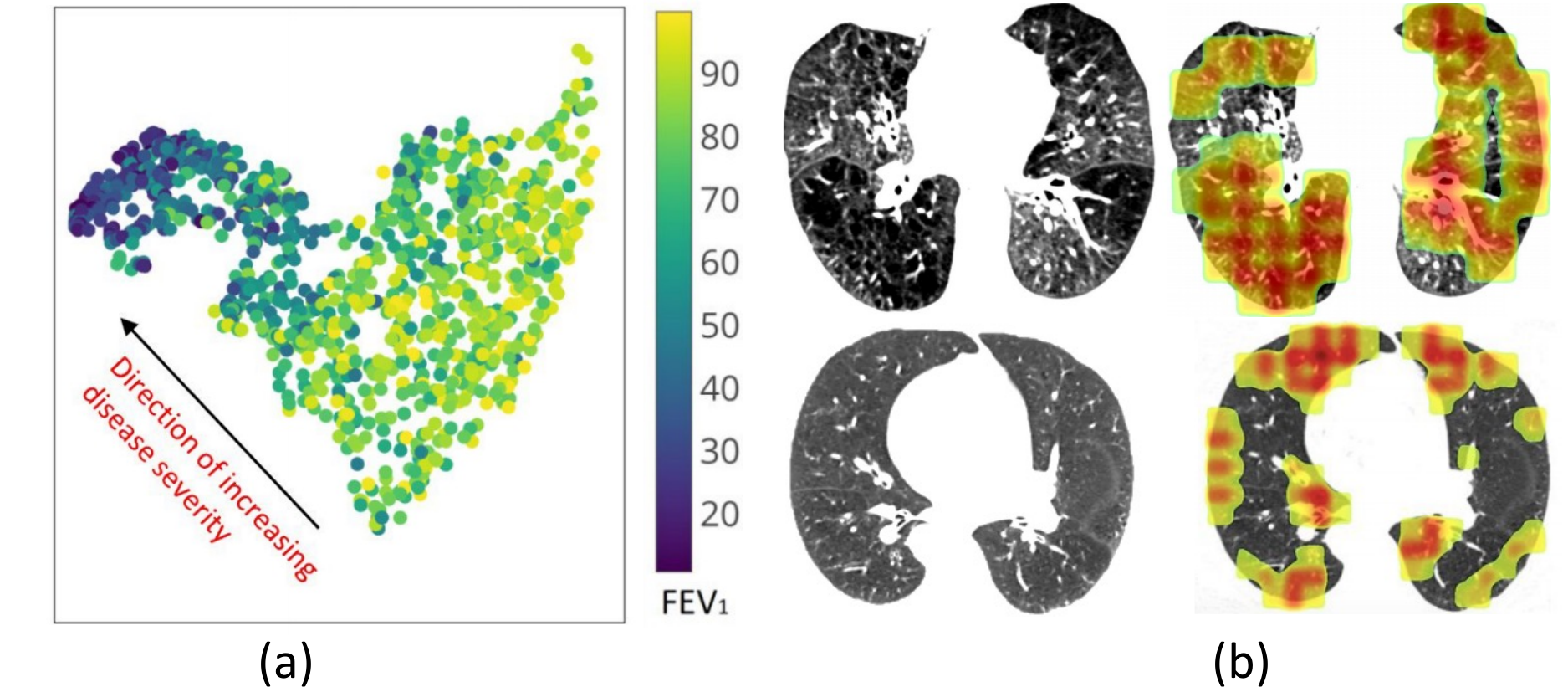}
     
   \caption{
   An axial view of the attention map on a subject. Red color indicate higher relevance to the disease severity.
    } 
    \label{fig_viz} 
\end{figure}

 Figure.~\ref{fig_viz}(b) visualizes the attention weights on a subject. The dark area on the left lung (tope-row), which is severely damaged, received high attention, while same regions in a different lung (bottom-row) have minimum attention.

 \vspace{0.9cm}

\vspace{-1.0cm}

\section{Discussion and Conclusion}
Our proposed Deep Learning-based method demonstrates the ability to predict multiple aspects of COPD disease pattern, severity, and future events. It does so by extracting the most relevant information from volumetric HRCT images of the subject. Unlike previous Deep Learning methods that process a collection of 2D slices, our method works on the entire 3D inspiratory scan of the subject. Deep Learning enables us to go beyond standard radiographic features such as LAA and construct data-driven radiological features that are optimal for a specific task. Our results show that large cohorts such as COPDGene enable DL methods to learn meaningful patterns and converge to reliable predictions. Another advantage of our method lies in its generalizability and flexibility to incorporate different aspects of COPD. Using the same DL model and architecture, we were not only able to predict spirometric obstruction but were also successful in predicting all-cause mortality and future exacerbations, quantifying emphysema burden and disease pattern, and evaluating symptom scores. 

In the direct approach, our model achieved high predictive strength by explicitly training to predict a target outcome. Our cross-validation experiments showed that the model was well-calibrated and achieved consistent performance over all folds. While in the in-direct approach, the model was trained only once, to predict respiratory measurements, this model performed well in predicting COPD outcomes including, acute exacerbations, and mortality.

Our predictions of spirometry measurements outperformed previously reported methods, including the previous DL method. Our method has a potential translational impact if it is utilized as a clinical screening tool, e.g., when obtained during routine cancer screening, to identify subjects with a high likelihood of COPD for further assessment. Our visualization of the COPDGene population colored by the FEV1 value shows subjects with high FEV1 clustered together and a progression of disease severity from low to high (\textbf{Figure.~\ref{ch_2_fig_2}(d)}). This population-level analysis may be helpful in prospectively identifying unique clinical subgroups or in quantifying disease severity across research cohorts.

This is the first study to use DL-based method to predicted various clinical outcomes associated with COPD like spirometric obstruction, emphysema severity, current and future exacerbation risk and mortality, using CT imaging alone. The results of our study  conclude that DL-based method can provide a holistic view of disease severity and progression from a single set of CT images. Further work toward developing interpretable DL models is essential for the development of standardized CT-based assessment of COPD.

High-resolution CT evaluation by a deep learning algorithm might provide low-cost, reproducible, near-instantaneous classification of fibrotic lung disease with human-level accuracy. These methods could be of benefit to centres at which thoracic imaging expertise is scarce, as well as for stratification of patients in clinical trials.

\chapter{Progressive Counterfactual Explainer}
\label{ch2}
\section{Introduction}
With the explosive adoption of deep learning for real-world applications,  explanation and model interpretability have received substantial attention from the research community~\cite{doshi2017towards,guidotti2019survey,kim2015interactive,molnar2018interpretable}. Primarily, DL is used for Computer-Aided Diagnosis~\cite{Hosny2018ArtificialRadiology} and other tasks in the medical imaging domain~\cite{Rajpurkar2018DeepRadiologists,Rodriguez-Ruiz2019Stand-AloneRadiologists}. However, for real-world deployment~\cite{Wang2020ShouldMedicine}, the decision-making process of these models should be explainable to humans to obtain their trust in the model ~\cite{Gastounioti2020IsAI,Jiang2018ToClassifier}. Explainability is essential for auditing the model~\cite{Winkler2019AssociationRecognition}, identifying failure modes~\cite{Eaton-Rosen2018TowardsPredictions,Oakden-Rayner2020HiddenImaging} or hidden biases in the data or the model~\cite{Larrazabal2020GenderDiagnosis}, and obtaining new insights from large-scale studies~\cite{Rubin2018LargeNetworks}. For example, consider evaluating a computer-aided diagnosis of Alzheimer's disease from medical
images. The physician should be able to assess whether or not the model pays attention to
age-related or disease-related variations in an image  to trust the system.

With the advancement of DL methods for medical imaging analysis, deep neural networks (DNNs)  have achieved near-radiologist performance in multiple image classification tasks~\cite{seah2021effect,Rajpurkar2017CheXNet:Learning}. However, DNNs are criticized for their ``black-box" nature, \ie they fail to provide a simple explanation as to why a given input image produces a corresponding output~\cite{Tonekaboni2019WhatUse}. To address this concern, multiple model explanation techniques have been proposed that aim to explain the decision-making process of DNNs~\cite{Selvaraju2017Grad-cam:Localization,cohen2021gifsplanation}. The most common form of explanation in medical imaging is a class-specific heatmap overlaid on the input image. It highlights the most relevant regions (\emph{where}) for the classification decisions~\cite{Bach2015OnPropagation,Lundberg2017APredictions,Selvaraju2017Grad-cam:Localization,Shrikumar2017LearningDifferences,Simonyan2013DeepMaps,Springenberg2015StrivingNet,Sundararajan2017AxiomaticNetworks}. The location information alone is insufficient for applications in medical imaging. Different diagnoses may affect the same anatomical regions, resulting in similar explanations for multiple diagnosis, resulting in inconclusive explanations. A thorough explanation should explain \emph{what} imaging features are present in those important locations, and \emph{how} changing such features modifies the classification decision.

Although  not  always  clear,  there  are  subtle  differences  between interpretability and \emph{explanation}~\cite{turner2016model}.  While the former mainly focuses on building or
approximating models that are locally or globally interpretable~\cite{ribeiro2016should}, the
latter aims at explaining a predictor a-posteriori.  The explanation approach does not compromise
the prediction performance. However, a rigorous definition for what is a good explanation is
elusive. Some researchers focused on providing feature importance (\eg in the form of a
heatmap~\cite{Selvaraju2017Grad-cam:Localization}) that influence the outcome of the predictor.  In
some applications (\eg diagnosis with medical images) the causal changes are spread out across a
large number of features (\ie large portions of the image are impacted by a disease). Therefore, a
heatmap may not be informative or useful, as almost all image features are highlighted.
Furthermore, those methods do not explain \emph{why} a predictor returns an outcome.  Others have
introduced local occlusion or perturbations to the
input~\cite{Zhou2014ObjectCnns,Fong2017InterpretablePerturbation} by assessing which manipulations
have the largest impact on the predictors. There is also recent interest in generating
counterfactual inputs that would change the black box classification decision with respect to the
query inputs~\cite{Goyal2019CounterfactualExplanations,Liu2019GenerativeLearning}.  Local
perturbations of a query are not guaranteed to generate realistic or plausible inputs, which
diminishes the usefulness of the explanation, especially for end users (\eg physicians).  We argue
that the explanation should depend not only on the predictor function but also on the data.
Therefore, it is reasonable to train a model that learns from data as well as the black-box
classifier
(\eg~\cite{Chang2019ExplainingGeneration,Dabkowski2017RealClassifiers,Fong2017InterpretablePerturbation}).

To address these gaps, we propose a novel explanation method to provide a counterfactual explanation. A counterfactual explanation is a perturbation of the input image such that the classification decision is flipped~\cite{ Joshi2019TowardsSystems,Liu2019GenerativeLearning,Mahajan2019PreservingClassifiers,SamangoueiPouyaandSaeedi2018ExplainGAN:Transformations}. By comparing, the input image and its corresponding counterfactual image, the end-users can visualize the difference in important image features that leads to a change in classification decision. Our proposed method falls into the local explanation paradigm. Our approach is model agnostic and
only requires access to the predictor values and its gradient with respect to the input. Given a
query input to a black-box, we aim at explaining the outcome by providing \emph{plausible} and
\emph{progressive} variations to the query that can result in a change to the output. The
plausibility
property ensures that perturbation is natural-looking. A user can employ our method as a ``tuning knob'' to progressively transform inputs, traverse the decision boundary from one side to the other, and gain
understanding about how the predictor makes a decision.

We adopted a conditional Generative Adversarial Network (cGAN) as our explanation framework.  However, using cGAN is challenging, as GANs with an encoder may ignore small or uncommon details during image generation~\cite{Bau2019SeeingGenerate}. This is particularly important in our application, as the missing information includes foreign objects such as a pacemaker that influence human users' perception. To address this issue, we stipulate when the input image has reconstructed the shape of the anatomy and that foreign objects are preserved. We achieve this by incorporating semantic segmentation and object detection into our loss function. 

We introduce three principles for
an explanation function that can be used beyond our application of interest. We evaluate our method
on a set of benchmarks as well as real medical imaging data. Our experiments show that the counterfactually generated samples are realistic-looking and in the real medical application, satisfy the
external evaluation. We also show that the method can be used to detect bias in training of the
predictor.  Our contributions are summarized as follows:
\begin{enumerate}
    \item We developed a progressive counterfactual explainer (PCE) that generates visual explanations for a black-box classifier. PCE explains the decision for a query image by gradually changing the image such that the classification decision is flipped. 
   \item Our method accounts for subtleties of medical imaging by  preserving the anatomical shape and foreign  objects such as support devices across generated images. The specialized reconstruction loss is proposed to incorporate  context from semantic segmentation and foreign object detection networks.
    \item We evaluated our method extensively on both natural and medical datasets.
      \item We proposed quantitative metrics based on clinical definition of two diseases (cardiomegaly and PE). We are one of the first methods to use such metrics for quantifying DNN model explanation. Specifically, we used these metrics to quantify statistical differences between counterfactual and query images. 
      \item We are one of the first methods to conduct a thorough human-grounded study to evaluate different counterfactual explanations for medical imaging task. Specifically, we collected and compared feedback from diagnostic radiology residents, on different aspects of explanations: (1) understandability,  (2) classifier's decision justification, (3) visual quality, (d) identity preservation, and (5) overall helpfulness of an explanation to the users.
      
      \item On the face images dataset, we show that our method successfully detects confounding bias in the classifier.
\end{enumerate}

\section{Method}
Consider a \emph{black box} classifier that maps an input space $\mathcal{X}$ (\eg images) to an output space $\mathcal{Y}$ (\eg labels). In this paper, we consider binary classification problems where $\mathcal{Y}=\{0, 1\}$. However, the proposed method is general and can be used for multi-class or multi-label settings. We use $f(\rvx)= \mathbb{P} (y|\rvx)\in[0,1]$ to denote the posterior probability of the classifier. We assume that $f$ is a differentiable function and we have access to its value as well as its gradient with respect to the input $\nabla_{\rvx} f(\rvx)$.

We view the (visual) explanation of the black-box as a generative process that produces a plausible and realistic perturbation of the query image such that the classification decision is changed to a desired value $\mathbf{c}$. By gradually changing the desired output $\mathbf{c}$ in range $[0,1]$, we can traverse the prediction space while visualizing the gradual exaggeration of the target class in generated images. 
We conceptualize this traversal from one side of the decision boundary to the other as walking across a data manifold, $\mathcal{M}_x$. Directly manipulating the high dimensional image space is very challenging. Hence, we assume that there is a low-dimensional embedding space ($\mathcal{M}_z$) that encodes the walk. An encoder, $E: \mathcal{M}_x \rightarrow \mathcal{M}_z$, maps an input, $\rvx$, from the data manifold, $\mathcal{M}_x$, to the embedding space. The desired output  $\mathbf{c}$  represents the step size of the walk. We gradually increase or decrease $\mathbf{c}$ to generate a new image to represent each step of this walk.
A generator, $G: \mathcal{M}_z \rightarrow \mathcal{M}_y$, takes both the embedding coordinate and the desired output  $\mathbf{c}$ (current step-size) and maps it back to the data manifold (see Figure. {\ref{Fig_Manifold}}).

The PCE is denoted as $\mathcal{I}_f(\cdot,\cdot)$. Formally, $\mathcal{I}_f(\rvx,\mathbf{c}): (\mathcal{X}, \mathbb{R}) \rightarrow \mathcal{X}$ is a function that takes two arguments: a query image $\rvx$ and the desired posterior probability $\mathbf{c}$ for some target class $y$. This function generates a perturbed image $\mathbf{x}_c$ such that $f(\rvx_c)[y] \approx \rvc$. This formulation allows us to view $\rvc$ as a ``knob'' that gradually perturb the input image to achieve visually perceptible differences in $\rvx$ while crossing the decision boundary given by function $f$. PCE function $\mathcal{I}_f$ should satisfy the following properties:

 \vspace{0.9cm}

\begin{figure}[!ht]
\centering
\includegraphics[width=0.3\linewidth]{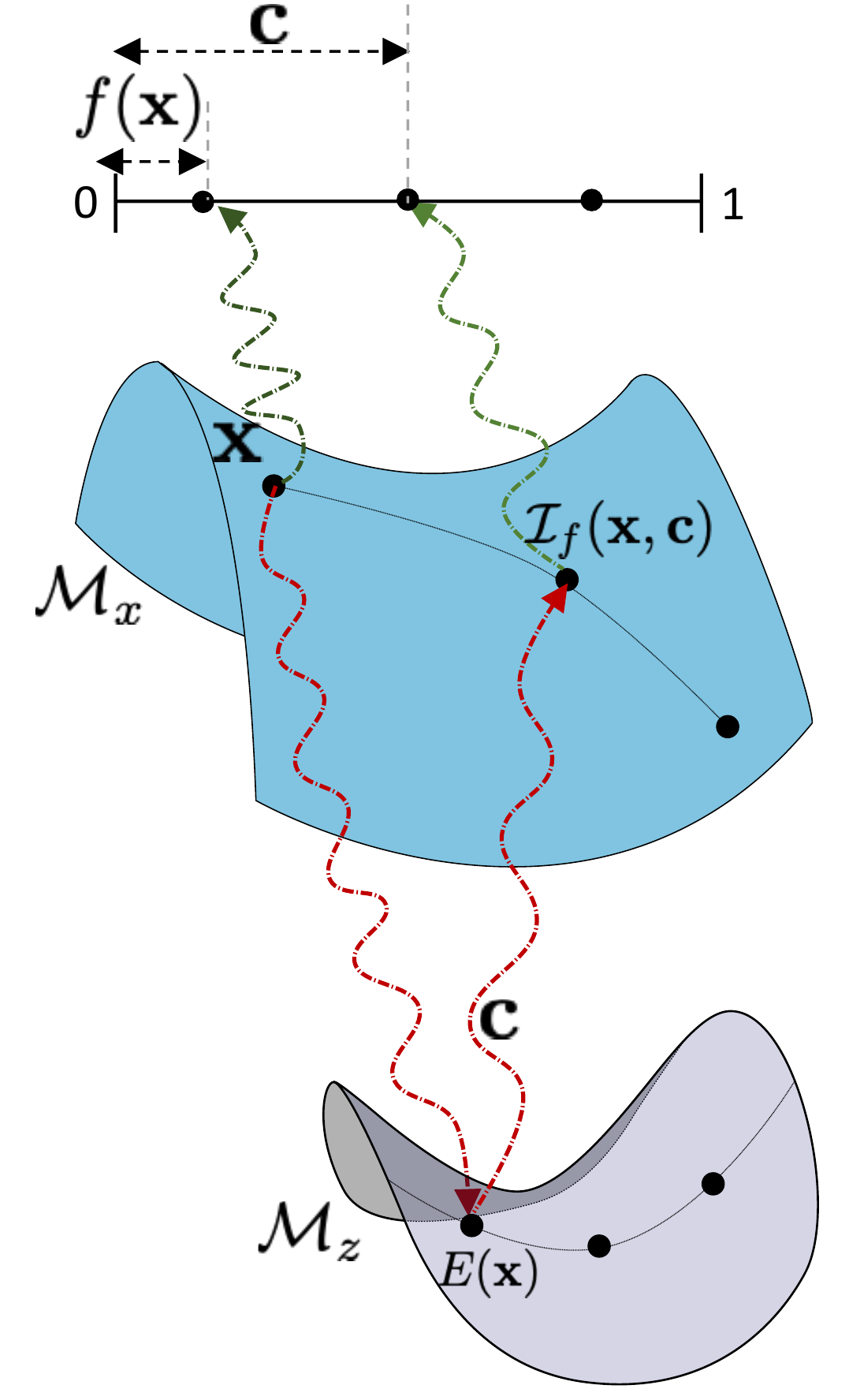}
 
\caption[The schematic of the method.]{The schematic of the method. $f$ is the black-box function producing the posterior probability $f(\rvx)$. $\rvc$ is the desired probability. $\mathcal{I}_f (\rvx,\rvc)$ is an explainer function for $f$, which creates a perturbation of $\rvx$ that produce a classifier's prediction of $\rvc$. The $E(\cdot)$ is an encoder that maps the data manifold $\mathcal{M}_x$ to the embedding manifold $\mathcal{M}_z$. Explanation function is generator that conditionally maps embedding back to the data manifold.}
\label{Fig_Manifold}
\end{figure}

\vspace{-0.2cm}

A.) \textbf{Data consistency}: The perturbed image, $\rvx_{\rvc}$ should resemble data instance from input space $\mathcal{X}$ \ie if input space comprises chest x-rays, $\rvx_{\rvc}$ should look like a chest x-ray with minimum artifacts or blurring.

B.) \textbf{Classification model consistency}: The perturbed image, $\rvx_{\rvc}$ should produce the desired output from the classifier $f$, \ie $f(\mathcal{I}_f(\rvx, \rvc)) \approx \rvc$.

C.) \textbf{Context-aware self-consistency}: To be self-consistent, the PCE should satisfy three criteria (1) Reconstructing the input image by setting $\rvc = f(\rvx)$ should return the input image, \ie $\mathcal{I}_f(\rvx, f(\rvx)) = \rvx$. (2) Applying a reverse perturbation on the explanation image $\rvx_{\rvc}$ should recover $\rvx$, \ie $\mathcal{I}_f(\rvx_{\rvc}, f(\rvx)) = \rvx$. (3) Achieving the aforementioned reconstructions while preserving anatomical shape and  foreign objects (\eg pacemaker) in the input image.
\newline

\begin{figure}[!ht]
\includegraphics[width=0.9\linewidth]{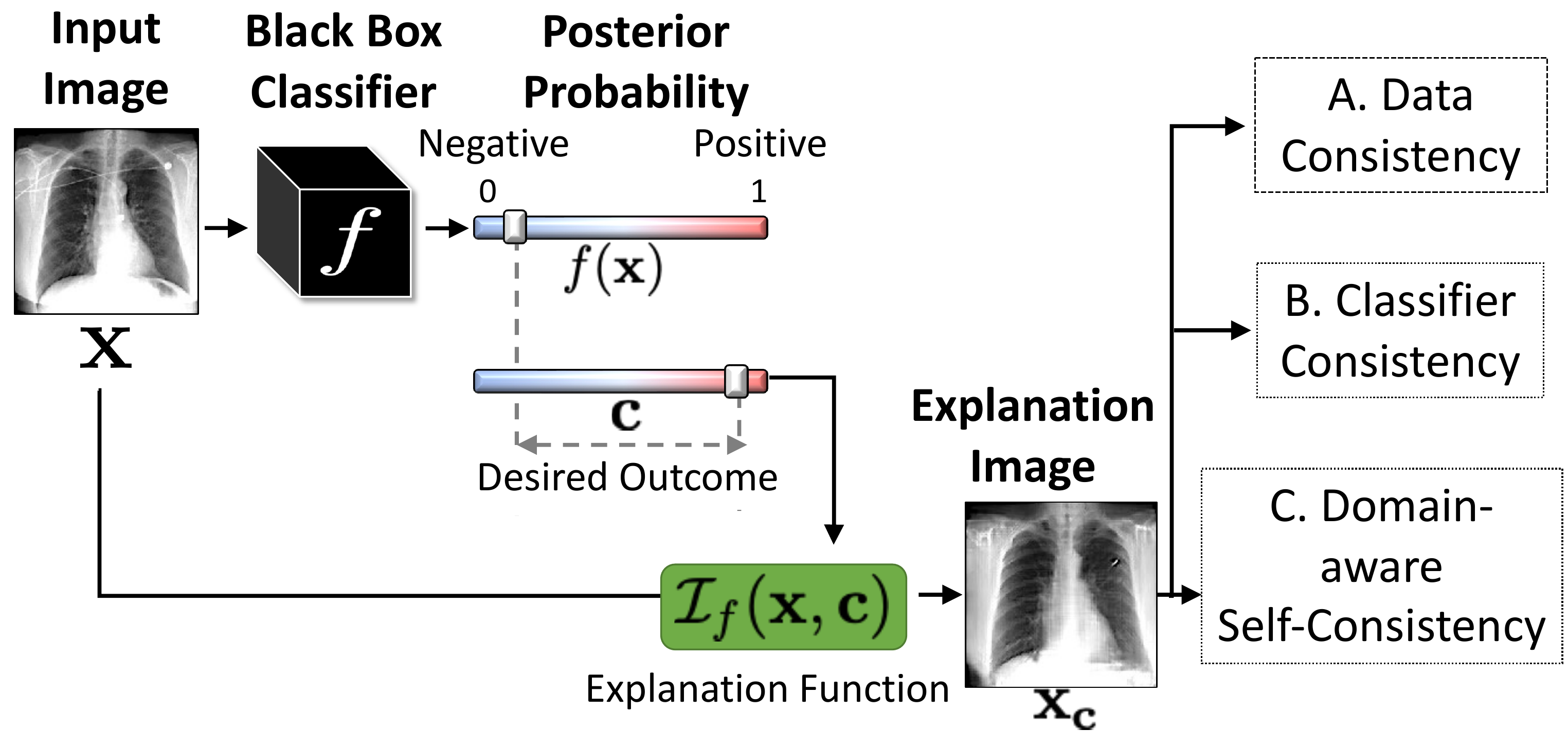}
 
\caption[PCE function $\mathcal{I}_f(\rvx, \rvc)$ for classifier $f$.]{PCE function $\mathcal{I}_f(\rvx, \rvc)$ for classifier $f$.
Given an input image $\rvx$, we  generates a perturbation of the input, $\rvx_{\rvc}$ as explanation, such that the posterior probability, $f$, changes from its original value, $f(\rvx)$, to a desired value $\rvc$ while satisfying the three consistency constraints.}
\label{Fig_Model}
\end{figure}

\vspace{-0.4cm}

We designed PCE as a novel deep learning (DL) framework, which is trained end to end to satisfy the three properties. It minimizing the following objective function:
\newline
\begin{equation}
    \min_{E, G} \max_{D} \lambda_{cGAN}  \mathcal{L}_{\text{cGAN}}(D,G) + \lambda_{f} \mathcal{L}_{f}(D,G) + \lambda_{rec}\mathcal{L}_{\text{rec}}(E,G)
\end{equation}

\vspace{0.5cm}

where $\mathcal{L}_{\text{cGAN}}$ is a conditional GAN-based loss function that enforces data-consistency, $\mathcal{L}_{f}$ enforces classification model consistency through a KullbackLeibler (KL) divergence loss and $\mathcal{L}_{\text{rec}}$ is a reconstruction loss that enforces self-consistency. The loss function is defined over three networks, an image encoder $E(\cdot)$, a conditional GAN generator $G(\cdot)$ and a discriminator $D(\cdot)$. $\lambda_{cGAN}, \lambda_{f}$ and $\lambda_{rec}$ controls the balance between the terms. In the following sections, we will discuss each property and the associated loss term in detail.

\subsection{Data consistency}
We formulated PCE, $\mathcal{I}_f(\rvx, \rvc)$,  as an image encoder $E(\cdot)$ followed by a conditional GAN (cGAN)~\cite{Miyato2018CGANsDiscriminator}, with $\rvc$ as the condition. The encoder enables transformation of a given image, while the GAN framework allows to generate realistic looking transformations as explanation image. The Generative Adversarial Network (GANs)~\cite{Goodfellow2014GenerativeNets} are implicitly models, that learn the underlying data distribution $p_{\text{data}}(\rvx)$ by setting up a min-max game between generative ($G)$ and discriminative ($D$) networks. The $G(\cdot)$ network learns to transform samples drawn from a canonical distribution such that $D(\cdot)$ network fails to distinguish the generated data from the real data. GANs optimizes the following objective function:
\newline
\begin{equation*}
    \mathcal{L}_{\text{GAN}}(D,G)  =  \E_{\rvx,c \sim P(\rvx)}\big[\log \big(D(\rvx)\big)\big] + 
    \E_{ \rvz \sim P_\rvz }\big[\log \big(1 - D(G(\rvz))\big)\big],
\end{equation*}
\newline
where $\rvz$ and $P_\rvz$ are the noise distribution and the corresponding canonical distribution.
There has been significant progress toward improving GANs stability as well as sample quality~\cite{Brock2019LargeSynthesis,Karras2019ANetworks}. The advantage of GANs is that they produce realistic-looking samples without an explicit likelihood assumption about the underlying probability distribution. This property is appealing for our application.

Furthermore, we use a Conditional GAN (cGAN) that allows the incorporation of a context as a condition to the GAN~\cite{Mirza2014ConditionalNets,Miyato2018CGANsDiscriminator}. We use the desired classification outcome as our condition $\rvc \in [0,1]$. The cGAN optimizes the following loss function:
\newline
\begin{equation}
    \mathcal{L}_{\text{cGAN}}(D,G)  =  \E_{\rvx,c \sim P(\rvx,c)}\big[\log \big(D(\rvx, \rvc)\big)\big] + 
    \E_{ \rvz \sim P_\rvz, c \sim P_c }\big[\log \big(1 - D(G(\rvz, \rvc), \rvc)\big)\big],
    \label{eq:cgan}
\end{equation}
\newline
where $\rvc$ denotes a condition. In our formulation, $\rvz$ is the latent representation of the input image $\rvx$, learned by the encoder $E(\cdot)$. Finally, the PCE is defined as,
\newline
\begin{equation}
 \mathcal{I}_f(\rvx, \rvc) =  G(E(\rvx), \rvc ).
\end{equation}
\newline
Our architecture is based on Projection GAN~\cite{Miyato2018CGANsDiscriminator}, a modification of cGAN. An advantage of the Projection GAN is that it scales well with the number of classes allowing to use very small bin size while discretizing $\rvc$.
The Projection GAN imposes the following structure on the discriminator loss function:
\newline
\begin{equation}
    \mathcal{L}_{\text{cGAN}}(D, \hat{G})(\rvx, \rvc) = \log \frac{p_{\text{data}}(\rvc| \rvx)}{ q ( \rvc| \rvx ) } + \log \frac{p_{\text{data}}(\rvx) }{q(\rvx)} := r(\rvc|\rvx) + \psi( \bm{\phi}(\hat{G}(\rvz))  ),
\end{equation}
\newline

\begin{figure}[!ht]
\centering
\includegraphics[width=0.9\linewidth]{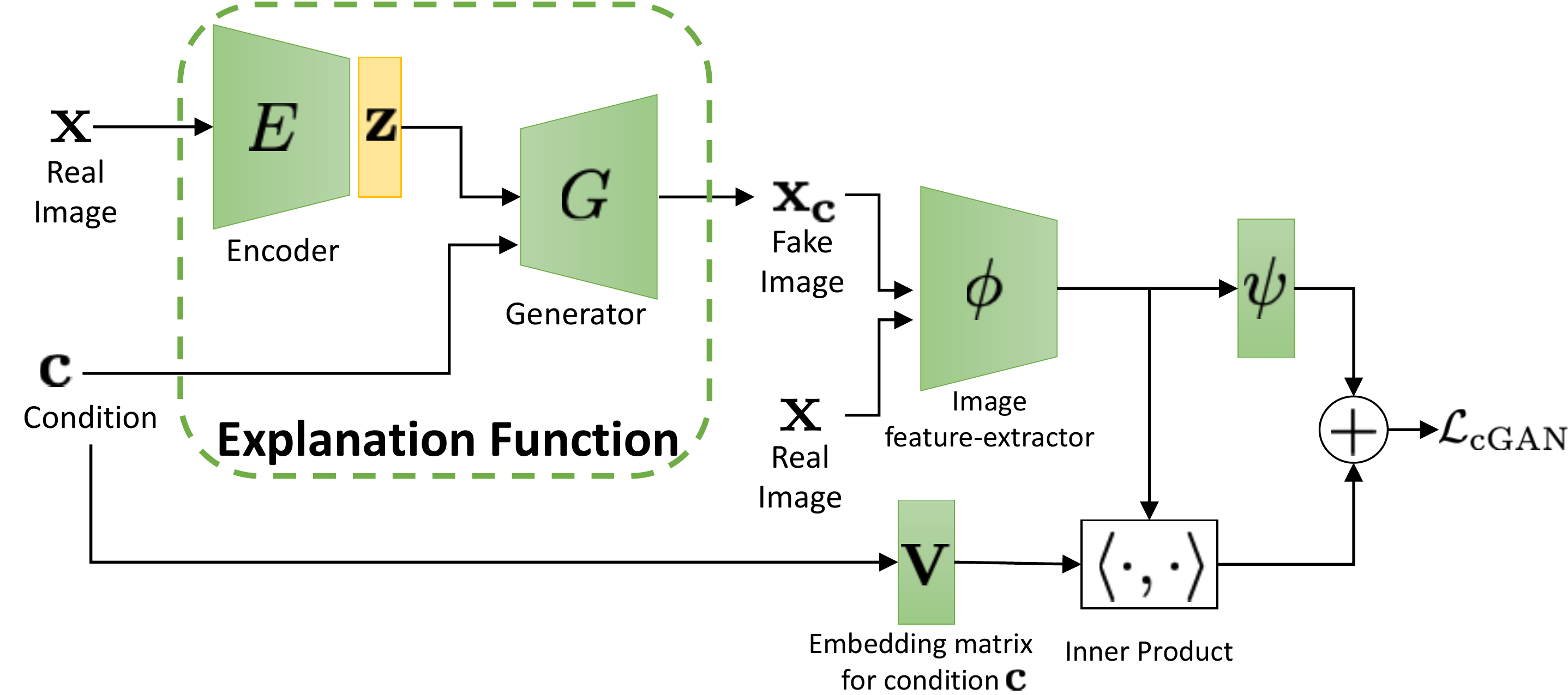}
 
\caption{Progressive counterfactual explainer (PCE) as a conditional-GAN with an encoder.} 
\label{Fig_GAN}
\end{figure}

\vspace{-0.4cm}

For the discriminator in cGAN, we adapted the loss function from Projection GAN~\cite{Miyato2018CGANsDiscriminator} based on our application, as shown in Figure.~\ref{Fig_GAN}. We can view $\rvc$ as a one-hot vector over $N$ classes. The loss function of projection cGAN has two terms. The first term is the distribution ratio between marginals \ie the real data distribution $p_{\text{data}}(\rvx)$ and the learned distribution of the generated data $q(\rvx)$.  The second term is the distribution ratio between conditionals. It evaluates the correspondence between the generated image and the condition. This formulation allows us to skip calculating $q$ as we are only interested in the ratio. The overall loss function is as follows,

\begin{equation}
\begin{split}
    \gL_{\text{cGAN}}(D, \hat{G})(\rvx, \rvc) & = \log \frac{p_{\text{data}}(\rvx) }{q(\rvx)} + \log \frac{p_{\text{data}}(\rvc| \rvx)}{ q ( \rvc| \rvx ) }  \\
    & := r(\rvx) + r(\rvc|\rvx)\\
\end{split}
\label{eq:disc}
\end{equation}

\vspace{0.5cm}

where $\gL_{\text{cGAN}}(D,\hat{G})$ indicates the loss function in Eq.~\ref{eq:cgan} when $\hat{G}$ is fixed. 
Further, $r(\rvx)$ is the discriminator logit that  evaluates the visual quality of the generated image. It is the discriminator's attempt to separate real images from the fakes images created by the generator. The second term evaluates the correspondence between the generated image $\rvx_{\rvc}$ and the condition $\rvc$. 

To represent the condition, the discriminator learns an embedding matrix $\mathbf{V}$ with $N$ rows, where $N$ is the number of conditions. The condition is encoded as an  $N$-dimensional one-hot vector which is multiplied by the embedding-matrix to extract the condition-embedding. When $\rvc = n$, the conditional embedding is given as the $n$-th row of the embedding-matrix ($\rvv_n$). The projection is computed as the dot product of the condition-embedding and the features extracted from the fake image, 
\newline
\begin{equation}
\label{eq_cgan}
     \gL_{\text{cGAN}}(D, \hat{G})(\rvx, \rvc) := r(\rvx) + \rvv_n^T\phi(\rvx),
\end{equation}
 \newline
where, $n$ is the current class for the conditional generation and $\phi$ is the feature extractor.

 In our use-case, the condition $\rvc$ is the desired posterior probability from the classification function $f$. $\rvc$ is a continuous variable with values in range $[0,1]$. Projection-cGAN requires the condition to be a discrete variable, to be mapped to the embedding matrix $\mathbf{V}$. Hence, we discretize the range $[0,1]$ into $N$ bins, where each bin is one condition. One can view change from $f(\rvx)$ to $\rvc$ as changing the bin index from the current value $C(f(\rvx))$ to $C(\rvc)$ where $C(\cdot)$ returns the bin index.

\subsection{Classification model consistency}

Ideally, cGAN should generate a series of smoothly transformed images as we change condition $\rvc$ in range $[0,1]$. These images, when processed by the classifier $f$ should also smoothly change the classification prediction between $[0,1]$. To enforce this, rather than considering bin-index $C(\rvc)$ as a scalar, we consider it as an ordinal-categorical variable, \ie $C(\rvc_1) < C(\rvc_2)$ when $\rvc_1 < \rvc_2$.   Specifically, rather than checking one condition that desired bin-index is equal to some value $n$, $C(\rvc) = n$, we check $n-1$ conditions that desired bin-index is greater than all bin-index which are less than $n$ \ie  $C(\rvc) > i \forall i \in [1, n)$~\cite{Frank2001AClassification}.

We adapted  Eq.~\ref{eq_cgan} to account for a categorical variable as the condition, by modifying the second term to support ordinal multi-class regression. The modified loss function is as follows:
\newline
\begin{equation}
    r(\rvc=n|\rvx) := \sum_{i<n} \rvv_i^{T} \bm{\phi} ( \rvx  ),
\end{equation}
\newline
Along with conditional loss for the discriminator, we need additional regularization for the generator to ensure that the actual classifier's outcome, \ie $f(\rvx_{\rvc})$, is very similar to the condition $\rvc$.  To ensure this compatibility with $f$, we further constrain the generator to minimize the KullbackLeibler (KL) divergence that encourages the classifier’s score for $\rvx_{\rvc}$ to be similar to $\rvc$ (\emph{see} Figure.~\ref{Fig_GAN}(b). Our final condition-aware loss is as follows,
\newline
\begin{equation}\label{eq:dkl}
\mathcal{L}_f(D,G) := r(\rvc|\rvx) + \KL(f(\rvx_{\rvc}) || \rvc),
\end{equation}
\newline
Here, the first term is a function of both $G$ and $D$, the second term influences only the $G$.

\subsection{Context-aware self consistency}
A valid explanation image is a small modification of the input image, and should preserve the inputs' identity \ie patient-specific information such as the shape of the anatomy. While images generated by a GAN is shown to be realistic looking~\cite{Karras2019ANetworks}, GAN with an encoder may ignore small or uncommon details in the input image~\cite{Bau2019SeeingGenerate}. To preserve these features, we propose a context-aware reconstruction loss (CARL) that exploits extra information from the input domain to refine the reconstruction results. This extra information comes in the form of semantic segmentation and detection of any foreign object present in the input image. The CARL is defined as,
\newline
\begin{equation}
    \gL_{\text{rec}}(\rvx, \rvx^\prime) = \sum_j \frac{S_j(\rvx) \odot || \rvx -  \rvx^{\prime} ||_1}{\sum S_j(\rvx)} +\KL(O(\rvx) || O(\rvx^{\prime})).
    \label{eq:rec_s}
\end{equation}

\vspace{0.5cm}

Here, $S(\cdot)$ is a pre-trained semantic segmentation network that produces a label map for different regions in the input domain. Rather than minimizing a distance such as $\ell_1$ over the entire image, we minimize the reconstruction loss for each segmentation label ($j$). Such a loss heavily penalizes differences in small regions to enforce local consistency.

$O(\rvx)$ is a pre-trained object detector that, given an input image $\rvx$, outputs a number of bounding boxes called region of interests (ROIs). For each bounding box, it outputs 2-d coordinates in the image where the box is located and an associated probability of presence of an object. Using the input image $\rvx$, we obtain the ROIs and associated $O(\rvx)$, which is a probability vector, stating probability of finding an object in each ROI. For reconstructed image $\rvx^{\prime}$, we reuse the ROIs obtained from image $\rvx$ and computed the associated probabilities for the reconstructed image as $O(\rvx^{\prime})$.  Next, we used KL divergence to quantify the difference between probability  vectors as $\KL(O(\rvx) || O(\rvx^{\prime}))$, in Eq.~\ref{eq:rec_s}.
 Finally, we used the CAR loss to enforce two important properties of the explanation function:
\begin{enumerate}
    \item If $\rvc = f(\rvx)$, the self-reconstructed image should resemble the input image.
    \item For $\rvc \neq f(\rvx)$, applying a reverse perturbation on the explanation image $\rvx_{\rvc}$ should recover the initial image \ie $\rvx \approx \mathcal{I}_f(\mathcal{I}_f(\rvx, \rvc), f(\rvx))$.
\end{enumerate}

We enforce these two properties by the following loss,

\begin{figure}[!ht]
\centering
\includegraphics[width=0.7\linewidth]{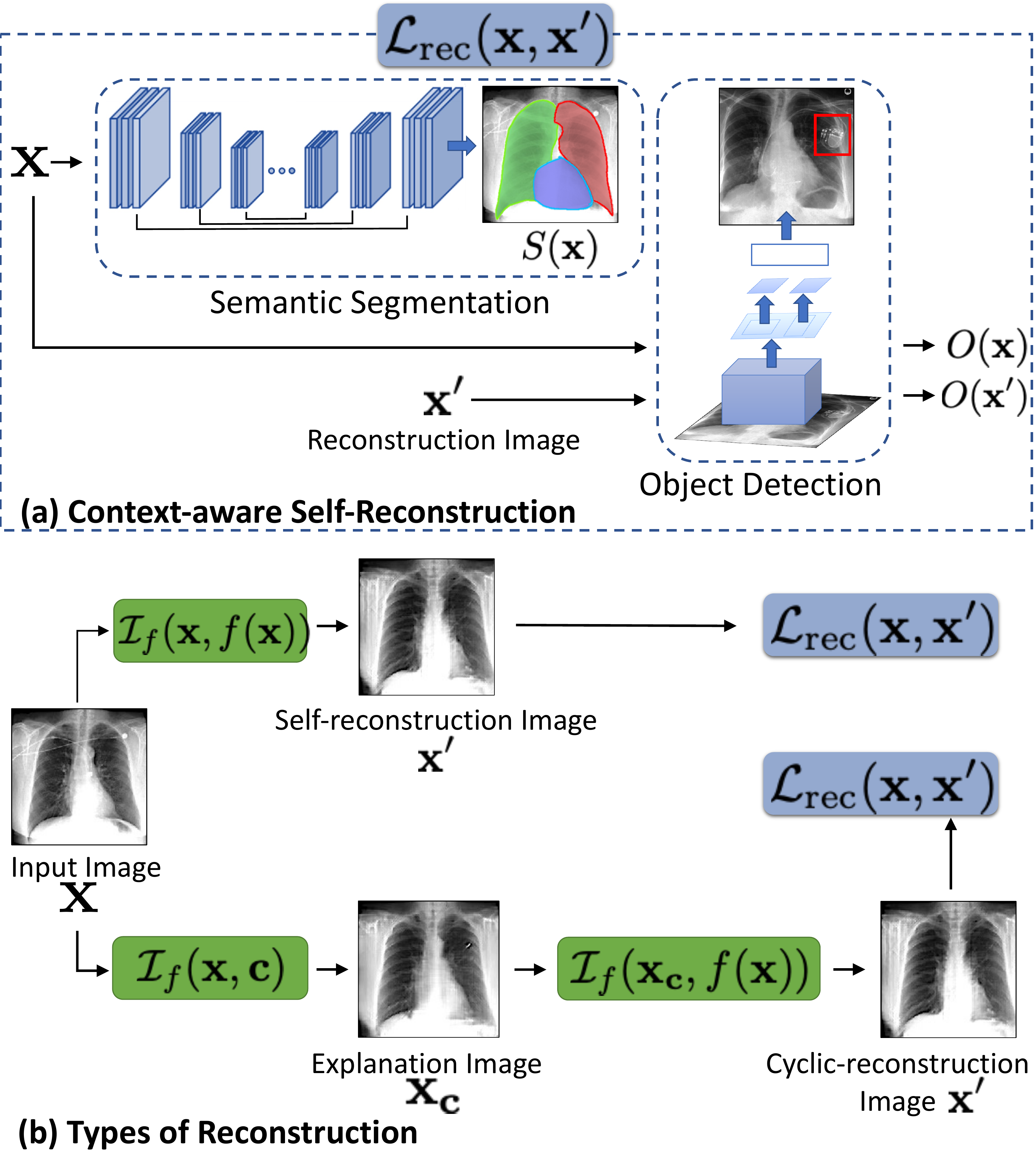}
 
\caption[Context-aware self consistency]
{(a) A context-aware self-reconstruction loss with pre-trained semantic segmentation $S(\rvx)$ and object detection $O(\rvx)$ networks. (b) The self and cyclic reconstruction should retain maximum information from $\rvx$. Note, explanation image $\rvx_{\rvc}$ may differ from input image, $\rvx$.}
\label{Fig_Rec}
\end{figure}

\begin{equation}
    \gL_{\text{rec}}(E,G) =    \gL_{\text{rec}}(\rvx, \mathcal{I}_f(\rvx, f(\rvx))) +
    \gL_{\text{rec}}(\rvx, \mathcal{I}_f(\mathcal{I}_f(\rvx, \rvc), f(\rvx))).
    \label{eq:rec}
\end{equation}

\vspace{0.5cm}

We minimize this loss only for  reconstruction of the input image.  Please note, the classifier $f$ and support networks $S(\cdot)$ and $O(\cdot)$ remained fixed during training.
%For the explanation image, $\rvx_{\rvc}$, with a bin-index different from the input image, we didn't enforce the reconstruction loss, to ensure that the explanation function is not biased towards foreign objects or region specific details. 

\section{Experiments and Results}
\subsection{Study cohort and imaging dataset}
Our experiments are conducted on the CelebA~\cite{Liu2015DeepWild} and
MIMIC-CXR~\cite{Johnson2019MIMIC-CXRReports} datasets. CelebA contains 200K celebrity face images, each with forty attribute labels.
MIMIC-CXR is a multi-modal dataset consisting of 473K chest X-ray images and 206K reports from 63K patients. The dataset provides binarized labels over fourteen radio-graphic observations, namely, enlarged cardiomediastinum, cardiomegaly, lung-lesion, lung-opacity, edema, consolidation, pneumonia, atelectasis, pneumothorax, pleural effusion, pleural other, fracture, support devices and no-finding.
The images are preprocessed using a standard pipeline involving cropping, re-scaling and intensity normalization. We consider a multi-label classifier that takes a frontal view chest x-ray image as input and outputs a posterior probability for the fourteen radio-graphic observations. 

\subsection{Experimental setup}

\emph{Classification model:} CelebA -  We considered two independently trained binary classifiers trained on the ``smiling'' and ``age'' attributes. The classifiers are deep learning models with a ResNet~\cite{He2016DeepRecognition} backbone. For training the classifier, we used the default test and train split as provided by the dataset. The classifiers are very accurate with a test AUC-ROC greater than 0.90. We also experimented with other attributes. 
%For extended results, please refer to the \textbf{Supplementary Material} (appendix ~\ref{SM-Extended-Results}).

MIMIC-CXR - Following the prior work on diagnosis classification~\cite{Irvin2019CheXpert:Comparison}, we used DenseNet-121~\cite{Huang2016DenselyNetworks} architecture as the baseline classification model.  The model is trained on 198K ($\sim$80\%) frontal view CXR from 51K patients and is test on a held-out set of 50K images from 12K non-overlapping patients.  We use the Adam optimizer with default $\beta$-parameters of $\beta_1 = 0.9$, $\beta_2 = 0.999$ and learning
rate $1 \times$ $10^{-4}$ which is fixed for the duration of the training. We used a batch size of 16 images and train for 3 epochs, saving checkpoints every 4800 iterations. The classifier have a mean AUC-ROC of 0.75. It is highly discriminative for three diagnosis: cardiomegaly (AUC-ROC = 0.87), pleural effusion (AUC-ROC = 0.95) and edema (AUC-ROC = 0.91). These results are comparable to performance of the published model~\cite{Irvin2019CheXpert:Comparison}.

\emph{Explanation function: } Our explanation function is implemented using TensorFlow version 2.0 and is trained on NVIDID P100 GPU. Before training the explanation function, we assume access to the pre-trained classification function, that we aim to explain. We also assume access to pre-trained segmentation and object detection networks, that are used to enforce CARL loss.

The explanation function is a cGAN with an encoder $E(\cdot)$ and generator $G(\cdot)$ network, that follows ResNet~\cite{He2016DeepRecognition} architecture. $G(\cdot)$ uses conditional batch normalization (cBN) to incorporate condition information. For discriminator $D(\cdot)$ network, we adapted the architecture from SNGAN~\cite{Miyato2018SpectralNetworks}.  We optimized the adversarial hinge loss for the cGAN training. We set the loss hyper-parameters as $\lambda_1 = 1.0$, $\lambda_2 = 1.0$ and $\lambda_3 = 0.5$. We used the Adam optimizer~\cite{Kingma2014Adam:Optimization}, with default hyper-parameters set to $\alpha = 0.0002, \beta_1 = 0, \beta_2 = 0.9$.

Using PCE, we can derive qualitative explanations for any target class.
 However, for chest imaging dataset currently we support quantitative metrics only for cardiomegaly and PE. Deriving such metrics for other diagnosis requires understanding of the clinical definition of the disease and is challenging. Previously, researcher have shown qualitative results on other diagnosis~\cite{cohen2021gifsplanation}, but quantitative evaluation is still missing. We train three independent cGANs to explain target classes: cardiomegaly, PE, and edema.

 To train PCE, we used 30K images from held-out set, which was not used in training the classifier. We divide  $f(\rvx)[y] \in [0,1]$ into $N = 10$ equally size bins. Here, $y$ is a target class. Each input image is mapped to a  bin-index depending on the prediction $f(\rvx)[y]$. We randomly sample images such that each bin has 2500 to 3000 images. After training the cGAN, we audited the classifier by deriving explanations on a held-out set of 20K images. Please refer appendix \ref{SM-EF} for further details.

 \emph{Semantic segmentation network function:} Semantic segmentation network $S(\cdot)$ is a 2D U-Net~\cite{Ronneberger2015U-NetSegmentation}. It marks the lung and the heart contour in a chest x-ray. The network is trained on 385 chest x-rays and masks from Japanese Society of Radiological Technology (JSRT)~\cite{vanGinneken2006SegmentationDatabase} and Montgomery~\cite{Jaeger2014TwoDiseases.} datasets.

  \emph{Object detector:} We trained a faster regional CNN~\cite{Ren2015FasterNetworks} network, for detecting FO such as pacemaker and hardware in a chest x-ray. The network learns to detect FO by placing a bounding box over them. To create the training dataset, we extracted 300 x-rays with a positive mention of these objects in the corresponding radiology reports, and collected bounding box annotations to mark the ground truth. We further trained two more detectors to evaluate our explanations. Specifically, we trained detectors for identifying normal and abnormal  costophrenic (CP) recess region in the chest x-ray. We associated an abnormal CP recess with the radiological finding of a blunt CP angle as identified by the  positive mention for \textit{``blunting of the costophrenic angle"} in the corresponding radiology report. For the normal-CP recess, we considered images with a positive mention for \emph{``lungs are clear"} in the reports. 

We compared our counterfactual explanations with closest existing methods such as xGEM\cite{Joshi2018XGEMs:Models} and CycleGAN~\cite{Narayanaswamy2020ScientificTranslation,DeGrave2020AISignal}. For proper comparison, we used the open-source implementation of these models and trained them on the MIMIC-CXR dataset, using the same training-set as our model. Please refer SM-Sec~\ref{SM-xGEM} for more details. We also compared against the saliency-based methods to provide \emph{post-hoc} model explanation.  We performed following experiments: 
\begin{enumerate}
    \item \textit{Desiderata of explanation function:} We compared our model with existing methods on the three desiderata of valid explanations and evaluated the following metrics: Fr\'echet Inception Distance (FID) score to assess visual quality,  counterfactual validity (CV) score to quantify compatibility with the classifier, and  face verification accuracy and foreign object preservation (FOP) score to evaluate the identity preservation in the explanations.
    \item \textit{Comparison with saliency-maps:} 
We compared the localization ability of our counterfactual explanations against the saliency maps generated by gradient-based methods.
\item \textit{Clinical evaluation:} We used two clinical metrics, namely, cardiothoracic ratio (CTR) and the Score for detecting a normal Costophrenic recess (SCP) to demonstrate the clinical relevance of our explanations.
\item \textit{Bias detection:} We trained two classifiers on biased and unbiased data and examined the performance of our method in identifying the bias.
\item \textit{Evaluating class discrimination:} We trained a multi-label classifier and demonstrate the sensitivity of PCE to the task being explained.
\item \textit{Human evaluation:} We evaluate the strength of PCE in explaining the classifier's decision to the end-user.
\end{enumerate}

\subsection{Desiderata of explanation function}
We evaluated our method on three desiderata of a valid counterfactual~\cite{Mothilal2019ExplainingExplanations}. First, \textit{Data consistency:} A counterfactual should be realistic-looking \ie it should be very similar to the input image. Second, \textit{Classifier consistency:} A counterfactual should flip the classification decision for the input image. Third, \textit{Identify preservation:} A counterfactual should preserve patient-specific details such as FOs.

\begin{figure}[!ht]
\begin{center}
\includegraphics[width=0.75\linewidth]{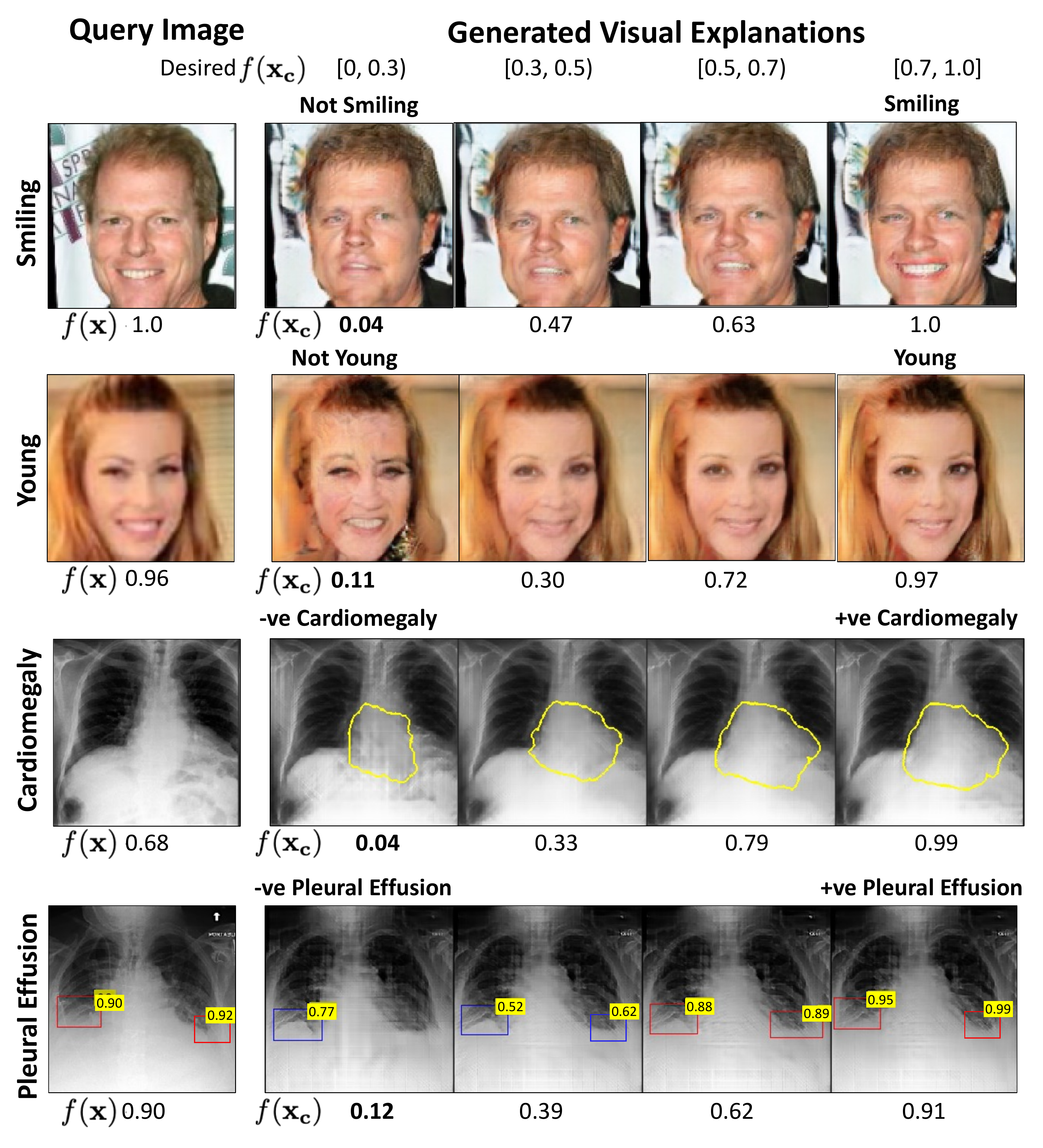}
 
\caption[Progressive counterfactual explanations generated for different prediction tasks.]{Progressive counterfactual explanations generated for different prediction tasks. The figure shows smiling/not-smiling
(first row), young/old face (second row), diagnosis of cardiomegaly (third row) and diagnosis of pleural effusion (last row). The first column shows the query image, followed by the corresponding generated explanations.
The bottom label is the output of the classifier $f$. For Cardiomegaly, we show the
segmentation of the heart (yellow edge). For PE, we show the bounding box (BB) for normal (blue) and abnormal (red) costophrenic (CP) recess.  Extended results in SM-Figure.~\ref{Fig_Quality_All}.}
\label{Fig_Quality_ch3}
\end{center}
\end{figure}

\subsubsection{Data consistency} 
Given an input image, our model generates a series of images $\rvx_{\rvc}$ as explanations by gradually changing $\rvc$ in range $[0, 1]$. Figure.~\ref{Fig_Quality_ch3} shows the qualitative results on CelebA dataset. We show results for two prediction tasks: smiling or not-smiling and young or old. Bottom two rows of Figure.~\ref{Fig_Quality_ch3} shows our result on MIMIC-CXR dataset. The left-most image is the input CXR.  For \textbf{cardiomegaly}, we highlight the heart contour (yellow). Its helps in visualizing enlargement of the cardiac silhouette. For \textbf{PE}, we showed the results of an object detector as bounding-box (BB) over the normal (blue) and abnormal (red) CP recess regions. The number on the top-right of the blue-BB is the Score for detecting a normal CP recess (SCP). The number on red-BB is 1-SCP. The CP recess is the potential area to be analyzed for PE~\cite{PleuralTomography}. From left to right, the normal CP recess changed into an abnormal CP recess with a high detection score.
We observed a gradual increase in posterior probability $f(\rvx_{\rvc})$ (bottom label) as we go from left to right.

\emph{Quantitatively evaluation:} 
We evaluated the visual quality of our explanations by computing Fr\'echet Inception Distance (FID) score~\cite{Heusel2017GansEquilibrium}. FID quantifies the visual similarity between the real images $\rvx$ and the synthetic counterfactuals $\rvx_{\rvc}$ by computing distance between their activation distributions as follow,
\newline
\begin{equation}
    \text{FID}(\rvx, \rvx_{\rvc}) = ||\mu_{\rvx} - \mu_{\rvx_{\rvc}}||_2^{2} + \text{Tr}(\Sigma_{\rvx} + \Sigma_{\rvx_{\rvc}} - 2(\Sigma_{\rvx}\Sigma_{\rvx_{\rvc}})^{\frac{1}{2}}),
\end{equation}
\newline
where $\mu$'s and $\Sigma$'s are mean and covariance of the activation vectors derived from the penultimate layer of a pre-trained Inception v3 network~\cite{Heusel2017GansEquilibrium}. The pre-trained network is trained on same dataset as the PCE.
We examined real and synthetic (\ie generated explanations) images on the two extreme of the decision boundary, \ie a normal group ($f(\rvx) < 0.2$) and an abnormal group  ($f(\rvx) > 0.8$). In Table.~\ref{FID-table}, we compared three counterfactual-generating algorithms: ours, xGEM, and cycleGAN, and reported the FID for each group. Our model creates natural-looking counterfactuals compared to xGEM. The cycleGAN model generates the most visually appealing images with the lowest FID score across the classification tasks.  However, a visually good image doesn't necessarily means a good counterfactual. It is also equally important to flip the classification decision as explained in next section.

\begin{table}[!ht]
\caption[The FID score quantifies the visual appearance of the explanations.]{The FID score quantifies the visual appearance of the explanations.
The counterfactual validity (CV) score is the fraction of explanations that have an opposite prediction compared to the input image. An ideal counterfactual explanation have low FID score and a high CV score. }
\label{FID-table}
\begin{center}
\begin{tabular}{c|ccc|ccc}
\multicolumn{1}{c|}{ }  &  \multicolumn{3}{c|}{\bf Negative ($f(\rvx), f(\rvx_{\rvc}) < 0.2$)} &  \multicolumn{3}{c}{\bf Positive ($f(\rvx), f(\rvx_{\rvc}) > 0.8$)} \\
\bf Target class & Ours & xGEM & CycleGAN & Ours & xGEM & CycleGAN \\
\hline
\multicolumn{7}{c}{\bf Fr\'echet Inception Distance (FID)} \\
\hline
CelebA:Smiling & 56.3 & 112.9& \bf30 & 46.9 &111.0  & \bf37 \\
CelebA:Young & 74.4 & 170.3 &\bf56 & 67.6 & 115.2& \bf35 \\
CXR:Cardiomegaly & 166 & 138 & \bf 30 & 137 & 316 & \bf 56\\
CXR:Pleural Effusion & 146 & 347 & \bf 37 & 122 & 355 & \bf 35\\
CXR:Edema & 149 & 376 & \bf 72 & 102 & 274 & \bf 77\\
\hline
\hline
\multicolumn{7}{c}{\bf Counterfactual Validity Score} \\
\hline
CXR:Cardiomegaly & \bf 0.91 & \bf 0.91 & 0.43\\
CXR:Pleural Effusion & \bf 0.97 & \bf 0.97 & 0.49\\
CXR:Edema & \bf 0.98 & 0.66 & 0.57 \\
\hline
\end{tabular}
\end{center}
\end{table}

\subsubsection{Classification model consistency}
By definition, a counterfactual image have an opposite classification decision as compared to the query image. A counterfactual provides explanation by showing how image-features should be modified to flip the classification decision. If the decision doesn't flip then the explanation is inconclusive. Counterfactual validity (CV) score is the fraction of counterfactual explanations that successfully flipped the classification decision \ie if the input image is negative (normal) then the generated explanation is predicted as positive (abnormal) for the target class. We compared different counterfactual-generating algorithms on CV score~\cite{Mothilal2019ExplainingExplanations} metric. The last rows of Table.~\ref{FID-table}, summarizes our result.  For all tasks, our model consistently achieved the highest CV score. CycleGAN achieved a low CV score, thus creating explanations that are frequently inconsistent with the classifier. 

\vspace{1.0cm}

\begin{figure}[!ht]
\begin{center}
\includegraphics[width=1.0\linewidth]{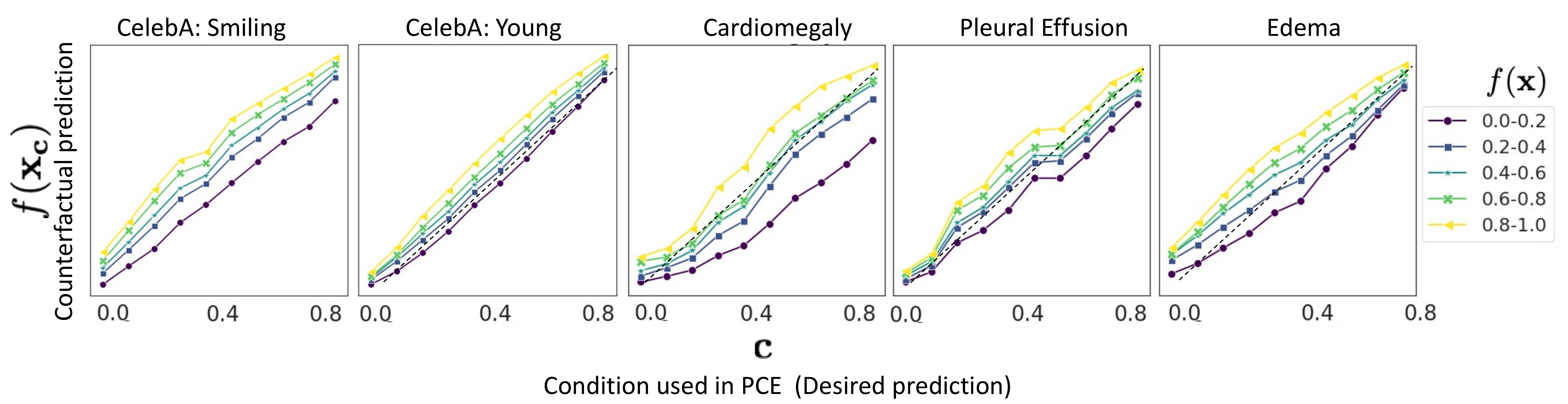}
 
\caption[Classifier consistency]{Plot of the desired outcome from the classifier, $\rvc$, against the
actual response of the classifier on generated explanations, $f(\rvx_{\rvc})$. The monotonically
increasing trend shows a positive correlation between $\rvc$ and $f(\rvx_{\rvc})$, and thus the generated explanations are consistent with the expected condition.  Each line represents a set of input images with prediction $f(\rvx)$ in a given range.}
\label{Fig_CC}
\end{center}
\end{figure}

\vspace{-0.5cm}

Next, we quantify this consistency at every step of the transformation. We divided the prediction range $[0,1]$ into $N=10$ equally sized bins. For each bin, we generated an explanation image by choosing an appropriate, $\rvc$. We further divided the input image space into five groups based on their initial prediction \ie $f(\rvx)$. In Figure.~\ref{Fig_CC}, we represented each group as a line and plotted the average response of the classifier \ie $f(\rvx_{\rvc})$ for explanations in each bin against the expected outcome \ie $\rvc$.  The positive slope of the line-plot, parallel to $y = x$ line confirms that starting from images with low $f(\rvx)$, our model creates fake images such that $f(\rvx_{\rvc})$ is high and vice-versa.

%Our model achieved a low FID score and a high CV score, producing realistic explanations that are consistent with classifier. In contrast, xGEM have a very high FID, with poor visual quality which  deemed  then  unsuited  for  clinical  applications.  CycleGAN produced the most visually appealing x-ray images, with a high FID score. However, during its training, the cycelGAN does not consider the classification function $f$ and follows a data-driven approach to learn the differences in positive and negative classes. Thus, it is not always necessary that the resulting counterfactual explanation flips the classifier's decision. As a result the counterfactual validity score (CVS) for cycleGAN is about 50\%. In contrast, 90\% of the explanations generated by our model produced the desired outcome from the classifier (\emph{see} Table~\ref{FID-table}).

\subsubsection{Identity preservation} 
The counterfactual explanations should differ only in semantic features associated with the target class while retaining the identity of the query image. For example, in CelebA, if the classifier is using image features near the lips to decide smiling or not, the other features such as hair and the person in the image should remain the same.
Similarly, in CXR, any foreign objects (FOs) such as pacemaker should be preserved. Furthermore, FO provides critical information to identify the patient in an x-ray. The disappearance of FO in explanation images may create confusion that explanation images show a different patient.

For PCE trained on CelebA dataset, we evaluate the identify preservation in counterfactual explanations through face verification. In face verification, we quantify the similarity between the faces in query image and corresponding fake counterfactual explanation image. We used state-of-the-art face recognition model trained on VGGFace2 dataset~\cite{Cao2018VGGFace2:Age} as feature extractor for both real images and their corresponding fake explanations. For face verification, we calculated the closeness between real and fake image as cosine distance between their feature vectors.  The faces were considered as verified \ie fake explanation have same identity as real image, if the distance is below 0.5. In Table~\ref{OD-table}, we report face verification accuracy as percentage of the verified query image and fake counterfactual image pairs. We evaluated this metric over a randomly sampled test set of 500 images.

\begin{table}[!ht]
\centering
\caption{Results on face-verification task to demonstrate that the identity of a person is preserved across counterfactual explanations.}
\label{OD-table}
\begin{tabular}{c|c}
\bf Target  Class & \bf Face Verification Accuracy\\
\hline
CelebA: Smiling & 85.3\%  \\
CelebA:Young & 72.2\%  \\
\hline
\end{tabular}
\end{table}

\begin{table}[!ht]
\centering
\caption{ The foreign object preservation (FOP) score with and without the context-aware reconstruction loss (CARL). FOP score depends on the performance of FO detector.}
\label{OD-table-FO}
\begin{tabular}{c|cc}
\bf Foreign  &  \multicolumn{2}{c}{\bf FOP score}\\
 \bf Objects & \bf Ours with CARL & \bf Ours with $\ell_1$ \\
\hline
 Pacemaker  & \bf 0.52 & 0.40   \\
Hardware  & \bf 0.63& 0.32   \\
\hline
\end{tabular}
\end{table}

For PCE trained on CXR dataset, we quantify the strength of our revised CARL loss in preserving FO in explanation images compared to an image-level $\ell_1$ reconstruction loss. We reported results on the FO preservation (FOP) score metric. FOP score is the fraction of real images, with successful detection of FO, in which FO was also detected in the corresponding explanation image $\rvx_{\rvc}$. Our model with CARL obtained a higher FOP score, as shown in Table~\ref{OD-table-FO}. The FO detector network has an accuracy of 80\%.

\begin{figure}[!ht]
    \centering
    \includegraphics[width = 0.65\linewidth]
    {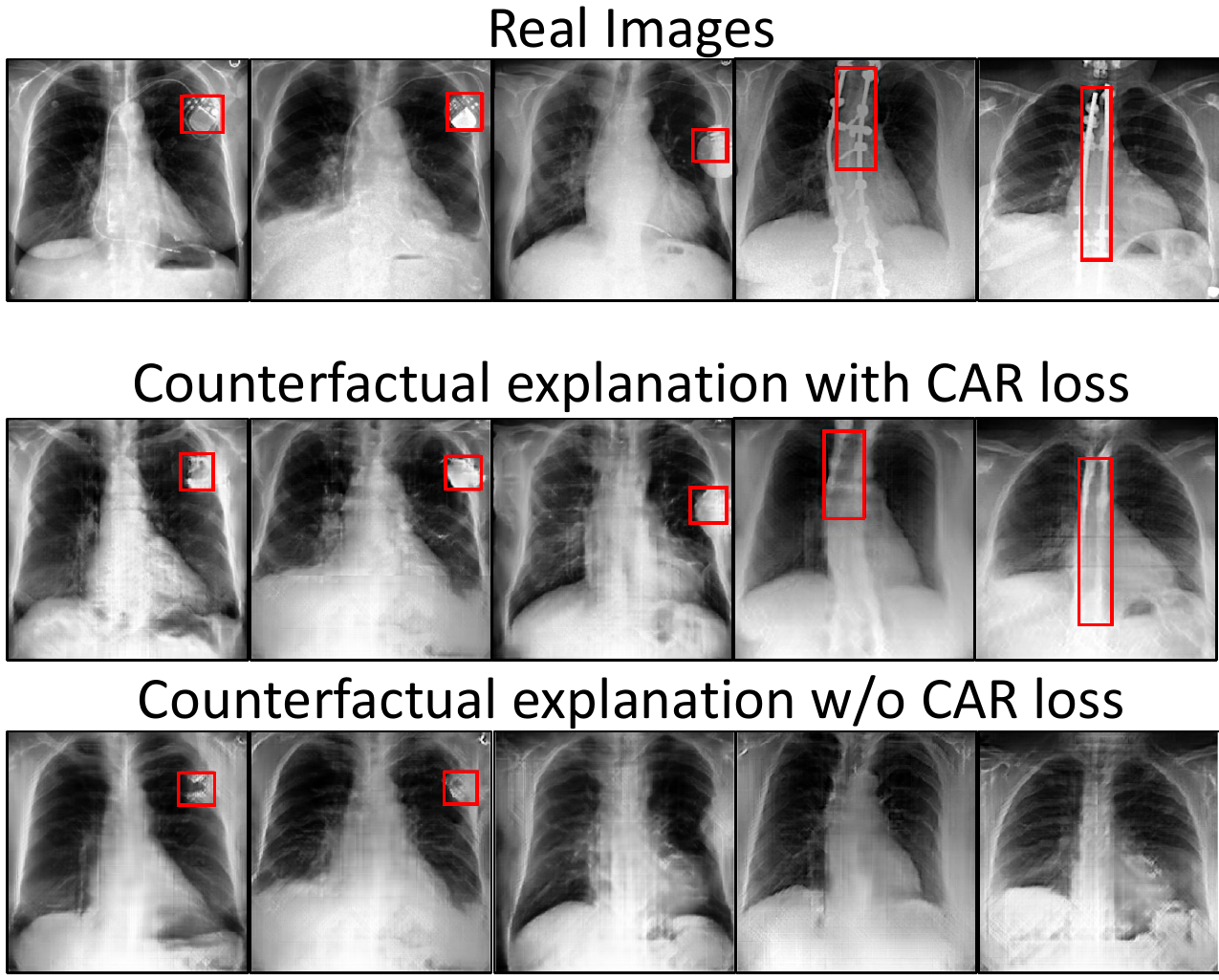}
    \caption{ Fidelity of generated images with respect to preserving FO.}
   \label{Fig_Pacemaker}
\end{figure}

Figure.~\ref{Fig_Pacemaker} presents examples of counterfactual explanations generated by our model with and without the CARL. Our results confirm that CARL is an improvement over $\ell_1$ reconstruction loss. We further provide a detailed ablation study over different components of our loss in appendix \ref{SM-ablation}.

\subsection{Comparison with saliency-maps}
Popular existing approaches for model explanation consist of gradient-based methods that provide a qualitative explanation in the form of saliency maps~\cite{Pasa2019EfficientVisualization,Irvin2019CheXpert:Comparison}. Saliency maps show the importance of each pixel of an image in the context of classification. Our
method is not designed to produce saliency maps as a continuous score for every feature of the
input. To compare against such methods, we approximated  a saliency map as an absolute difference map between the explanations generated for the two extremes; negative decision with $f(\rvx_{\rvc}) < 0.2$ and positive decision with $f(\rvx_{\rvc}) > 0.8$. For proper comparison, we considered the absolute values of the saliency maps and normalized them in the range $[0, 1]$.

 \begin{figure}[!ht]
\centering
\includegraphics[width=0.9\linewidth]{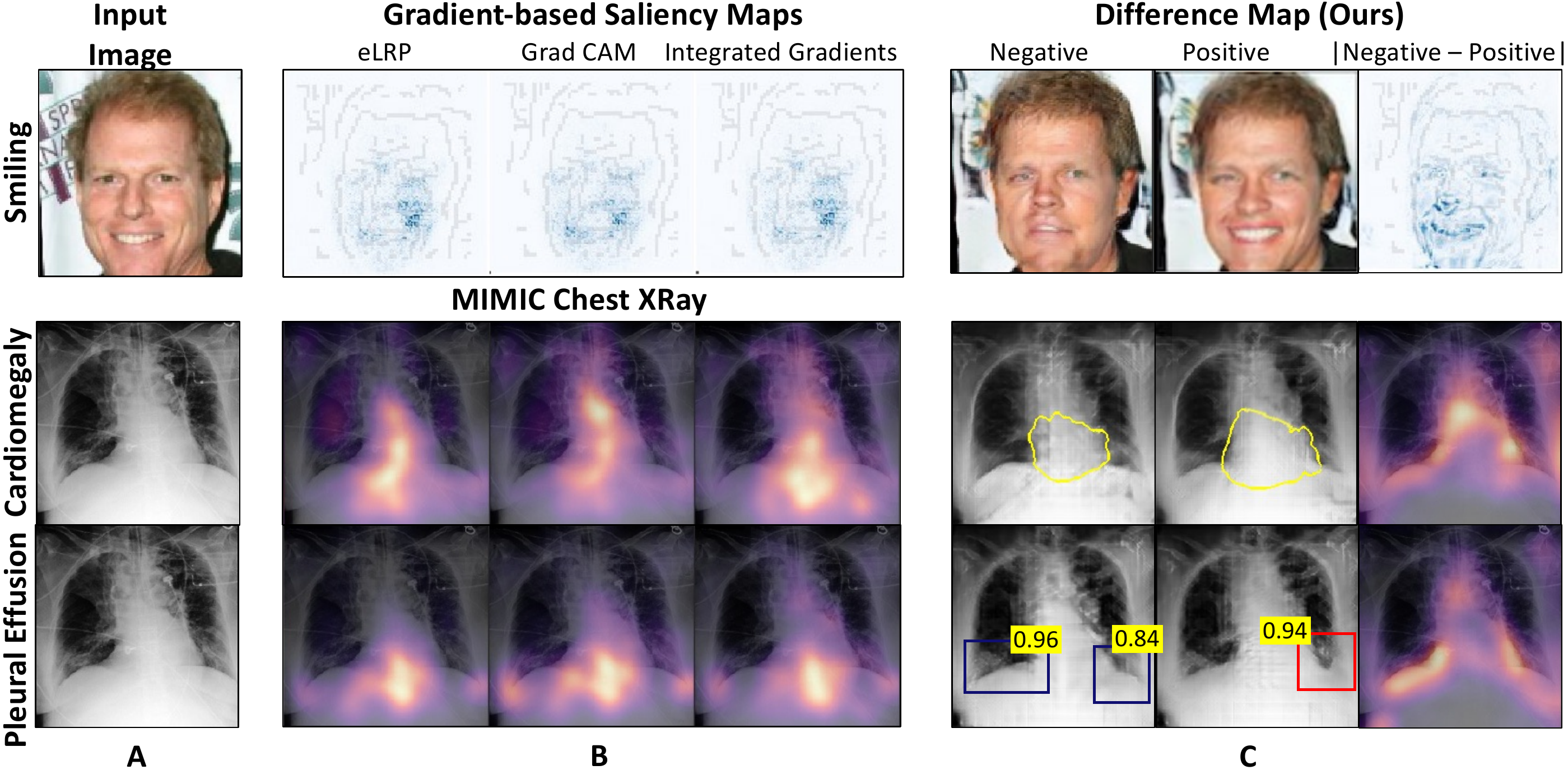}
 
\caption[Comparison of our method against different gradient-based methods.]{Comparison of our method against different gradient-based methods.
A: Input image; B: Saliency maps from existing works; C: Our simulation of saliency map as difference map between the normal and abnormal explanation images. More examples are shown in SM-Figure.~\ref{Fig_c}.}
\label{Fig_saliency}
\end{figure}

\vspace{-0.2cm}

Figure.~\ref{Fig_saliency} shows the saliency map obtain from our method and its comparison with popular gradient based methods. For CelebA, we compare the explanations derived for the ``smiling" classifier. For CXR dataset,  we show an example of an input image, where the gradient-based saliency maps highlight almost the same region for two different target tasks. In contrast, our difference map localized disease to specific regions in the chest. Figure.~\ref{Fig_saliency}.C shows the two extreme explanation images and the corresponding difference map, derived for input images shown in Figure.~\ref{Fig_saliency}.A.

For quantitative evaluation,  we used the \textit{deletion} evaluation metric to  compare our difference map with saliency maps produced by different gradient-based methods~\cite{Petsiuk2018RISE:Models}.  The deletion metric quantifies how the probability of the target-class changes as important pixels are removed from an image. A sharp drop in in prediction accuracy, resulting in a low area under the probability curve (AUC) (as a function of
the percentage of the salient pixels removed), represents a good explanation. To remove salient pixels from an image, 
%in face images from CelebA dataset, we replace the removed pixels with random values sampled from a uniform distribution. While 
in CXR images, we selectively impaint the removing regions based on its surrounding.

\begin{table}[!ht]
\centering
\caption[Quantity comparison of our method against gradient-based methods.]{Quantity comparison of our method against gradient-based methods. Mean area under the probability curve (AUC), plotted as a function of the fraction of removed pixels. A low AUC shows a sharp drop in prediction accuracy as fraction of removed pixels increases.}
\label{table}
\begin{tabular}{c|ccc}
Method & \bf Cardiomegaly &  \bf Pleural Effusion &  \bf Edema\\
\hline
Ours &  \bf 0.040$\pm$0.04 &  \bf0.023$\pm$0.02 &  0.083$\pm$0.05 \\
eLRP & 0.071$\pm$0.05&  0.033$\pm$0.02 & 0.055$\pm$0.03 \\
Grad-CAM & 0.045$\pm$0.04&   0.058$\pm$0.05 &  \bf 0.035$\pm$0.02 \\
Integrated Gradients & 0.058$\pm$0.06&  0.046$\pm$0.05  & 0.077$\pm$0.04\\
\hline
\end{tabular}
\end{table}

%Figure.~\ref{Fig_SM_celebA} shows the plot of the drop in classification accuracy vs the percentage of the relevant pixels that were removed for ``smiling" classifier.  
%All the methods experienced a drop in the accuracy of the classifier with increase in the percentage of perturb pixels. The saliency maps produced by our model are comparable to the other saliency map methods.

% \begin{figure}[ht]
%\centering
%\includegraphics[width=0.4\linewidth]{Images/ch3/ch3_sm_1.jpg}
%\caption{\footnotesize  Plot to show the drop in accuracy of the classifier as we perturb the most relevant pixels (relevance calculated from saliency map) in the image.}
%\label{Fig_SM_celebA}
%\end{figure}

\begin{figure}[!ht]
    \centering
    \includegraphics[width = 1.0\linewidth]
    {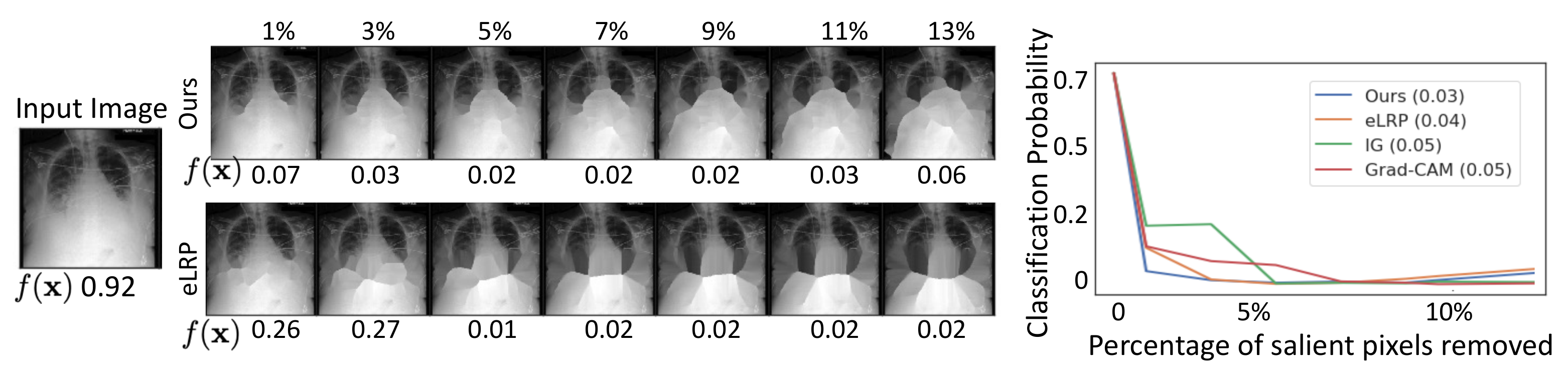}
     
    \caption[Evaluation using deletion metric.]{Evaluation using deletion metric.
    The plot shows the drop in classification probability for  pleural effusion as important pixels are removed from the input image. Top label shows the percentage of removed pixels. The bottom label shows the classification prediction. }
    \label{Fig_SM}
\end{figure}

For CXR dataset, in Table~\ref{table}, we report the mean AUC over a sample of 500 images. The images were selected such that the $f(\rvx) > 0.9$ for the target-disease. Our model achieved the lowest AUC in deletion-by-impainting for cardiomegaly and pleural effusion. In Figure.~\ref{Fig_SM}, we show an example of deletion-by-impainting.  The results show that the regions modified by our explanation model are important for the classification decision. 
\newline

\vspace{-0.5cm}

\subsection{Clinical evaluation}
In this experiment, we demonstrate the clinical relevance of our explanations. First, we translate the clinical definition of two diseases (cardiomegaly and pleural effusion) into quantitative metrics.
 Next, we used these clinical metrics to quantify the counterfactual changes between normal and abnormal populations, as identified by the given classifier. If the change in classification decision is associated with the corresponding change in clinical-metric, we can conclude that the classifier considers clinically relevant information in its diagnosis prediction. We considered the following two  metrics:
 \subsubsection{Cardio Thoracic Ratio (CTR)}
 The CTR is the ratio of the cardiac diameter to the maximum internal diameter of the thoracic cavity.  A CTR ratio greater than 0.5 indicates cardiomegaly~\cite{Mensah2015EstablishingScreening,2017EvaluatingEchocardiography,Dimopoulos2013CardiothoracicDisease}. We followed the approach in ~\cite{Chamveha2020AutomatedApproach} to calculate CTR from a CXR. In the absence of ground truth lung and heart segmentation on the MIMIC-CXR dataset, we used a segmentation network trained on open-sourced supervised datasets~\cite{Maduskar2013AutomatedRadiographs,Jaeger2014TwoDiseases.}. We calculated heart diameter as the distance between the leftmost and rightmost points from the lung centerline on the heart segmentation.    The thoracic diameter is calculated as the horizontal distance between the widest points on the lung mask. Please refer appendix \ref{SM-SS} for details on segmentation network.
\newline
 
\begin{figure}[!ht]
\centering
\includegraphics[width = 0.95\linewidth]
{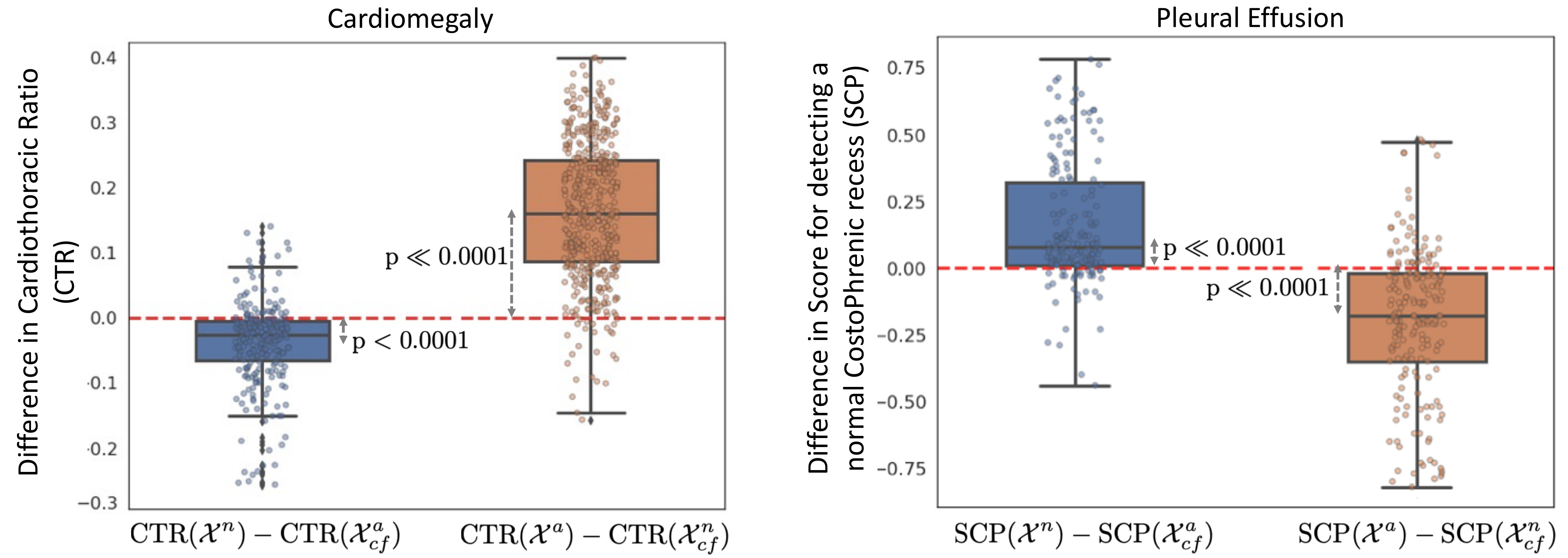}
 
\caption[Box plots to show distributions of pairwise differences in clinical-metrics.]{Box plots to show distributions of pairwise differences in clinical-metrics.
We consider clinical metrics such as CTR for cardiomegaly and the Score of normal CP recess (SCP) for pleural effusion, before (real) and after (counterfactual) our generative counterfactual creation process. The mean value corresponds to the average causal effect of the clinical-metric on the target disease. The low p-values for the dependent t-test statistics confirms the statistically significant difference in the distributions of metrics for real and counterfactual images. Further numbers are summarized in SM-Table~\ref{t-table}. }
\label{Fig_Box}
\end{figure}

 \vspace{-0.8cm}

 \subsubsection{Costophrenic recess}
The fluid accumulation in costophrenic (CP) recess may lead to the diaphragm's flattening and the associated blunting of the angle between the chest wall and the diaphragm arc, called costophrenic angle (CPA). The blunt CPA is an indication of pleural effusion~\cite{Maduskar2016AutomaticRadiographs}.

Marking the CPA angle on a CXR requires expert supervision, while annotating the CP region with a bounding box is a much simpler task (\emph{see} SM-Figure.~\ref{FIG_CPA}). We learned an object detector to identify normal or abnormal CP recess in the CXRs and used the Score for detecting a normal CP recess (SCP) as our evaluation metric. Further details on the training of the object detector are provided in appendix ~\ref{SM-OD}.

 We performed a statistical test to quantify the differences in real images and their corresponding counterfactuals based on the above two metrics. We randomly sample two groups of real images (1) a \emph{real-normal} group defined as $\mathcal{X}^{n} = \{\rvx; f(\rvx) < 0.2\}$. It consists of real CXR that are predicted as normal by the classifier $f$. (2) A \emph{real-abnormal} group defined as $\mathcal{X}^{a} = \{\rvx; f(\rvx) > 0.8\}$. For $\mathcal{X}^{n}$ we generated a counterfactual group as,  $\mathcal{X}_{cf}^{a} = \{\mathcal{I}_f(\rvx, \rvc); \rvx \in \mathcal{X}^n; \rvc > 0.8\}$. Similarly for $\mathcal{X}^a$, we derived a counterfactual group as   $\mathcal{X}_{cf}^{n} = \{\mathcal{I}_f(\rvx,\rvc); \rvx \in \mathcal{X}^a; \rvc< 0.2\}$. 

In Figure.~\ref{Fig_Box}, we showed the distribution of differences in CTR for cardiomegaly and SCP for PE in a pair-wise comparison between real (normal/abnormal) images and their respective counterfactuals. Patients with cardiomegaly have higher CTR as compared to normal subjects. Hence, one should expect CTR($\mathcal{X}^n$) $<$ CTR($\mathcal{X}_{cf}^a$) and likewise CTR($\mathcal{X}^a$) $>$ CTR($\mathcal{X}_{cf}^n$). Consistent with clinical knowledge, in Figure.~\ref{Fig_Box}, we observe a negative mean difference for CTR($\mathcal{X}^n$) $-$ CTR($\mathcal{X}_{cf}^a$) (a p-value of $< 0.0001$) and a positive mean difference for CTR($\mathcal{X}^a$) $-$ CTR($\mathcal{X}_{cf}^n$) (with a p-value of $\ll 0.0001$). The low p-value in the dependent t-test statistics supports the alternate hypothesis that the difference in the two groups is statistically significant, and this difference is unlikely to be caused by sampling error or by chance.

By design, the object detector assigns a low SCP to any indication of blunting CPA or abnormal CP recess. Hence, SCP($\mathcal{X}^n$) $>$ SCP($\mathcal{X}_{cf}^a$) and likewise SCP($\mathcal{X}^a$) $<$ SCP($\mathcal{X}_{cf}^n$). Consistent with our expectation, we observe a positive mean difference for SCP($\mathcal{X}^n$) $-$ SCP($\mathcal{X}_{cf}^a$) (with a p-value of $\ll 0.0001$) and a negative mean difference for SCP($\mathcal{X}^a$) $-$ SCP($\mathcal{X}_{cf}^n$) (with a p-value of $\ll 0.0001$). A low p-value confirmed the statistically significant difference in SCP for real images and their corresponding counterfactuals. For further details and visual examples of samples in normal and abnormal groups, please refer appendix ~\ref{Extended-R-CE}.

\subsection{Human Evaluation}
\subsubsection{CelebA dataset}
\begin{figure}[!ht]
\centering
\includegraphics[width=0.95\linewidth]{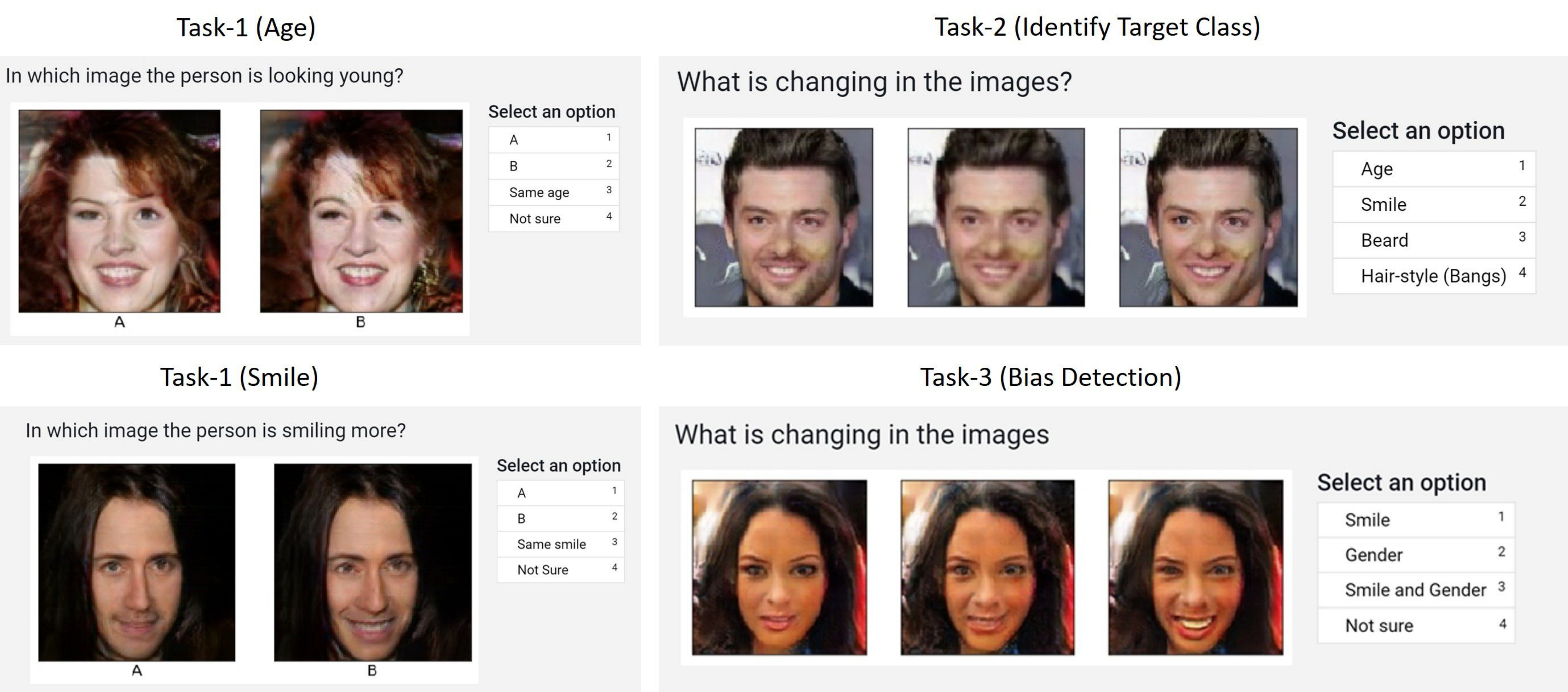}
 
\caption[The interface for the human evaluation done using Amazon Mechanical Turk (AMT).]{The interface for the human evaluation done using Amazon Mechanical Turk (AMT). Task-1 evaluated if humans can detect the relative order between two explanations. Task-2 evaluated if humans can identify the target class for which our model has provided the explanations. Task-3 demonstrated that our model can help the user to identify problems like possible bias in the black-box training.}
\label{Fig_HT}
\end{figure}

We used Amazon Mechanical Turk (AMT) to conduct human experiments to demonstrate that the progressive exaggeration produced by our model is visually perceivable to humans. We presented AMT workers with three tasks. In the first task, we evaluated if humans can detect the relative order between two explanations produced for a given image. We ask the AMT workers, ``Given two images of the same person, in which image is the person younger (or smiling more)?" (\textit{see } Figure~\ref{Fig_HT}).  We experimented with 200 query images and generated two pairs of explanations for each query image (\ie 400 hits). The first pair (\textit{easy}) imposed the two images are samples from opposite ends of the explanation spectrum (counterfactuals), while the second pair (\textit{hard}) makes no such assumption. 
\newline

\vspace{-0.4cm}
In the second task, we evaluated if humans can identify the target class for which our model has provided the explanations.  We ask the AMT workers, ``What is changing in the images? (age, smile, hair-style or beard)". We experimented with 100 query images from each of the four attributes (\ie 400 hits). In the third task, we demonstrate that our model can help the user to identify problems like possible bias in the black-box training. Here, we used the same setting as in the second task but also showed explanations generated for a biased classifier. We ask the AMT workers, ``What is changing in the images? (smile or smile and gender)" (\textit{see } Figure~\ref{Fig_HT}).  We generated explanations for 200 query images each, from a biased-classifier ($f_{\text{Biased}}$) explainer from Section \ref{bias} and an unbiased classifier ($f_{\text{No-biased}}$) explainer (\ie 400 hits). In all the three tasks, we collected eight votes for each task, evaluated against the ground truth, and used the majority vote for calculating accuracy.

We summarize our results in Table~\ref{HE-table}. In the first task, the annotators achieved high accuracy for the \textit{easy} pair when there was a significant difference among the two explanation images, as compared to the \textit{hard} pair when the two explanations can have very subtle differences. Overall, the annotators were successful in identifying the relative order between the two explanation images.
\begin{table}[!ht]
\centering
\caption[Summarizing the results of human evaluation.]{Summarizing the results of human evaluation. The $\kappa$ -statistics measure inter-rater agreement for qualitative classification of items into some mutually exclusive categories. One possible interpretation of $\kappa$ as given in ~\cite{Viera2005UnderstandingStatistic} is $< 0.0 $: Poor, $0.01 - 0.2$: Slight, $0.21 - 0.40$: Fair, $0.41 - 0.60$: Moderate, $0.61 - 0.80$: Substantial and $0.81 - 1.00$: Almost perfect agreement.}
\label{HE-table}
\begin{tabular}{c|cc|ccc}
\bf Annotation Task &  \multicolumn{2}{c}{\bf Overall} & \multicolumn{3}{c}{\bf Sub categories}\\
 & Accuracy  & $\kappa$-statistic   & Category & Accuracy & $\kappa$-statistic \\

\hline
\bf Task-1 (Age) &  83.5\% & 0.41 (Moderate) & Hard & 73\% & 0.31 (Fair) \\
 &  &  & Easy & \bf 94\% & 0.51 (Moderate) \\
 
\bf Task-1 (Smile) & 77.5\% & 0.28 (Fair) & Hard & 66\% & 0.23 (Fair) \\
 &  &  & Easy & \bf 89.5\% & 0.32 (Fair) \\
 \hline
 \centering \bf Task-2   & 77\% & 0.35 (Fair) & Age & 72\% & -\\
 \bf (Identify &  & & Smile & \bf 99\% & - \\
\bf Target Class)  &  & & Bangs & 50\% & - \\
 &  & & Beard & 87\% & - \\
  \hline
 \bf Task-3 (Bias & 93.75\% & 0.14 (Slight) & $f_{\text{Biased}}$  & 87.5\% & 0.09 (Slight) \\
 \bf Detection) &  & & $f_{\text{No-biased}}$ & \bf 100\% & 0.02 (Slight) \\
    \hline
\end{tabular}
\end{table}

In the second task, the annotators were generally successful in correctly identifying the target class. The target class ``bangs'' proved to be the most difficult to identify, which was expected. The generated images for ``bangs'' were qualitatively, the most subtle. For the third task, the correct answer was always the target class \ie ``smile''. In the case of biased classifier explainer, the annotators selected ``Smile and Gender'' 12.5\% of the times. The gradual progression made by the explainer for a biased classifier was very subtle and was changing large regions of the face as compared to the unbiased explainer. The difference is much more visible when we compare the explanation generated for the same query image for a biased and no-biased classifier, as in Figure~\ref{Fig_Bias_Detection}. But in a realistic scenario, the no-biased classifier would not be available to compare against. Nevertheless, the annotators detected bias at roughly the same level of accuracy as our classifier (Table~\ref{Table_Bias}). Future work could improve upon bias detection.

\subsubsection{MIMIC CXR Dataset}
We conducted a human-grounded experiment with diagnostic radiology residents to compare different styles of explanations (no explanation, saliency map, cycleGAN explanation, and our counterfactual explanation) by evaluating different aspects of explanations: (1) understandability,  (2) classifier's decision justification, (3) visual quality, (d) identity preservation, and (5) overall helpfulness of an explanation to the users.

Our results show that our counterfactual explanation  was the only explanation method that significantly improved  the users' understanding of the classifier's decision compared to the no-explanation baseline. In addition, our counterfactual explanation had a significantly higher classifier's decision justification than the cycleGAN explanation, indicating that the participants found a good evidence for the classifier's decision more frequently in our counterfactual explanation as compared to cycleGAN explanation.

Further, cycleGan explanation performed better in terms of visual quality and identity preservation. However, at times the cycleGAN explanations were identical to the query image, thus providing inconclusive explanations.
Overall the participants found our explanation method the most helpful method in understanding the assessment made by the AI system in comparison to other explanation methods. 
%Even though cycleGAN explanations were visually similar to the query CXR, on an average our counterfactual explanations were more aligned with the user's reasoning for the diagnosis as compared to cycleGAN explanation. 
Below, we describe the design of the study, the data analysis methods, along with the results of the experiment in detail.

\textbf{Experiment Design: }We conducted an online survey experiment with 12 diagnostic radiology residents. Participants first reviewed an instruction script, which described the AI system developed to provide an autonomous diagnosis for CXR findings such as cardiomegaly. The study comprised of the radiologists evaluating six CXR images which were presented in random order to them. For selecting these siz CXR, we first, divided the  test-set of the explanation function for cardiomegaly  into three groups, positive ($f(\rvx) \in [0.8, 1.0]$), mild ($f(\rvx) \in [0.4,0.6]$) and negative ($f(\rvx) \in [0.0, 0.2]$). Next, we randomly selected two CXR images from each group. The six CXR images were anonymized as part of the MIMIC-CXR dataset protocol.

For each image, we had the same procedure consisted of a diagnosis tasks, followed by four explanation conditions, and ended by a final evaluation question between the explanation conditions. Further details of the study design are includes in appendix ~\ref{sm-he}.

\emph{Diagnosis:} For each CXR image, we first asked a participant to provide their diagnosis for cardiomegaly. This question ensures that the participants carefully consider the imaging features that helped them diagnose. Subsequently, the participants were presented with the classifier's decision and were asked to provide feedback on whether they agreed.

\emph{Explanation Conditions:}
Next, the study provides the classifier's decision  with the following explanation conditions: 
\begin{enumerate}
    \item \textbf{No explanation (Baseline)}: This condition simply provides the classifier decision without any explanation, and is used as the control condition.
    \item \textbf{Saliency map}: A heat map overlaid on the query CXR, highlighting essential regions for the classifier's decision.
    \item \textbf{CycleGAN explanation}: A video loop over two CXR images, corresponding to the query CXR transformation with a negative and a positive decision for cardiomegaly.
    \item \textbf{Our counterfactual explanation}: A video showing a series of CXR images gradually changing the classifier's decision from negative to positive.
\end{enumerate}

Please note that after showing the baseline condition, we provided the other explanation conditions in random order to avoid any learning or biasing effects.

\emph{Evaluation metrics:} Given the classifier's decision and corresponding explanation, we consider the following metrics to compare different explanation conditions:

\begin{enumerate}

\item \textbf{Understandability}: For each explanation condition, the study included a question to measure whether the end-user understood the classifier's decision, when explanation was provided. The participants were asked to rate agreement with \emph{``I understand how the AI system made the above assessment for Cardiomegaly"}.

\item \textbf{Classifier's decision justification}: Human user's may perceive explanations as the reason for the classifier's decision. For the cycleGAN and our counterfactual explanation conditions, we quantify whether the provided explanation were actually related to the classification task by measuring  the participants' agreement with \emph{``The changes in the video are related to Cardiomegaly"}.
\item \textbf{Visual quality}: The study quantifies the proximity between the explanation images and the query CXR by measuring the participants' agreement with "\emph{Images in the video look like a chest x-ray.}".

\item \textbf{Identity preservation}: The study also measures the extent to which participants think the explanation images correspond to the same subject as the query CXR by measuring the participants' agreement with \emph{``Images in the video look like the chest x-ray from a given subject"}.

\item {\textbf{Helpfulness:}} For each CXR image, we asked the participants to select the most helpful explanation condition in understanding the classifier's decision, \emph{``Which explanation helped you the most in understanding the assessment made by the AI system?"}. This evaluation metric directly compares the different explanation conditions. 
\end{enumerate}
All metrics, but the helpfulness metric were evaluated for agreement  on a 5-point Likert scale, where one means ``\emph{strongly disagree}" and five means ``\emph{strongly agree}".

\emph{Free-form Response:} After each question, we also asked the participants a free-form question: ``\emph{Please explain your selection in a few words.}" We used answers to these questions to triangulate our findings and complement our quantitative metrics by understanding our participants' thought-processes and reasoning.

\emph{Participants.} Our participants include 12 diagnostic radiology residents who have completed medical school and have been in the residency program for one or more years. On average, the participants finished the survey in 40 minutes and were paid \$100 for their participation in the study.

\textbf{Data analysis: }For each evaluation metric, the study asked the same question to the participants while showing them different explanations. For each question, we gather 72 responses (6 - number of CXR images $\times$ 12 - number of participants).

For the understandability and helpfulness metrics, we conducted a one-way ANOVA test to determine if there is a statistically significant difference between the mean metric scores for the four explanation conditions. Specifically, we built a one-way ANOVA with the metric as our dependent variable and analyzed agreement rating as the independent variable. If we found a significant difference in the ANOVA method, we ran Tukey’s Honestly Significant Difference (HSD) posthoc test to perform a pair-wise comparison between different explanation conditions.

We measured the classifier's  decision justification,  visual quality and identity preservation metrics only for the cycleGAN and our counterfactual explanations. We conducted  paired t-tests to compare these evaluation metrics between these two explanation conditions.  We also qualitatively analyzed the participants' free-form responses to find themes and patterns in their responses.

\begin{figure}[!ht]
    \centering
    \includegraphics[width = 0.95\linewidth]
    {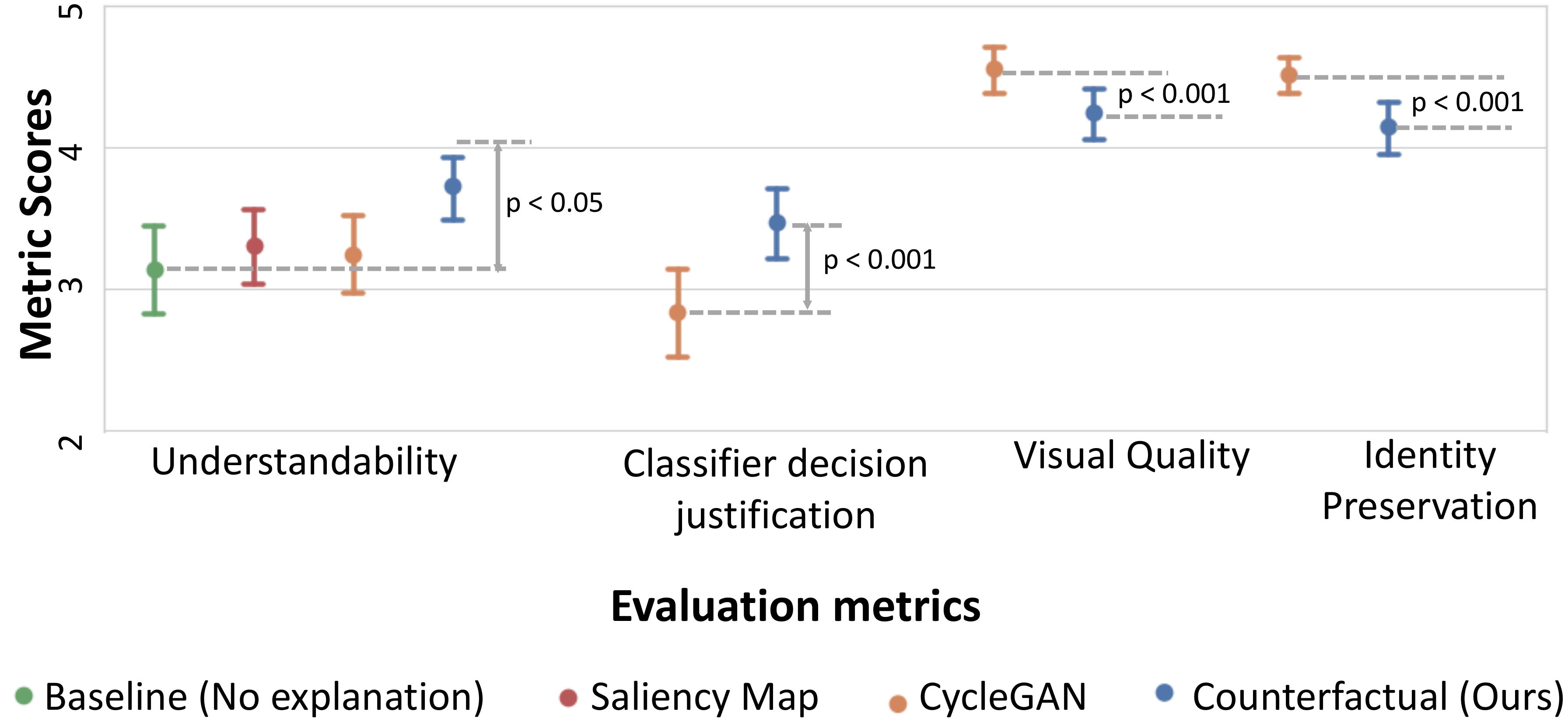}
    \caption{ Comparing the different metrics in human evaluation study.}
   \label{fig_he}
\end{figure}

\textbf{Results: }Fig.~\ref{fig_he} shows the mean score for the evaluation metrics of understandability, classifier's decision justification, visual quality, and identity preservation among the different explanation conditions. Below, we report the statistical analysis for these results, followed by analysis of the participants' free-form responses to understand the reasons behind these results.

\vspace{1.0cm}

\vspace{-0.4cm}

\emph{Understandability:} The results show that our counterfactual explanation was the most understandable explanation to the participants. A one-way ANOVA revealed that there was a statistically significant difference in the understandability metric between at least two explanation conditions (F(3, 284) = [3.39], p=0.019). The Tukey post-hoc test showed that the  understandability metric for our counterfactual explanation was significantly higher than the no-explanation baseline (p = 0.018). However, there was no statistically significant difference in mean scores between other pairs of explanations (refer to Table~\ref{table_anova}, ``Understandability" column). 
This finding indicates that providing our counterfactual explanations along with the classifier's decision made the algorithm most understandable to our clinical participants,  while other explanation conditions, 
 saliency map and cycleGAN failed to achieve significant difference from no-explanation baseline on the understandability metric. Next, we use responses from free-text question to supplement our findings.

 For the no-explanation baseline, the primary reason for poor understanding was the absence of explanation (n=30), (\eg  they stated that \emph{``there is no indication as to how the AI made this decision"}). Interestingly, many responses (n=23) either associated their high understanding with the correct classification decision \ie participants understood the decision as the decision is correct (\emph{``I agree, it is small and normal"}) or they assumed the AI-system is using similar reasoning as them to arrive at its decision (\emph{``I assume the AI is just measuring the width of the heart compared to the thorax", ``Assume the AI measured the CT ratio and diagnosed accordingly."}).
 
 Participants' mostly found saliency maps to be correct but incomplete  (n=23), (\emph{``Unclear how assessment can be made without including additional regions"}). Specifically, for cardiomegaly, the saliency maps were highlighting parts of the heart and not its border (\emph{``Not sure how it gauges not looking at the border"}) or thoracic diameter (\emph{``thoracic diameter cannot be assessed using highlighted regions of heat map"}). We observe a similar result in Fig.~\ref{Fig_saliency}, where the heatmap focuses on the heart but not its border. Further, some participants expressed a concern that they didn't understand how relevant regions were used to derive the decision (\emph{``i understand where it examined but not how that means definite cardiomegaly"}).
 
 For cycleGAN explanation, the primary reason for poor understanding was the minimal perceptible changes between the negative and positive images (n=3), (\emph{``There is no change in the video."}). In contrast, many participant's explicitly reported an improved understanding of the classifier's decision in the presence of our counterfactual explanations (n=33), (\emph{``I think the AI looking at the borders makes sense.", ``i can better understand what the AI is picking up on with the progression video"}).
 
 \emph{Classifier's decision justification}:
 Our counterfactual explanation (M=3.46; SD=1.12) achieved a positive mean difference of 0.63 on this metric as compared to cycleGAN (M=2.83; SD=1.33), with t(71)=3.55 and $p<0.001$. This result indicates that the  participants found a good evidence for the predicted class (cardiomegaly), much frequently  in our counterfactual explanations as compared to cycleGAN.
  
 Most responses (n=25) explicitly mentioned visualizing changes related to cardiomegaly such as an enlarged heart in our explanation video as compared to cycleGAN (n=17). In cycleGAN, many reported that changes in the explanation video was not perceptible (n=23). Further, the participants reported changes in density, windowing level or other attributes which were not related to cardiomegaly (\emph{``Decreasing the density does not impact how I assess for cardiomegaly.", ``they could be or just secondary to windowing the radiograph"}). Such responses were observed in both cycleGAN (n=17) and our explanation (n=17). This indicates that the classifier may have associated such secondary information (short-cuts) with cardiomegaly diagnosis. A more in-depth analysis is required to quantify the classifiers' behaviour.

 \emph{Visual quality and identity preservation}: We observe a negative mean difference of 0.31 and 0.37 between our and cycleGAN explanation methods in visual quality and identity preservation metrics, respectively. The mean score for visual quality was higher for cycleGAN (M=4.55; SD=0.71) as compared to our method (M=4.24; SD=0.80) with t(71)=3.49 and $p<0.001$. Similarly, the mean  score for identity preservation was also higher for cycleGAN (M=4.51; SD=0.56) as compared to our method (M=4.14; SD=0.78) with t(71)=3.96 and $p<0.001$.

  \begin{table}[!ht]
\caption[Results for one-way ANOVA for understandability metric, followed by Tukey's HSD post-hoc test between different levels of agreement.]{Results for one-way ANOVA for understandability metric, followed by Tukey's HSD post-hoc test between different levels of agreement.
Note that the mean value for E4 (our counterfactual explanation) is the highest, indicating that our explanations helped users the most in understanding the classifier's decision. *$p < 0.05$; ***$p<0.0001$.}
\centering
\label{table_anova}
\begin{tabular}{c|c|c|c|c|c}
\multicolumn{3}{c|}{\bf Understandability} & \multicolumn{3}{c}{\bf Helpfulness} \\
\multicolumn{3}{c|}{F(3, 284) = 3.39} & \multicolumn{3}{c}{F(3, 284) = 21.5}\\
\multicolumn{3}{c|}{$p < 0.05$} & \multicolumn{3}{c}{$p < 0.001$}\\
\hline
\multicolumn{2}{c|}{ Explanation method}  &  $p$ & \multicolumn{2}{c|}{ Explanation method}  &  $p$\\
\hline
E1 (No explanation) & E2 & & E1 & E2 & \\
M=3.14  & E3 &  & M=0.05 & E3 & \\
SD=1.39 & E4 & * & SD=0.23 & E4 & ***\\
\hline
E2 (Saliency Map) & E1 & & E2 & E1 & \\
M=3.31 & E3 & & M=0.18 & E3 & \\
SD=1.13 & E4 & & SD=0.39& E4& *** \\
\hline
E3 (CycleGAN) & E1 & & E3 & E1&\\
M=3.24 & E2 & &M=0.16&E2&\\
SD=1.19 & E4 & &SD=0.37&E4& ***\\
\hline
E4 (Our counterfactual  & E1 & * & E4 & E1&***\\
explanation) \bf M=3.72 & E2 & & \bf M=0.24 & E2 & *** \\
SD=0.97 & E3 & &SD=0.42&E3&***\\
\hline
\end{tabular}
\end{table}

 Most of the responses (n=69) agreed that the CycleGAN explanation were marked as highly similar to the query CXR image. These results are consistent with our earlier results, that cycleGAN has better visual quality with a lower FID score (\emph{see} Table.~\ref{FID-table}). However, in some responses, the participants pointed out that the explanation images were almost identical to the query image (\emph{``There's virtually no differences. This is within the spectrum of a repeat chest x-ray for instance."}). An explanation image identical to the query image provides no information about the classifier's decision.  Further, similar looking CXR will also result in similar classification decision, and hence will fail to flip the classification decision. As a result, we also observed a lower agreement in the classifier consistency metric and a lower counterfactual validity score in Table.~\ref{FID-table} for cycleGAN.

 \emph{Helpfulness:} In our concluding question, \emph{``Which explanation helped you the most in understanding the assessment made by the AI system?"}, ~57\% of the responses selected our counterfactual explanation as the most helpful method. A one-way ANOVA revealed that there was a statistically significant difference in the helpfulness metric between at least two explanation conditions (F(3, 284) = [21.5], $p<0.0001$). In pair-wise Tukey's HSD posthoc test,  we found that the mean  usefulness metric for our counterfactual explanations was significantly different from all the rest explanation conditions($p<0.0001$). Table~\ref{table_anova} ( ``Helpfulness" column) summarizes these results.

 These results indicates that the participant's selected our counterfactual explanations as the most helpful form of explanation for understanding the classifier's decision.

\subsection{Bias detection}
\label{bias}
Our model can discover confounding bias in the data used for training the black-box classifier. Confounding bias provides an alternative explanation for an association between the data and the target label. For example, a classifier trained to predict the presence of a disease may make decisions based on hidden attributes like gender, race, or age. In a simulated experiment, we trained two classifiers to identify smiling vs not-smiling images in the CelebA dataset. The first classifier $f_{\text{Biased}}$ is trained on a biased dataset, confounded with gender such that all smiling images are of male faces. We train a  second classifier $f_{\text{No-biased}}$ on an unbiased dataset, with data uniformly distributed with respect to gender. Note that we evaluate both the classifiers on the same validation set. Additionally, we assume access to a proxy Oracle classifier $f_{\text{Gender}}$ that perfectly classifies the confounding attribute \ie gender.

\vspace{0.8cm}

\begin{figure}[!ht]
\centering
\includegraphics[width=0.95\linewidth]{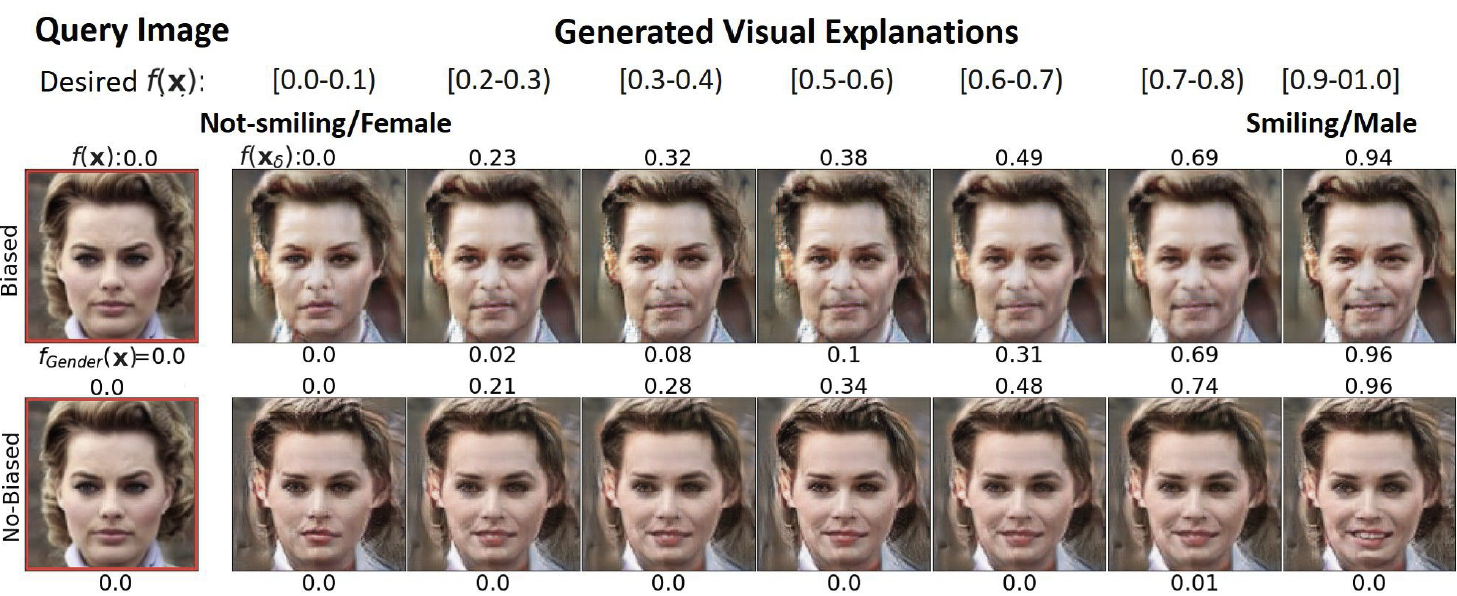}
 
\caption[Explanations for two classifiers, both trained to classify ``Smiling" attribute on CelebA dataset.]{ Explanations for two classifiers, both trained to classify ``Smiling" attribute on CelebA dataset.
For each example, the top row shows results from ``Biased" classifier  whose data distribution is confounded with ``Gender". The bottom row shows explanations from ``No-Biased" classifier  with uniform data distribution w.r.t gender.   The top label indicates output of the classifier and the bottom label is the output of an oracle classifier for the con-founding attribute gender. The visual explanations for the ``Biased" classifier changes the gender as it adds smile on the face. }
\label{Fig_Bias_Detection}
\end{figure}

\vspace{-0.2cm}

As shown in~\cite{Cohen2018DistributionTranslation}, if the training data for the GAN is biased, then the inference would reflect that bias. In
Figure~\ref{Fig_Bias_Detection}, we compare the explanations generated for the two classifiers. The visual explanations for the biased classifier change gender as it increases the amount of smile. We adapted the confounding metric proposed in~\cite{Joshi2018XGEMs:Models} to summarize our results in
Table~\ref{Table_Bias}. Given the data $\mathcal{D} = \{(\rvx_i, y_i, a_i), \rvx_i \in \mathcal{X}, y_i, a_i \in \mathcal{Y}\}$, we quantify that a classifier is confounded by an attribute $a$ if the generated explanation $x_{\rvc}$ has a different attribute $a$, as compared to query image $\rvx$, when processed through the Oracle classifier $f_{\text{a}}$. The metric is formally defined as $\E_{\mathcal{D}}[1(f_{\text{a}}(\rvx_{\rvc} ) \neq f_{\text{a}}(\rvx ))]/|\mathcal{D}|$.  For a biased classifier, the Oracle function predicted the female class for the majority of the images, while the unbiased classifier is consistent with the true distribution of the validation set for gender. Thus, the fraction of generated explanations that changed the confounding attribute ``gender’' was found to be high for the biased classifier.

\begin{table}[!ht]
\begin{center}
\caption[Confounding metric for biased detection.]{Confounding metric for biased detection. For target label ``Smiling'' and  ``Not-Smiling'', the explanations are generated using condition $\rvc > 0.9$  and $\rvc < 0.1$ respectively. The Male and Female values quantifies the fraction of the generated explanations classifier as male or female, respectively by oracle classifier $f_{\text{Gender}}$. The overall value quantifies the fraction of the generated explanations who have different gender as compared to the query image. A small overall value shows least bias.}
\label{Table_Bias}
\begin{tabular}{lll}
{} & \multicolumn{2}{c}{\bf Target Label} \\ 
Black-box classifier & Smiling & Not-Smiling \\
\hline
$f_{\text{Biased}}$  & Male: 0.52  & Male: 0.18\\
 & Female: 0.48 & Female: 0.82\\
 & Overall: \textbf{0.12} & Overall: \textbf{0.35} \\
\hline
$f_{\text{No-biased}}$  & Male: 0.48  & Male: 0.47 \\
 & Female:  0.52 & Female: 0.53\\
 & Overall: 0.07 & Overall: 0.08 \\
 \hline
\end{tabular}
\end{center}
\end{table}

\subsection{Evaluating class discrimination}
In multi-label settings, multiple labels can be true for a given image. A multi-label setting is common in CXR diagnosis. For example, cardiomegaly and pleural effusion are associated with cardiogenic edema and frequently co-occur in a CXR. Please note that our classification model is also trained in a multi-label setting where the fourteen radiological findings may co-occur in a CXR. In this evaluation, we demonstrate the sensitivity of our generated explanations to the task being explained. Ideally, an explanation model trained to explain a given task should produce explanations consistent with the query image on all the other classes besides the given task. Specifically, if we are training a model to explain ``cardiomegaly" then the counterfactual image should flip classification decision only for ``cardiomegaly" class and not for any other class. 
\newline

\begin{figure}[!ht]
    \centering
    \includegraphics[width = 0.95\linewidth]
    {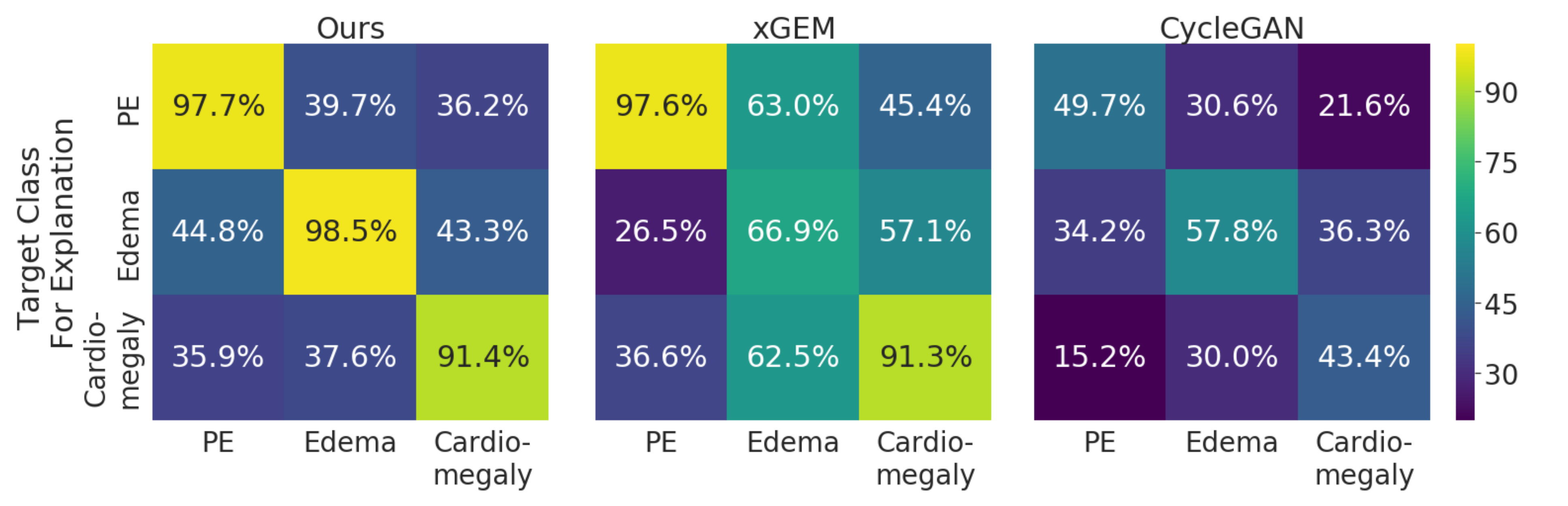}
     
    \caption[Evaluating class discrimination.]{Evaluating class discrimination. Each cell is the fraction of the generated explanations, that have flipped in a class as compared to the query image. The x-axis shows the classes in a multi-label setting, and the y-axis shows the target class for
which an explanation is generated. Note: This is not a confusion matrix. }
    \label{Fig_multi}
\end{figure}

\vspace{-0.4cm}

We considered three diagnosis tasks, cardiomegaly, pleural effusion, and edema. For each task, we trained one explanation model.  Figure.~\ref{Fig_multi} plots the fraction of the generated explanations, that have flipped in other classes as compared to the query image.  In Figure.~\ref{Fig_multi}, each column represents one task, and each row is one run of our method to explain a given task. The diagonal values also represent the counterfactual validity (CV) score reported in Table.~\ref{FID-table}.

\section{Discussion and Conclusion}
We provided  a BlackBox a 
\emph{Progressive Counterfactual Explainer} designed to explain image classification models for medical applications.
 Our framework explains the decision by gradually transforming the input image to its counterfactual, such that the classifier's prediction is flipped. We have formulated and evaluated our framework on three properties of a valid counterfactual transformation: data consistency, classifier consistency, and self-consistency. Our results showed that our framework adheres to all three properties. 
 
 \emph{Comparison with xGEM and cycleGAN: }Our model satisfy all three essential properties of a valid counterfactual explanation. Our model creates natural-looking explanations that produce a desired outcome from the classification model while retaining maximum patient-specific information.  In comparison, both xGEM and cycleGAN failed on at least one essential property. xGEM model fails to create realistic images with a high FID score ($> 300$). Furthermore, the cycleGAN model fails to flip the classifier's decision with a low CV score $(< 60\%)$.

 \emph{xGEM: }The visual quality of images generated by xGEM, is limited by the expressiveness of its generator. xGEM adopted a variational autoencoder (VAE) as the generator. VAE uses a Gaussian likelihood ($\ell_2$ reconstruction), an unrealistic assumption for image data, and is known to produce over-smoothed images~\cite{Huang2018IntroVAE}. In contrast, our model uses an implicit likelihood assumption of GAN~\cite{Miyato2018CGANsDiscriminator}, resulting in realistic explanation images.
 
 \emph{CycleGAN: }The cycleGAN model learns two generator networks to transform an input image into a positive or a negative sample for a given target class.    However, during training,  cycleGAN loss function does not incorporate the external black-box classifier. It primarily follows a data-driven approach to learn all the differences between positive and negative samples. Hence, the cycleGAN model learns to explain the data and not the classification model. As a result, the counterfactual explanations derived from cycleGAN model frequently fails to flip the classification decision, despite their high visual quality, resulting in  inconclusive images that are not counterfactual.
 
 Further, we present a thorough comparison between cycleGAN and our explanation in a human evaluation study.  The clinical experts' expressed high agreement that explanation images from cycleGAN were of high quality and they resembles the query CXR. But at the same time, users found the explanation images to be too similar to query CXR, and the cycleGAN explanations failed to provide the counterfactual reasoning for the decision. In comparison, our explanation were most helpful in understanding the classification decision. Though the users reported inconsistencies in the visual appearance, but the overall sentiment looks positive and they selected our method as their preferred explanation method for improved understandability.

 \emph{Comparison with saliency maps: }
 As compared to saliency maps, counterfactual explanations provide extra information to the end-user to understand the classification decision. Our quantitative experiments show that the region modified by the PCE to create counterfactual image, frequently matches the salient regions highlighted by the saliency-map based explanations models. Also, saliency-map-based explanations may highlight almost the same region for different tasks, resulting in misleading and inconclusive explanations (\emph{see} Figure.~\ref{Fig_saliency}). In contrast,  our counterfactual explanations provide additional information to clarify \emph{how} input features in the important regions could be modified to change the prediction decision.  Our difference map localizes disease to specific regions in the chest, and these regions align with the clinical knowledge of the disease. In Figure.~\ref{Fig_saliency}, our difference map focused on the heart region for cardiomegaly and the CP recess region for PE.

 \emph{Clinical relevance of the explanations: } From a clinical perspective, we demonstrated that the counterfactual changes associated with normal (negative) or abnormal (positive) classification decisions are also associated with corresponding changes in disease-specific metrics such as CTR and SCP.  For example, changes associated with an increased posterior probability for cardiomegaly also resulted in an increased CTR. Similarly, for PE, a healthy CP recess with a high SCP score  transformed into an abnormal CP recess with blunt CPA, as the posterior probability for PE increases (\emph{see} Figure.~\ref{Fig_Quality_ch3} and Figure.~\ref{Fig_Box}).  
 
 To the best of our knowledge, ours is the first attempt to quantify model explanations using clinical metrics.  At the same time, our evaluation has certain limitations. Our automatic pipeline to compute CTR and SCP suffers from inaccuracies.  This contributed to the large variance in difference plots in Figure.~\ref{Fig_Box}. These inaccuracies are due to the sub-optimal performance of the segmentation and object detector networks. In the absence of ground truth annotations for lung and heart segmentation and limited annotations for the CP recess region, these networks have sub-optimal performance.
Nevertheless, our goal is not to compute these metrics correctly for each image but to perform a population-level analysis. In our experiments, CTR and SCP successfully captured the difference between normal and abnormal CXR for cardiomegaly and PE, respectively. One may argue using CTR and SCP to perform disease classification. However, models based on these features will also suffer from similar inaccuracies, resulting in poor performance and generalization compared to the DL methods.

Defining clinical metrics for different diseases is a challenging task. For example, consider edema. It may appear as different radiographic concepts (\eg cephalization, peribronchial cuffing, perihilar batwing appearance, and opacities \etc) in different patients~\cite{Milne1985TheEdema}. Transforming a healthy CXR to a counterfactual image for edema introduce changes in multiple such concepts.  Future research should determine appropriate metrics to quantify and understand these concepts. Manual annotation is one solution for obtaining ground truth to train models that can identify concepts. Efforts should be made to reduce the dependency on manual labelling as it is expensive and not scalable. 

 \emph{Usability of explanations: } Counterfactual explanations can help in model auditing and recovering hidden bias in the classifier's training. In our experiments, we visualize counterfactual explanations from a biased classifier and contrast it with a classifier without any data bias. Further, using human evaluation we demonstrate the prospective use-case for counterfactual explanations.

We acknowledge that our GAN-generated counterfactual explanations may have missing details such as small wires. In our extended experiments, we found that the foreign objects such as pacemaker have minimal importance in the classification decision (\emph{see} appendix ~\ref{SM-ASPM}). We attempted to improve the preservation of such information through our revised context-aware reconstruction loss (CARL). However, even with CARL, the FO preservation score is not perfect.  A possible reason for this gap is the limited capacity of the object detector used to calculate the FOP score. Training a highly accurate FO detector is outside the scope of this study.

Further, a resolution of $256 \times 256$ for counterfactually generated images is smaller than a standard CXR. Small resolution limits the evaluation for fine details by both the algorithm and the interpreter. Our formulation of cGAN uses conditional-batch normalization (cBN) to encapsulate condition information while generating images. For efficient cBN, the mini-batches should be class-balanced. To accommodate high-resolution images with smaller batch sizes, we must decrease the number of conditions to ensure class-balanced batches. Fewer conditions resulted in a coarse transformation with abrupt changes across explanation images. In our experiments, we selected the smallest bin width, which created a class-balanced batch that fits in GPU memory and resulted in stable cGAN training.  However, with the advent of larger-memory GPUs, we intend to apply our methods to higher resolution images in future work; and assess how that impacts interpretation by clinicians.

To conclude, this study uses counterfactual explanations as a way to audit a given black-box classifier and evaluate whether the radio-graphic features used by that classifier have any clinical relevance. In particular, the proposed model did well in explaining the decision for cardiomegaly and pleural effusions and was corroborated by an experienced radiology resident physician.
By providing visual explanations for deep learning decisions, radiologists better understand the causes of its decision-making. This is essential to lessen physicians' concerns regarding the ``BlackBox" nature by an algorithm and build needed trust for incorporation into everyday clinical workflow.  As an increasing amount of artificial intelligence algorithms offer the promise of everyday utility, counterfactually generated images are a promising conduit to building trust among diagnostic radiologists.

By providing counterfactual explanations, our work opens up many ideas for future work. Our framework showed that valid counterfactual can be learned using an adversarial generative process, that is regularized by the classification model. However, counterfactual reasoning is incomplete without a causal structure and explicitly modeling of the interventions. An interesting next step should explore incorporating or discovering plausible causal structures and creating explanations that are grounded with them.

  \chapter{Concept-based Counterfactual Explanation}
\label{ch3}
\section{Introduction}

Machine Learning, specifically, Deep Learning (DL) methods are increasingly adopted in healthcare applications. Model explainability is essential to build trust in the AI system~\cite{glass2008toward} and to receive clinicians' feedback. Standard explanation methods for image classification delineates regions in the input image that significantly contribute to the model's outcome~\cite{Selvaraju2017Grad-cam:Localization,Lundberg2017APredictions,ribeiro2016should}.  However, it is challenging to explain \emph{how} and \emph{why} variations in identified regions are relevant to the model's decision. Ideally, an explanation should resemble the decision-making process of a domain expert. This paper aims to map a DL model's neuron activation patterns to the radiographic features and constructs a simple rule-based model that partially explains the Black-box.

Methods based on feature attribution have been commonly used for explaining DL models for medical imaging~\cite{basu2020deep}. However, an alignment between feature attribution and radiology concepts is difficult to achieve, especially when a single region may correspond to several radiographic concepts. Recently, researchers have focused on providing explanations in the form of human-defined concepts~\cite{Kim2017InterpretabilityTCAV,bau2017network,Zhou2018InterpretableExplanation}. In medical imaging, such methods have been adopted to derive an explanation for breast mammograms ~\cite{yeche2019ubs}, breast histopathology~\cite{graziani2020concept} and cardiac MRIs~\cite{gloabl_local}. A major drawback of the current approach is their dependence on explicit concept-annotations, either in the form of a representative set of images~\cite{Kim2017InterpretabilityTCAV} or semantic segmentation~\cite{bau2017network}, to learn explanations. Such annotations are expensive to acquire, especially in the medical domain. 
We use weak annotations from radiology reports to derive concept annotations. Furthermore, these methods measure correlations between concept perturbations and classification predictions to quantify the concept's relevance. However, the neural network may not use the discovered concepts to arrive at its decision.  We borrow tools from causal analysis literature to address that drawback~\cite{NEURIPS2020_92650b2e}.

In this work, we used radiographic features mentioned in radiology reports to define concepts. Using a National Language Processing (NLP) pipeline, we extract weak annotations from text and classify them based on their positive or negative mention~\cite{irvin2019chexpert}. Next, we use sparse logistic regression to identify sets of hidden-units correlated with the presence of a concept. To quantify the causal influence of the discovered concept-units on the model's outcome, we view concept-units as a \emph{mediator} in the treatment-mediator-outcome framework~\cite{imai2011commentary}. Using measures from mediation analysis, we provide an effective ranking of the concepts based on their causal relevance to the model's outcome. Finally, we construct a low-depth decision tree to express discovered concepts in simple decision rules, providing the global explanation for the model. The rule-based nature of the decision tree resembles many decision-making procedures by clinicians.

\section{Method}
We consider a pre-trained \emph{black-box} classifier $f: \rvx \rightarrow \rvy$ that takes an image $\rvx$ as input and process it using a sequence of hidden layers to produce a final output $\rvy \in \mathbb{R}^D$. Without loss of generality, we decompose function $f$ as $\Phi_2 \circ \Phi_1(\rvx)$, where $\Phi_1(\rvx) \in \mathbb{R}^L$ is the output of the initial few layers of the network and $\Phi_{2}$ denotes the rest of the network. 
We assume access to a dataset $\mathcal{X} = \{(\rvx_n, \rvy_n, \rvc_n)\}^N$, where $\rvx_n$ is input image, $\rvy_n$ is a $d$-dimensional one-hot encoding of the class labels and $\rvc_n \in \mathbb{R}^K$ is a $k$-dimensional concept-label vector. We define concepts as the radiographic observations mentioned in radiology reports to describe and provide reasoning for a diagnosis. We used a NLP pipeline~\cite{irvin2019chexpert} to extract concept annotations. The NLP pipeline follows a rule-based approach to extract and classify observations from the free-text radiology report. The extracted $k^{th}$ concept-label $\rvc_n [k]$ is either 0 (negative-mention), 1(positive-mention) or -1 (uncertain or missing-mention).

\begin{figure}[!ht]
  \centering
        \includegraphics[width=0.89\textwidth]{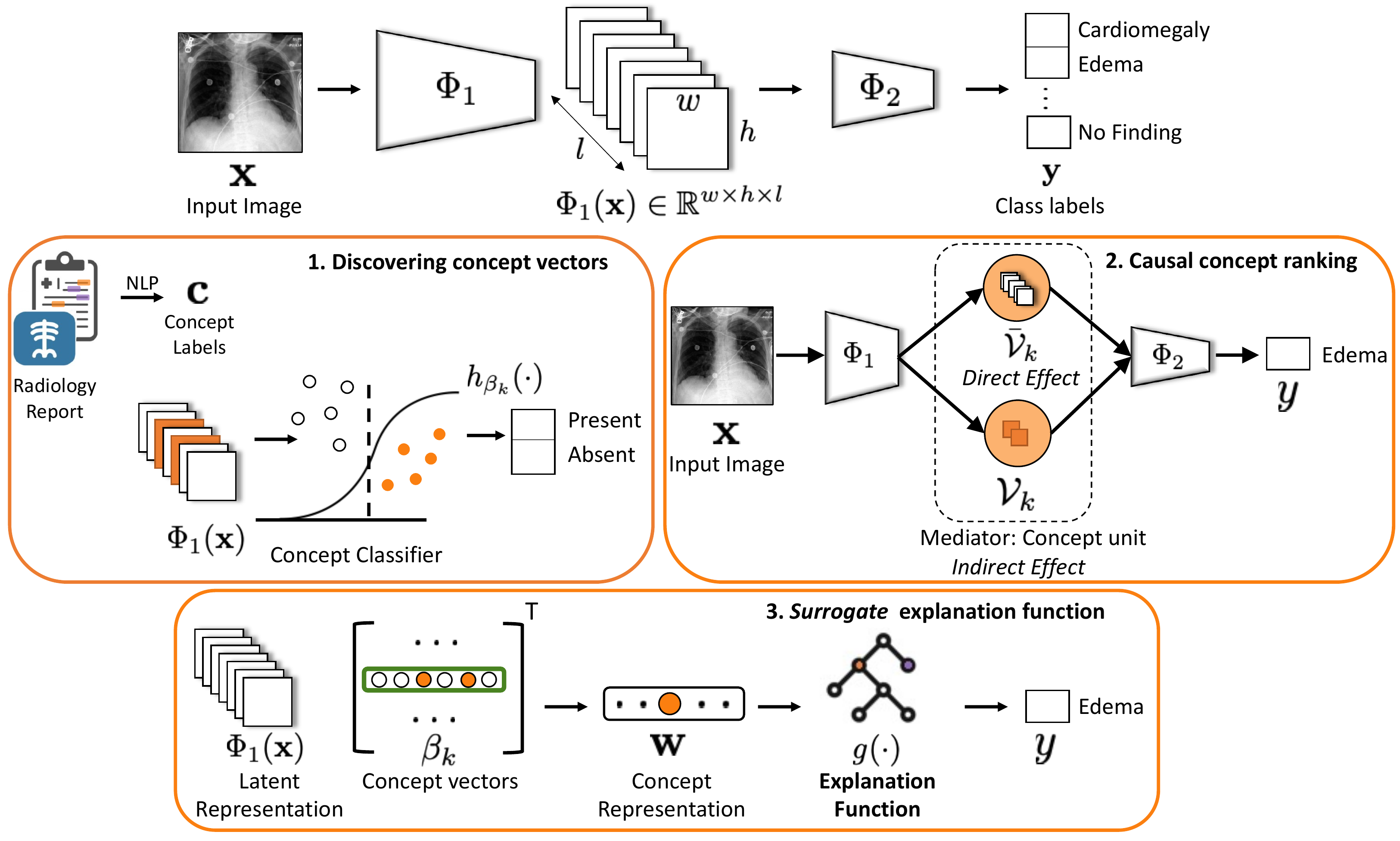}
        \caption[Method overview for concept-based counterfactual explanations.]{Method overview for concept-based counterfactual explanations.
        We provide explanation for the black-box function $f(\rvx)$ in-terms of concepts, that are radiographic observations mentioned in radiology reports. 1) The intermediate representation  $\Phi_1(\rvx)$ is used to learn a sparse logistic regression $h_{\rvv_k}(\cdot)$ to classify $k^{th}$ concept. 2) The non-zero coefficients of $\rvv_k$ represents a set of concept units $\mathcal{V}_k$ that serves as a mediator in the causal path connecting input $\rvx$ and outcome $y$. 3) A decision tree function is learned to map concepts to class labels.}
        \label{fig:model}
    \end{figure}

An overview of our method is shown in Fig.~\ref{fig:model}. Our  method consists of three sequential steps: 

(1) \emph{Concept associations}: We seek to discover sparse associations between concepts and the hidden-units of $f(\cdot)$. We express $k^{th}$ concept as a sparse vector $\rvv_{k} \in \mathbb{R}^L$ that represents a linear direction in the
intermediate space $\Phi_1(\cdot)$.

  (2) \emph{ Causal concept ranking}: Using tools from causal inference, we find an effective ranking of the concepts based on their relevance to the classification decision. Specifically, we consider each concept as a mediator in the causal path between the input and the outcome. We measure concept relevance as the effect of a counterfactual intervention on the outcome that passes indirectly through the concept-mediator.
       
  (3) \emph{Surrogate explanation function}: We learn an  easy-to-interpret function $g(\cdot)$  that mimics function $f(\cdot)$ in its decision. Using $g(\cdot)$, we seek to learn a global explanation for $f(\cdot)$ in terms of the concepts.

  \vspace{0.8cm}

    \subsection{Concept associations}
    We discover concept associations with intermediate representation $\Phi_1(\cdot)$ by learning a binary classifier that maps $\Phi_1(\rvx)$ to the concept-labels~\cite{Kim2017InterpretabilityTCAV}. We treat each concept as a separate binary classification problem and extract a representative set of images $\mathcal{X}^{k}$, in which concept $c_n[k]$ is present and a random negative set. We define concept vector ($\rvv_{k}$) as the solution to the logistic regression model $ c_n[k] = \sigma(\rvv_k^T \vect(\Phi_1(\rvx_n))) + \epsilon$, where $\sigma(\cdot)$ is the sigmoid function. For a convolutional neural network, $\Phi_1(\rvx) \in \mathbb{R}^{w \times h \times l}$ is the output activation of a convolutional layer with width $w$, height $h$ and number of channels $l$. We experimented with two vectorization for $\Phi_1$. In first, we flatten $\Phi_1(\rvx)$ to be a $whl$-dimensional vector. In second, we applied a spatial aggregation by max-pooling along the width and height to obtain $l$-dimensional vector. Unlike TCAV~\cite{Kim2017InterpretabilityTCAV} that uses linear regression, we  used lasso regression to enable sparse feature selection and minimize the following loss function,
    \newline
 \begin{equation}\label{eq:matrixeqn2}
\min_{\rvv_k} \sum_{\rvx_n \in \mathcal{X}_k}  \ell(h_{\rvv_k}(\rvx), c_n[k]) + \lambda ||\rvv_{k}||_1
   \end{equation}  
   \newline
    where $\ell(\cdot,\cdot)$ is the cross entropy loss, $h_{\rvv_k}(\rvx) = \sigma(\rvv_k^T \vect(\Phi_1(\rvx_n)))$ and $\lambda$ is the regularization parameter. 
We performed 10-fold nested-cross validation to find $\lambda$ with least error. The non-zero elements in the concept vector $\rvv_{k}$ forms the set of hidden units ($\mathcal{V}_{k}$) that are most relevant to the $k^{th}$ concept. 
%e call these hidden units as concept-units .

% repeated lasso regression multiple times with different samples from $\mathcal{X}_{c_k}$ and choose the final concept-units through majority voting.
%\[\mathcal{V}_{c_k} = \{m; v_{c_k}^{(m)} \neq 0 \}\]

  \subsection{Causal concept ranking}
% We quantify the relevance of a concept to the classification outcome, based measuring the causal effect of a concept on the outcome
 Concept associations identified hidden units that are strongly correlated with a concept. However, the neural network may or may not use the discovered concepts to arrive at its decision. We use tools from causal inference, to quantify what fraction of the outcome is mediated through the discovered concepts. 
 
 To enable causal inference, we first define  \textit{counterfactual} $\rvx'$ as a perturbation of the input image $\rvx$ such that the decision of the classifier is flipped. Following the approach proposed in Chapter 3, we used the Progressive Counterfactual Explainer (PCE), a conditional generative adversarial network (cGAN) to learn the counterfactual perturbation. We conditioned  on the output of the classifier, to ensure that cGAN learns a classifier-specific perturbation for the given image $\rvx$. Next, we used theory from causal mediation analysis to causally relate a concept with the classification outcome. Specifically, %, while using the desired change in classifier outcome as the condition.} 
 %\textit{do}-operation on the input image. \textit{do}($\rvx$) denotes an intervention on input $\rvx$ to create a counterfactual 
 %such that prediction of the classifier flips as a result of the intervention. 
 %We obtain counterfactual for $\rvx$, using the counterfactual generation process proposed in \cite{Singla2020ExplanationExaggeration}. They trained a conditional generative adversarial network (cGAN) to learn a  perturbation of the input image such that the resulting image produces a desired outcome from the black-box classifier. 
 we consider concept as a mediator in the causal pathway from the input $\rvx$ to the outcome $\rvy$. We specify following effects to quantify the causal effect of the counterfactual perturbation and the role of a mediator in transferring such effect, 
 
 \begin{enumerate}
     \item Average treatment effect (ATE): ATE  is the  total change in the classification outcome $\rvy$ as a result of the counterfactual perturbation. 
     \item Direct effect (DE): DE is the effect of the counterfactual perturbation that comprises of any causal mechanism that \textit{do not} pass through a given mediator. It captures how the perturbation of input image changes classification decision directly, without considering a given concept.
      \item Indirect effect (IE): IE is the effect of the  counterfactual perturbation which is mediated by a set of mediators. It captures how the  perturbation of input image changes classification decision indirectly through a given concept.
 \end{enumerate}
 
\begin{figure}[!ht]
        \centering
        \includegraphics[width=0.9\textwidth]{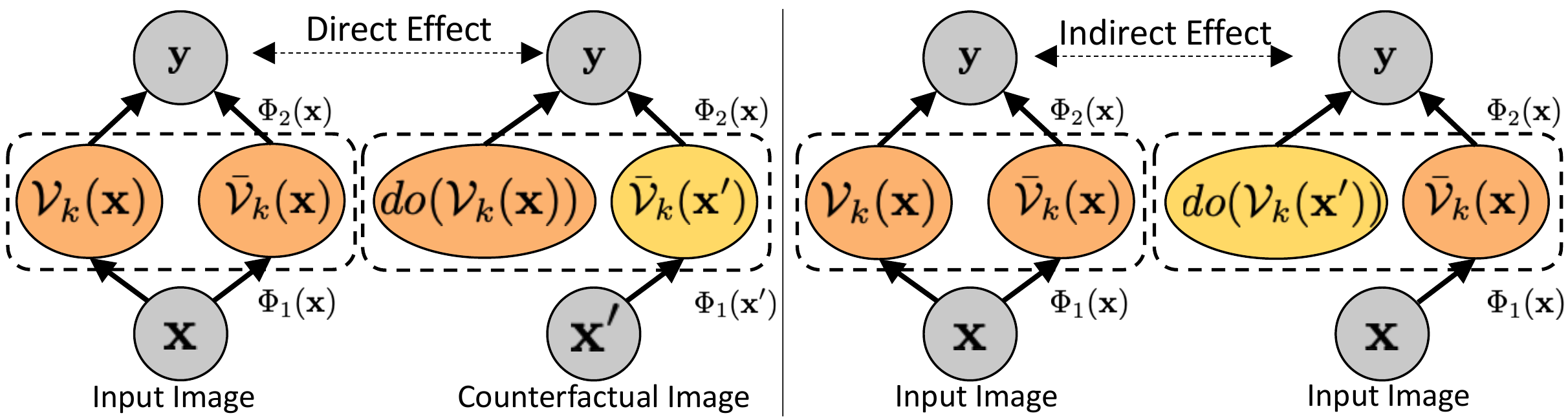}
        \caption{ Illustration of direct and indirect effects in causal mediation analysis.}
        \label{fig:cma1}
    \end{figure}

 Following the potential outcome framework from \cite{rubin1974estimating,NEURIPS2020_92650b2e}, we define the ATE as the proportional difference between the factual and the counterfactual classification outcome,
 \newline
  \begin{equation}\label{eq:ate}
\textbf{ATE} =  \mathbb{E}\big[\frac{f(\rvx')}{f(\rvx)}-1\big].
   \end{equation}   
 \newline
  To enable causal inference through a mediator, we  borrow Pearl’s definitions of natural direct and indirect effects~\cite{pearl2001direct} (\emph{ref} Fig.~\ref{fig:cma1}). We consider set of concept-units $\mathcal{V}_{k}$ as a mediator, representing the $k^{th}$ concept. We decompose the latent representation $\Phi_1(\rvx)$ as concatenation of response of concept-units $\mathcal{V}_k(\rvx)$ and rest of the hidden units $\Bar{\mathcal{V}}_k(\rvx)$ \ie   $\Phi_1(\rvx) = [\mathcal{V}_k (\rvx) , \Bar{\mathcal{V}_k}(\rvx)]$. We can re-write classification outcome as $f(\rvx) = \Phi_2(\Phi_1(\rvx)) = \Phi_2([\mathcal{V}_k(\rvx), \Bar{\mathcal{V}_k}(\rvx)])$.
 To disentangle the direct effect from the indirect effect, we use the concept of \textit{do}-operation on the unit level of the learnt network. Specifically, we use 
   \textit{do}($\mathcal{V}_k(\rvx)$) to denote that we set the value of the concept-units to the value obtained by using the original image as input. By intervening on the network and setting the value of the concept units, we can compute the direct effect as the proportional difference between the factual and the counterfactual classification outcome, while holding mediator \ie $\mathcal{V}_k$ fixed to its value before the perturbation,
 \newline
    \begin{equation}\label{eq:de}
\textbf{DE} =  \mathbb{E} \big[\frac{\Phi_2([do(\mathcal{V}_k(\rvx)), \Bar{\mathcal{V}}_k(\rvx')])}{\Phi_2([\mathcal{V}_k(\rvx) , \Bar{\mathcal{V}}_k(\rvx)])}-1\big].
   \end{equation} 
   \newline
   
    \vspace{-0.5cm}

  We compute indirect effect as the expected change in the outcome, if we change  the mediator from its original value to its value using counterfactual, while holding everything else fixed to its original value,
  \newline
  \begin{equation}\label{eq:ide}
\textbf{IE} = \mathbb{E} \big[\frac{\Phi_2([do(\mathcal{V}_k(\rvx')) , \Bar{\mathcal{V}}_k(\rvx)]) }{\Phi_2([\mathcal{V}_k(\rvx) , \Bar{\mathcal{V}}_k(\rvx)])}-1\big].
 \end{equation}
 \newline
 %Specifically, we used causal mediation analysis (CMA)~\cite{robins1992identifiability,pearl2001direct} to measure the causal effect of a concept on an outcome. 
If the perturbation has no effect on the mediator, then the causal indirect effect will be zero. Finally, we use the indirect effect associated with a concept, as a measure of its relevance to the classification decision. 
 
 \subsection{Surrogate explanation function}
We aim to learn a surrogate function $g(\cdot)$, such that it reproduces the outcome of the function $f(\cdot)$  using an interpretable and straightforward function.
We formulated $g(\cdot)$ as a decision tree as many clinical decision-making procedures follow a rule-based pattern. We summarize the internal state of the function $f(\cdot)$ using output of $k$ concept regression functions $h_{\rvv_k}(\cdot)$ as follows, 
\newline
 \begin{equation}\label{eq:matrixeqn}
 \rvw_n = [\text{logit}(h_{\rvv_1}(\rvx_n)) , \text{logit}(h_{\rvv_2}(\rvx_n)) , \cdots ].
   \end{equation}
   \newline
Next, we fit a decision tree function, $g(\cdot)$, to mimic the outcome of the function $f(\cdot)$ as,
\newline
 \begin{equation}\label{eq:matrixeqn1}
 g^* = \arg\min_g \sum_n\mathcal{L}(g(\rvw_n), f(\rvx_n)),
   \end{equation}
   \newline
where $\mathcal{L}$ is the splitting criterion based on minimizing entropy for  highest information gain from every split. 
 
% We further quantify the extend to which our explanation function $g(\rvw)$ approximates the decision of $f(\rvx)$. We adapt concept completeness score to quantify the part of the decision function $f(\cdot)$ that is expressible in-terms of concepts~\cite{yeh2019completeness} as, 
%\begin{equation}
 %   \eta_f(\rvv_{1}, \cdots, %\rvv_{k}) = \frac{ %\mathbb{P}_{\rvx, \rvy \sim %\mathcal{X}_{test}}[y = %\arg \max_{y'}g(\rvw)]-\alp%ha_r}{\mathbb{P}_{\rvx, %\rvy \sim %\mathcal{X}_{test}}[y = %\arg \max_{y'}f(\rvx)]-\alp%ha_r},
%\end{equation}
%where $\mathbb{P}_{\rvx, \rvy \sim \mathcal{X}_{test}}[y = \arg \max_{y'}g(\rvw)]$ is the accuracy achieved by the surrogate model $g(\cdot)$ and $\alpha_r$ is the accuracy of random prediction, which forms the lower bound of the completeness score to 0. 

\section{Experiments and Results}
\subsection{Study cohort and imaging dataset}
We perform experiments on the MIMIC-CXR~\cite{Johnson2019MIMIC-CXRReports} dataset, which is a multi-modal dataset consisting of 473K chest X-ray images and 206K reports. The dataset is labeled for 14 radiographic observations, including 12 pathologies. We used state-of-the-art DenseNet-121~\cite{Huang2016DenselyNetworks} architecture for our classification function~\cite{irvin2019chexpert}. DenseNet-121 architecture is composed of four dense blocks. We experimented with three versions of $\Phi_1(\cdot)$ to represent the network until the second, third, and fourth dense block. For concept annotations, we considered radiographic features that are frequently mentioned in radiology reports in the context of labeled pathologies. Next, we used Stanford CheXpert~\cite{irvin2019chexpert} to extract and classify these observations from free-text radiology reports.

\subsection{Experimental setup}

We first evaluated the concept classification performance and visualized  concept-units to demonstrate their effectiveness in localizing a concept.   Next, we summarized the indirect effects associated with different concepts across different layers of the classifier. We evaluated a proposing ranking of the concepts based on their causal contribution to the classification decision. Finally, we used  the top-ranked concepts to learn a surrogate explanation function in the form of a decision tree.

\subsection{Evaluation of concept classifiers}

\begin{figure}[!ht]
        \centering
        \includegraphics[width=0.9\textwidth]{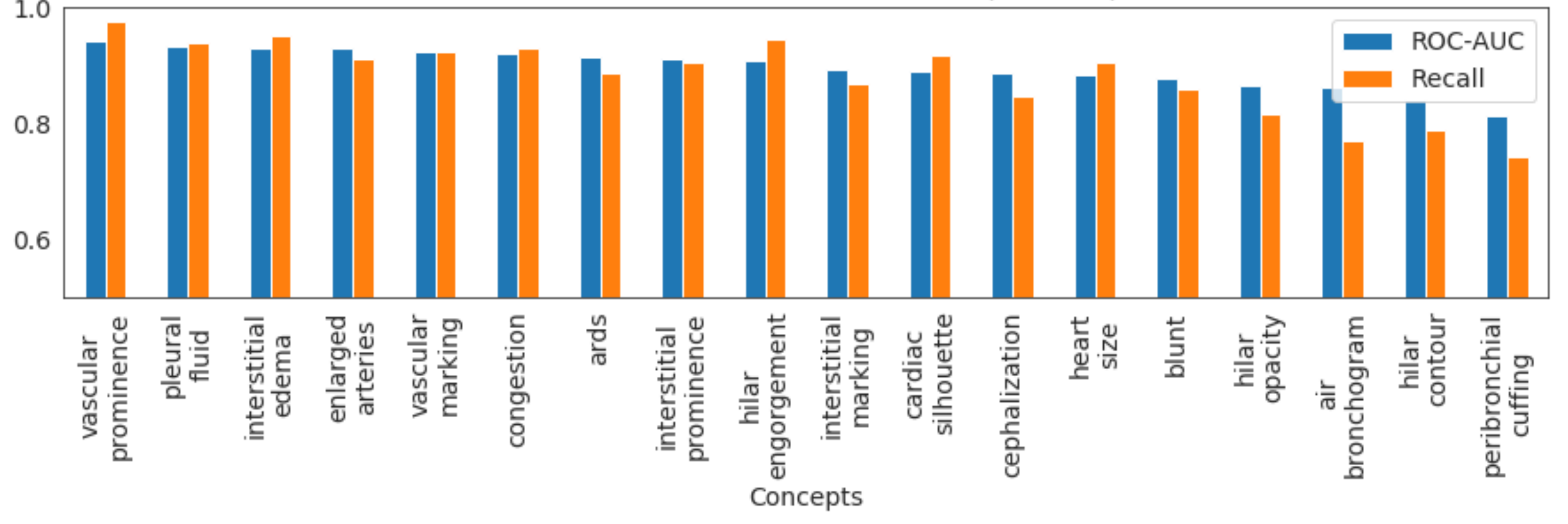}
        
        \caption{ AUC-ROC and recall metric for different concept classifiers.}
        \label{fig:lasso}
    \end{figure}

The intermediate representations from third dense-block consistently outperformed other layers in concept classification. In Fig.~\ref{fig:lasso}, we show the testing-ROC-AUC and recall metric for different concept classifiers. All the concept classifiers achieved high recall, demonstrating a low false-negative (type-2) error. 

\vspace{0.8cm}

    \vspace{-0.5cm}
    
\begin{figure}[!ht]
        \centering
        \includegraphics[width=0.95\textwidth]{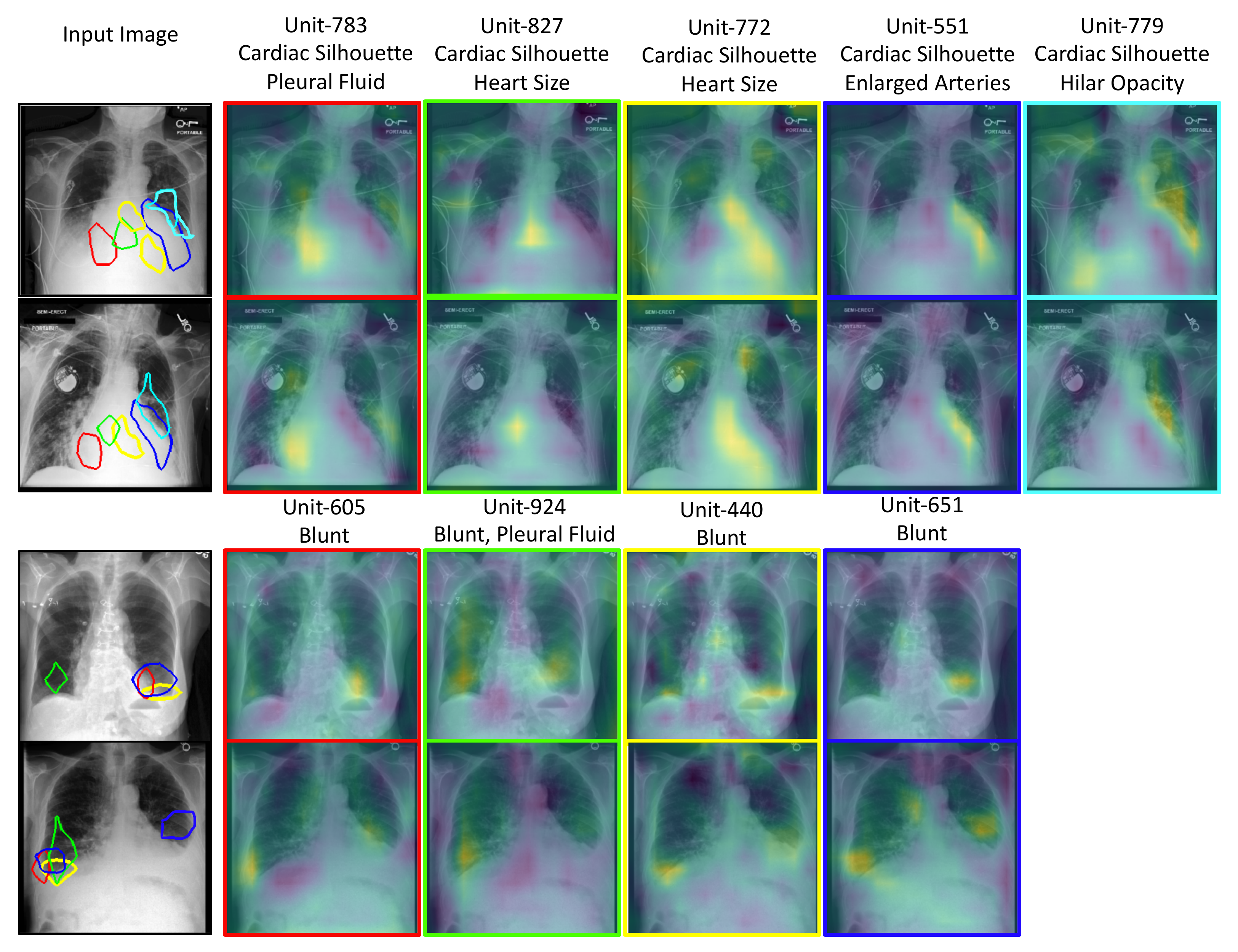}
       
        \caption[A qualitative demonstration of the activation maps of the hidden units that act as visual concept detectors.]{A qualitative demonstration of the activation maps of the hidden units that act as visual concept detectors. Each column represents one hidden unit identified as part of concept vector $\mathcal{V}_k$. Top two rows show $k=$ \textit{cardiac-silhouette} and bottom rows have $k=$\textit{blunt costophrenic angle}.}
        \label{fig:as}
    \end{figure}

 In Fig.~\ref{fig:as},  we visualize the activation map of hidden units associated with the concept vector $\mathcal{V}_k$. For each concept, we visualize hidden units that have large logistic regression-coefficient ($\beta_k$). To highlight the most activated region for a unit, we threshold activation map by the top 1\% quantile of the distribution of the selected units' activations~\cite{bau2017network}. Consistent with prior work \cite{bau2020understanding}, we observed that several hidden units have emerged as concept detectors, even though concept labels were not used while training $f$. For \textit{cardiac-silhouette}, different hidden units highlight different regions of the heart and its boundary with the lung. For localized concept such as \textit{blunt costophrenic angle}, multiple relevant units were identified that all focused on the lower-lobe regions. Same hidden unit can be relevant for multiple concepts. The top label in Fig.~\ref{fig:as}. shows the top two important concepts for each hidden unit.
 
 \vspace{0.5cm}
 
 %associated with concepts.
 
 %for a given concept, we visualize the activation map of hidden units with logistic regression-coefficient ($\beta_k$).

 %For a given concept, we visualize top hidden units sorted based on logistic regression-coefficient ($\beta_k$).
 
 %that have with large coefficient learned in logistic regression. We show the top-activated images for the selected hidden unit. 

  \vspace{-1.3cm}
  
\subsection{Evaluating causal concepts using decision tree as surrogate function}

\begin{figure}[!ht]
        \centering
        \includegraphics[width=1.0\textwidth]{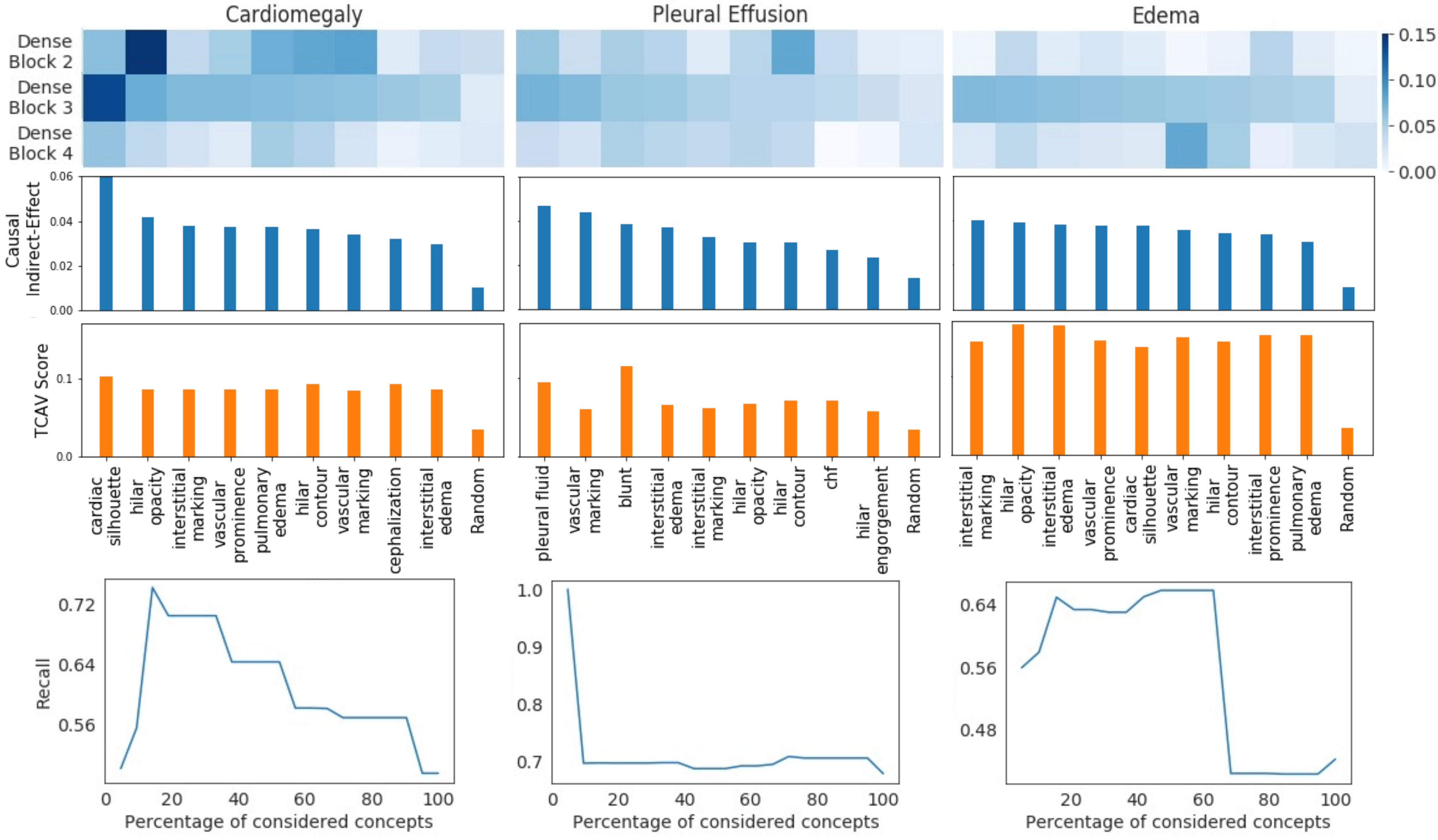}
       
        \caption[Evaluating concept vectors and their causal effect.]{Evaluating concept vectors and their causal effect.
        Indirect effects of the concepts, calculated over different layers of the DenseNet-121 architecture (heat-map).
        The derived  ranking of the concepts based on their causal relevance to the diagnosis (bar-graph). A comparative ranking based on concept sensitivity score from TCAV~\cite{Kim2017InterpretabilityTCAV}. The trend of recall metric for the decision tree function $g(\cdot)$, while training using top x\% of top-ranked concepts (trend-plot).}
        \label{fig:cma}
    \end{figure}

%In-direct effect captures the strength of the causal relationship between a mediator (concept) and the outcome (class-label). 
We evaluate the success of the counterfactual intervention by measuring average total effect (ATE). High values for ATE confirms that counterfactual image generated by~\cite{Singla2020ExplanationExaggeration} successfully flips the classification decision. We achieved an ATE of 0.97 for cardiomegaly, 0.89 for pleural effusion and 0.96 for edema. In Fig.~\ref{fig:cma} (heat-map), we show the distribution of the indirect effect associated with concepts, across different layers. The middle layer demonstrates a large indirect effect across all concepts. This shows that the hidden units in dense-block 3 played a significant role in mediating the effect of counterfactual intervention. 
\newline

 \vspace{-0.5cm}

 In Fig.~\ref{fig:cma} (bar-graph), we rank the concepts based on their indirect effect. The top-ranked concepts recovered by our ranking are consistent with the radiographic features that clinicians associates with the examined three diagnoses~\cite{karkhanis2012pleural,Milne1985TheEdema,nakamori1991effect}. Further, we used the concept sensitivity score from TCAV~\cite{Kim2017InterpretabilityTCAV} to rank concepts for each diagnosis. The top-10 concepts identified by our in-direct effect and TCAV are the same, while their order is different. The top-3 concepts are also the same, with minor differences in ranking. Both the methods have low importance score for random concept. This confirms that the trend in importance score is unlikely to be caused by chance. For our approach, random concept represents an ablation of the concept-association step. Here, rather than performing lasso regression to identify relevant units, we randomly select units. 
  \newline
     \begin{figure}[!ht]
\centering
        \includegraphics[width=0.75\textwidth]{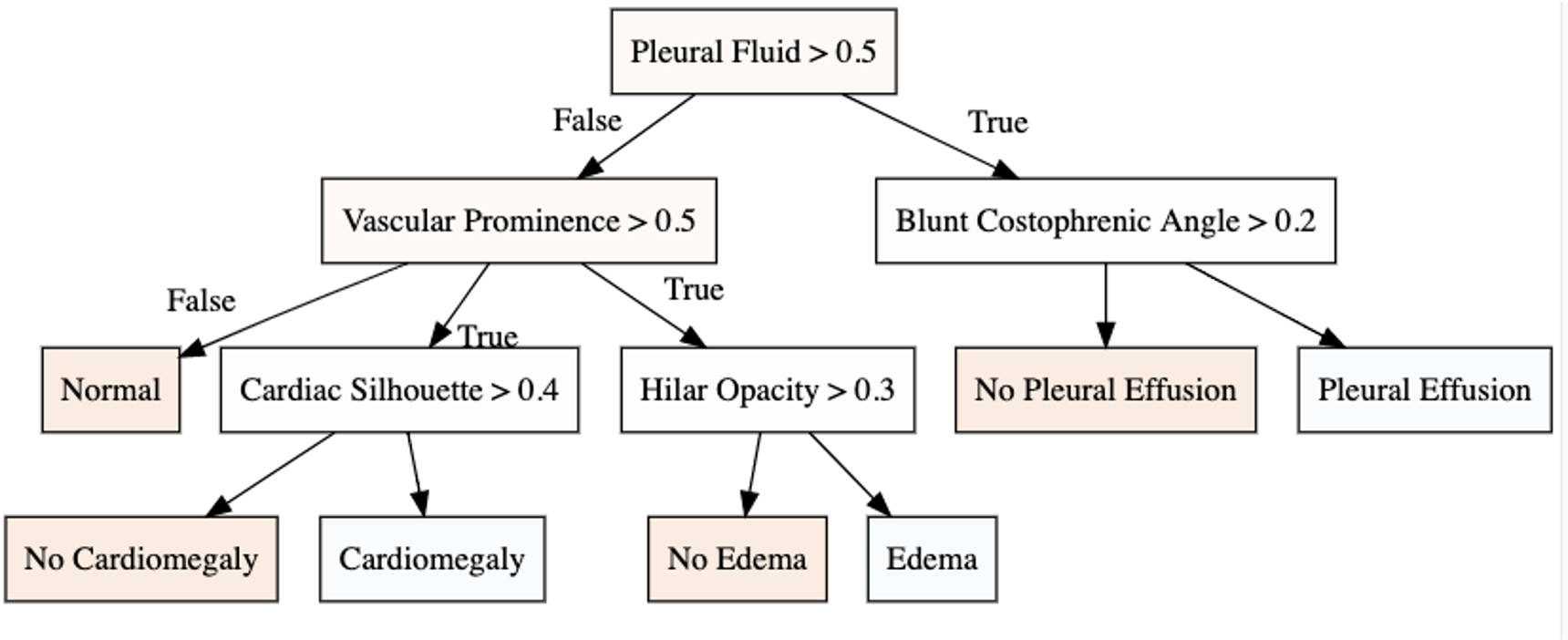}
        \caption{ The decision tree for the three diagnosis with best performance on recall metric.}
        \label{fig:dt1}
\end{figure}

\vspace{-0.5cm}

 To quantitatively demonstrate the effectiveness of our ranking, we iteratively consider $x\%$ of top-ranked concepts and retrain the explanation function $g(\rvw)$. In Fig.~\ref{fig:cma} (bottom-plot), we observe the change in recall metric for the classifier $g(\cdot)$ as we consider more concepts. In the beginning, as we add relevant concepts, the true positive rate increases resulting in a high recall. However, as less relevant concepts are considered, the noise in input features increased, resulting in a lower recall. Fig.~\ref{fig:dt1} visualize the decision tree learned for the best performing model. %Finally, we summarize our explanation framework in Fig.~\ref{fig:dt_ex}.  

 %We achieved a concept completeness score ($\eta_f$) of 0.17 for Cardiomegaly, 0.25 for Pleural effusion and 0.26 for Edema. In future work, we would consider more concepts to improve $\eta_f$.

    % The decision tree provides global explanation for a class label. $\rvv_c$ provides local explanation for a sample $\rvx$ in terms of important concepts for current sample.

 \section{Discussion and Conclusion}
 Model explainability is essential for the creation of trustworthy Machine Learning models in healthcare. An ideal explanation resembles the decision-making process of a domain expert and is expressed using concepts or terminology that is meaningful to the clinicians.
To provide such explanation, we grounded our explanation in terms of clinically relevant concepts that are causally influencing the model's decision. We first associate the hidden units of the classifier to clinically relevant concepts. We take advantage of radiology reports accompanying the chest X-ray images to define concepts. We discover sparse associations between concepts and hidden units using a linear sparse logistic regression. To ensure that the identified units truly influence the classifier's outcome, we adopt tools from Causal Inference literature and, more specifically, mediation analysis through counterfactual interventions. Finally, we construct a low-depth decision tree to translate all the discovered concepts into a straightforward decision rule, expressed to the radiologist. We evaluated our approach on a large chest x-ray dataset, where our model produces a global explanation consistent with clinical knowledge.  We successfully discovered highly discriminative neurons associated with fine-grains concepts that clinicians uses to explain their decision.

 \textit{Acknowledgement} This work was partially supported by NIH Award Number 1R01HL141813-01, NSF 1839332 Tripod+X, SAP SE, and Pennsylvania's Department of Health. We are grateful for the computational resources provided by Pittsburgh SuperComputing grant number TG-ASC170024.

 \chapter{Augmentation by Counterfactual Explanation - 
Fixing an Overconfident Classifier}
\label{ch4}
\section{Introduction}

A highly accurate but overconfident model is ill-suited for decision-making pipelines, especially in critical applications such as healthcare and autonomous driving. The classification outcome should reflect a high uncertainty on ambiguous in-distribution samples that lie close to the decision boundary. The model should also refrain from making overconfident decisions on samples that lie far outside its training distribution, far-out-of-distribution (far-OOD), or on  unseen samples from novel classes that lie near its training distribution (near-OOD). This paper proposes a method to fine-tune a given pre-trained classifier to fix its uncertainty characteristics while retaining its predictive performance.

We propose using a Progressive Counterfactual Explainer (PCE) to generate counterfactually augmented data (CAD) for fine-tuning the classifier. The PCE is a form of conditional Generative Adversarial Networks (cGANs) trained to generate samples that visually traverse the separating boundary of the classifier. The discriminator of the PCE serves as a density estimator to identify and reject OOD samples. We perform extensive experiments with detecting far-OOD, near-OOD, and ambiguous samples. Our empirical results show that our model improves the uncertainty of the baseline, and its performance is competitive to other methods that require a significant change or a complete re-training of the baseline model.

Deep neural networks (DNN) are increasingly being used in \emph{decision-making} pipelines for real-world high-stake applications such as medical diagnostics~\cite{esteva2017dermatologist} and autonomous driving~\cite{Autonomous}. For optimal decision making, the DNN should produce accurate predictions as well as quantify uncertainty over its predictions~\cite{Gal2016UncertaintyID,christian2017}. While substantial efforts are made to engineer highly accurate architectures~\cite{Huang2016DenselyNetworks}, many existing state-of-the-art DNNs do not capture the uncertainty correctly~\cite{pmlr-v48-gal16}. This hinders the re-use of openly available pre-trained models for real-world applications. We proposed to fine-tune the given \emph{pre-trained} DNN on counterfactually augmented data, to improve its uncertainty quantification  while retaining its original predictive accuracy. 

Any classification model is essentially learning a hyperplane to separate samples from different classes. 
Accuracy only captures the proportion of samples that are on the correct side of the decision boundary. However, it ignores the relative distance of the sample from the decision boundary~\cite{krishnan2020improving}. Ideally, samples closer to the boundary should have high uncertainty. The actual predicted value from the classifier should reflect this uncertainty via a low confidence score~\cite{Hllermeier2021AleatoricAE}. Conventionally, DNNs are trained on hard-label datasets to minimize a negative log-likelihood (NLL) loss.  Such models tend to over-saturate on NLL and end-up learning very sharp decision boundaries~\cite{guo2017temperaturescaling,mukhoti2020calibrating}. The resulting classifiers extrapolate over-confidently on ambiguous, near boundary samples, and the problem amplifies as we move to OOD regions~\cite{Gal2016UncertaintyID}.

\begin{figure}[!ht]
  \centering
  \includegraphics[width=1.0\textwidth]{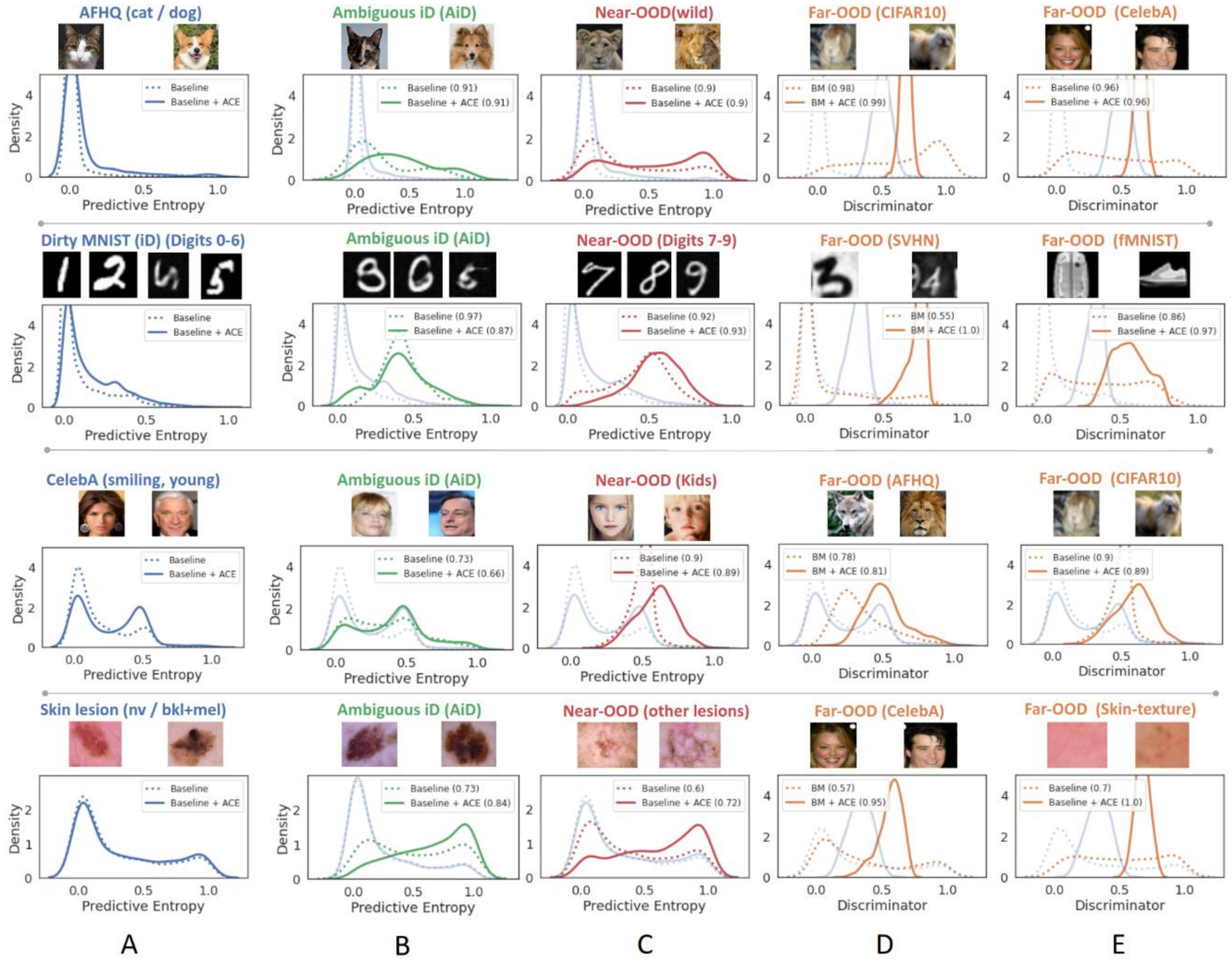}

  \caption[Comparison of the uncertainty estimates from the baseline, before and after fine-tuning with ACE.]{Comparison of the uncertainty estimates from the baseline, before and after fine-tuning with ACE.  Each row represents a different dataset. A) Fine-tuning has little effect on the predicted entropy (PE) of \textbf{\uid{in-distribution (iD)}} samples. We use PE to identify \textbf{\aid{ambiguous iD (AiD)}} samples (B) and \textbf{\nood{near-OOD}} samples (C). D-E) We use the discriminator of the PCE to identify \textbf{\food{far-OOD}} samples (last two columns). The legend shows the AUC-ROC for binary classification over uncertain samples and iD samples. Our method improved the uncertainty estimates across the spectrum.}
  \label{fig:all}
\end{figure}

We consider two types of uncertainty: \emph{epistemic uncertainty}, caused due to limited data and knowledge of the model, and \emph{aleatoric uncertainty}, caused by inherent noise or ambiguity in the data~\cite{KIUREGHIAN2009105}.  We evaluate these uncertainties with respect to three test  distributions (\emph{see} Fig~\ref{fig:all}):

\begin{itemize}
\item \aid{\bf{Ambiguous in-Distribution (AiD)}}: These are the samples within the training distribution that have an inherent ambiguity in their class labels. Such ambiguity represents high aleatoric uncertainty arising from class overlap or noise~\cite{smith2018understanding}, \eg an image of a `5' that is similar to a `6'.

\item \nood{\bf{Near-OOD}}: Near-OOD represents a label shift where label
space is different between ID and OOD data.  It has high epistemic uncertainty arising from the classifier's limited information on unseen data. We use samples from unseen classes of the training distribution as near-OOD.

\item \food{\bf{Far-OOD}}: Far-OOD represents data distribution that is significantly different from the training distribution. It has high epistemic uncertainty arising from mismatch between different data distributions.

\end{itemize}

Earlier work focuses on threshold-based detectors that use information from a pre-trained DNN to identify OOD samples~\cite{guo2017temperaturescaling,hendrycks17baseline,huang2021on,wang2021can,9156473}. Such methods predominantly focus on far-OOD detection and often do not address the over-confidence problem in DNN.

\vspace{-0.8cm}

In another line of research, variants of Bayesian models~\cite{neal2012bayesian,pmlr-v48-gal16} and ensemble learning~\cite{Snapshot,NIPS2017_9ef2ed4b} were explored to provide reliable uncertainty estimates. Recently, there is a shift towards designing generalizable DNN that provide robust uncertainty estimates in a single forward pass~\cite{van2021feature,chen2021atom,mukhoti2021deterministic}. Such methods usually propose changes to the DNN architecture~\cite{sun2021react}, training procedure~\cite{zhang2017mixup} or loss functions~\cite{Mukhoti2020CalibratingDN} to encourage separation between ID and OOD data. Popular methods include, training deterministic DNN with a distance-aware feature space~\cite{van2020uncertainty,Liu2020SimpleAP} and  regularizing DNN training with a generative model that simulates OOD data~\cite{lee2018training}. However, these methods require a DNN model to be trained from scratch and are not compatible with an existing pre-trained DNN. Also, they may use auxiliary data to learn to distinguish OOD inputs~\cite{liu2020energy}.

In this work, we introduce an augmentation by counterfactual explanation (ACE) strategy to fine-tune an existing \emph{pre-trained} DNN. Fine-tuning improves the uncertainty estimates without changing the network's architecture or compromising on its predictive performance. ACE uses a progressive counterfactual explainer (PCE) similar to Lang \etal~\cite{explaining_in_style} and Singla \etal~\cite{singla2021explaining} to generate counterfactually augmented data (CAD). The discriminator of the PCE is a density estimator and is used in a threshold-based selection function to identify and reject far-OOD samples.

The PCE is a conditional-Generative Adversarial Network (cGAN)-based explanation function that explains the decision of a DNN by gradually perturbing a query image to flip its classification decision. We used PCE to generate augmented samples closer to the decision boundary. We assign soft labels to these generated samples that mimic their distance from the boundary. Fine-tuning on such augmented data helps the DNN to recover from the over-saturation on NLL loss, thus making the decision boundary smoother. Smooth decision boundary facilitates improved uncertainty estimates for AiD and near-OOD samples and also makes the classifier robust to adversarial attacks.

Our contributions are as follows: (1)
We propose a novel strategy to fine-tune an existing \textit{pre-trained} DNN to improve its uncertainty estimates and facilitate its deployment in real-world applications. (2) Our approach generates counterfactual augmentations near the decision boundary, allowing the classifier to widen its boundary, to successfully capture the uncertainty over ambiguous-iD and near-OOD samples. (3) Our GAN-based augmenter provides a density estimator (the discriminator) to detect far-OOD samples.

We provide a comprehensive evaluation of our method on specifically defined test datasets to capture different uncertainties. Furthermore, our fine-tuned classifier exhibits better robustness to popular adversarial attacks.

\section{Method}

In this paper, we consider a pre-trained DNN classifier, $f_{\theta}$, with good prediction accuracy but sub-optimal uncertainty estimates. We assume $f_{\theta}$ is a differentiable function and we have access to its gradient with respect to the input, $\nabla_\mathbf{x}f_{\theta}(\mathbf{x})$, and to its final prediction outcome $f_{\theta}(\mathbf{x})$.  We also assume access to either the training data for $f_{\theta}$, or an equivalent dataset with competitive prediction accuracy. 
We further assume that the training dataset for $f_{\theta}$ has hard labels $\{0,1\}$ for all the classes. 
\newline

\begin{figure}[!ht]
  \centering
  \includegraphics[width=1.0\textwidth]{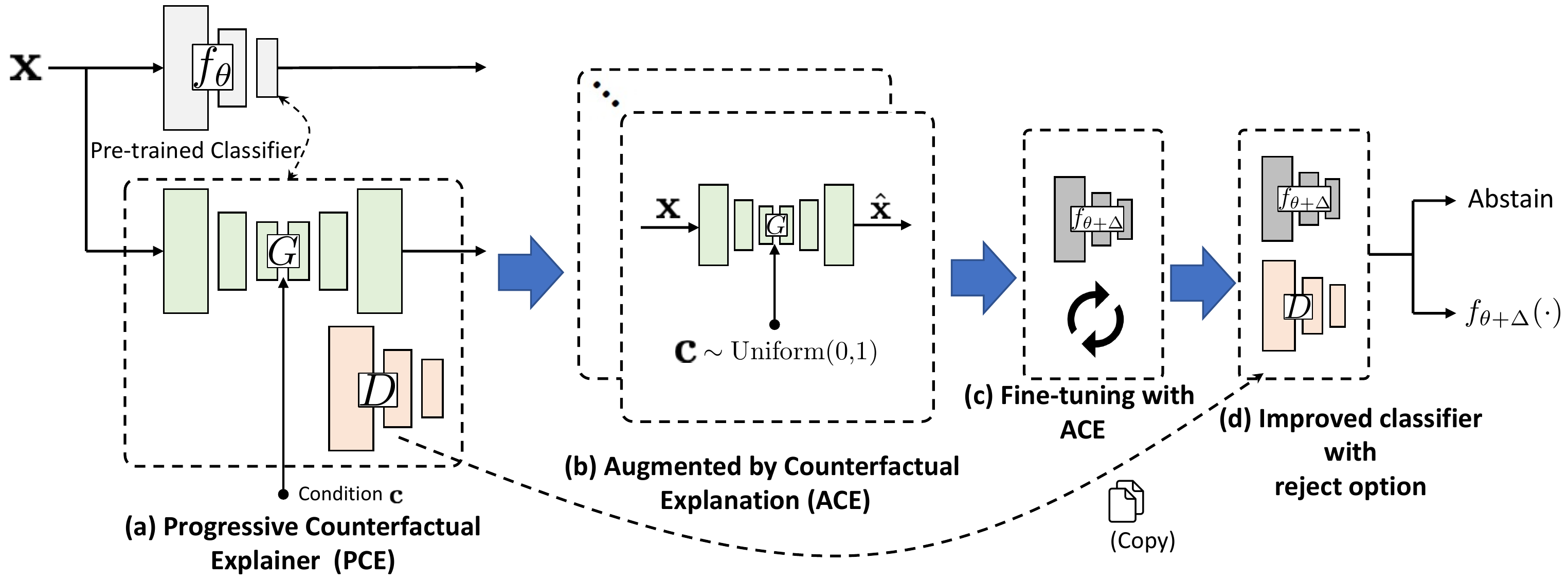}

  \caption[Overview of the method.]{Overview of the method. (a) Given a \emph{pre-trained} classifier $f_{\theta}$, we learn a c-GAN based progressive counterfactual explainer (PCE) $G(\mathbf{x}, \mathbf{c})$, while keeping $f_{\theta}$ fixed. (b) The trained PCE creates counterfactually augmented data. (c) A combination of original training data and augmented data is used to fine-tune the classifier, $f_{\theta + \Delta}$. (d) The discriminator from PCE serves as a selection function to detect and reject OOD data. }
  \label{fig:2}
\end{figure} 

\vspace{-0.4cm}

Our goal is to \emph{fine-tune} $f_{\theta}$ such that the revised model provides better uncertainty estimates, while retaining its original predictive accuracy. More specifically, we aim to improve uncertainty estimates for OOD and ambiguous samples.  We use a progressive counterfactual explainer (PCE) to generate counterfactually augmented data. This data is then used to apply a few updates to $f_{\theta}$, to gradually widen its decision boundary, resulting in improved uncertainty estimates on ambiguous and near-OOD samples. The PCE generates realistic perturbations of a given query image while gradually traversing the decision boundary between the classes, as defined by $f_{\theta}$\cite{singla2021explaining,explaining_in_style}. We used the PCE with a conditional-GAN backbone that is trained with respect to $f_{\theta}$. The discriminator of the cGAN-based PCE models is a density estimator that provides essential information to enhance  $f_{\theta}$ far-OOD detection. More specifically, we used the discriminator as a \emph{selection function} to abstain $f_{\theta}$ from making prediction on far-OOD samples. Fig.~\ref{fig:2} summarizes our approach. 
\newline

\begin{figure}[!ht]
  \centering
  \includegraphics[width=1.0\textwidth]{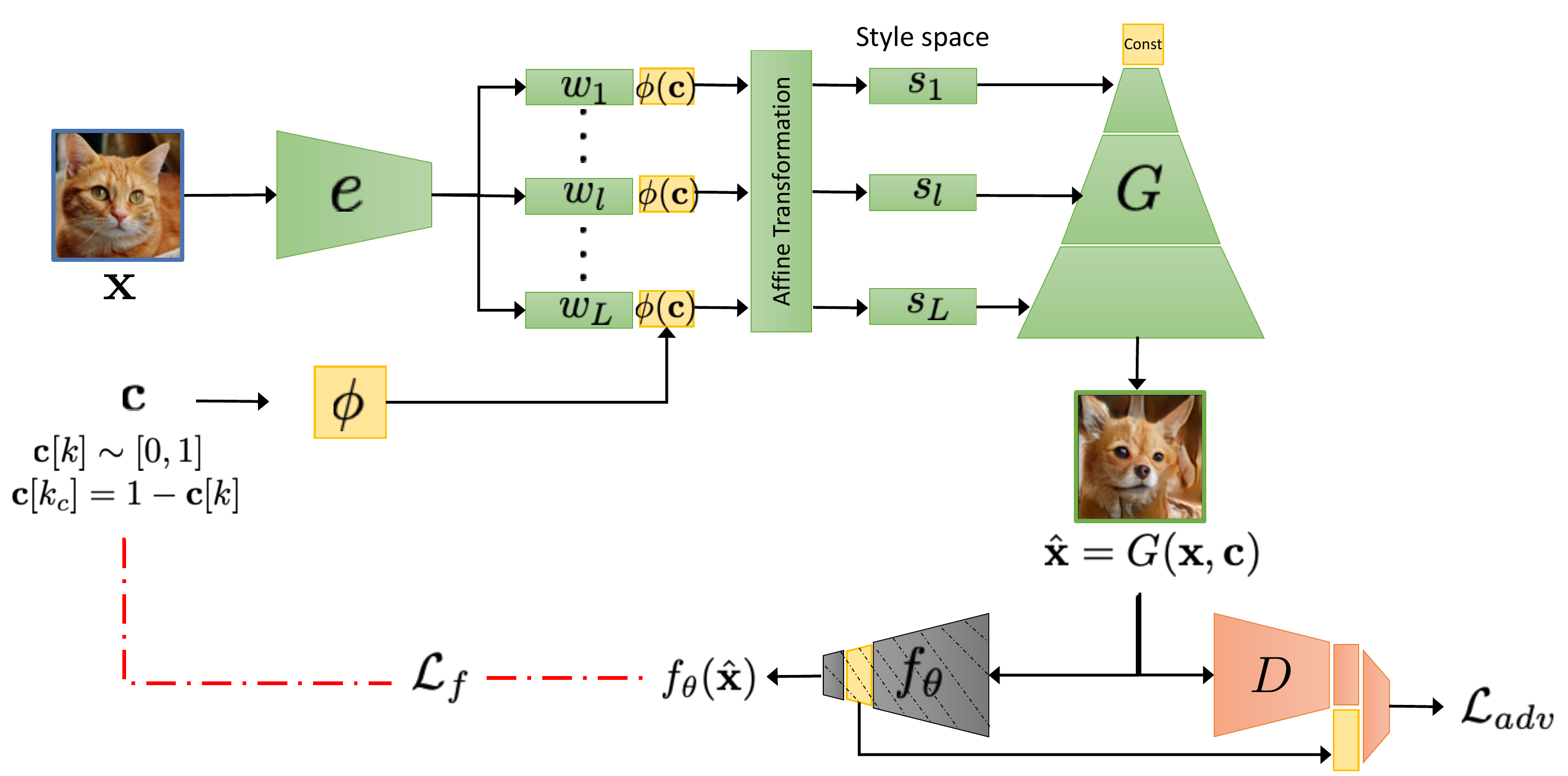}
  \caption{ PCE: The encoder-decoder architecture to create counterfactual augmentation for a given query image.  }
  \label{fig:PCE}
\end{figure} 

\vspace{-0.4cm}

The remaining sections are structured as follow: we first formulate the cGAN-based PCE model in Section~\ref{sec:AN}. In Section~\ref{sec:ACE}, we describe our novel Augmentation by Counterfactual Explanation (ACE) strategy that uses the trained PCE to generate CDA and fine-tune $f_{\theta}$. Finally, in Section~\ref{sec:SC}, we combine the discriminator from the PCE with the fine-tuned $f_{\theta}$ to provide our final classifier.

\emph{Notation:} The classification function is defined as $f_{\theta}: \mathbb{R}^d \rightarrow \mathbb{R}^K$, where $\theta$ represents model parameters. The training dataset for $f_{\theta}$ is defined as $\mathcal{D} = \{\mathcal{X}, \mathcal{Y}\}$, where $\mathbf{x} \in \mathcal{X} $ represents an input space and  $y \in \mathcal{Y} = \{1,2, \cdots, K\}$ is a label set over $K$ classes.  The classifier produces point estimates to approximate the posterior probability $\mathbb{P}(y|\mathbf{x}, \mathcal{D})$.

\subsection{Progressive Counterfactual Explainer (PCE) v2.0 }
\label{sec:AN}
 We designed the PCE network to take a query image ($\mathbf{x} \in \mathbb{R}^d$) and a desired classification outcome ($\mathbf{c} \in \mathbb{R}^K$) as input, and create a perturbation of a query image ($\hat{\mathbf{x}}$) such that $f_{\theta}(\hat{\mathbf{x}}) \approx \mathbf{c}$. Our formulation, $\hat{\mathbf{x}}= G(\mathbf{x}, \mathbf{c})$ allows us to use $\mathbf{c}$ to traverse through the decision boundary of $f_{\theta}$ from the original class to a counterfactual class. Following previous work~\cite{explaining_in_style,singla2021explaining}, we design the PCE to satisfy the following three properties:

\begin{enumerate}
     \item \textbf{Data consistency: } The perturbed image, $\hat{\mathbf{x}}$ should be realistic and should resemble samples in $\mathcal{X}$.
     \item \textbf{Classifier consistency:} The perturbed image, $\hat{\mathbf{x}}$ should produce
the desired output from the classifier $f_{\theta}$ \ie $f_{\theta}(G(\mathbf{x}, \mathbf{c})) \approx \mathbf{c}$.

\item  \textbf{Self consistency:} Using the original classification decision $f_{\theta}(\mathbf{x})$ as condition, the PCE should produce a perturbation that is very similar to the query image, \ie

$G(G(\mathbf{x}, \mathbf{c}), f_{\theta}(\mathbf{x})) = \mathbf{x}$ and $G(\mathbf{x}, f_{\theta}(\mathbf{x})) = \mathbf{x}$.
 \end{enumerate}

  \emph{Data Consistency:} 
 We formulate the PCE as a cGAN that learns the underlying data distribution of the input space $\mathcal{X}$ without an explicit likelihood assumption. The GAN model comprised of two networks -- the generator $G(\cdot)$ and the discriminator $D(\cdot)$. The $G(\cdot)$ learns to generate fake data, while the $D(\cdot)$  is trained to distinguish between the real and fake samples.  We jointly train $G, D$ to optimize the following logistic adversarial loss~\cite{Goodfellow2014GenerativeNets},
 \newline
\begin{equation}
\begin{split}
    \mathcal{L}_{\text{adv}}(D,G) = \mathbb{E}_{\mathbf{x}}[\log D(\mathbf{x}) + \log (1 - D(G(\mathbf{x}, \mathbf{c})))]
\end{split}
\end{equation}
\newline
Our architecture for the cGAN is adapted from StyleGANv2~\cite{abdal2019image2stylegan}. We formulate the generator as $G(\mathbf{x}, \mathbf{c}) = g(e(\mathbf{x}), \mathbf{c})$, a composite of two functions, an image encoder $e(\cdot)$ and a conditional decoder $g(\cdot)$~\cite{abdal2019image2stylegan}. The encoder function $e: \mathcal{X} \rightarrow \mathcal{W}^+$, learns a mapping from the input space $\mathcal{X}$ to an extended latent space $\mathcal{W}^+$. The  $\mathcal{W}^+$ represents a concatenation of $L$ different latent representations ($w_l$), one for each layer of the decoder $g(\cdot)$. The conditional decoder $g(\cdot)$, maps the embedding back to the input space $\mathcal{X}$ while respecting the condition $\mathbf{c}$. We provide condition information to the decoder by concatenating the condition $\mathbf{c}$ to each $w_l \in \mathcal{W}^+$. The decoder further transforms the layer-specific latent representation into a layer-specific style-vector as $s_l = A_l([w_l , \phi(\mathbf{c})])$ where, $A_l$ is an affine transformation and $\phi(\mathbf{c})$ is an embedding for $\mathbf{c}$. Further, we also extended the discriminator network $D(\cdot)$ to have auxiliary information from the classifier $f_{\theta}$. Specifically, we concatenate the penultimate activations from the $f_{\theta}(\mathbf{x})$ with the penultimate activations from the $D(\mathbf{x})$, to obtain a revised representation  before the final fully-connected layer of the discriminator.
The detailed architecture is summarized in Fig.~\ref{fig:PCE}.

We also borrow the concept of path-length regularization $\mathcal{L}_{\text{reg}}(G)$ from StyleGANv2 to enforce smoother latent space interpolations for the generator. 
\newline
\begin{equation}
\mathcal{L}_{\text{reg}}(G) = \mathbb{E}_{\mathbf{w}\sim e(\mathbf{x}), \mathbf{x}\sim \mathcal{X}} (||J^T_\mathbf{w}\mathbf{x}||_2 - a)^2
\end{equation}
\newline
where $\rvx$ denotes random images from the training data, $J_{\mathbf{w}}$ is the Jacobian matrix, and $a$ is a constant that is set dynamically during optimization.

\emph{Classifier consistency: }By default, GAN training is independent of the classifier $f_{\theta}$. We add a classifier-consistency loss to regularize the generator and ensure that the actual classification outcome for the generated image $\hat{\mathbf{x}}$, is similar to the condition $\mathbf{c}$ used for generation. We enforce classification-consistency by a
KullbackLeibler (KL) divergence loss as follow\cite{singla2021explaining},
\newline
\begin{equation}
    \mathcal{L}_{f}(G) = D_{KL}(f_{\theta}(\hat{\mathbf{x}})|| \mathbf{c})
\end{equation}
\newline
\emph{Self consistency: }We define the following reconstruction loss to regularize and constraint the Generator to preserve maximum information between the original image $\mathbf{x}$ and its reconstruction $\bar{\mathbf{x}}$,
\newline
\begin{equation}
    \mathcal{L}(\mathbf{x}, \bar{\mathbf{x}}) = ||\mathbf{x} - \bar{\mathbf{x}}||_1 + ||e(\mathbf{x}) - e(\bar{\mathbf{x}})||_1 
\end{equation}
\newline
Here, first term is a distance loss between the input and the reconstructed image, and the second term is a style reconstruction loss adapted from StyleGANv2~\cite{abdal2019image2stylegan}. We minimize the reconstruction loss to satisfy the identify constraint on self reconstruction using $\bar{\mathbf{x}}_{self} = G(\mathbf{x}, f_{\theta}(\mathbf{x}))$. We further insure that the PCE learns a reversible perturbation by recovering the original image from a given perturbed image $\hat{\mathbf{x}}$ as  $\bar{\mathbf{x}}_{\text{cyclic}} = G(\hat{\mathbf{x}}, f_{\theta}(\mathbf{x}))$, where $\hat{\mathbf{x}} = G(\mathbf{x}, \mathbf{c})$ with some condition $\mathbf{c}$. Our final reconstruction loss is defined as,
\newline
\begin{equation}
    \mathcal{L}_{\text{rec}}(G) = \mathcal{L}(\mathbf{x}, \bar{\mathbf{x}}_{\text{self}}) + \mathcal{L}(\mathbf{x}, \bar{\mathbf{x}}_{\text{cyclic}})
\end{equation}
\newline
\emph{Objective function: }Finally, we trained our model in an end-to-end fashion to learn parameters for the two networks,  while fixing $f_{\theta}$. Our overall objective function is:
\newline
\begin{equation}
    \min_{G} \max_{D} \lambda_{\text{adv}}\left( \mathcal{L}_{\text{adv}}(D,G) +     \mathcal{L}_{\text{reg}}(G)\right)
    + \lambda_{f} \mathcal{L}_f(G) + \lambda_{\text{rec}} \mathcal{L}_{\text{rec}}(G),
\label{eq:5}
\end{equation}
\newline
where, $\lambda$'s are the hyper-parameters to balance each of the loss terms.

\subsection{Augmentation by Counterfactual Explanation}
\label{sec:ACE}
\begin{figure}[!ht]
  \centering
  \includegraphics[width=0.8\linewidth]{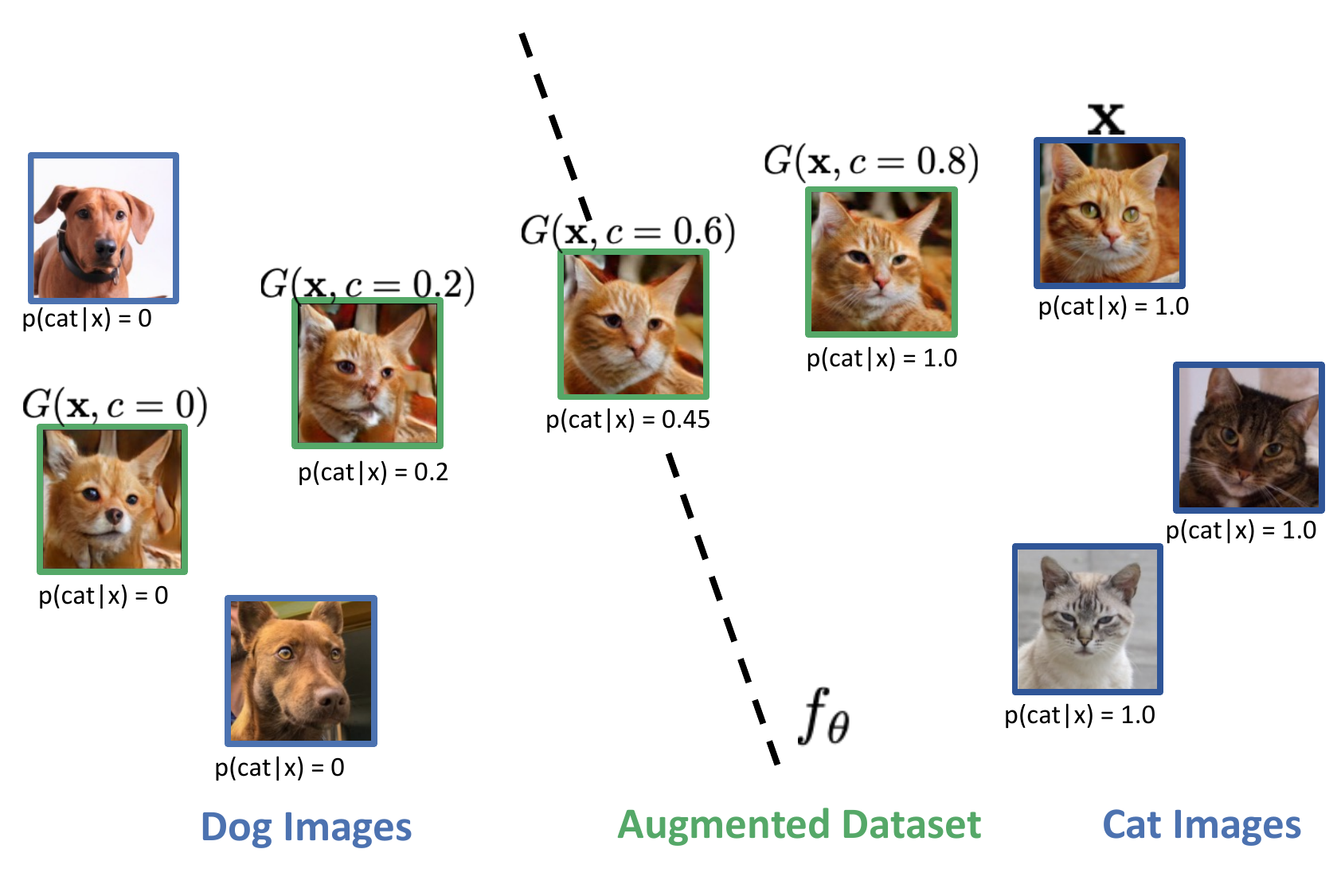}

  \caption[Augmentation by counterfactual explanation.]  {Augmentation by counterfactual explanation. Given a query image, the trained PCE generates a series of perturbations that gradually traverse the decision boundary of $f_{\theta}$ from the original class to a counter-factual class,  while still remaining plausible and realistic-looking.}
  \label{fig:ace}
\end{figure}

Given a query image $\mathbf{x}$, the trained PCE generates a series of  perturbations of $\mathbf{x}$ that gradually traverse the decision boundary of $f_{\theta}$ from the original class to a counterfactual class, while still remaining plausible and realistic-looking. 
This series of  perturbations is essentially mimicking a traversal on a latent manifold, as guided by the condition $\mathbf{c}$. Our trained AN enables conditional generation of an image at any point on the manifold as $G(\mathbf{x}, \mathbf{c})$ (\emph{see} Fig.\ref{fig:ace}). 
We modify $\mathbf{c}$ to represent different steps in this traversal.  We start from a high data-likelihood region for original class $k$ ($\mathbf{c}[k] \in [0.8, 1.0]$), walk towards the decision hyper-plane ($\mathbf{c}[k] \in [0.5, 0.8)$), and eventually cross the decision boundary ($\mathbf{c}[k] \in [0.2, 0.5)$) to end the traversal in a high data-likelihood region for the counterfactual class $k_c$ ($\mathbf{c}[k] \in [0.0, 0.2)$). Accordingly, we  set $\mathbf{c}[k_{c}] = 1- \mathbf{c}[k]$.

Ideally, the predicted confidence from NN should be  indicative of the distance from the decision boundary. Samples that lies close to the decision boundary should have low confidence, and confidence should increase as we move away from the decision boundary. We used $\mathbf{c}$ as a pseudo indicator of confidence to generate synthetic augmentation. Our augmentations are essentially showing how the query image $\mathbf{x}$ should be modified to have low/high confidence. 
\newline 

\vspace{-0.4cm}

To generate CAD, we randomly sample a subset of real training data as $\mathcal{X}_r {\displaystyle \subset }\mathcal{X}$. Next, for each $\mathbf{x} \in \mathcal{X}_r$ we generate multiple augmentations ($\hat{\mathbf{x}} = G(\mathbf{x}, \mathbf{c})$) by randomly sampling $\mathbf{c}[k] \in [0,1]$. We used $\mathbf{c}$ as soft label for the generate sample while fine-tuning the $f_{\theta}$. The  $\mathcal{X}_{c}$ represents our pool of generated augmentation images. Finally, we create a new dataset by randomly sampling images from $\mathcal{X}$ and $\mathcal{X}_c$. We fine-tune the $f_{\theta}$ on this new dataset, for only a few epochs, to obtain a revised classifier  given as $f_{\theta + \Delta}$. In our experiments, we show that the revised decision function $f_{\hat{\theta}}$ provides improved confidence estimates for AiD and near OOD samples and demonstrate robustness to adversarial attacks, as compared to given classifier $f_{\theta}$.

\subsection{Discriminator as a Selection Function}
\label{sec:SC}
A selection function $g: \mathcal{X} \rightarrow \{0,1\}$ is an addition head on top of a classifier that decides when the classifier should  abstain from making a prediction. We propose to use the discriminator network $D(\mathbf{x})$ as a selection function for $f_{\theta}$. Upon the convergence of the PCE training, the generated samples resemble the in-distribution training data. Far-OOD samples are previously unseen samples which are very different from the training input space. Hence, $D(\cdot)$ can help in detecting such samples.  Our final improved classification function is represented as follow,
\newline
\begin{equation}
    (f,D)(\mathbf{x})= 
\begin{cases}
    f_{\theta + \Delta} (\mathbf{x}),& \text{if } D(\mathbf{x}) \geq h\\
    \texttt{Abstain},              & \text{otherwise}
\end{cases}
\end{equation}
\newline
where, $f_{\theta + \Delta}$ is the fine-tuned classifier and $D(\cdot)$ is a discriminator network from the PCE which serves as a selection function that permits $f$ to make prediction if $D(\mathbf{x})$ exceeds a threshold $h$ and abstain  otherwise

 During inference, the discriminator first uses its density estimates to quantify the similarity between the learned data distribution and the query image. These estimates provide a pseudo signal to quantify epistemic uncertainty. Using a strict threshold, the discriminator may reject any sample that lies outside its learned data distribution as far-OOD. The fine-tuned classifier then processes the accepted samples. We use the fine-tuned classifier's predictive entropy (PE) to quantify the sample's uncertainty. This uncertainty can be epistemic (associated with near-OOD samples) or aleatoric (associated with AID samples). Our model cannot differentiate between the two.

\section{Experiments and Results}
We set up three experiments to compare the baseline model before and after fine-tuning with our proposed counterfactual augmentation:
First, we assess if the models can correctly capture aleatoric uncertainty by identifying ambiguous test samples. Second, we evaluate the models on OOD detection tasks. We consider a standard OOD task over separate datasets and a challenging task to detect near-OOD samples from previously unseen classes from the same dataset. Finally, we compare the model's behaviour under adversarial attacks.
%We also visually compare the changes to the loss landscape of the model after fine-tuning. 
In SM, we show further experiments on more datasets and an ablation study over different components of the PCE. 

\subsection{Imaging dataset}
In our experiments, we consider classification models trained on following datasets:

\begin{enumerate}
\item AFHQ~\cite{choi2020afhq}: Animal face high quality (AFHQ) dataset is a high resolution dataset of animal faces with 16K images from cat, dog and wild labels. The classifier is trained at an image resolution of $256\times256$.

\item Dirty MNIST~\cite{mukhoti2021deterministic}: The dataset is a combination of original MNIST~\cite{LeCun2010MNISTDatabase} and simulated Ambiguous-MNIST dataset. Each sample in Ambiguous-MNIST is constructed by decoding a linear combination of latent representations of two different MNIST digits from a pre-trained VAE~\cite{NIPS2015_bc731692}. The samples are generated by combining latent representation of different digits, to simulate ambiguous samples, with multiple plausible labels~\cite{mukhoti2021deterministic}.
The training dataset of the classifier comprises of 60K clean-MNIST and 60K Ambiguous-MNIST samples, with one-hot labels.  The original dataset consists of grayscale images of size 28$\times$28 pixels. We consider a classification model trained on 64$\times$64 resolution.

 \item CelebA~\cite{Liu2015DeepWild}: Celeb Faces Attributes Dataset (CelebA) is a large-scale face attributes dataset with more than 200K celebrity images, each with 40 binary attributes annotations per image.  Our \aid{AiD} samples comprises of middle-aged people who are arguably neither young nor old. To obtain such data, we use aleatoric uncertainty estimates from MC-Dropout averaged across 50 runs on test-set of CelebA. The classifier is trained at an image resolution of $256\times256$. We center-crop the images as a pre-processing step.

  \item Skin lesion (HAM10K)~\cite{tschandl2018ham10000}:  The HAM10000 is a dataset of 100K dermatoscopic images of pigmented skin lesions. It contains seven different lesion types -- Melanocytic Nevi (nv), Melanoma (mel), Benign Keratosis (bkl), Actinic Keratoses and Intraepithelial Carcinoma (akiec), Basal Cell Carcinoma (bcc), Dermatofibroma (df), Vascular skin lesions (vasc). In our experiments, we consider classifier trained to distinguish the majority class nv from mel and bkl. We consider images from rest of the lesions as near-OOD.  The classifier is trained at an image resolution of $256\times256$.
 \end{enumerate}

\emph{Classification tasks: }We consider four classification problems, in increasing level of difficulty: 
 \begin{enumerate}
  \item AFHQ~\cite{choi2020afhq}: We consider binary classification over well separated classes, cat vs dog. We consider images with ``wild" label as near-OOD.
    \item Dirty MNIST~\cite{mukhoti2021deterministic}: We consider multi-class classification over seven classes of hand-written digits `0' - `6'. We consider images from digits `7' - `9' as near-OOD samples.
    \item CelebA~\cite{Liu2015DeepWild}: we consider a two-class classifier over attributes ``Young" and ``Smiling" trained on CelebA dataset. Without age labels, identifying 'young' faces is a challenging task. 
    \item Skin lesion (HAM10K)~\cite{tschandl2018ham10000}:  We consider a binary classification to separate Melanocytic nevus (nv) from Melanoma (mel) and Benign Keratosis (bkl) lesions.  Skin lesion classification is a challenging task as different lesions may exhibit similar features~\cite{nachbar1994abcd}. 
   
 \end{enumerate}

\subsection{Experimental setup}
\emph{Classification Model: } We consider state-of-the-art DenseNet~\cite{Huang2016DenselyNetworks} architecture for the baseline. The \emph{pre-trained} DenseNet model followed the training procedures as described in~\cite{Huang2016DenselyNetworks}. In DenseNet, each layer implements a non-linear transformation based on composite functions such as Batch Normalization (BN), rectified linear unit (ReLU), pooling, or convolution. The resulting feature map at each layer is used as input for all the subsequent layers, leading to a highly convoluted multi-level multi-layer non-linear convolutional neural network.  We aim to improve such a model in a post-hoc manner without accessing the parameters learned by any layer or knowing the architectural details. Our proposed approach can be used for any DNN architecture.

\emph{Progressive Counterfactual Explainer v2.0: }
We formulate the progressive counterfactual explainer (PCE) as a composite of two functions, an image encoder $e(\cdot)$ and a conditional decoder ($g(\cdot)$)~\cite{abdal2019image2stylegan}. Our architecture for the conditional decoder is adapted from StyleGANv2~\cite{abdal2019image2stylegan}. In order to keep the architecture and training procedure of PCE simple, we consider the default training parameters from~\cite{abdal2019image2stylegan} for training the StyleGANv2. This encourages reproducibility as we didn't do hyper-parameter tuning for each dataset and classification model. Specifically, we use Adam optimizer to train StyleGANv2 at 256$^2$ resolution for $\sim$200K iterations with a batch size of 8 over 8 GPUs. For training  StyelGANv2, we use a randomly sampled subset $(\sim50\%)$ of the baseline model's training data. For multi-class classification, we consider all pairs of classes while creating counterfactual augmentations. Further, given an input image, the predicted class $k$ and a counterfactual class $k_c$, we initialize the condition $\mathbf{c}$ with all zeros and then set $\mathbf{c}[k] \sim \text{Uniform}(0,1) $ and $\mathbf{c}[k_c] = 1 -  \mathbf{c}[k]$. In all our experiments, we used $\lambda_{adv} = 10$, $\lambda_{rec} = 100$ and $\lambda_{f} = 10$.

To generate CAD, we randomly sample a subset of real training data as $\mathcal{X}_r {\displaystyle \subset }\mathcal{X}$. Next, for each $\mathbf{x} \in \mathcal{X}_r$ we generate four augmentations ($\hat{\mathbf{x}} = G(\mathbf{x}, \mathbf{c})$) by randomly sampling $\mathbf{c}[k] \in [0,1]$. We used $\mathbf{c}$ as soft label for the generate sample while fine-tuning the $f_{\theta}$. The  $\mathcal{X}_{c}$ represents our pool of generated augmentation images.

For fine-tuning the given baseline with consider a combination of the original training dataset $\mathcal{X}$ and the augmented data $\mathcal{X}_c$. We randomly selected a subset of samples from the two distributions and fine-tune the baseline for 5 to 10 epochs. We used the expected calibration error and the test-set accuracy to choose the final checkpoint. Our model does not require access to OOD or AiD dataset during fine-tuning. During evaluation we compute predicted entropy (PE) for original test-set and OOD samples and   measure for a range of thresholds how well
the two are separated. We report the AUC-ROC and the true negative rate (TNR) at 95\% true positive rate (TPR) (TNR@TPR95) in our results (\emph{see Table~1~and~2}). We will release the GitHub for the project after the review process.

 \emph{Comparison methods: } Our baseline is a standard DNN classifier $f_{\theta}$ trained with cross-entropy loss. For baseline and its post-hoc variant with temperature-scaling (\textbf{TS})~\cite{guo2017temperaturescaling}, we used threshold over predictive entropy (PE) to identify OOD. PE is defined as $-\sum[f_{\theta}(\mathbf{x})]_k \log[f_{\theta}(\mathbf{x})]_k$. We compared against three methods that changes network training or architecture to learn better uncertainty estimates: 
 \begin{enumerate}
     \item \textbf{Mixup}: Baseline model with mixup training using $\alpha = 0.2$~\cite{zhang2017mixup}.
     \item \textbf{DUQ}: Baseline model with radial basis function as the end-layer. Here, we use the closest kernel distance to quantify uncertainties~\cite{van2020uncertainty}.
     \item \textbf{DDU}: A ResNet-18~\cite{He2016DeepRecognition} model with spectral normalisation and Gaussian Mixture Model (GMM) for density estimation~\cite{mukhoti2021deterministic}.
 \end{enumerate}
 We also compared against methods that obtain uncertainty estimates from a pre-trained DNN output using threshold-based scoring functions.
 \begin{enumerate}
 \setcounter{enumi}{3}
     \item \textbf{Energy-based} scoring function: Baseline with an energy function. We experimented with two variants, in first we compute energy score in a post-hoc manner and use it to identify OOD samples. In the second, we fine-tune the pre-trained DNN using the energy-score based loss and then identify OOD samples~\cite{liu2020energy}.
     \item \textbf{Outlier exposure}: Baseline model with extra regularization from known outliers. While training the DNN, we assume access to outlier data and we force softmax output  to be a uniform distribution for the outlier data~\cite{hendrycks2018deep}.
     \item \textbf{ODIN}: Baseline with TS followed by the post-hoc approach of Out-of-DIstribution detector for Neural networks (ODIN). In ODIN, we added  small perturbations to the input to effectively separate OOD images from the in-distribution ones~\cite{liang2018enhancing}.
     
 \end{enumerate}
 
Further, we also compared against two ensemble approaches:  
\begin{enumerate}
 \setcounter{enumi}{6}
    \item \textbf{MC Dropout}: Baseline model trained with dropout. At inference time we took 20 dropout samples~\cite{pmlr-v48-gal16} to compute PE.
    \item \textbf{5-Ensemble}: Baseline model trained five times with different seeds using the same dataset, shuffled using different seed~\cite{NIPS2017_9ef2ed4b}. We view the ensemble approaches as an upper bound for uncertainty quantification.  
\end{enumerate}

 \subsection{Identifying AiD samples}
 AiD samples have an inherent ambiguity in their class label, arising from the overlapping class definitions. 
 Curated datasets for image classification are well-defined and provide binarized labels as ground truth. Assigning a label to an image removes ambiguity about its class membership. In the absence of ground truth uncertainty labels, we use the PE estimates from an MC Dropout classifier to label AiD samples. Specifically,   we sort the test set using PE and consider the top 5 to 10\% samples as AiD.  
 
 \begin{table}[!ht]
 \begin{center}
\caption[Performance of different methods on identifying \textbf{\aid{ambiguous in-distribution (AiD)}} samples.]{Performance of different methods on identifying \textbf{\aid{ambiguous in-distribution (AiD)}} samples.
For all metrics, higher is better. The best results from the methods that require a single forward pass at inference time are highlighted. The ensemble approaches form an upper-bound and are for reference only and not comparison. Given a baseline model, our results are averaged over 10 runs of fine-tuning with different augmentation by counterfactual explanation (ACE) datasets. }
\label{AiD-table1}

\small
% \resizebox{\textwidth}{!}{%
    \begin{tabular}{cccccc}
    \hline
    \multicolumn{1}{c}{\bf Train }  &  \multicolumn{1}{c}{\bf Method/ }  & \multicolumn{1}{c}{\bf Test-Set} &  \multicolumn{2}{c}{\bf Identifying \aid{AiD}}  \Tstrut{}\\
    \bf Dataset & \bf Model  & Accuracy & AUC-ROC & TNR@TPR95  \Bstrut{}\\
    \hline
    & Baseline  &   99.44$\pm$0.02  &  0.87$\pm$0.04  & 48.93$\pm$10 \Tstrut{}\\ 
     & Baseline+TS~\cite{guo2017temperaturescaling} & 99.45$\pm$0.00  &   0.85$\pm$0.07  &  48.77$\pm$9.8 \\ 
    & Mixup~\cite{zhang2017mixup} & 99.02$\pm$0.10    & 0.80$\pm$0.05  &  35.66$\pm$6.7 \\ 
    AFHQ & DUQ~\cite{van2020uncertainty} & 94.00$\pm$1.05  & 0.67$\pm$0.01    & 26.15$\pm$4.5 \\ 
    & DDU~\cite{mukhoti2021deterministic} & 97.66$\pm$1.10  &  0.74$\pm$0.02  &  19.65$\pm$4.5 \\
    & Baseline+Energy~\cite{liu2020energy}  &   99.44$\pm$0.02  &  0.87$\pm$0.06  & 49.00$\pm$1.64\\ 
    & Energy w/ fine-tune~\cite{liu2020energy}  &  99.45$\pm$0.11   & 0.69$\pm$1.28   & 30.36$\pm$2.52 \\
    & Outlier Exposure~\cite{hendrycks2018deep}  &  99.50$\pm$0.14   &  0.85$\pm$0.01  & 41.07$\pm$0.75 \\ 
    & Baseline+TS+ODIN~\cite{liang2018enhancing}  &  99.45$\pm$0.00   &  0.85$\pm$0.06  & 35.72$\pm$1.26 \\
    & \bf Baseline+ACE & \bf 99.52$\pm$0.21  & \bf 0.91$\pm$0.02  &  \bf 50.75$\pm$3.9 \Bstrut{}\\
    \cline{2-5}
    & MC Dropout~\cite{pmlr-v48-gal16} & 98.83$\pm$1.12 & 0.87$\pm$0.04 & 51.56$\pm$1.2 \Tstrut{}\\
    & 5-Ensemble~\cite{NIPS2017_9ef2ed4b} &  99.79$\pm$0.01  &  0.98$\pm$0.01  &  51.93$\pm$2.7 \Bstrut{} \\
    \hline 
     & Baseline  &    95.68$\pm$0.02 & \bf 0.96$\pm$0.00  & 28.5$\pm$2.3 \Tstrut{}\\ 
     & Baseline+TS~\cite{guo2017temperaturescaling} &  95.74$\pm$0.02 &  0.94$\pm$0.01  & 27.90$\pm$1.3\\ 
     & Mixup~\cite{zhang2017mixup} & 94.66$\pm$0.16   &  0.94$\pm$0.02  & 25.78$\pm$2.1 \\ 
    Dirty & DUQ~\cite{van2020uncertainty} & 89.34$\pm$0.44  & 0.67$\pm$0.01   & 23.89$\pm$1.2 \\ 
    MNIST & DDU~\cite{mukhoti2021deterministic}& 93.52$\pm$1.12   &     0.65$\pm$0.12  &   20.78$\pm$4.0 \\ 
    & Baseline+Energy~\cite{liu2020energy}  &  95.68$\pm$0.02   & 0.80$\pm$0.03   & 17.60$\pm$0.55 \\ 
    & Energy w/ fine-tune~\cite{liu2020energy}  & 96.17$\pm$0.02    &  0.39$\pm$0.04  & 11.59$\pm$0.25 \\
    & Outlier Exposure~\cite{hendrycks2018deep}  &  \bf 96.30$\pm$0.07   & 0.63$\pm$0.07   & 17.6$\pm$2.88 \\ 
    & Baseline+TS+ODIN~\cite{liang2018enhancing}  &  95.74$\pm$0.02   & 0.79$\pm$0.03   & 13.25$\pm$4.88 \\
    & \bf Baseline+ACE &   95.36$\pm$0.45  &  0.86$\pm$0.01  &  \bf 34.12$\pm$2.6 \Bstrut{}\\
    \cline{2-5}
    & MC Dropout~\cite{pmlr-v48-gal16} & 89.50$\pm$1.90  &     0.75$\pm$0.07  &  36.10$\pm$1.8 \Tstrut{}\\ 
    & 5-Ensemble~\cite{NIPS2017_9ef2ed4b} &  95.90$\pm$0.12   &     0.98$\pm$0.02  &  34.87$\pm$3.4 \Bstrut{}\\
    \hline
   
    \end{tabular}
\end{center}
\end{table}

 In Fig.~\ref{fig:all}, we qualitatively compare the PE distribution from the given baseline and its fine-tuned version (baseline + ACE). Fine-tuning resulted in minor changes to the PE distribution of the iD samples (Fig.~\ref{fig:all}.A). 
 We observe a significant separation in the PE distribution of AiD samples and the rest of the test set (Fig.~\ref{fig:all}.B), even on the baseline. This suggests that the PE correctly captures the aleatoric uncertainty. Fine-tuning with counterfactual augmentation further enhanced this separation by shifting the PE distribution of AiD samples to the right and assigning a higher PE value to the uncertain samples. 
 
 The Table~\ref{AiD-table1} and Table~\ref{AiD-table2}  compare our model to several methods.  We report the test set accuracy,  the AUC-ROC for the binary task of identifying AiD samples and the  true negative rate (TNR) at 95\% true positive rate (TPR) (TNR@TPR95), which simulates
an application requirement that the recall of in-distribution
data should be 95\%~\cite{9156473}. For all metrics higher value is better. Our model outperformed other deterministic models, in identifying AiD samples with a high AUC-ROC and TNR@TPR95 across all datasets. 
Also, the fine-tuned model retained the predictive accuracy of the baseline in AFHQ and Dirty MNIST datasets. We observe a little drop in test accuracy in  complex classification problems, where multiple classes may look similar, e.g. medical datasets (HAM10K). Fine-tuning makes the decision boundary broader. As a result, the samples near the decision boundary may flip their decision, decreasing accuracy. We consider this drop more like a flag to indicate possible label noise. Please note, in the tables, we highlighted the best results from the methods that require a single forward pass at inference time.

 \begin{table}[!ht]
 \begin{center}
\caption[Performance of different methods on identifying \textbf{\aid{ambiguous in-distribution (AiD)}} samples.]{ Performance of different methods on identifying \textbf{\aid{ambiguous in-distribution (AiD)}} samples.
For all metrics, higher is better. The best results from the methods that require a single forward pass at inference time are highlighted. The ensemble approaches form an upper-bound and are for reference only and not comparison. Given a baseline model, our results are averaged over 10 runs of fine-tuning with different augmentation by counterfactual explanation (ACE) datasets. }
\label{AiD-table2}
\small
% \resizebox{\textwidth}{!}{%
    \begin{tabular}{cccccc}
    \hline
    \multicolumn{1}{c}{\bf Train }  &  \multicolumn{1}{c}{\bf Method/ }  & \multicolumn{1}{c}{\bf Test-Set} &  \multicolumn{2}{c}{\bf Identifying \aid{AiD}}  \Tstrut{}\\
    \bf Dataset & \bf Model  & Accuracy & AUC-ROC & TNR@TPR95  \Bstrut{}\\
    \hline
    & Baseline  &  89.36$\pm$0.96 & 0.73$\pm$0.01 & 17.18$\pm$1.6 \Tstrut{}\\  
    & Baseline+TS~\cite{guo2017temperaturescaling} & 89.33$\pm$0.01 & 0.72$\pm$0.02 & 17.21$\pm$1.5   \\  
    & Mixup~\cite{zhang2017mixup} & 89.04$\pm$0.47 & \bf 0.74$\pm$0.02 & 15.09$\pm$1.9 \\ 
    CelebA & DUQ~\cite{van2020uncertainty} & 71.75$\pm$0.01 & 0.65$\pm$0.01 & 14.20$\pm$1.0 \\ 
     & DDU~\cite{mukhoti2021deterministic} & 70.15$\pm$0.02 & 0.67$\pm$0.06 & 11.39$\pm$0.4 \\ 
     & Baseline+Energy~\cite{liu2020energy}  &  89.36$\pm$0.96   & 0.57$\pm$0.28   & 4.87$\pm$0.32\\
      & Energy w/ fine-tune~\cite{liu2020energy}  & \bf 90.22$\pm$0.96   & 0.53$\pm$1.25   & 5.06$\pm$0.28 \\
      & Outlier Exposure~\cite{hendrycks2018deep}  &   86.65$\pm$1.22  &  0.53$\pm$0.46  & 5.06$\pm$0.19 \\ 
      & Baseline+TS+ODIN~\cite{liang2018enhancing}  & 89.33$\pm$0.01    &0.57$\pm$0.01    & 6.34$\pm$0.38 \\ 
     & \bf Baseline+ACE &  86.8$\pm$0.79 & \bf 0.74$\pm$0.06 & \bf 22.36$\pm$ 2.3 \Bstrut{}\\
    \cline{2-5}
    & MC Dropout~\cite{pmlr-v48-gal16} & 89.86$\pm$0.33 & 0.73$\pm$0.03 & 19.78$\pm$0.7 \Tstrut{}\\
    & 5-Ensemble~\cite{NIPS2017_9ef2ed4b} &  90.76$\pm$0.00 & 0.84$\pm$0.11 & 17.79$\pm$0.6 \Bstrut{}\\
    \hline
     & Baseline  &    85.88$\pm$0.75   & 0.82$\pm$0.06  & 20.52$\pm$3.7 \Tstrut{}\\ 
     & Baseline+TS~\cite{guo2017temperaturescaling} &  \bf 86.27$\pm$0.40  &     \bf 0.84$\pm$0.03  & 23.34$\pm$2.8 \\ 
    & Mixup~\cite{zhang2017mixup} &  85.81$\pm$0.61  &  \bf  0.84$\pm$0.04& 31.29$\pm$7.0 \\ 
    Skin-Lesion & DUQ~\cite{van2020uncertainty} & 75.47$\pm$5.36  &  0.81$\pm$0.02     &  30.12$\pm$4.4\\ 
    (HAM10K) & DDU~\cite{mukhoti2021deterministic}& 75.84$\pm$2.34   &     0.79$\pm$0.03  &   26.12$\pm$6.6 \\ 
    & Baseline+Energy~\cite{liu2020energy}  & 85.88$\pm$0.75    & 0.77$\pm$0.12   & 18.40$\pm$0.51 \\
    & Energy w/ fine-tune~\cite{liu2020energy}  &  \bf 86.56$\pm$0.53   &0.64$\pm$0.06    & 17.45$\pm$1.78 \\
    & Outlier Exposure~\cite{hendrycks2018deep}  &  86.37$\pm$0.46   & 0.73$\pm$0.02   & 13.21$\pm$2.70 \\ 
    & Baseline+TS+ODIN~\cite{liang2018enhancing}  & 86.27$\pm$0.40    & 0.78$\pm$0.01   & 15.87$\pm$4.33 \\ 
    & \bf Baseline+ACE & 81.21$\pm$1.12  & \bf 0.84$\pm$0.05  & \bf 71.60$\pm$3.8 \Bstrut{}\\
    \cline{2-5}
    & MC Dropout~\cite{pmlr-v48-gal16} & 84.90$\pm$1.17   &  0.85$\pm$0.06  &  43.78$\pm$1.9 \Tstrut{}\\ 
    & 5-Ensemble~\cite{NIPS2017_9ef2ed4b} &  87.89$\pm$0.13   &     0.86$\pm$0.02  &  40.49$\pm$5.1 \Bstrut{}\\
    \hline
    \end{tabular}
\end{center}
\end{table}

\subsection{Detecting OOD samples}

We consider two tasks to evaluate the model’s OOD detection performance. First, a standard OOD task where OOD samples are derived from a separate dataset. Second, a difficult near-OOD detection task where OOD samples belongs to novel classes from the same dataset, which are not seen during training. We consider the following OOD datasets:
 
\begin{enumerate}
 
\item AFHQ~\cite{choi2020afhq}: We consider ``wild" class from AFHQ to define near-OOD samples. For the far-OOD detection task, we use the CelebA dataset, and also cat/dog images from CIFAR10~\cite{Krizhevsky2009LearningML}.
  
    \item Dirty MNIST~\cite{mukhoti2021deterministic}: We consider digits 7-9 as near-OOD samples. For far-OOD detection, we use SVHN~\cite{37648} and fashion MNIST~\cite{Xiao2017FashionMNISTAN} datasets.

 \item CelebA~\cite{Liu2015DeepWild}: We consider images of kids in age-group: 0-11 from the UTKFace~\cite{zhifei2017cvpr} dataset to define the near-OOD samples. For far-OOD detection task, we use the AFHQ and CIFAR10 datasets.
 
    \item Skin lesion (HAM10K)~\cite{tschandl2018ham10000}: We consider samples from lesion types: Actinic Keratoses and Intraepithelial Carcinoma (akiec), Basal Cell Carcinoma (bcc), Dermatofibroma (df) and Vascular skin lesions (vasc) as near-OOD. For far-OOD, we consider CelebA and an additional simulated dataset with different skin textures/tones.
 \end{enumerate}

We summarize our qualitative results in Fig.~\ref{fig:all}. We observe much overlap between the predictive entropy (PE) distribution of the near-OOD samples  and in-distribution samples in Fig.~\ref{fig:all}.C. Fine-tuning with counterfactual augmentation helped in reducing this overlap, by amplifying the uncertainty associated with OOD data. Further, in Fig.~\ref{fig:all}.D-E, we observe that the PE distribution from the baseline model does not capture epistemic uncertainty associated with far-OOD samples, but our model successfully disentangles OOD samples from the in-distribution samples by using density estimates from the discriminator of the PCE.

  \begin{table}[!ht]
  \begin{center}
\caption[OOD detection performance for different baselines.]{OOD detection performance for different baselines. \textbf{\nood{Near-OOD}} represents label shift, with samples from the unseen classes of the same dataset. \textbf{\food{Far-OOD}} samples are from a separate dataset.  The numbers are averaged over five runs. }
%\textcolor{red}{The ensemble approaches form an upper-bound and are for reference only and not comparison.
\label{OOD-table1}
\scriptsize
\resizebox{\textwidth}{!}{%
    \begin{tabular}{cccccccc}
    \hline 
    \multicolumn{1}{c}{\bf Train }  &  \multicolumn{1}{c}{\bf Method }  &   \multicolumn{2}{c}{\bf \nood{Near-OOD} (Wild)} &  \multicolumn{2}{c}{\bf \food{Far-OOD} (CIFAR10)} &  \multicolumn{2}{c}{\bf \food{Far-OOD} (CelebA)} \Tstrut{}\\
    \bf Dataset & \bf   & AUC-ROC & TNR@TPR95 & AUC-ROC & TNR@TPR95 &   AUC-ROC & TNR@TPR95 \Bstrut{}\\
    \hline
     & Baseline  &  0.88$\pm$0.04 & 47.40$\pm$5.2    & 0.95$\pm$0.04  & 73.59$\pm$9.4 &   0.95$\pm$0.03 & 70.69$\pm$8.9 \Tstrut{}\\
     
     & Baseline+TS~\cite{guo2017temperaturescaling} & 0.88$\pm$0.03   & 45.53$\pm$9.8 &   0.95$\pm$0.04  & 71.77$\pm$8.9& 0.95$\pm$0.03 & 65.89$\pm$8.3 \\

    & Mixup~\cite{zhang2017mixup} & 0.86$\pm$0.06 &   53.83$\pm$6.8  &  0.82$\pm$0.11  & 57.01$\pm$8.6 & 0.88$\pm$0.13 &70.51$\pm$9.8 \\ 
     
    AFHQ & DUQ~\cite{van2020uncertainty}       & 0.78$\pm$0.05    & 20.98$\pm$2.0 & 0.67$\pm$0.59  & 16.23$\pm$1.5 & 0.66$\pm$0.55 & 15.34$\pm$2.6  \\ 
     
     & DDU~\cite{mukhoti2021deterministic} & 0.83$\pm$0.02   & 23.19$\pm$2.6 & 0.90$\pm$0.02    &  32.98$\pm$10& 0.75$\pm$0.02 & 10.32$\pm$5.6 \\ 
     
     & Baseline+Energy~\cite{liu2020energy}  & 0.88$\pm$0.03 & 47.77$\pm$1.10 & 0.94$\pm$0.05  & 72.68$\pm$2.69 & 0.96$\pm$0.04   & 74.75$\pm$2.89 \\
     
     & Energy w/ fine-tune~\cite{liu2020energy}  &    \bf 0.93$\pm$3.06 & 45.97$\pm$2.78 & \bf 0.99$\pm$0.00  & 0.66$\pm$0.01 & 0.94$\pm$1.86 & 68.38$\pm$3.03  \\ 
     & Outlier Exposure~\cite{hendrycks2018deep} & 0.92$\pm$0.01  & \bf 73.99$\pm$2.62 & 0.99$\pm$0.20 & \bf 99.54$\pm$0.79 & 0.96$\pm$0.01 & 78.69$\pm$3.02\\ 
     
     & Baseline+TS+ODIN~\cite{liang2018enhancing} & 0.87$\pm$0.05 &45.02$\pm$1.51    & 0.95$\pm$0.05   & 69.42$\pm$2.38 & 0.95$\pm$0.03 & 67.18$\pm$2.16\\ 
    
    & \bf Baseline+ACE &  0.89$\pm$0.03 &  51.39$\pm$4.4 &  0.98$\pm$0.02    &  88.71$\pm$5.7 &  \bf0.97$\pm$0.03 & \bf 88.87$\pm$9.8 \Bstrut{}\\
    
     \cline{2-8}
    & MC-Dropout~\cite{pmlr-v48-gal16} & 0.84$\pm$0.09  & 30.78$\pm$2.9 & 0.94$\pm$0.02  & 73.59$\pm$2.1& 0.95$\pm$0.02 & 71.23$\pm$1.9 \Tstrut{}\\

    & 5-Ensemble~\cite{NIPS2017_9ef2ed4b} & 0.99$\pm$0.01   & 65.73$\pm$1.2 &   0.97$\pm$0.02  & 89.91$\pm$0.9& 0.99$\pm$0.01 & 92.12$\pm$0.7 \Bstrut{}\\ 
    
    \hline
     &  \multicolumn{1}{c}{ }  &   \multicolumn{2}{c}{\bf \nood{Near-OOD} (Digits 7-9) } &  \multicolumn{2}{c}{\bf \food{Far-OOD} (SVHN)} &  \multicolumn{2}{c}{\bf \food{Far-OOD} (fMNIST)} \Tstrut{}\\
     & \bf   & AUC-ROC & TNR@TPR95 & AUC-ROC & TNR@TPR95 &   AUC-ROC & TNR@TPR95 \Bstrut{}\\
    \cline{3-8}
     & Baseline  & 0.86$\pm$0.04  &  28.23$\pm$2.9 &  0.75$\pm$0.15 &  51.98$\pm$0.9 & 0.87$\pm$0.02 & 58.12$\pm$1.5 \Tstrut{}\\  
    
     & Baseline+TS~\cite{guo2017temperaturescaling} & 0.86$\pm$0.01  &  30.12$\pm$2.1 &  0.73$\pm$0.07 &  48.12$\pm$1.5 & 0.89$\pm$0.01 & 61.71$\pm$2.8\\
     
    Dirty & Mixup~\cite{zhang2017mixup}  & 0.86$\pm$0.02 &  35.46$\pm$1.0 & 0.95$\pm$0.03 &  65.12$\pm$3.1 & 0.94$\pm$0.05 & 66.00$\pm$0.8\\
    
    MNIST & DUQ~\cite{van2020uncertainty}       &  0.78$\pm$0.01   &  15.26$\pm$3.9 & 0.73$\pm$0.03  & 45.23$\pm$1.9 & 0.75$\pm$0.03 & 50.29$\pm$3.1  \\
    
     & DDU~\cite{mukhoti2021deterministic} & 0.67$\pm$0.07   &  10.23$\pm$0.9 &   0.68$\pm$0.04  &  39.31$\pm$2.2 & 0.85$\pm$0.02 & 53.76$\pm$3.7 \\ 
     
     & Baseline+Energy~\cite{liu2020energy}  & 0.87$\pm$0.04 & 40.30$\pm$1.05  &  0.86$\pm$0.12 & 43.92$\pm$2.30 &  0.91$\pm$0.02  & 62.10$\pm$5.17 \\
     
     & Energy w/ fine-tune~\cite{liu2020energy}  & 0.60$\pm$0.08 & 37.43$\pm$0.93 & \bf 1.00$\pm$0.00  & \bf 99.99$\pm$0.00 & \bf 1.00$\pm$0.00 & 99.06$\pm$0.01  \\ 
     & Outlier Exposure~\cite{hendrycks2018deep} &  \bf 0.94$\pm$0.01 & \bf 65.58$\pm$1.64 & \bf 1.00$\pm$0.00 & \bf 99.99$\pm$0.00 & \bf 1.00$\pm$0.00 & \bf 99.56$\pm$0.12 \\ 
     
     & Baseline+TS+ODIN~\cite{liang2018enhancing} & 0.83$\pm$0.04 & 34.13$\pm$12.07   & 0.77$\pm$0.13   & 21.59$\pm$19.62 & 0.89$\pm$0.02 & 46.43$\pm$4.31\\

     & \bf Baseline+ACE & \bf0.94$\pm$0.02   &  37.23$\pm$1.9   &  0.98$\pm$0.02  &   67.88$\pm$3.1 &  0.97$\pm$0.02 &  70.71$\pm$1.1 \Bstrut{}\\

    \cline{2-8}
    & MC-Dropout~\cite{pmlr-v48-gal16} & 0.97$\pm$0.02  &  40.89$\pm$1.5 & 0.95$\pm$0.01 &  62.12$\pm$5.7 & 0.93$\pm$0.02 &65.01$\pm$0.7 \Tstrut{}\\  
    
    & 5-Ensemble~\cite{NIPS2017_9ef2ed4b} & 0.98$\pm$0.02  &  42.17$\pm$1.0 &  0.82$\pm$0.03&  55.12$\pm$2.1 & 0.94$\pm$0.01 &64.19$\pm$4.2 \Bstrut{}\\ 
    \hline
    \end{tabular}}
\end{center}
\end{table}

In Table~\ref{OOD-table1} and Table~\ref{OOD-table2}, we report the AUC-ROC and TNR@TPR95 scores on detecting the two types of OOD samples. We first use the discriminator from the PCE to detect far-OOD samples. The discriminator achieved near-perfect AUC-ROC for detecting far-OOD samples. It consistently outperformed the deep ensemble, MC Dropout, and other deterministic methods across all datasets. The near-OOD samples are relatively similar to the training distribution of the discriminator. Hence, the discriminator performed sub-optimally on the near-OOD detection task. We used the PE estimates from the fine-tuned model (baseline + ACE) to detect near-OOD samples. We outperformed all other deterministic methods in identifying near-OOD samples. Overall our model performed better on both near and far-OOD detection tasks with high TNR@TPR95.

 \begin{table}[!ht]
 \begin{center}
\caption[OOD detection performance for different baselines.]{OOD detection performance for different baselines. \textbf{\nood{Near-OOD}} represents label shift, with samples from the unseen classes of the same dataset. \textbf{\food{Far-OOD}} samples are from a separate dataset.  The numbers are averaged over five runs. }
%\textcolor{red}{The ensemble approaches form an upper-bound and are for reference only and not comparison.}
\label{OOD-table2}
\scriptsize
\resizebox{\textwidth}{!}{%
    \begin{tabular}{cccccccc}
    \hline 
    \multicolumn{1}{c}{\bf Train }  &  \multicolumn{1}{c}{\bf Method }  &     \multicolumn{2}{c}{\bf \nood{Near-OOD} (Kids) } &  \multicolumn{2}{c}{\bf \food{Far-OOD} (AFHQ)} &  \multicolumn{2}{c}{\bf \food{Far-OOD} (CIFAR10)} \Tstrut{}\\
     & \bf   & AUC-ROC & TNR@TPR95 & AUC-ROC & TNR@TPR95 &   AUC-ROC & TNR@TPR95 \Bstrut{}\\
\hline
    & Baseline  & 0.84$\pm$0.02 & 1.25$\pm$0.1 & 0.86$\pm$0.03 & 88.57$\pm$0.9 & 0.79$\pm$0.02 & 29.01$\pm$5.1 \Tstrut{}\\ 
    & Baseline+TS~\cite{guo2017temperaturescaling} & 0.82$\pm$0.04 & 1.24$\pm$0.1 & 0.87$\pm$0.06 & 88.75$\pm$0.9 & 0.78$\pm$0.04 & 29.01$\pm$5.1  \\ 
    & Mixup~\cite{zhang2017mixup} & 0.82$\pm$0.08 & 22.18$\pm$2.7 & 0.95$\pm$0.02 & 82.96$\pm$2.5 & 0.79$\pm$0.13 & 30.54$\pm$1.3\\ 
    CelebA & DUQ~\cite{van2020uncertainty} & 0.80$\pm$0.03 & 14.68$\pm$3.1 & 0.72$\pm$0.07 & 26.62$\pm$7.7 & 0.86$\pm$0.04 & 28.70$\pm$4.1\\ 
    & DDU~\cite{mukhoti2021deterministic} & 0.73$\pm$0.15  & 7.9$\pm$0.4 & 0.74$\pm$0.13  & 8.18$\pm$0.4 & 0.81$\pm$0.15 & 25.45$\pm$1.4 \\ 
    & Baseline+Energy~\cite{liu2020energy}  & 0.76$\pm$0.51 & 9.40$\pm$0.01 & 0.94$\pm$0.08  & 32.08$\pm$1.70 &  0.85$\pm$0.76  & 17.10$\pm$0.72 \\
    & Energy w/ fine-tune~\cite{liu2020energy}  & 0.85$\pm$1.27    & 32.81$\pm$1.92 & \bf 0.99$\pm$0.00  & \bf 99.99$\pm$0.00 & 0.91$\pm$0.77 & \bf 84.35$\pm$1.29  \\ 
    & Outlier Exposure~\cite{hendrycks2018deep} & 0.66$\pm$0.69  & 8.44$\pm$0.45 & 0.75$\pm$0.70 & 26.09$\pm$0.51 & 0.69$\pm$0.53 & 16.63$\pm$0.90\\ 
    & Baseline+TS+ODIN~\cite{liang2018enhancing} & 0.65$\pm$0.01 & 8.75$\pm$2.21   &    0.55$\pm$0.01& 23.03$\pm$0.16 & 0.54$\pm$0.01 & 5.00$\pm$0.07\\ 
    & \bf Baseline+ACE & \bf 0.87$\pm$0.03 & \bf 34.37$\pm$2.5 &     0.96$\pm$0.01 &  96.35$\pm$2.5 & \bf 0.92$\pm$0.05 &   63.51$\pm$1.5  \Bstrut{}\\
    \cline{2-8}
    & MC-Dropout~\cite{pmlr-v48-gal16} & 0.70$\pm$0.10 & 25.62$\pm$1.7 & 0.86$\pm$0.1 & 91.72$\pm$7.5 & 0.74$\pm$0.12 & 64.79$\pm$1.8 \Tstrut{}\\ 
    & 5-Ensemble~\cite{NIPS2017_9ef2ed4b} & 0.93$\pm$0.03 & 10.35$\pm$0.2 & 0.99$\pm$0.0 & 98.31$\pm$1.2 & 0.92$\pm$0.10 & 61.88$\pm$1.2 \Bstrut{}\\ 
    
    \hline
      &  \multicolumn{1}{c}{ }  &   \multicolumn{2}{c}{\bf \nood{Near-OOD} (other lesions) } &  \multicolumn{2}{c}{\bf \food{Far-OOD} (CelebA)} &  \multicolumn{2}{c}{\bf \food{Far-OOD} (Skin-texture)} \Tstrut{}\\
     & \bf   & AUC-ROC & TNR@TPR95 & AUC-ROC & TNR@TPR95 &   AUC-ROC & TNR@TPR95 \Bstrut{}\\
    \cline{3-8}
     & Baseline  &  0.67$\pm$0.04  &8.70$\pm$2.5    &  0.66$\pm$0.06  & 10.00$\pm$3.6 &  0.65$\pm$0.10 & 5.91$\pm$2.8 \Tstrut{}\\ 
     
      & Baseline+TS~\cite{guo2017temperaturescaling} & 0.67$\pm$0.05   & 8.69$\pm$2.0 & 0.63$\pm$0.06  & 9.24$\pm$4.3 & 0.68$\pm$0.07 & 5.70$\pm$3.2 \\ 
     
    Skin& Mixup~\cite{zhang2017mixup} & 0.67$\pm$0.01 & 8.52$\pm$2.8   & 0.64$\pm$0.08  & 10.21$\pm$4.0 &  0.72$\pm$0.05 &5.26$\pm$3.1 \\ 
    
    Lesion & DUQ~\cite{van2020uncertainty} & 0.67$\pm$0.04  &   3.12$\pm$1.8 & 0.89$\pm$0.09 &   11.89$\pm$2.5 & 0.64$\pm$0.03 & 4.8$\pm$1.5\\
    
     & DDU~\cite{mukhoti2021deterministic} & 0.65$\pm$0.03  &   3.45$\pm$1.9 & 0.75$\pm$0.04 &  15.45$\pm$2.9 & 0.71$\pm$0.05 & 4.19$\pm$1.3 \\
    & Baseline+Eenergy~\cite{liu2020energy}  & 0.70$\pm$0.04 & 10.85$\pm$0.08 & 0.70$\pm$0.14  & 7.90$\pm$0.29 & 0.65$\pm$0.20   & 2.83$\pm$1.33\\
    & Energy w/ fine-tune~\cite{liu2020energy}  & 0.62$\pm$0.02 & 9.80$\pm$1.81 &  \bf 1.00$\pm$0.00 & \bf 99.77$\pm$0.33 & 0.76$\pm$0.13 & 16.04$\pm$1.08  \\ 
    & Outlier Exposure~\cite{hendrycks2018deep} & 0.67$\pm$0.09 & 10.38$\pm$3.30 & 0.99$\pm$0.00 & 97.17$\pm$2.37 & 0.81$\pm$0.08 & 22.64$\pm$4.30 \\ 
    & Baseline+TS+ODIN~\cite{liang2018enhancing} & 0.68$\pm$0.01 & 9.43$\pm$0.33   & 0.67$\pm$0.07   & 11.32$\pm$4.66 & 0.68$\pm$0.07 & 6.60$\pm$0.29\\

    & \bf Baseline+ACE & \bf0.72$\pm$0.04   & \bf 11.00$\pm$2.8 & 0.97$\pm$0.02   & 66.77$\pm$1.4 &  \bf 0.96$\pm$0.03 & \bf 95.83$\pm$5.0 \Bstrut{} \\
     
     \cline{2-8}
    & MC-Dropout~\cite{pmlr-v48-gal16} & 0.67$\pm$0.05   & 9.45$\pm$3.9 &  0.80$\pm$0.07  & 30.00$\pm$3.2&0.56$\pm$0.03  & 10.87$\pm$2.3 \Tstrut{}\\

    & 5-Ensemble~\cite{NIPS2017_9ef2ed4b} & 0.88$\pm$0.01   & 11.23$\pm$1.7 &  0.91$\pm$0.03  & 27.89$\pm$5.9 & 0.76$\pm$0.02 & 17.89$\pm$3.5 \Bstrut{}\\

    \hline
    \end{tabular}}
\end{center}
\end{table}

\subsection{Toy-Setup - Two Moons}
\begin{figure}[!ht]
  \centering
  \includegraphics[width=0.95\linewidth]{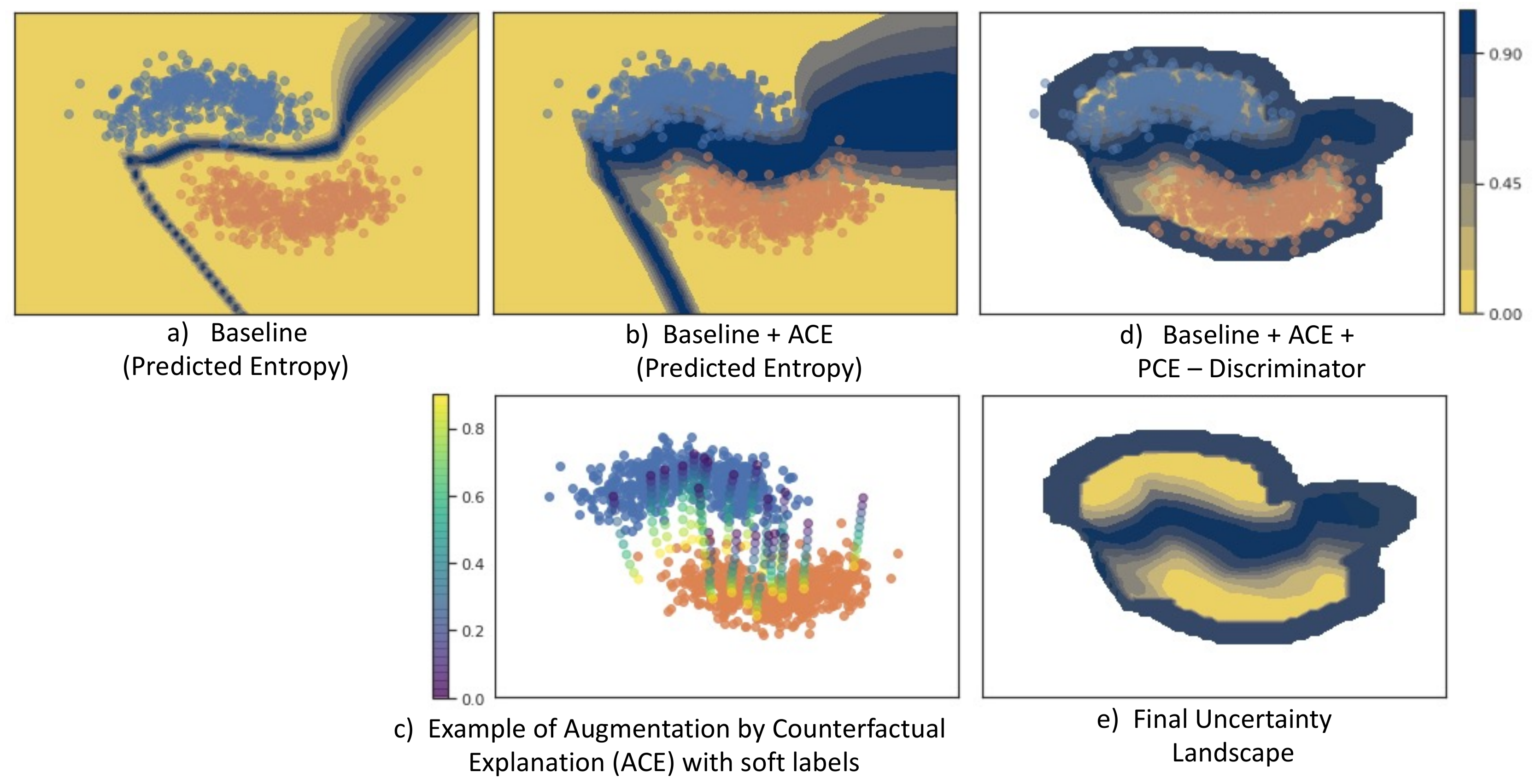}

  \caption[Uncertainty results on the Two Moons dataset.]{Uncertainty results on the Two Moons dataset.
  Yellow indicates low uncertainty, while blue indicates uncertainty. a) The baseline classifier is uncertain only along the decision boundary, and certain elsewhere. b) Fine-tuning baseline model on ACE data  improves uncertainty estimates near the decision boundary. c) An example of augmented data and corresponding soft labels. d) The discriminator from PCE rejects OOD samples, hence the rejected space have no uncertainty values (white color). e) The final uncertainty landscape, the improved classifier is certain on in-distribution regions and rejects OOD data.}
  \label{fig:2D}
\end{figure}

This section demonstrates the over-confidence problem in a classifier trained on the two moons dataset. Using the scikit-learn’s datasets package, we generated 2000 samples with a noise rate of 0.1. Our baseline classification model is a 2-layer MLP. In Fig.~\ref{fig:2D}.a, we visualize the uncertainty estimates from this classifier. A classifier optimized for cross-entropy loss learns a very sharp decision boundary with low uncertainty only near the decision boundary and high uncertainty everywhere else.
\newline

In Fig.~\ref{fig:2D}.b, we visualize the revised decision boundary after fine-tuning the classifier with counterfactually augmented data (CAD). The decision boundary is much broader, and uncertainty is high near the decision boundary, decreasing as one moves away from it. In Fig.~\ref{fig:2D}.c, we show examples of CAD. Using the progressive counterfactual explainer (PCE), we created samples resembling a walk from one class to the other while crossing the decision boundary.

This simple example demonstrates that fine-tuning the classifier with augmented data near the decision boundary with soft labels helped the classifier recover from the over-saturation on the negative log-likelihood (NLL) loss. Hence, the fine-tuned classifier has better uncertainty estimates near the decision boundary and is not over-confident on ambiguous in-distribution samples in the class over-lapping regions.

Further, in Fig.~\ref{fig:2D}.d, we show the hard threshold used by the discriminator of the PCE as the selection function. We processed all the samples through the discriminator of the PCE and used a pre-defined threshold to separate in-distribution samples from the OOD samples. The white colour in the plot is the samples the discriminator rejects as OOD. Fig.~\ref{fig:2D}.e shows the final uncertainty landscape. The near-OOD around the in-distribution samples all have high uncertainty. The ambiguous in-distribution samples are assigned a high uncertainty near the decision boundary.

\subsection{Robustness to Adversarial Attacks}

\begin{figure}[!ht]
\centering
      \includegraphics[width=1.0\textwidth]{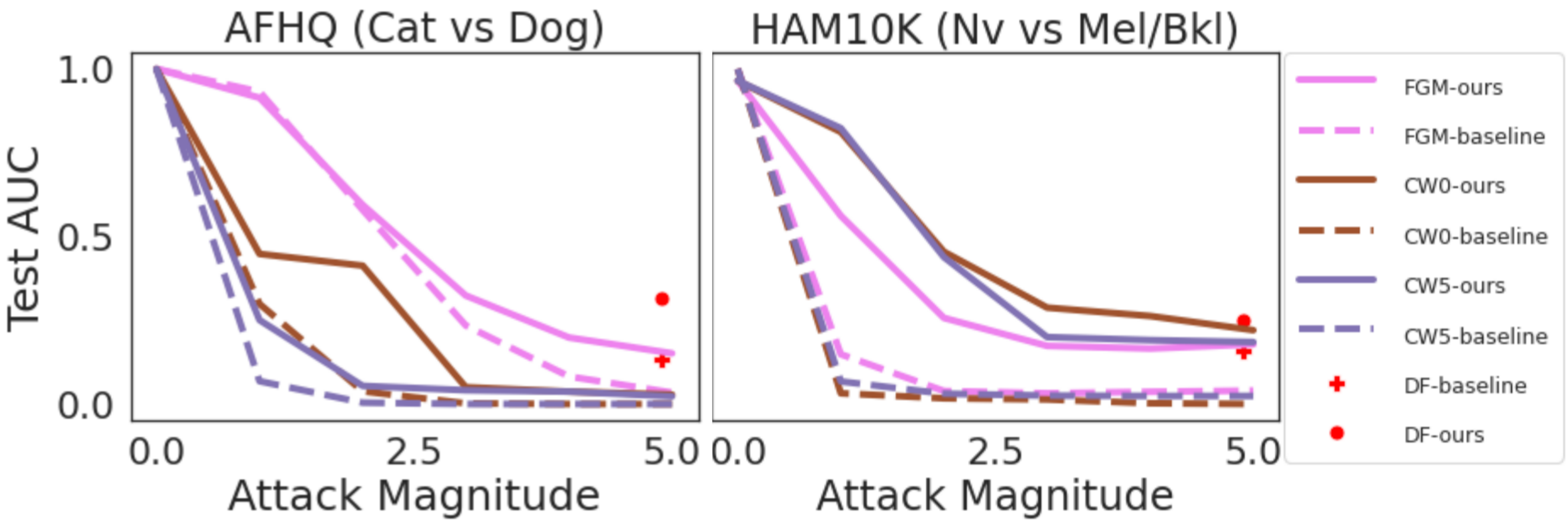}
     
    \caption[Plots comparing baseline model before and after fine-tuning (ACE) for different magnitudes of adversarial attack.]{Plots comparing baseline model before and after fine-tuning (ACE) for different magnitudes of adversarial attack. The figure shows three different attacks -- FGSM~\cite{fgsm_attack}, CW~\cite{carlini_wagner_attack}, DeepFool~\cite{deepfool_attack}, on three different datasets -- HAM10K, AFHQ, MNIST. The x-axis denotes maximum perturbation ($\epsilon$) for FGSM, and iterations in multiples of $10$ for CW and DeepFool. Attack magnitude of $0$ indicates no attack. For CW we used $\kappa=0$ and $5$. (All results are reported on the test-set of the classifier).}
    \label{fig:adv_attacks}
  \end{figure}

In this experiment, we compared the baseline model before and after fine-tuning (baseline + ACE) in their robustness to three adversarial attacks: Fast Gradient Sign Method (FGSM)~\cite{fgsm_attack}, Carlini-Wagner (CW)~\cite{carlini_wagner_attack}, and DeepFool~\cite{deepfool_attack}. For each attack setting, we transformed the test set into an adversarial set.  In Fig.~\ref{fig:adv_attacks}, we report the AUC-ROC over the adversarial set as we gradually increase the magnitude of the attack. For FGSM,  we use the maximum perturbation ($\epsilon$) to specify the attack's magnitude. For CW, we gradually increase the number of iterations to an achieve a higher magnitude attack. We set box-constraint parameter as $c=1$, learning rate $\alpha=0.01$ and confidence $\kappa=0,5$. For  DeepFool ($\eta=0.02$), we show results on the best attack. Our improved model (baseline + ACE) consistently out-performed the baseline model in test AUC-ROC.  
  
   % We investigate this further in the supplementary section.

\subsection{Ablation Study}

We conducted an ablation study over the three-loss terms of PCE in Eq.~\ref{eq:5}. The three terms of the loss function enforce three properties of counterfactual explanation, data consistency: explanations should be realistic looking images, classifier consistency: explanations should produce the desired outcome from the classifier and self-consistency: explanation image should retain the identity of the query image. For the ablation study, we consider the cat and dog classifier. We train three PCE; in each run, we ablate one term from the final loss function. 
\newline

  \vspace{-0.4cm}

\begin{figure}[!ht]
  \centering
  \includegraphics[width=1.0\textwidth]{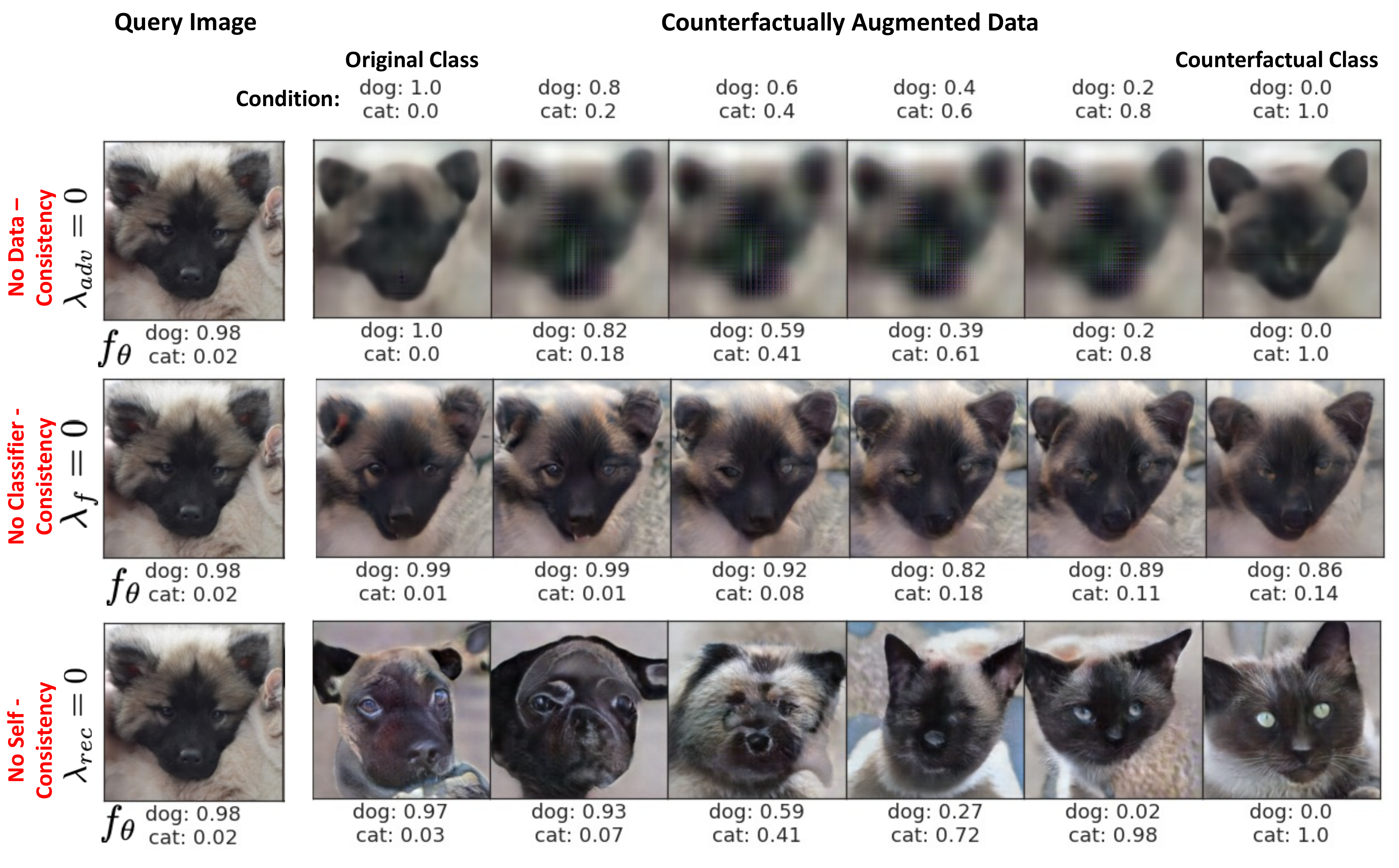}
  \caption { Examples of data augmentation while ablating different loss terms.}
  \label{fig:ablation}
\end{figure}

In Fig.~\ref{fig:ablation}, we show a qualitative example of the counterfactual data augmentation generated through each PCE. Without data consistency, the images are blurry and are no longer realistic. Without classifier consistency loss, though the images are natural, the classifier's output is not changing with the condition. Hence such PCE won't generate augmented samples near the decision boundary, which is the goal of our proposed strategy. With self-consistency, the generated images are not a gradual transformation of a given query image.

\begin{figure}[!ht]
  \centering
  \includegraphics[width=1.0\textwidth]{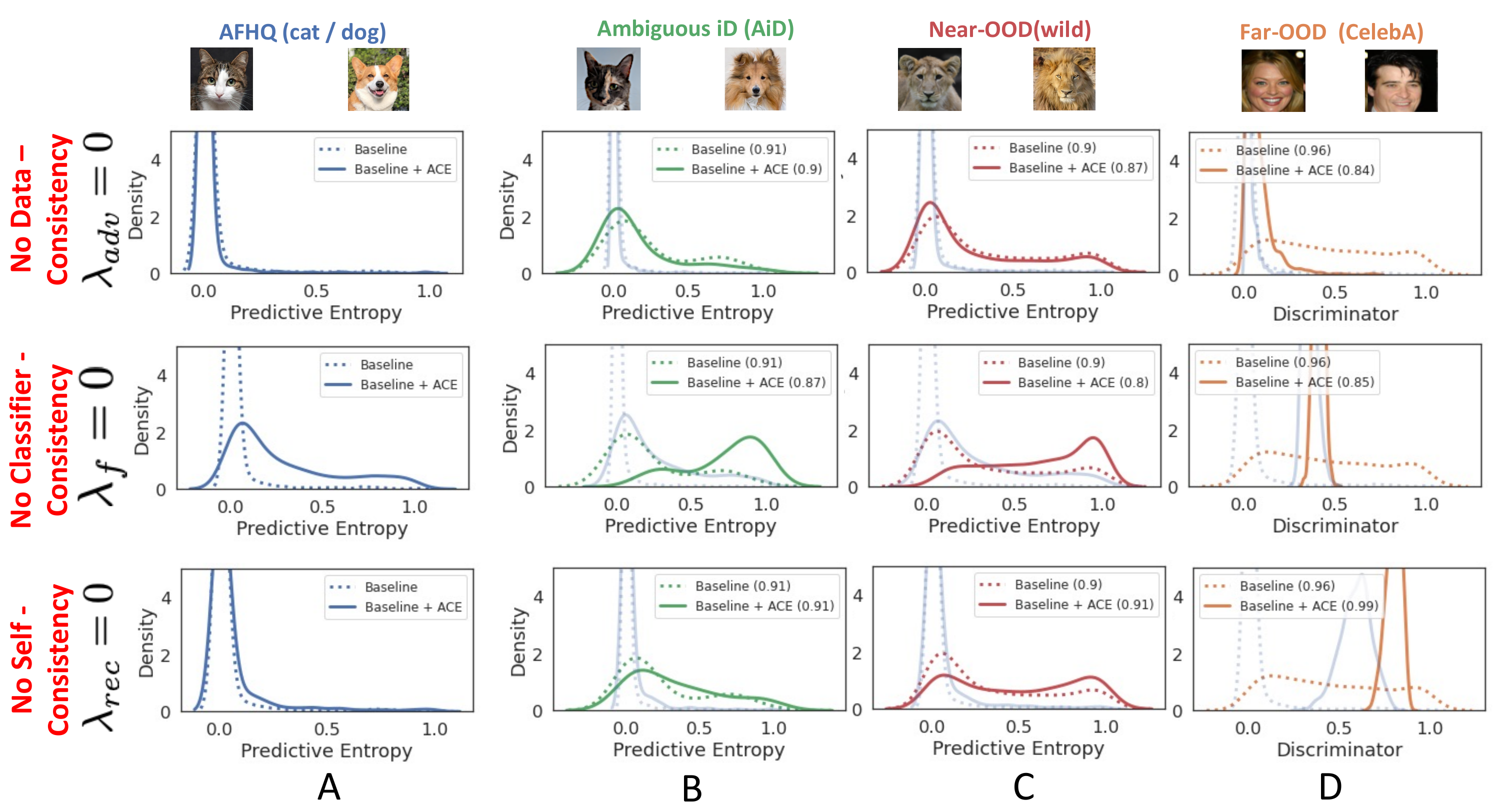}
  
  \caption[Comparison of the uncertainty estimates from the baseline, before and after the fine-tuning with ACE.]{Comparison of the uncertainty estimates from the baseline, before and after the fine-tuning with ACE.
  Each row represents a different ablation over the three loss terms. A) Predicted entropy (PE) of \textbf{\uid{in-distribution (iD)}} samples. Ideally, fine-tuning should minimally effect the PE distribution over iD samples. Without classification consistency loss (second row), the PE distribution of iD samples changed significantly. B)  PE distribution over \textbf{\aid{ ambiguous in-distribution (AiD)}} samples. C) PE distribution over \textbf{\nood{near-OOD}} samples.   The data augmentation derived from PCE without adversarial loss or reconstruction loss, is not able to separate AiD samples or near-OOD from rest of the test set. D)  We use the discriminator of the PCE to identify \textbf{\food{far-OOD}} samples. In all three rows, we observe sub-optimal performance of the discriminator in identifying and rejecting far-OOD samples.  The legend shows the AUC-ROC for binary classification over uncertain samples and iD samples. Hence, all three loss terms are important to improve the uncertainty estimates of the baseline over all samples across the uncertainty spectrum.} 
  \label{fig:ablation_all}
\end{figure}

Further, in Fig.~\ref{fig:ablation_all} we present quantitatively compare the uncertainty estimates from the baseline, before and after the fine-tuning with ACE. We represent a different ablation over the three-loss terms in each row. 
\newline 

\vspace{-0.4cm}

Fig.~\ref{fig:ablation_all}.A. shows the predicted entropy (PE) of \textbf{\uid{in-distribution (iD)}} samples. Ideally, fine-tuning should minimally affect the PE distribution over iD samples. Without classification consistency loss (second row), the PE distribution of iD samples changed significantly. Fig.~\ref{fig:ablation_all}.B and  Fig.~\ref{fig:ablation_all}.C shows the PE distribution over \textbf{\aid{ ambiguous in-distribution (AiD)}} samples and  \textbf{\nood{near-OOD}} samples, respectively.   The data augmentation derived from PCE without adversarial loss or reconstruction loss cannot separate AiD samples or near-OOD from the rest of the test set. In Fig.~\ref{fig:ablation_all}.D, we use the discriminator of the PCE to identify \textbf{\food{far-OOD}} samples. In all three rows, we observe the sub-optimal performance of the discriminator in identifying and rejecting far-OOD samples. The legend shows the AUC-ROC for binary classification over uncertain and iD samples. Hence, all three loss terms are important to improve the uncertainty estimates of the baseline over all samples across the uncertainty spectrum.
\newline

 \section{Discussion and Conclusion}
 
We propose a novel method to improve the uncertainty quantification of an existing  \emph{pre-trained} DNN by fine-tuning it on counterfactually augmented data. We used a cGAN-based counterfactual explainer to generate the data augmentation. Our fine-tuned model, combined with the discriminator of the GAN, can successfully capture uncertainty over ambiguous samples, unseen near-OOD samples with label shift and far-OOD samples from independent datasets. Comparative post-hoc methods such as thresholding softmax outputs and temperature scaling cannot recover a pre-trained model from over-saturation on log-likelihood loss. Other deterministic methods significantly change the classification model design to enable better uncertainty quantification over OOD samples. These methods require a network to be trained from scratch and are not compatible with a pre-trained classifier. Our proposed strategy reuses the counterfactual explanation model for the given classifier to fix its over-confidence problem. We out-performed state-of-the-art methods for uncertainty quantification on four datasets with varying difficulty levels.
Furthermore, our improved model also exhibits robustness to prevalent adversarial attacks. We recognize that our proposed strategy involves training a GAN and fine-tuning the classifier with augmented data, which creates a one-time computational overhead. But, once we have a fine-tuned classifier, it requires only a single forward pass, with fast inference. The trained GAN has the added benefit of explaining the given classification model that can help in making it more user-accessible. Our work opens up a new direction for improving uncertainty quantification in existing classification models.

 \chapter{Conclusions and Future Directions}
\label{ch5}
\section{Conclusion}
In this thesis, we developed new deep learning architectures and post-hoc explainability techniques that improved the application of DL methods for medical image classification.  This dissertation proposes models to account for subtleties of medical imaging and add support for specific clinical needs. The major contribution of this work is to tackle model explanation from different perspectives. We started with building an interpretable model, that not only provides accurate predictions but also use an attention mechanism to show important/relevant regions of the input. We designed the model to handle the subtleties of medical images, by have an input processing pipeline that can process the entire 3D volume with minimum resizing, thus reserving the spatial integrity of the imaging data (Chapter-3).

Moving on, we developed a progressive counterfactual explainer, that provides visual explanation to explain the decision of a pre-trained classifier in a post-hoc manner. The design of our explainer is highly motivated by the clinical use-cases. For instance, most of the lung diseases are developed progressively and hardy have a sudden appearance. Out explanation, shows a gradual transformation of the query image, where the input CXR gradually becomes positive for a diagnosis. This aligns with the clinical expectation of how decision for a diagnosis should change. Further, counterfactual explanations are superior than saliency-map based methods as they not only show where in the image the classifier is paying attention to make its prediction, but also shows what image features in those salient regions are essential for the positive or negative decision. We supported our methods with through experiments, on both natural image datasets and medical image datasets. From a clinical perspective, we demonstrated that the counterfactual changes associated with normal (negative) or abnormal (positive)  classification decisions are also associated with corresponding changes in disease-specific metrics such as CTR and SCP. For example, changes associated with an increased posterior probability for cardiomegaly also resulted in an increased CTR. Similarly, for PE, a healthy CP recess with a high SCP score transformed into an abnormal CP recess with blunt CPA, as the posterior probability for PE increase. Further, we evaluated our methods through a human evaluation study. The results of our human evaluation study confirms that the counterfactual explanations obtain from our method helped the clinicians better understand the classification decision (Chapter-4).

Chapter-5 presents an application of the counterfactual explainer in obtaining concept-based explanations. This method is also motivated by the need of the domain expert, to receive model explanation in a terminology that is meaningful to them. To fulfill this requirement, we provide explanation in terms of clinical concepts that are  used in radiology reports to support the presence of a diagnosis. Specifically, we associate the internal structure of the deep neural network with clinically relevant concepts and used our counterfactual explanations to measure the causal effect of these concepts on the model's prediction. We adopted tools from Causal Inference
literature and, more specifically, mediation analysis through counterfactual interventions. Using measures from mediation analysis, we provide an effective ranking of the concepts based on their causal relevance to the model’s outcome. Finally, we construct a low-depth decision tree to express discovered concepts in simple decision rules, providing the
global explanation for the model. We presented a through experiment of our proposed method on a clinical dataset.

For a more comprehensive interpretation of the deep learning models by their end-users, in Chapter-7 we demonstrate how to improve the uncertainty estimates from a pre-trained classifier, by fine-tuning the classifier with counterfactually augmented data. Counterfactual data lies near the decision boundary between two classes, Fine-tuning with such data helps in making the decision boundary wider and thus preventing the classifier from making over confident predictions on the sample near the decision boundary. Further, we show that the discriminator from the counterfactual explanier is a good proxy to the data distribution. The likelihood estimates from this model thus can identify and reject OOD samples. The experiments with the natural and medical images showed that our proposed technique is helpful in learning more reliable classifiers.  We out-performed state-of-the-art methods for uncertainty quantification on four datasets with varying difficulty levels. Furthermore, our improved model also exhibits robustness to prevalent adversarial attacks

Overall, this thesis is a summarization of different ways to explain the deep learning model decision. The methods proposed in this thesis provided a set of tools to the deep learning model designers to better design and explain DNNs, while satisfying the clinical needs. Making progress in this direction will ensure the path to deployment for DNN models. For all the proposed methods we provide thorough comparisons with existing baselines and in each case we demonstrate reliable and superior performance.

%There are also several avenues for improvement which are left for future work:

\section{Future directions}

There are also several avenues for improvement which are left for future work:

\begin{enumerate}
    \item We lose the context information when representing a volumetric image as a set of patches. There is no notion of spatial context when elements of a set are processed in a format invariant to their order. Future work should explore adding a positional encoding to the patches to incorporate spatial information. Much of this is inspired by recent advanced DNN architectural designs. Highly complex DNN designs such as vision transformers also take tokens as input. These patches, along with positional encoding, can become an essential way of processing 3D volumetric data for transformer-based architectures~\cite{fan2021multiscale}.
    
    \item Diversity is an essential aspect of counterfactual explanations. Diversity among the generated counterfactuals provides different ways of changing the outcome decision. Diverse counterfactuals offer users multiple options to understand which input features are important for the classification decision. Diverse counterfactuals may include changes to a particular concept or several concepts. Current work is restrictive as it creates only a single counterfactual image. Future work should explore generating diverse counterfactual explanations showing all possible ways of changing the classification decision~\cite{Rodriguez_2021_ICCV,Mothilal2019ExplainingExplanations}. Further, disentangling the counterfactual changes into human-understandable concepts can enrich the quality and usability of counterfactual explanations.
    
    \item A counterfactual explanation is incomplete without causal analysis~\cite{mandel2007counterfactual}. The current work uses neuron activations as a weak signal to capture concept information in the network. Future work should build on it to identify concepts in counterfactual explanations~\cite{Goyal2019Explainingcace}. A possible next step is to use the concept vectors to navigate the direction in which counterfactual perturbation should be made to achieve the desired change in the classification decision.
    
    \item The current definition of the concept in the concept-based explanations is restrictive. It only captures a few radiological terms and provides no intuition behind what part of the prediction decision is not explainable using the current definition of the concepts. Future work should explore adding the notion of completeness, which quantifies how sufficient a particular set of concepts is in explaining a model’s prediction behavior based on the assumption that complete concept scores are sufficient statistics of the model predictions~\cite{yeh2019completeness}. Researchers have already started exploring this direction. However, there is a limited progress in terms of identifying extensive concepts for medical tasks. Also, further research is required to quantify \emph{completeness}. 
    
    \item Uncertainty quantification is a challenging task. Our current work focuses on improving uncertainty estimates from a pre-trained classifier. The training of PCE is a computationally expensive and time intensive overhead. The future work should explore removing this overhead and proposing an augmentation that can be quickly achieved. One direction would be to use a pre-trained generative model for getting augmented data.
    
    \item The clinical concepts and their definitions are blurred and uncertain~\cite{NguyenICDM11}. Another prospective direction is to explore for learning a DNN with improved uncertainty quantification is to train the DNN with soft labels instead of strict hard labels~\cite{NguyenAMIA11}. Some preliminary work in this direction have found that the clinical annotators prefer to label clinical data with soft labels which can come in form of probability estimates or discrete categories defining different degrees (strength) of labels~\cite{NguyenJAMIA13,ValizadeganJBI13}. Further, the soft labels help to both reduce the sample size from which we can train the high quality models and also make these methods less sensitive to highly unbalanced data.
    
\end{enumerate}

%==========================================================================================%
% APPENDIX
%==========================================================================================%
\appendix     
%After this command, chapters will be formatted as appendices. 

\chapter{Progressive Counterfactual Explanation}

\section{Human evaluation}
\label{sm-he}
In our human evaluation study, we asked the following 15 questions for each CXR:
\newline

\begin{figure}[!ht]
    \centering
    \includegraphics[width = 0.8\linewidth]
    {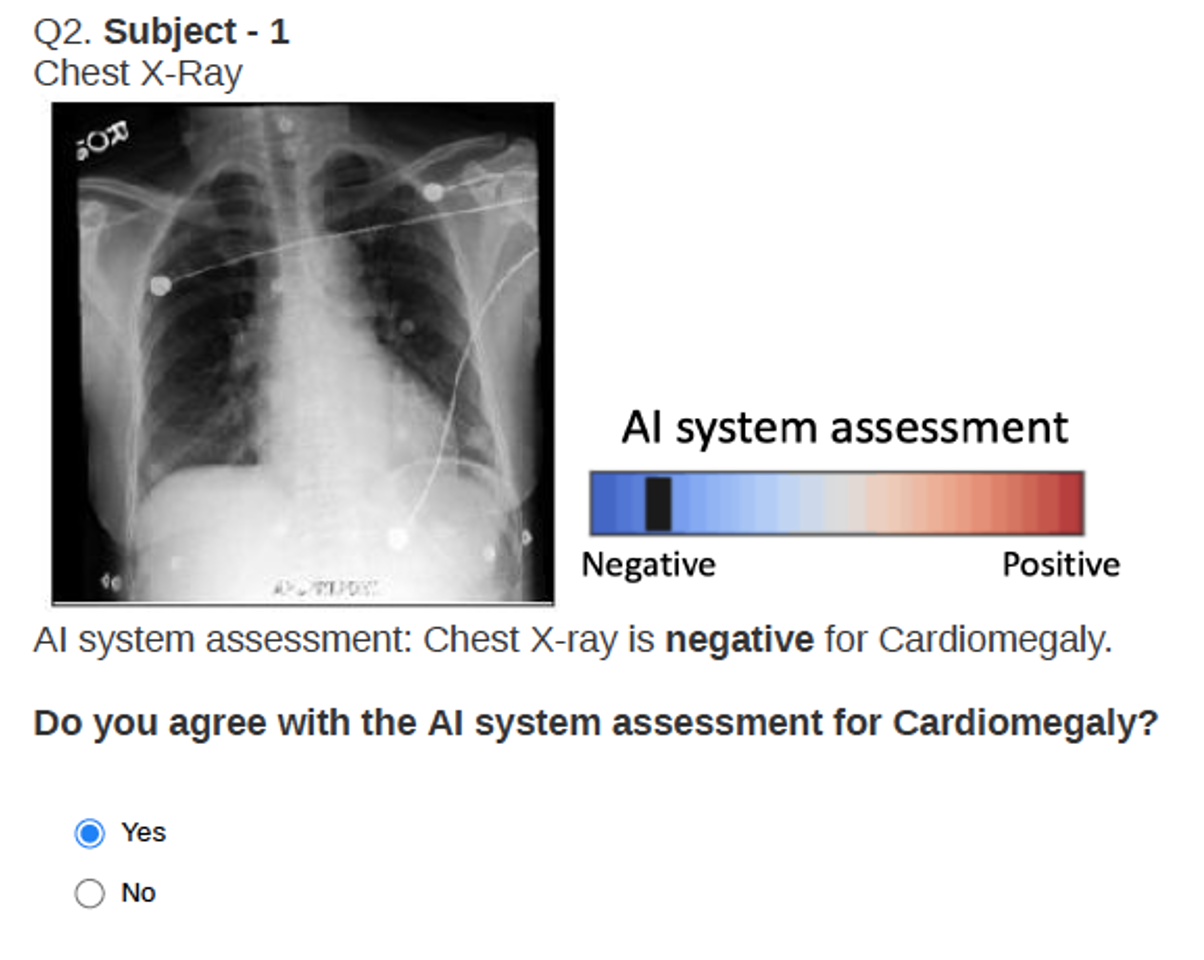}
    \caption{ Question 2-3 showing the query CXR and the classifier's decision.}
\end{figure}

\vspace{-0.5cm}
\begin{enumerate}
    \item Please provide your diagnosis for Cardiomegaly. Answers: Negative, mild, positive, not sure.
    \item (Only assessment) Do you agree with the AI system assessment for Cardiomegaly? Answers: yes, no
    \item (Only assessment)  I understand how the AI system made the above assessment for Cardiomegaly. Answers: 5-point Likert scale.
   
   \begin{figure}[!ht]
    \centering
    \includegraphics[width = 0.8\linewidth]
    {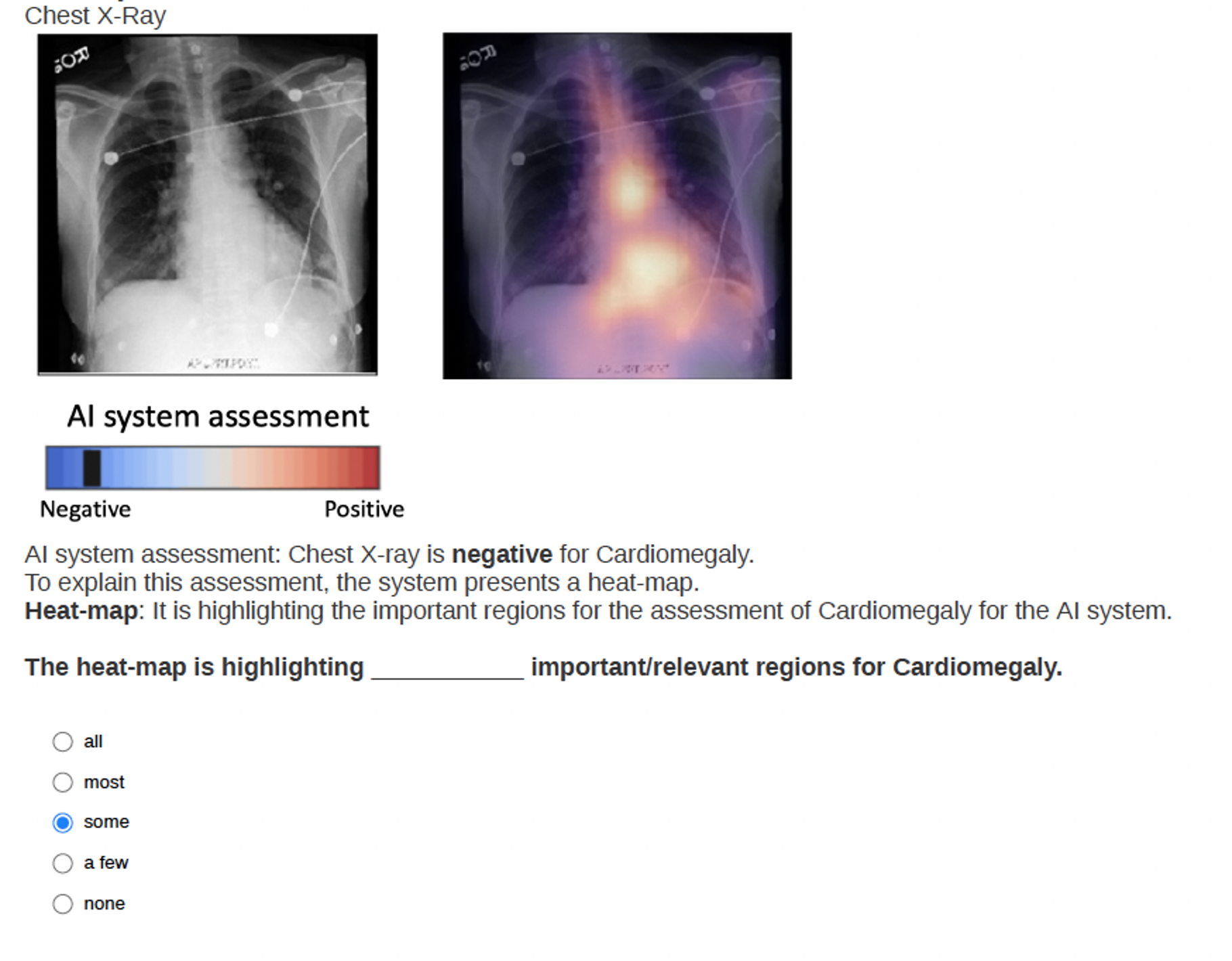}
    \caption{ Question 4-5 showing the query CXR, the classifier's decision and the saliency map explanation.}
\end{figure}

    \item (Assessment + SM) The heat-map is highlighting $<$blank$>$ important/relevant regions for Cardiomegaly. Answers: all, most, some, a few, none.
    \item (Assessment + SM) I understand how the AI system made the above assessment for Cardiomegaly. Answers: 5-point Likert scale.
    
    \begin{figure}[!ht]
    \centering
    \includegraphics[width = 0.8\linewidth]
    {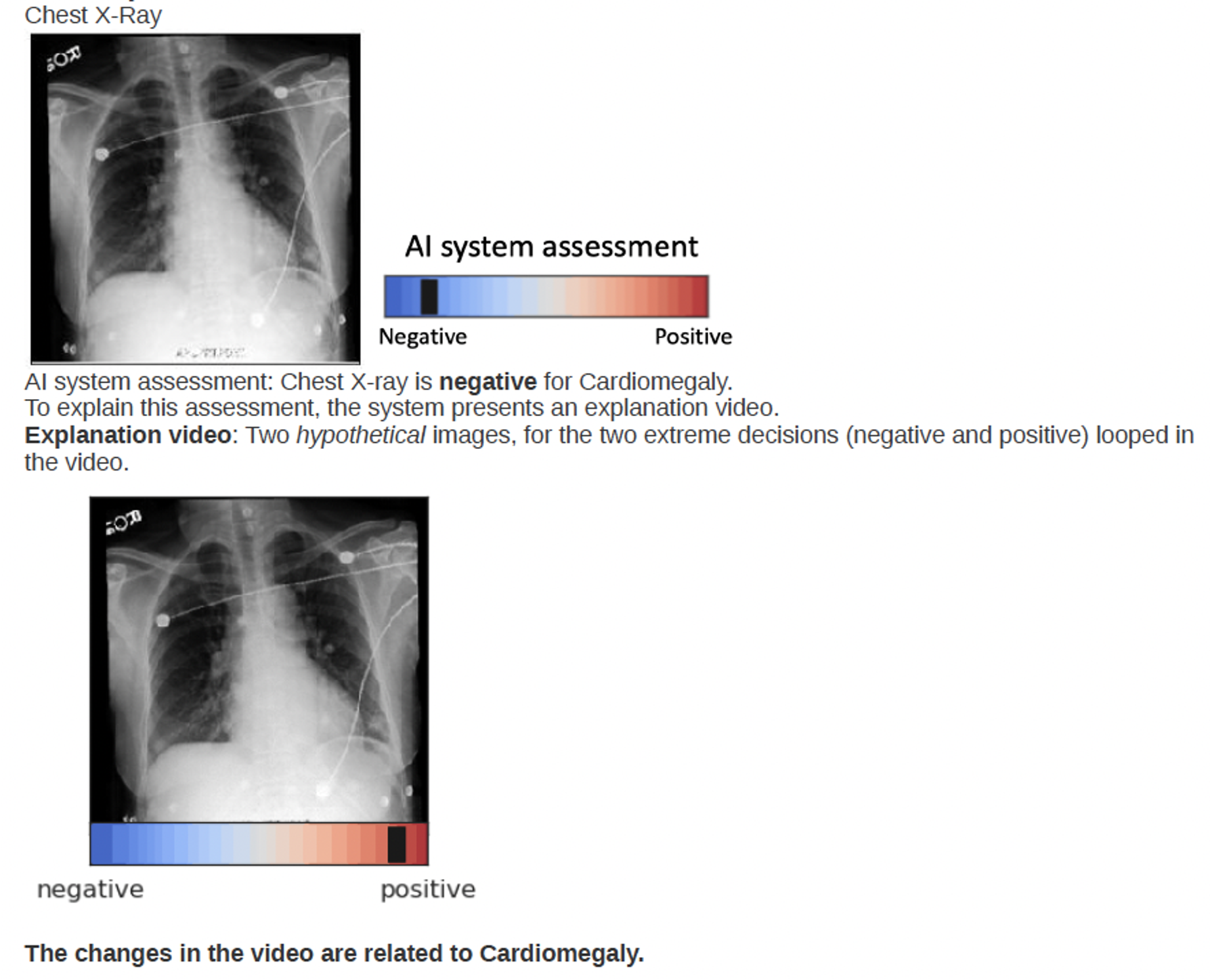}
    \caption{ Question 6-9 showing the query CXR, the classifier's decision and the cycleGAN explanation.}
\end{figure}

    \item (Assessment + cycleGAN) The changes in the video are related to Cardiomegaly. Answers: 5-point Likert scale.
    \item (Assessment + cycleGAN)  I understand how the AI system made the above assessment for Cardiomegaly. Answers: 5-point Likert scale.
    \item (Assessment + cycleGAN) Images in the video look like a chest x-ray. Answers: 5-point Likert scale.
    \item (Assessment + cycleGAN) The images in the video look like the chest x-ray from the subject. Answers: 5-point Likert scale.

\begin{figure}[!ht]
    \centering
    \includegraphics[width = 0.8\linewidth]
    {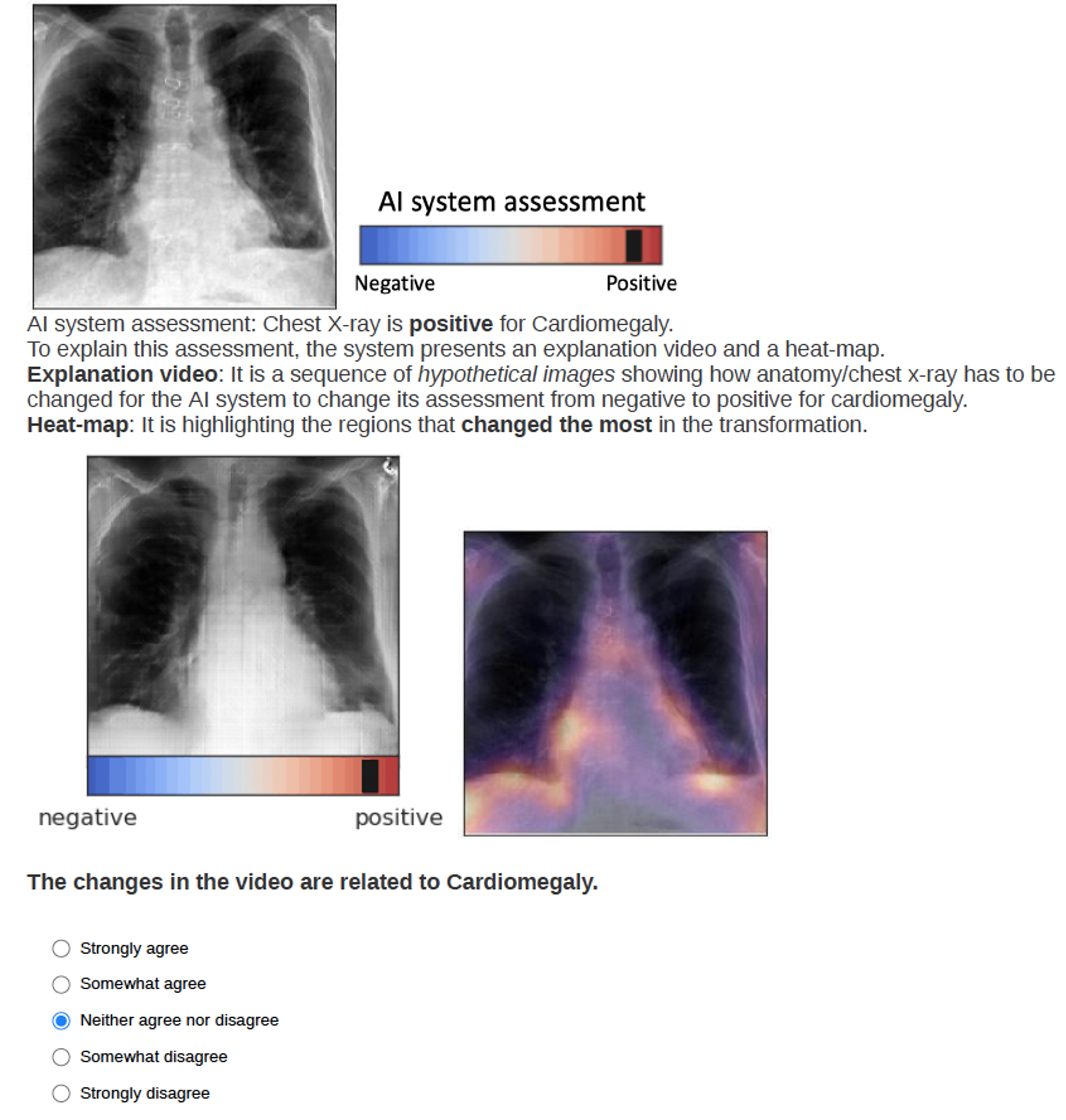}
    \caption{ Question 10-14 showing the query CXR, the classifier's decision and our counterfactual explanation.}
\end{figure}

    \item (Assessment  + ours) The changes in the video are related to Cardiomegaly. Answers: 5-point Likert scale.
    \item (Assessment  + ours) Changes in the anatomy in the highlighted regions in the heat-map will change the assessment
of Cardiomegaly. Answers: 5-point Likert scale.
\item (Assessment + ours)  I understand how the AI system made the above assessment for Cardiomegaly. Answers: 5-point Likert scale.
\item (Assessment + ours) Images in the video look like a chest x-ray. Answers: 5-point Likert scale.
\item (Assessment + ours) The images in the video look like the chest x-ray from the subject. Answers: 5-point Likert scale.
\item Which explanation helped you the most in understanding the assessment made by the AI system
    Answers: Explanation-1: Heat-map highlighting important regions for assessment, Explanation-2: A video showing the transformation from negative to positive decision, Explanation-3: Two images at the two extreme ends of the decision (positive and negative), none.
\end{enumerate}

\section{Summarizing the notation}
Table.~\ref{Notation-table} summarizes the notation used in the manuscript.

\begin{table}[!ht]
\centering
\caption{ Summarizing the notation.}
\label{Notation-table}
\begin{tabular}{l|l}
Notation &  Description\\
\hline
$\mathcal{X}$ & Input image space\\
$\rvx \in \mathcal{X}$ & Input image\\
$f:\mathcal{X}\rightarrow \mathcal{Y}$ & Pre-trained classification function\\
$f(\rvx)[k] \in [0,1]$ & Classifier's output for $k^{\text{th}}$ class \\
$\rvc$ & The condition used in cGAN, the desired classifier's output\\
&  for the $k^{\text{th}}$ class\\
$\rvx_{\rvc}$ & Explanation image\\
$f(\rvx_{\rvc})$ & Classifier's output for the explanation image\\
$\mathcal{I}_f(\rvx, \rvc)$ & Explanation function\\
$E(\cdot)$ & Image encoder\\
$\mathbf{z}$ & Latent representation of the input image\\
$C(\rvc)$&Discretizing function that maps $\rvc$ to an integer\\
$G(\mathbf{z}, \mathbf{c})$ & Generator of cGAN\\
$D(\rvx, \mathbf{c})$ & Discriminator of cGAN\\
$p_{\textnormal{data}}(\rvx)$ & Real image data distribution\\
$q(\rvx)$ & Learned data distribution by cGAN\\
$r(\rvx)$ & Loss term of cGAN that measures similarity between real\\
&  and learned data distribution \\
$r(\mathbf{c}|\rvx)$ & Loss term of cGAN that evaluates correspondence between \\
&  generated images and condition\\
$\phi(\rvx)$& Image feature extractor; part of the discriminator function\\
\hline
\end{tabular}
\end{table}

\section{MIMC-CXR Dataset}
We focus on explaining classification models based on deep convolution neural networks (CNN); most state-of-the-art performance models fall in this regime. We used large, publicly available datasets of chest x-ray (CXR) images, MIMIC-CXR~\cite{Johnson2019MIMIC-CXRReports}. MIMIC-CXR dataset is a multi-modal dataset consisting of 473K CXR, and 206K reports from 63K patients.  We considered only frontal (posteroanterior PA or anteroposterior AP) view CXR. The datasets provide image-level labels for fourteen radio-graphic observations.  These labels are extracted from the radiology reports associated with the x-ray exams using an automated tool called the Stanford CheXpert labeler~\cite{Irvin2019CheXpert:Comparison}. The labeller first defines some thoracic observations using a radiology lexicon~\cite{Hansell2008FleischnerImaging}. It extracts and classifies (positive, negative, or uncertain mentions) these observations by processing their context in the report. Finally, it aggregates these observations into fourteen labels for each x-ray exam. For the MIMIC-CXR dataset, we extracted the labels ourselves, as we have access to the reports. 

\section{Classification Model}
\label{SM-CM}
To train the classifier, we considered the uncertain mention as a positive mention. We crop the original images to have the same height and width, then downsample them to 256 $\times$ 256 pixels. The intensities were normalized to have values between 0 and 1.  Following the approach in prior work~\cite{Rajpurkar2017CheXNet:Learning,Rubin2018LargeNetworks,Irvin2019CheXpert:Comparison} on diagnosis classification, we used DenseNet-121~\cite{Huang2016DenselyNetworks} architecture as the classification model.  In DenseNet, each layer implements a non-linear transformation based on composite functions such as Batch Normalization (BN), rectified linear unit (ReLU), pooling, or convolution. The resulting feature map at each layer is used as input for all the subsequent layers, leading to a highly convoluted multi-level multi-layer non-linear convolutional neural network.  We aim to explain such a model post-hoc without accessing the parameters learned by any layer or knowing the architectural details. Our proposed approach can be used for explaining any DL based neural network.

\section{Progressive Counterfactual Explainer}
\label{SM-EF}
\begin{figure}[!ht]
    \centering
    \includegraphics[width = 0.8\linewidth]
    {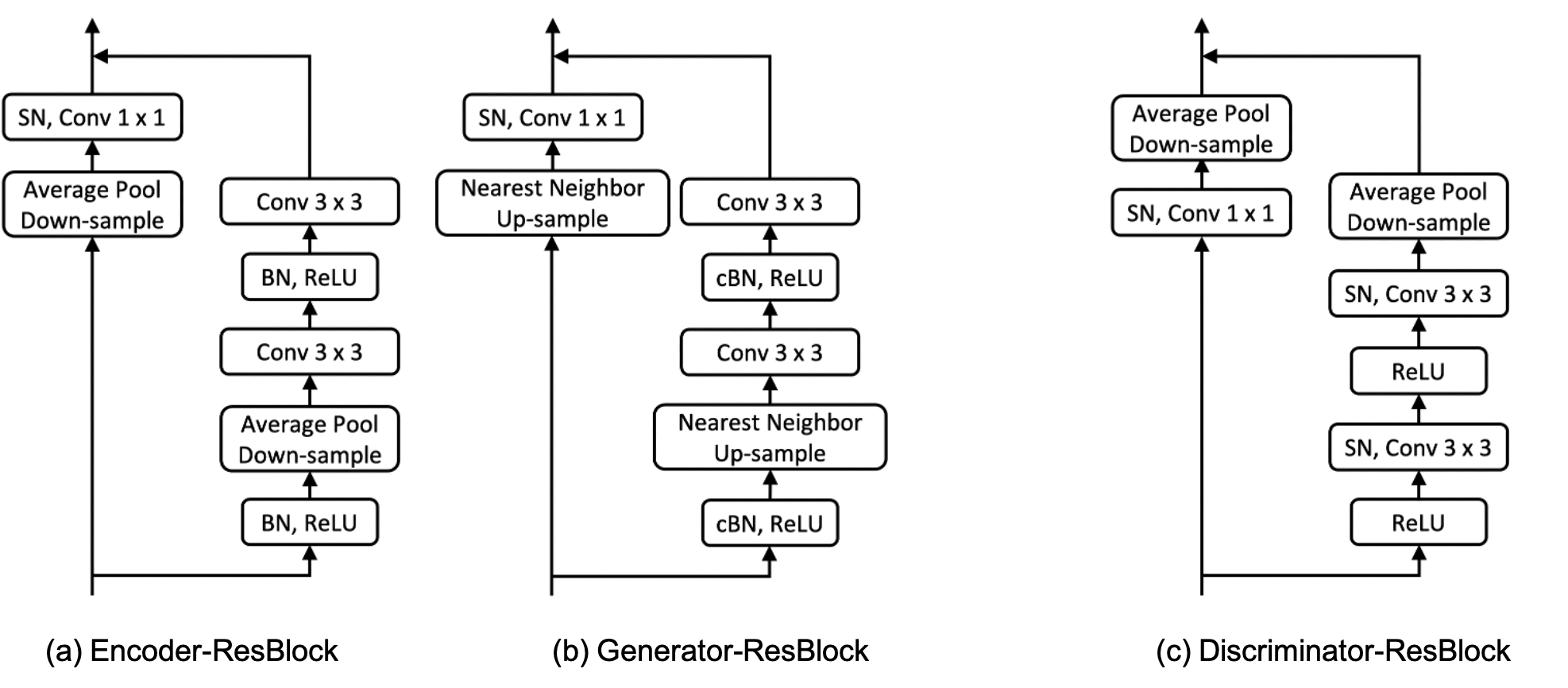}
    \caption{ Architecture of the ResBlocks used in all experiments.
    }
    \label{Fig_resblock}
\end{figure}

The PCE function is a conditional GAN with an encoder. We used a ResNet~\cite{He2016DeepRecognition} architecture for the Encoder, Generator, and Discriminator. The details of the architecture are given in Table~\ref{Tab-arch}.  For the encoder network, we used five ResBlocks with the standard batch normalization layer (BN). In encoder-ResBlock, we performed down-sampling (average pool) before the first \textit{conv} of the ResBlock as shown in Figure.~\ref{Fig_resblock}.a. For the  generator network, we follow the details in~\cite{Miyato2018SpectralNetworks} and replace the BN layer in encoder-ResBlock with  conditional BN (cBN) to encode the condition (\textit{see} Figure.~\ref{Fig_resblock}.b.). The architecture for the generator has five ResBlocks; each ResBlock performed up-sampling through the nearest neighbour interpolator.  For the discriminator, we used spectral normalization (SN)~\cite{Miyato2018CGANsDiscriminator} in Discriminator-ResBlock and performed down-sampling after the second \textit{conv} of the ResBlock as shown in Figure.~\ref{Fig_resblock}.c. For the optimization, we used Adam optimizer~\cite{Kingma2014Adam:Optimization}, with hyper-parameters set to $\alpha = 0.0002, \beta_1 = 0, \beta_2 = 0.9$ and updated the discriminator five times per one update of the generator and encoder.

\begin{table}[!ht]
\centering
\caption{ Explanation Model (cGAN) Architecture}
\label{Tab-arch}
\begin{tabular}{ c c}
(a) Encoder  & \\  
% leftmost table of the top level table
\hline
\hline
Grayscale image $\rvx \in \mathbb{R}^{256 \times 256 \times 1}$ \\
BN, ReLU, 3$\times$3 conv 64 \\
Encoder-ResBlock down 128 \\
Encoder-ResBlock down 256 \\
Encoder-ResBlock down 512 \\
Encoder-ResBlock down 1024 \\
Encoder-ResBlock down 1024 \\
(b) Generator & (c) Discriminator\\
\hline
\hline
Latent code $\rvz \in \mathbb{R}^{1024}$ & Grayscale image $\rvx \in \mathbb{R}^{256 \times 256 \times 1}$\\
Generator-ResBlock up 1024, $\rvc$ & Discriminator-ResBlock down 64\\
Generator-ResBlock up 512, $\rvc$ & Discriminator-ResBlock down 128\\
Generator-ResBlock up 256, $\rvc$ & Discriminator-ResBlock down 256\\
Generator-ResBlock up 128, $\rvc$ & Discriminator-ResBlock down 512\\
Generator-ResBlock up 64, $\rvc$ & Discriminator-ResBlock down 1024\\
BN, ReLU, 3$\times$3 conv 1 & ReLU, Global Sum Pooling (GSP) $|$ Embed($\rvc$)\\
Tanh & Inner Product (GSP, Embed($\rvc$)) $\rightarrow \mathbb{R}^1$\\
& Add(SN-Dense(GSP) $\rightarrow \mathbb{R}^1$, Inner Product)\\
\end{tabular} 
\end{table}

For creating the training dataset, we divide the posterior distribution  for the target class, $f(\rvx) \in [0,1]$ into $N$ equally-sized bins. For efficient training, cBN requires class-balanced batches. A large $N$ results in more conditions for training cGAN, increasing cGAN complexity and training time. Also,  we have to increase the batch size to ensure each condition is well represented in a batch. Hence, the GPU memory size bounds the upper value for $N$. A small value of $N$ is equivalent to fewer bins, resulting in a coarse transformation which leads to abrupt changes across explanation images. In our experiments, we used $N = 10$, with a batch size of 32. We experimented with different values of $N$ and selected the largest $N$, which created a class-balanced batch that fits in GPU memory and resulted in stable cGAN training.

\section{Semantic Segmentation}
\label{SM-SS}
We adopted a 2D U-Net~\cite{Ronneberger2015U-NetSegmentation} to perform semantic segmentation, to mark the lung and the heart contour in a CXR. The network optimizes a multi-categorical cross-entropy loss function, defined as,
\newline
\begin{equation}\label{eq:ss}
\mathcal{L_{\theta}} := \sum_s \sum_i \mathbbm{1}(y_i = s) \log(p_{\theta}(x_i)) ,
\end{equation}
\newline
where $\mathbbm{1}$ is the indicator function, $y_i$ is the ground truth label for i-th pixel. $s$ is the segmentation label with values (background, the lung or the heart). $p_{\theta}(x_i)$ denotes the output probability for pixel $x_i$ and $\theta$ are the learned parameters. The network is trained on 385 CXRs and corresponding masks from Japanese Society of Radiological Technology (JSRT) ~\cite{vanGinneken2006SegmentationDatabase} and Montgomery~\cite{Jaeger2014TwoDiseases.} datasets.

\section{Object Detection}
\label{SM-OD}
We trained an object detector network to identify medical devices in a CXR. For the MIMIC-CXR dataset, we pre-processed the reports to extract keywords/observations that correspond to medical devices, including pacemakers, screws, and other hardware.  Such foreign objects are easy to identify in a CXR and do not requires expert knowledge for manual labelling.  Using the CheXpert labeller, we extracted 300 CXR images with positive mentions for each observation.  The extracted x-rays are then manually annotated with bounding box annotations marking the presence of foreign objects using the LabelMe~\cite{Wada2016LabelmePython} annotation tool. Next, we trained an object detector based on Fast Region-based CNN ~\cite{Ren2015FasterNetworks}, which used VGG-16 model~\cite{Simonyan2014VeryRecognition}, trained on the MIMIC-CXR dataset as its foundation. We used this object detector to enforce our novel context-aware reconstruction loss (CARL). 
\newline

\begin{figure}[!ht]
    \centering
    \includegraphics[width = 0.4\linewidth]
    {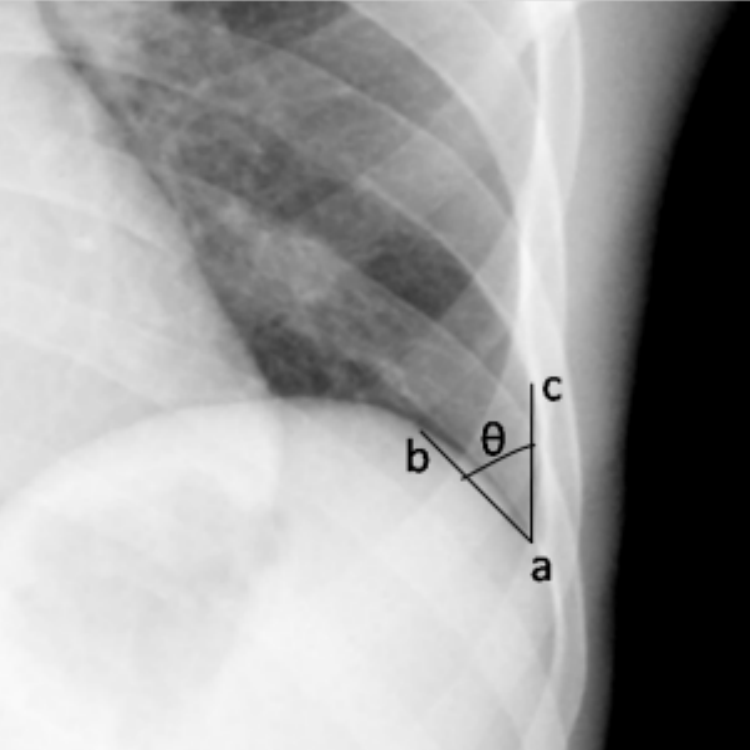}
 
    \caption[The costophrenic angle (CPA) on a CXR]{The costophrenic angle (CPA) on a CXR is marked as the angle formed by, (a) costophrenic angle point, (b) hemidiaphragm point and (c) lateral chest wall point, as shown by Maduskar \etal in~\cite{Maduskar2016AutomaticRadiographs}.}
    \label{FIG_CPA}
\end{figure}

\vspace{-0.4cm}

We trained similar detectors for identifying normal and abnormal CP recess regions in a CXR. We associated an abnormal CP recess with the radiological finding of a blunt CP angle as identified by the positive mention for \textit{``blunting of costophrenic angle"} in the corresponding radiology report. For the normal-CP recess, we considered images with a positive mention for \emph{``lungs are clear"} in the reports. We extracted 300 CXR images with positive mention of respective terms for normal and abnormal CP recess to train the object detector.

Please note that the object detector for CP recess is only used for evaluation purposes, and they were not used during the training of the explanation function. In literature, the blunting of CPA is an indication of pleural effusion~\cite{Maduskar2013AutomatedRadiographs,Maduskar2016AutomaticRadiographs}. The angle between the chest wall and the diaphragm arc is called the costophrenic angle (CPA). Marking the CPA angle on a CXR requires an expert to mark the three points, (a) costophrenic angle point, (b) hemidiaphragm point and (c) lateral chest wall point and then calculate the angle as shown in Figure.~\ref{FIG_CPA}.  Learning automatic marking of CPA angle requires expert annotation and is prone to error. Hence, rather than marking the CPA angle, we annotate the CP region with a bounding box which is a much simpler task. We then learned an object detector to identify normal or abnormal CP recess in a CXR and used the Score for detecting a normal CP recess (SCP) as our evaluation metric.

\section{xGEM}
\label{SM-xGEM}
We refer to work by Joshi \etal~\cite{Joshi2019TowardsSystems} for the implementation of xGEM. 
xGEM iteratively traverses the input image's latent space and optimizes the traversal to flip the classifier's decision to a different class.  Specifically, it solves the following optimization
\newline
\begin{equation}
    \Tilde{\rvx} = \mathcal{G}_{\theta}(\arg \min_{\mathbf{z}\in \mathbb{R}^d}\mathcal{L}(\rvx,\mathcal{G}_{\theta}(\mathbf{z})) + \lambda\ell(f(\mathcal{G}_{\theta}(\mathbf{z})),y^{'}))
\end{equation}
\newline
where the first terms is an $\ell_2$ distance loss for comparing real and generated data. The second term ensures that the classification decision for the generated sample is in favour of class $y^{'}$ and $y^{'} \neq y$ is a class other than original decision. Unless explicitly imposed, the explanation image does not look realistic. The explanation image is generated from an updated latent feature, and the expressiveness of the generator limits its visual quality.  xGEM adopted a variational autoencoder (VAE) as the generator. VAE uses a Gaussian likelihood ($\ell_2$ reconstruction), an unrealistic assumption for image data. Hence, vanilla VAE is known to produce over-smoothed images~\cite{Huang2018IntroVAE}.
 The VAE used is available at \href{https://github.com/LynnHo/VAE-Tensorflow}{https://github.com/LynnHo/VAE-Tensorflow}. All settings and architectures were set to default values. The original code generates an image of dimension 64x64. We extended the given network to produce an image with dimensions 256$\times$256. 
 
 \section{cycleGAN}
\label{SM-CG}
We refer to the work by Narayanaswamy \etal~\cite{Narayanaswamy2020ScientificTranslation} and DeGrave \etal~\cite{DeGrave2020AISignal} for the implementation details of cycleGAN. The network architecture for cycleGAN is replicated from the GitHub repository \href{https://github.com/junyanz/pytorch-CycleGAN-and-pix2pix}{https://github.com/junyanz/pytorch-CycleGAN-and-pix2pix}. For training cycleGAN, we consider two sets of images. The first set comprises 2000 images from the MIMIC-CXR dataset such that the classifier has a strong positive prediction for the presence of a target disease \ie $f(\rvx) > 0.9$, and the second set has the same number of images but with strong negative prediction \ie  $f(\rvx) < 0.1$. We train one such model for each target disease.

\begin{table}[!ht]
\caption[Results for six prediction tasks on CelebA dataset.]{Results for six prediction tasks on CelebA dataset. FID (Fr\'echet Inception Distance) score measures the quality of the generated explanations. Lower FID is better.  FVA (Face verification accuracy) measures percentage of the times the query image and generated explanation have same face identity as per model trained on VGGFace2. Higher LSC and FVA is better.}
\label{Extra-table}
\begin{center}
\begin{tabular}{c|cc}
\bf Prediction Task &  \multicolumn{2}{c}{\bf Data Consistency (FID)} \\
 & Negative ($f(\rvx), f(\rvx_{\rvc}) < 0.2$)  & Positive ($f(\rvx), f(\rvx_{\rvc}) > 0.8$)    \\
\hline
CelebA-Smiling   & 56.3 & 46.9  \\
CelebA-Young        &74.4 & 67.5   \\
CelebA-No beard &  72.3 &  79.2 \\
CelebA-Heavy makeup &  98.2  & 64.9 \\
CelebA-Black hair  & 72.8  & 55.8 \\
CelebA-Bangs  & 57.8 & 54.1  \\
\hline
\end{tabular}
\end{center}
\end{table}

\section{Extended results on the three desiderata of explanation function}
\label{SM-Extended-Results}
\begin{figure}[!ht]
\centering
\includegraphics[width=1\linewidth]{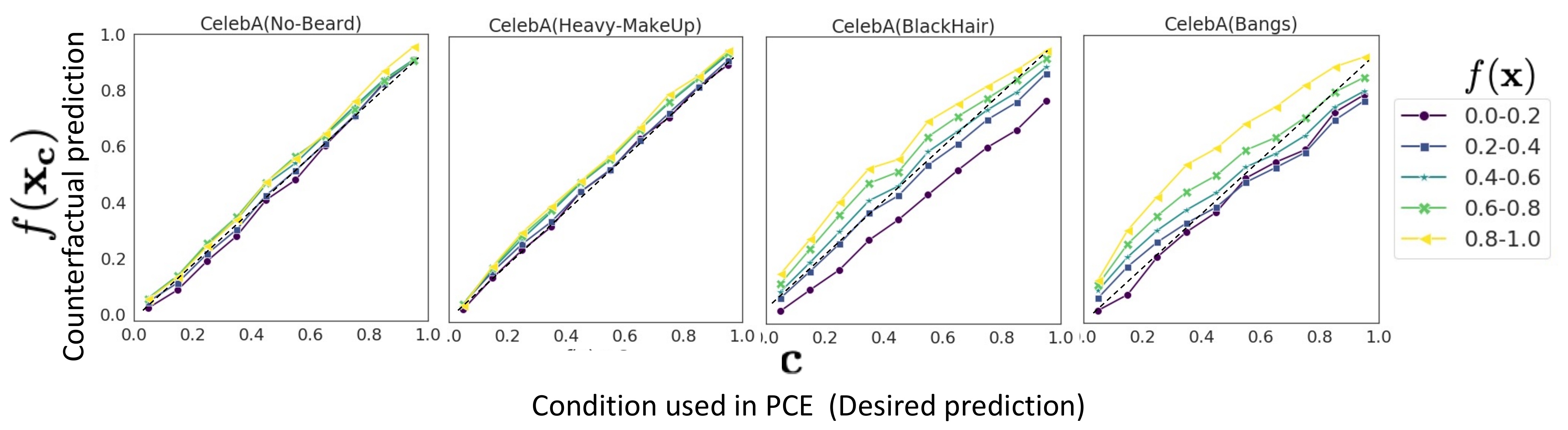}
\caption[Plot of the expected outcome from the classifier, $\rvc$, against the
actual response of the classifier on generated explanations, $f(\rvx_{\rvc})$.]{Plot of the expected outcome from the classifier, $\rvc$, against the
actual response of the classifier on generated explanations, $f(\rvx_{\rvc})$. The monotonically increasing trend shows a positive correlation between $\rvc$ and $f(\rvx_{\rvc})$. Hence, the explanations are consistent with the given condition.}
\label{fig_CC_SM}
\end{figure}

\begin{figure}[!ht]
\centering
\includegraphics[width=0.99\linewidth]{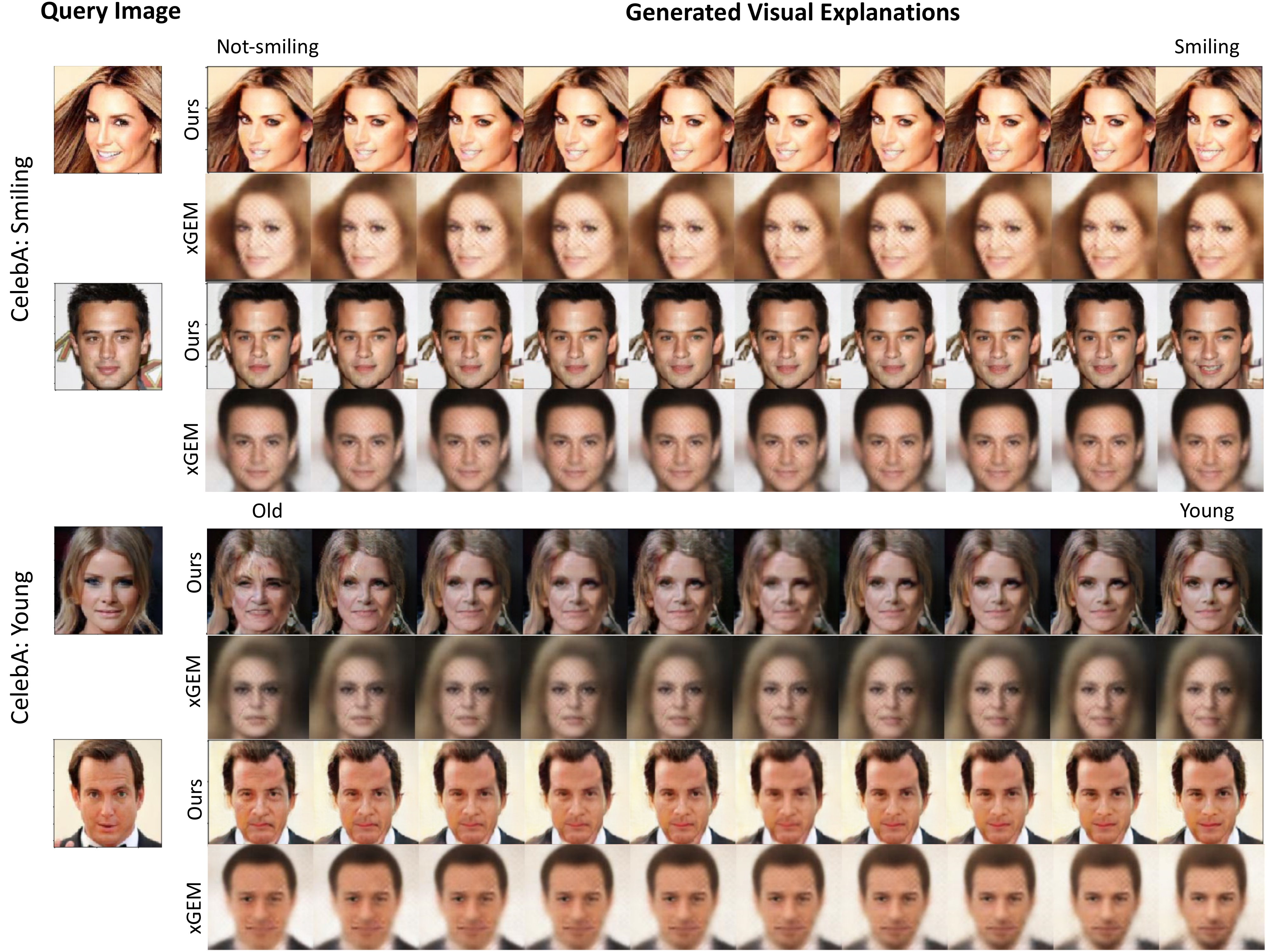}
\caption{ Visual explanations generated for ``smiling" and ``young" attribute classification on CelebA dataset. }
\label{Fig_xGEM}
\end{figure}

\begin{figure}[!ht]
\centering
\includegraphics[width=0.75\linewidth]{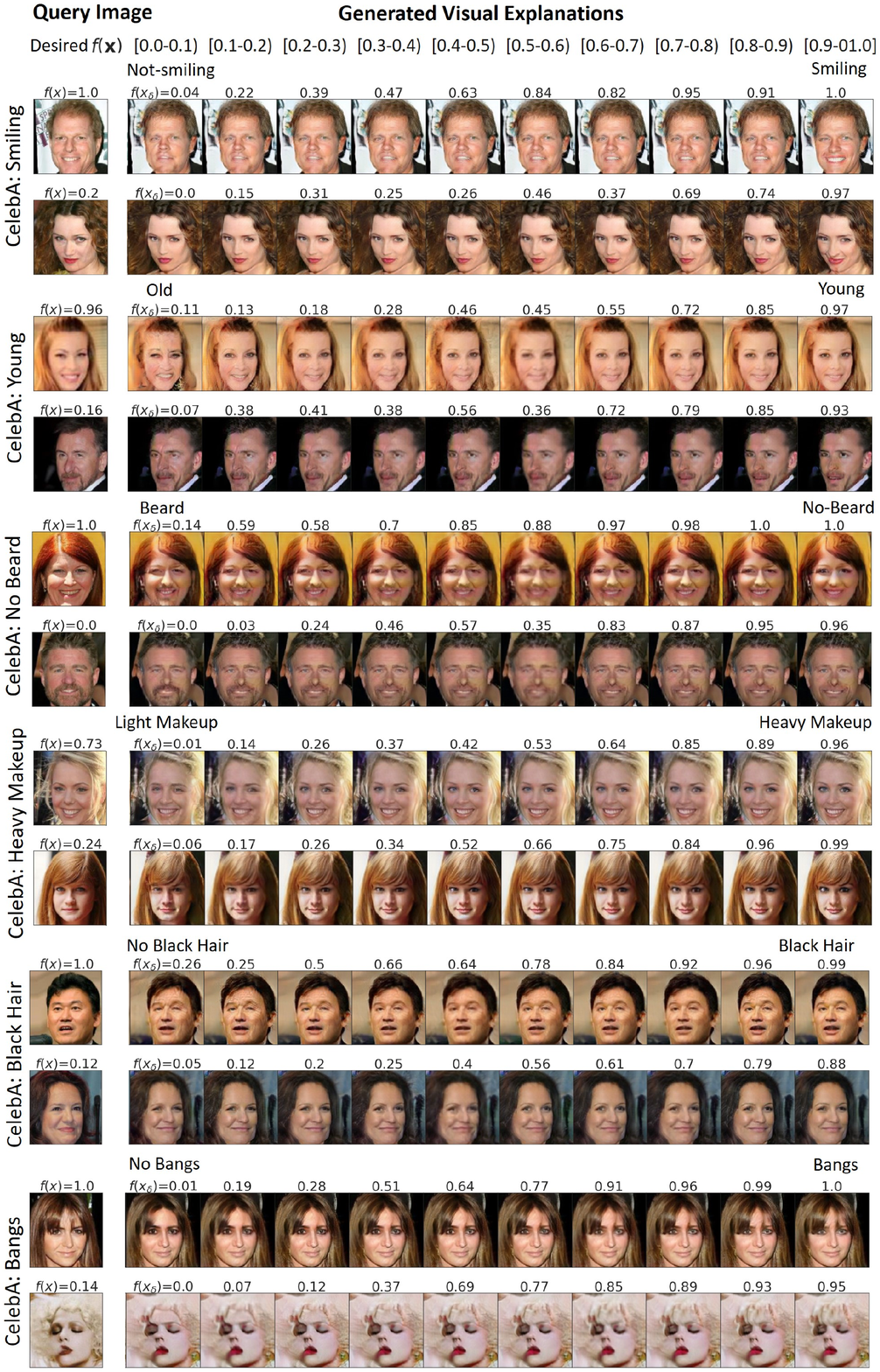}
\caption{ Visual explanations generated for different prediction tasks on CelebA dataset. }
\label{Fig_Quality_All}
\end{figure}

Here, we provide results for four more prediction tasks on celebA dataset: no-beard or beard, heavy makeup or light makeup,  black hair or not back hair, and bangs or no-bangs.
Figure~\ref{Fig_Quality_All} shows the qualitative results, an extended version of results in Figure~\ref{Fig_Quality_ch3}. In Table~\ref{Extra-table}, we summarize the FID scores for each PCE trained to explain a specific classification task.  To demonstrate classifier consistency, similar to Figure~\ref{Fig_CC}, we plotted the average response of the classifier \ie $f(\rvx_{\rvc})$ for explanations in each bin against the expected outcome \ie $\rvc$ (\textit{see} Figure.~\ref{fig_CC_SM}).  The positive slope of the line-plot, parallel to $y = x$ line confirms that starting from images with low $f(\rvx)$, our model creates fake images such that $f(\rvx_{\rvc})$ is high and vice-versa. Further, in Figure.~\ref{Fig_xGEM}, we provide a qualitative comparison between counterfactual images generated by our method and from xGEM. The explanation images generated by xGEM are blurred and lacks the natural-looking appeal of a face or an x-ray image. Consistent with this observation, earlier in our results  Table.~\ref{FID-table}, xGEM has a high FID score, validating that the xGEM explanation images are significantly different from the real images. 
\newline

\vspace{-0.4cm}

\begin{figure}[!ht]
    \centering
    \includegraphics[width=0.70\linewidth]
    {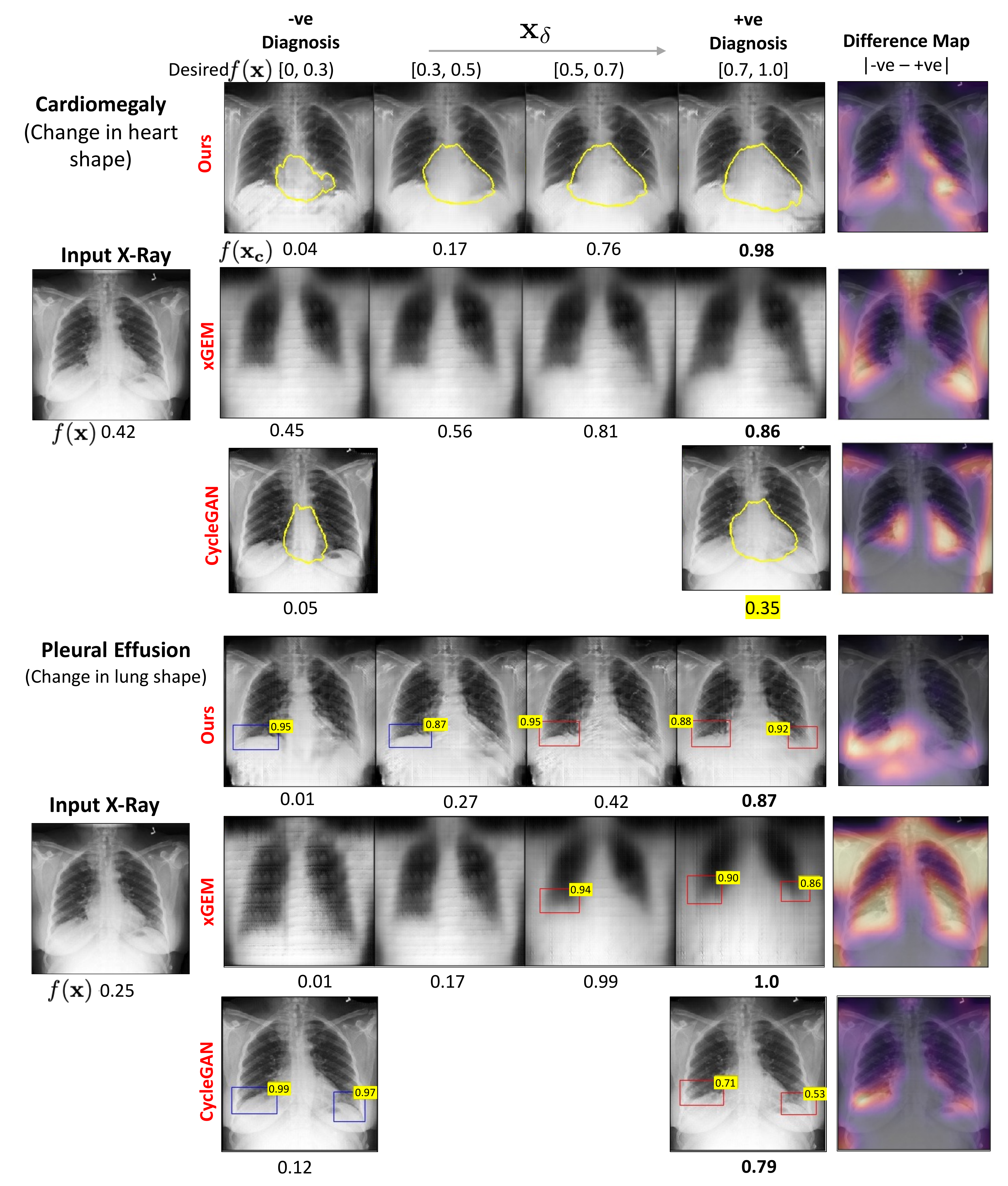}
  
    \caption[Example of counterfactual explanations.]{ 
    The transformation of an input chest x-ray into the counterfactual explanations for two diagnosis, cardiomegaly (first row) and pleural effusion (PE) (last row). The bottom labels are the classifier's prediction for the specific class. The yellow color highlight the prediction where counterfactual fails to flip the decision.  The last column shows the difference map between negative and positive explanation. For cardiomegaly, we are highlight the heart segmentation (yellow). For PE, we show the bounding-box (BB) for  normal (blue) and abnormal (red) costophrenic (CP) recess. The number on blue-BB is the Score for detecting a normal CP recess (SCP). The number on red-BB is 1-SCP.}
    \label{Extend_fig_quality_xgem}
\end{figure}

\begin{figure}[!ht]
    \centering
    \includegraphics[width=0.85\linewidth]
    {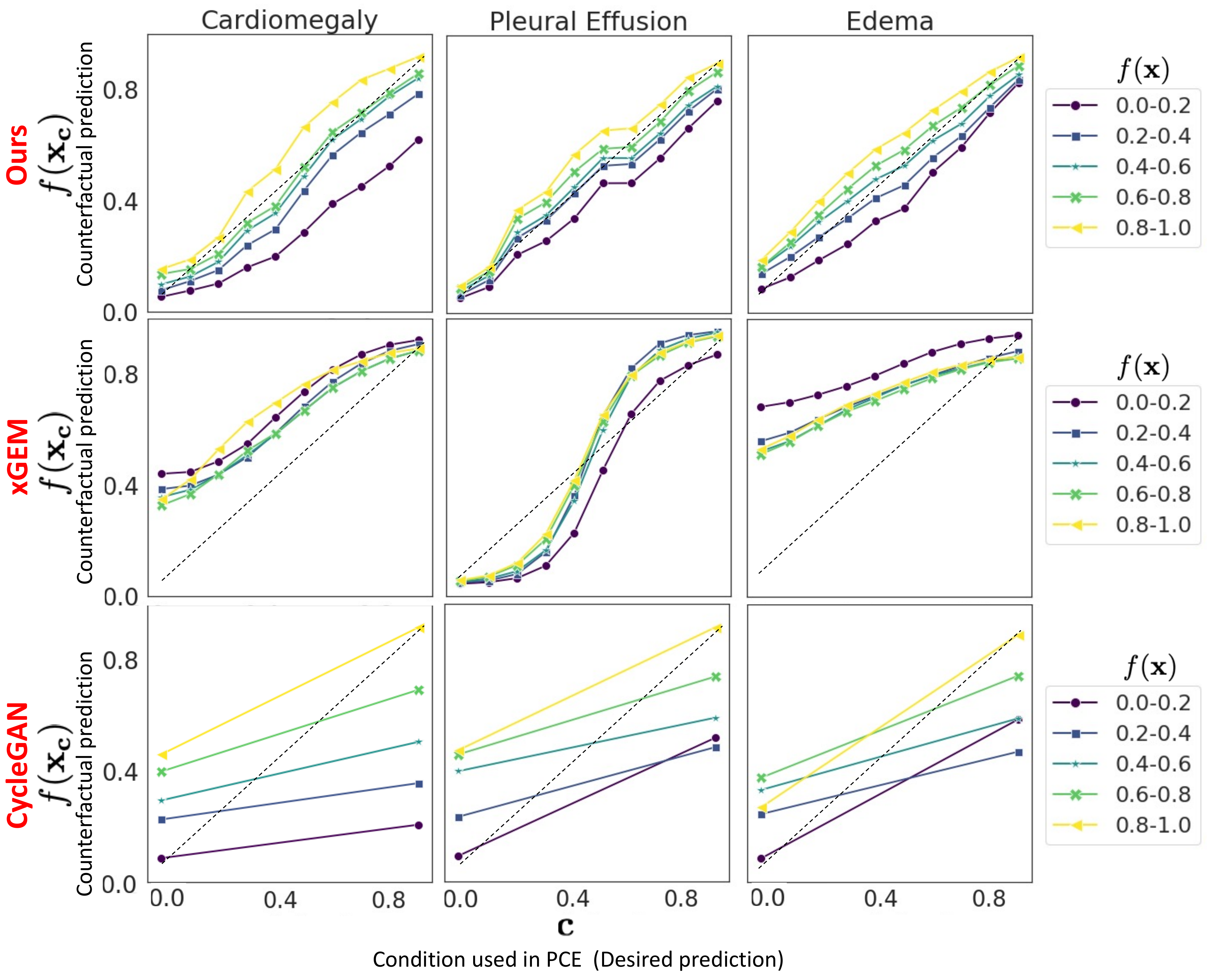}
    \caption[The plot of desired prediction, $\rvc$, against actual response of the classifier
on generated explanations, $f(\rvx_{\rvc})$.]
{The plot of desired prediction, $\rvc$, against actual response of the classifier
on generated explanations, $f(\rvx_{\rvc})$.
Each line represents a set of input images with classification prediction $f(\rvx)$ in a given range.  Dashed line represents $y = x$ line. A good explanation should cover the entire range of y-axis $[0,1]$ for all set of images ( lines of different colors).}
    \label{Fig_Cls_ext}
\end{figure}

Next, in Figure~\ref{Extend_fig_quality_xgem}, we provide similar results on CXR dataset. The bottom labels are the classifier's prediction for specific class. We also show the corresponding difference map, obtained by taking an absolute difference between explanations generated for the two extreme ends, negative (second column) and positive (fifth column) diagnosis. For cardiomegaly  the counterfactual image obtained by cycleGAN failed to flip the classification decision. Further, in Figure.~\ref{Fig_Cls_ext} we provide the classifier consistency results. For cycleGAN, starting with input images with $f(\rvx) \in [0.0, 0.2]$ (purple line), when we create explanation images with the desired predictions \ie x-axis with $\rvc \in [0.8,1.0]$, the resulting images ($\rvx_{\rvc}$) doesn't satisfy $f(\rvx_{\rvc}) \in [0.8, 1.0]$. This shows that on-an-average the CycleGAN counterfactual images doesn't flip the classification decision. This finding is consistent with the low counterfactual validity score in Table.~\ref{FID-table}. 
\newline

\section{Extended results on clinical evaluation}
\label{Extended-R-CE}
For quantitative analysis, 
we randomly sample two groups of real images (1) a \emph{real-normal} group defined as $\mathcal{X}^{n} = \{\rvx; f(\rvx) < 0.2\}$. It consists of real CXR images that are predicted as normal by the classifier $f$. (2) A \emph{real-abnormal} group defined as $\mathcal{X}^{a} = \{\rvx; f(\rvx) > 0.8\}$. For $\mathcal{X}^{n}$ we generated a counterfactual group as,  $\mathcal{X}_{cf}^{a} = \{\rvx \in \mathcal{X}^n; f(\mathcal{I}_f(\rvx, \rvc)) > 0.8\}$. Similarly for $\mathcal{X}^a$, we derived a counterfactual group as   $\mathcal{X}_{cf}^{n} = \{\rvx \in \mathcal{X}^a; f(\mathcal{I}_f(\rvx, \rvc)) < 0.2\}$.

Next, we quantify the differences in real and counterfactual groups by performing statistical tests on the distribution of clinical metrics such as cardiothoracic ratio (CTR) and the Score of normal Costophrenic recess (SCP). Specifically, we performed the dependent t-test statistics on clinical metrics for paired samples ($\mathcal{X}^n$ and $\mathcal{X}^a_{cf}$), ($\mathcal{X}^a$ and $\mathcal{X}^n_{cf}$) and the independent two-sample t-test statistics for normal ($\mathcal{X}^n$, $\mathcal{X}^n_{cf}$) and abnormal ($\mathcal{X}^a$, $\mathcal{X}^a_{cf}$) groups. The two-sample t-tests are statistical tests used to compare the means of two populations. A low p-value $< $ 0.0001 rejects the null hypothesis and supports the alternate hypothesis that the difference in the two groups is statistically significant and that this difference is unlikely to be caused by sampling error or by chance. For paired t-test, the mean difference corresponds to the average causal effect of the intervention on the variable under examination. In our setting, intervention is a \textit{do} operator on input image ($\rvx$), before intervention, resulting in a counterfactual image ($\rvx_{\rvc}$), after intervention. 

\begin{table}[!ht]
\caption[Results of independent t-test.]{Results of independent t-test. We compared the difference distribution of cardiothoracic ratio (CTR) for cardiomegaly and the Score for normal Costophrenic recess (SCP) for pleural effusion. CI: confidence interval; CF: counterfactual.
}
\label{t-table}
\footnotesize
\centering
\begin{tabular}{c|cc|cccc|c|c|c}
Target  & & & \multicolumn{4}{c|}{\bf Paired Differences} &  & & \\
Disease & Real & CF &Mean & & \multicolumn{2}{c|}{95\% CI} & & & \\
& Group & Group &  Difference & Std & Lower & Upper & t & df & p-value \\
\hline
Cardiomegaly & $\mathcal{X}^n$ & $\mathcal{X}_{cf}^a$ & \bf -0.03 & 0.07 & -0.03 & -0.01 & -4.4 & 304 & $<$ 0.0001 \\
(CTR) & $\mathcal{X}^a$ & $\mathcal{X}_{cf}^n$ & \bf 0.14 & 0.12 & 0.13 & 0.15 & 24.7 & 513 & $\ll$ 0.0001 \\
Pleural effusion & $\mathcal{X}^n$ & $\mathcal{X}_{cf}^a$ & \bf 0.13 & 0.22 & 0.06 & 0.13 & 5.9 & 217 & $\ll$ 0.0001 \\
(SCP)  & $\mathcal{X}^a$ & $\mathcal{X}_{cf}^n$ & \bf -0.19 & 0.27 & -0.18 & -0.09 & -6.7 & 216 & $\ll$ 0.0001 \\
\hline
\hline
  & & & \multicolumn{4}{c|}{\bf Un-Paired Differences} &  & & \\
 &  &  & Mean Real & Mean CF & \multicolumn{2}{c|}{95\% CI} & & & \\
& & & Group & Group & Lower & Upper & t & df & p-value \\
\hline
Cardiomegaly & $\mathcal{X}^n$ & $\mathcal{X}_{cf}^n$ & \bf 0.46 & 0.42 & 0.02 & 0.06 & 5.2 & 817 & $<$ 0.0001 \\
(CTR) & $\mathcal{X}^a$ & $\mathcal{X}_{cf}^a$ & \bf0.56 & 0.50 & 0.04 & 0.07 & 9.9 & 817 & $\ll$ 0.0001 \\
Pleural effusion & $\mathcal{X}^n$ & $\mathcal{X}_{cf}^n$ & \bf0.69 & 0.61 & 0.18 & 0.27 & 9.3 & 433 & $\ll$ 0.0001 \\
(SCP)  & $\mathcal{X}^a$ & $\mathcal{X}_{cf}^a$ & 0.42 & \bf0.56 & -0.32 & -0.21 & -9.7 & 433 & $\ll$ 0.0001 \\
\hline
\end{tabular}
\end{table}

\vspace{0.5cm}

Table~\ref{t-table} provides the extended results for the Fig.~\ref{Fig_Box}. Patients with cardiomegaly have higher CTR as compared to normal subjects. Hence, one should expect CTR($\mathcal{X}^n$) $<$ CTR($\mathcal{X}_{cf}^a$) and likewise CTR($\mathcal{X}^a$) $>$ CTR($\mathcal{X}_{cf}^n$). Consistent with clinical knowledge, in Table.~\ref{t-table}, we observe a negative mean difference of -0.03 for CTR($\mathcal{X}^n$) $-$ CTR($\mathcal{X}_{cf}^a$) (a p-value of $< 0.0001$) and a positive mean difference of 0.14 for CTR($\mathcal{X}^a$) $-$ CTR($\mathcal{X}_{cf}^n$) (with a p-value of $\ll 0.0001$). On  a  population-level  CTR  was  successful  in capturing  the  difference  between  normal  and  abnormal  CXRs.  Specifically in un-paired differences, we observe a low mean CTR values for normal subjects \ie mean CTR($\mathcal{X}^n$) = 0.46 as compared to mean CTR for abnormal patients \ie mean CTR($\mathcal{X}^a$) = 0.56. The low p-values supports the alternate hypothesis that the difference in the two groups is statistically significant.

\begin{figure}[!ht]
    \centering
    \includegraphics[width = 0.9\linewidth]
    {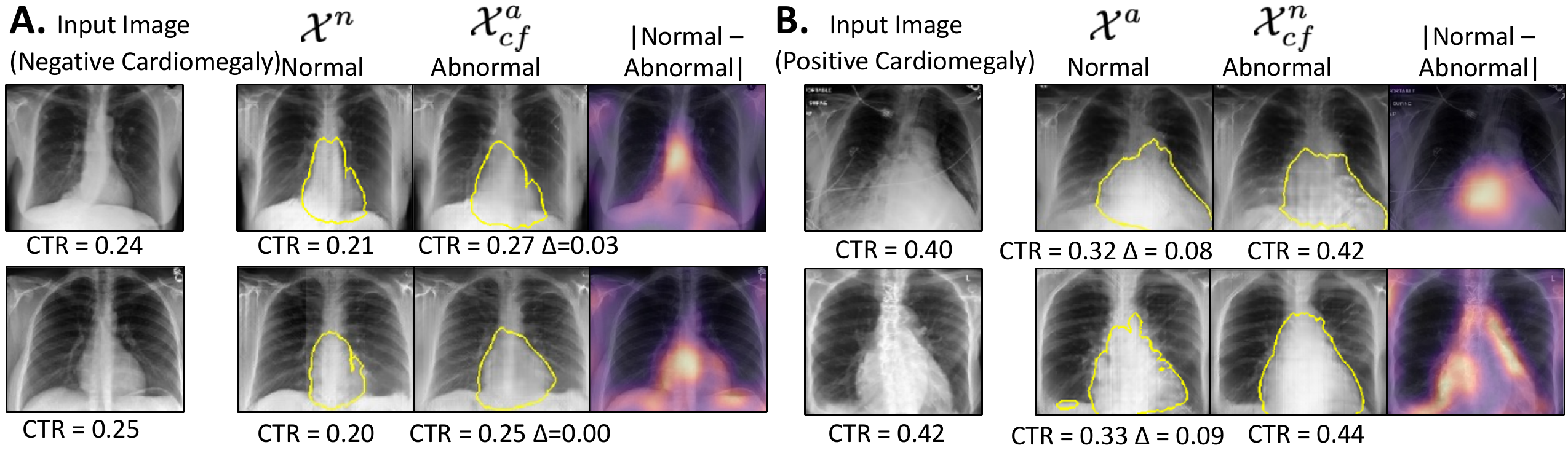}

    \caption[
    Extended results for explanation produced by our model for \textbf{Cardiomegaly}.]{Extended results for explanation produced by our model for \textbf{Cardiomegaly}. For each image, we generate a normal and an abnormal explanation image. We show pixel-wise difference of the two generated images as the saliency map. In column A.(B.), we show input images negatively (positively) classified for Cardiomegaly. The yellow contour shows the heart boundary learned by a segmentation network. CTR is the cardiothoracic ratio. For column A, we observe a relatively minor change in CTR ($\Delta$) between real and counterfactual images than in column B. 
    \label{Fig_c}.}
\end{figure}

Further, in Fig~\ref{Fig_c}.A, we show samples from input images that were predicted as negative for cardiomegaly ($\mathcal{X}^{n}$). In their counterfactual abnormal images (third column), we observe small changes in CTR are sufficient to flip the classification decision. This is consistent with a small mean difference CTR($\mathcal{X}^{n}$) - CTR($\mathcal{X}^{a}_{cf}) = -0.03$.
 In contrast, when we generate counterfactual normal (sixth column) from real abnormal images (positive for cardiomegaly, Fig~\ref{Fig_c}.B), significant changes in CTR lead to flipping of the prediction decision. This observation is consistent with a large mean difference CTR($\mathcal{X}^{a}$) - CTR($\mathcal{X}^{n}_{cf}) = 0.14$. 
\newline

\vspace{-0.5cm}

By design, the object detector assigns a low SCP to any indication of blunting CPA or abnormal CP recess. Hence, SCP($\mathcal{X}^n$) $>$ SCP($\mathcal{X}_{cf}^a$) and likewise SCP($\mathcal{X}^a$) $<$ SCP($\mathcal{X}_{cf}^n$). Consistent with our expectation, in Table.~\ref{t-table}, we observe a positive mean difference of 0.13 for SCP($\mathcal{X}^n$) $-$ SCP($\mathcal{X}_{cf}^a$) (with a p-value of $\ll 0.0001$) and a negative mean difference of -0.19 for SCP($\mathcal{X}^a$) $-$ SCP($\mathcal{X}_{cf}^n$) (with a p-value of $\ll 0.0001$). On  a  population-level  SCP  was  successful  in capturing  the  difference  between  normal  and  abnormal  CXR for pleural effusion.  Specifically in un-paired differences, we observe a high mean SCP values for normal subjects \ie mean SCP($\mathcal{X}^n$) = 0.69 as compared to mean SCP for abnormal patients \ie mean SCP($\mathcal{X}^a$) = 0.42. 
%A low p-value confirmed the statistically significant difference in SCP for real images and their corresponding counterfactuals. 
\newline

\begin{figure}[!ht]
\includegraphics[width=1\linewidth]{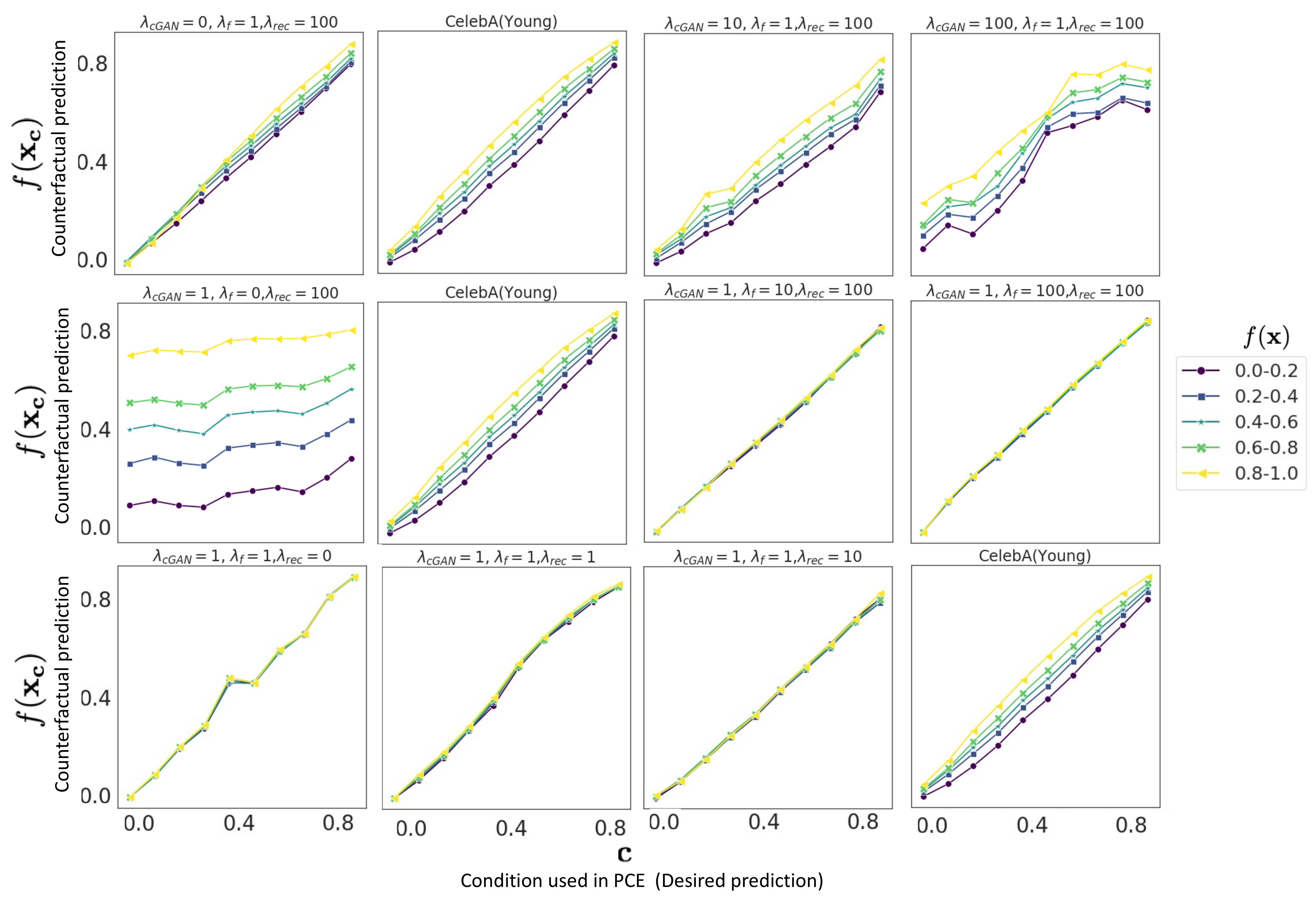}

\caption[Ablation study to show the effect of KL loss term.]{Ablation study to show the effect of KL loss term. Plot of the expected outcome from the classifier, $\rvc$, against the actual response of the classifier on generated explanations, $f(\rvx_{\rvc})$. }
\label{Fig_Quatitative_Ablation_All}
\end{figure}

\begin{table}[!ht]
\begin{center}

\caption[Our model with ablation on prediction task of young vs old on CelebA dataset.]{Our model with ablation on prediction task of young vs old on CelebA dataset. FID (Fr\'echet Inception Distance) score measures the quality of the generated explanations. Lower FID is better. FVA (Face verification accuracy) measures percentage of the times the query image and generated explanation have same face identity as per model trained on VGGFace2. }
\label{Ablation-table-celebA}
\begin{tabular}{ccc|ccc|c}
\multicolumn{3}{c}{\bf Configuration} &  \multicolumn{3}{c}{\bf Data Consistency (FID)} & \multicolumn{1}{c}{\bf Self Consistency}\\
$\lambda_{cGAN}$ & $\lambda_{f}$ & $\lambda_{rec}$ & Present  & Absent  & Overall &  FVA \\
\hline
\cellcolor{blue!25} 0 & 1 & 100 & 69.7   & 105.7 & 67.2    &  99.8\\
\cellcolor{blue!25}1 & 1 & 100 & \bf 67.5   & \bf 74.4  & \bf 53.4    &  72.2\\
\cellcolor{blue!25}10 & 1 & 100& 89.4   & 105.2 &  63.0   &  82.7\\
\cellcolor{blue!25}100 & 1 & 100& 71.6  & 80.6  &  44.26  &  18.0\\
\hline
1 & \cellcolor{blue!25}0 & 100   & 66.2   & 66.2  &  44.9    & 99.4\\
1 & \cellcolor{blue!25}1 & 100   & 67.5   & 74.4  & 53.4   & 72.2\\
1 & \cellcolor{blue!25}10 & 100  & 95.5   & 90.4  & 62.4   &  96.8\\
1 & \cellcolor{blue!25}100 & 100 & 77.4   & 73.1  & 71.2   &  42.23\\
\hline
1 & 1 & \cellcolor{blue!25}0   & 116.2    & 118.9       & 72.2    &  0.0\\
1 & 1 & \cellcolor{blue!25}1   & 63.0     & 78.6        & 61.6    &  5.5\\
1 & 1 & \cellcolor{blue!25}10  & 87.6     & 83.6        & 65.7    & \bf 88.8\\
1 & 1 & \cellcolor{blue!25}100 & 67.5    & 74.4      & 53.4      & 72.2\\
\hline
\end{tabular}
\end{center}
\end{table}

\section{Ablation Study}
\label{SM-ablation}
Our proposed model has three types of loss functions: adversarial loss from cGAN $\mathcal{L}_{\text{cGAN}}(D,G)$, KL loss $\gL_{f}(D,G)$, and CARL reconstruction loss $\gL_{\text{rec}}(E,G)$. The three losses enforce the three properties of our proposed explainer function: data consistency, compatibility with $f$, and self-consistency, respectively. In the ablation study, we quantify the importance of each of these components by training different models, which differ in one hyper-parameter. For \textbf{data consistency}, we evaluate Fr\'echet Inception Distance (FID). FID  score measures the visual quality of the generated explanations by comparing them with the real images. We show results for two groups. In the first group, we consider real and fake images where the classifier has high confidence in \textit{presence} of the target label $y$ \ie  $f(\rvx_{\rvc})[y], f(\rvx)[y] \in [0.8, 1.0]$. In second group, the target label $y$ is \textit{absent} \ie $f(\rvx_{\rvc})[y], f(\rvx)[y] \in [0.0, 0.2)$.  For \textbf{compatibility with $f$}, we plotted the desired output of the classifier \ie $\rvc$ against the actual output of the classifier $f(\rvx_{\rvc})$ for the generated explanations. For \textbf{self consistency}, we calculated the   Face verification accuracy (FVA) for celebA dataset and the foreign object preservation (FOP) score for CXR dataset. FVA measures the percentage of the instances in which the query image and generated explanation have the same face identity as per the model trained on VGGFace2. FOP score is the fraction of real images, with successful detection of FO, in which FO was also detected in the corresponding explanation image $\rvx_{\rvc}$.

\begin{table}[!ht]
\begin{center}
\caption[Evaluation metrics for ablation study.]{Evaluation metrics for ablation study.
FID score quantifies the visual appearance of the explanations. CV score is the fraction of explanations that have an opposite prediction compared to the input image. FOP score is the fraction of real images with FO, in which FO was also detected in the corresponding explanation image. In configuration with $\lambda_{cGAN} = 0$ there is no adversarial loss from cGAN, in  $\lambda_{f} = 0$ there is no KL-loss for classifier consistency and in $\lambda_{rec} = 0$ there is no context-aware self reconstruction loss. }
\label{Ablation-table}
\begin{tabular}{c|cccc|cccc}
\multicolumn{1}{c}{ }  &  \multicolumn{4}{c}{\bf Cardiomegaly} &  \multicolumn{4}{c}{\bf Pleural Effusion} \\
& Baseline & $\lambda_{cGAN}$=0 & $\lambda_{f}$=0 &$\lambda_{rec}$=0  & Baseline & $\lambda_{CGAN}$=0 & $\lambda_{f}$=0 &$\lambda_{rec}$=0    \\
\hline
\multicolumn{9}{c}{\bf FID score} \\
\hline
Normal&  166 & 200& 174 & 160 &146  & 210 & 150 & 149\\
Abnormal & 137 & 189 & 138 & 140 & 122& 178 & 120 & 130\\
\hline
\hline
\multicolumn{9}{c}{\bf Counterfactual Validity (CV) Score} \\
\hline
Overall&  0.91&  0.89 & 0.43 & 0.92 &  0.97& 0.93 & 0.43 & 0.97 \\
\hline
\multicolumn{9}{c}{\bf Foreign Object Preservation (FOP) score} \\
\hline
Pacemaker&  0.52&  0.2 & 0.55 & 0.19\\
\hline
\end{tabular}
\end{center}
\end{table}

For celebA, we consider the prediction task of young vs old.
Figure~\ref{Fig_Quatitative_Ablation_All} shows the results for compatibility with $f$.
Table~\ref{Ablation-table-celebA} summarizes the results for data consistency and self-consistency. For MIMIC-CXR, Table~\ref{Ablation-table} summarizes our results. In the absence of adversarial loss from cGAN ($\lambda_{cGAN} = 0$), FID score is very high as the generated images looks very different from the real images.  On removing the KL loss for classifier consistency ($\lambda_{f} = 0$), the CV score is poor as the generated explanations are derived without considering the classification function and hence they failed to flip the classification decision. In the absence of reconstruction loss ($\lambda_{rec} = 0$), the generated explanations are no longer for the same person as in query image. This results in a low FVA score. In CXR dataset, FO in query CXR are absent in generated explanations, resulting in low FOP score.

\section{Ablation study over pacemaker}
\label{SM-ASPM}
We performed an ablation study to investigate if a pacemaker is influencing the classifier's prediction for cardiomegaly. We consider 300 subjects that are positively predicted for cardiomegaly and have a pacemaker. We used our pre-trained object detector to find the bounding-box annotations for these images. Using the bounding-box, we created a perturbation of the input image by masking the pacemaker and in-filling the masked region with the surrounding context.

\begin{figure}[!ht]
    \centering
    \includegraphics[width = 0.7\linewidth]
    {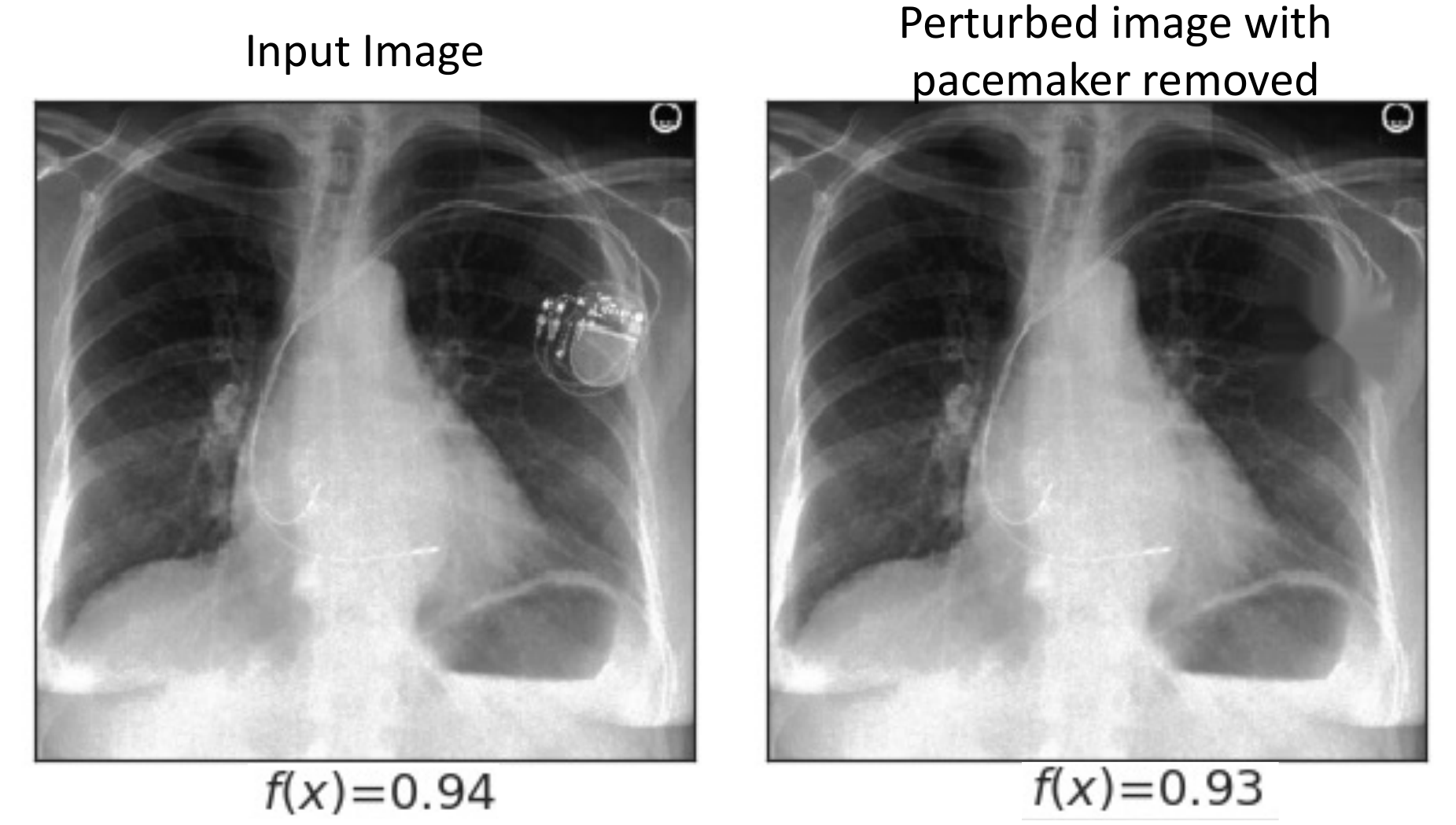}
    \caption{ An example of input image before and after removing the pacemaker.}
    \label{Fig_pacemaker-1}
\end{figure}

    An example of the perturbation image is shown in Fig.~\ref{Fig_pacemaker-1}. We passed the perturbed image through the classifier and calculated the difference in the classifier's prediction before and after removing the pacemaker. The average change in prediction was negligible ($0.03$). Hence, pacemaker is not influencing classification decisions for cardiomegaly. 

We performed an ablation study to investigate if a pacemaker is influencing the classifier's prediction for cardiomegaly. We consider 300 subjects that are positively predicted for cardiomegaly and have a pacemaker. We used our pre-trained object detector to find the bounding-box annotations for these images. Using the bounding-box, we created a perturbation of the input image by masking the pacemaker and in-filling the masked region with the surrounding context.

%==========================================================================================%
% BIBLIOGRAPHY
%==========================================================================================%
\safebibliography{main}
\bibliographystyle{plain}

%==========================================================================================%
%==========================================================================================%

\end{document}